\documentclass[10pt]{article}
\usepackage[margin=.8in,left=.8in]{geometry}

\usepackage{amsthm}
\usepackage{amssymb}
\usepackage{amsmath}
\usepackage{mathtools}
\usepackage{adjustbox}

\usepackage[table]{xcolor}

\usepackage{stmaryrd}    % for \sslash
\usepackage{rotating}
\usepackage{xfrac}

\usepackage{enumitem}
\setlist{
  itemsep=-3pt,
  topsep=1.5pt,
  leftmargin=.6cm,
}

\usepackage[safe]{tipa} % for \esh

\usepackage{accents}
\newlength{\dhatheight}
\newcommand{\doublehat}[1]{%
    \hspace{1.3pt}
    \settoheight{\dhatheight}{\ensuremath{\widehat{#1}}}%
    \addtolength{\dhatheight}{-0.23ex}%
    \widehat{\vphantom{\rule{1pt}{\dhatheight}}%
    \hspace{-1.3pt}
    \smash{\widehat{#1}}}}

\usepackage{tikz}
\usetikzlibrary{
  backgrounds, 
  snakes
}
\usetikzlibrary{cd}

\usepackage{mathrsfs} %mathscr
\DeclareMathAlphabet{\mathpzc}{OT1}{pzc}{m}{it} %for math script

\usepackage{hyperref}
\hypersetup{
  colorlinks = true,
  linkcolor=darkblue, citecolor=darkblue,
  urlcolor=black
}

\definecolor{darkblue}{rgb}{0.05,0.25,0.65}
\definecolor{darkgreen}{RGB}{20,140,10}
\definecolor{lightgray}{rgb}{0.9,0.9,0.9}
\definecolor{darkorange}{RGB}{200,100,5}
\definecolor{darkyellow}{rgb}{.91,.91,0}
\definecolor{lightolive}{RGB}{225, 220, 185}

\theoremstyle{definition}

\newcommand{\acts}{\raisebox{1.4pt}{\;\rotatebox[origin=c]{90}{$\curvearrowright$}}\hspace{.5pt}}

\newcommand{\HilbertSpace}[1]{\mathcal{#1}}

\newcommand{\ZTwo}{\mathbb{Z}_2}

\newcommand{\hotype}[1]{\mathcal{#1}}

\newcommand{\filling}{K}

\newcommand{\unit}[2]{\mathrm{ret}^{\scalebox{.7}{$#1$}}_{\scalebox{.7}{$#2$}}}

\let\PLAINthebibliography\thebibliography
\renewcommand\thebibliography[1]{
  \PLAINthebibliography{#1}
  \setlength{\parskip}{1.5pt}
  \setlength{\itemsep}{1.5pt plus .3ex}
}

\newcommand{\shape}{
  \raisebox{1pt}{\rm\normalfont\textesh}
}

\newcommand{\defneq}{\equiv}

\newcommand\bosonic[1]{\mathstrut\mkern2.5mu#1\mkern-14mu\raise1.7ex%
  \hbox{$\scriptstyle\rightsquigarrow$}}

\newcommand{\grayunderbrace}[2]{\mathcolor{gray}{\underbrace{\mathcolor{black}{#1}}}_{\mathcolor{gray}{#2}}}

\newcommand{\grayoverbrace}[2]{\mathcolor{gray}{\overbrace{\mathcolor{black}{#1}}}^{\mathcolor{gray}{#2}}}

\newcommand{\plus}{{\sqcup \{\infty\}}}
\newcommand{\cpt}{\hspace{.8pt}{\adjustbox{scale={.5}{.77}}{$\cup$} \{\infty\}}}

\newcommand{\vortex}[2]{
\begin{scope}[shift={(#1,#2)}]

\clip
  (-.2, .08) rectangle (.2,-.365);

\begin{scope}[yscale=.9, shift={(-.32,0)}]
\draw 
  (0,0)
    .. controls (.25,0) and (.3,-.4) ..
  (.3,-.6);
\end{scope}

\begin{scope}[
  shift={(+.32,0)},
  xscale=-1, yscale=.9
]
\draw 
  (0,0)
    .. controls (.25,0) and (.3,-.4) ..
  (.3,-.6);
\end{scope}

\begin{scope}[shift={(0,-.335)}, scale=.11]
\draw[gray, fill=white]
  (0,0)
    ellipse
      (.32 and .11);
\end{scope}

\begin{scope}[shift={(0,-.314)}, scale=.123]
\draw[gray, fill=white]
  (0,0)
    ellipse
      (.32 and .11);
\end{scope}

\begin{scope}[shift={(0,-.29)}, scale=.139]
\draw[gray, fill=white]
  (0,0)
    ellipse
      (.32 and .11);
\end{scope}

\begin{scope}[shift={(0,-.265)}, scale=.156]
\draw[gray, fill=white]
  (0,0)
    ellipse
      (.32 and .11);
\end{scope}

\begin{scope}[shift={(0,-.24)}, scale=.17]
\draw[gray, fill=white]
  (0,0)
    ellipse
      (.32 and .11);
\end{scope}

\begin{scope}[shift={(0,-.21)}, scale=.2]
\draw[gray, fill=white]
  (0,0)
    ellipse
      (.32 and .11);
\end{scope}

\begin{scope}[shift={(0,-.175)}, scale=.24]
\draw[gray, fill=white]
  (0,0)
    ellipse
      (.32 and .11);
\end{scope}

\begin{scope}[scale=.324, shift={(0,-.4)}]
\draw[gray, fill=white]
  (0,0)
    ellipse
      (.32 and .11);
\end{scope}

\begin{scope}[scale=.424, shift={(0,-.19)}]
\draw[gray, fill=white]
  (0,0)
    ellipse
      (.32 and .11);
\end{scope}

\begin{scope}[scale=.63, shift={(0,-.015)}]
\draw[gray, fill=white]
  (0,0)
    ellipse
      (.32 and .11);
\end{scope}

\draw[gray] 
 (0,0)  to (0,-.07); 
\draw[gray] 
 (0,-.086)  to (0,-.12); 
\draw[gray] 
 (0,-.135)  to (0,-.155); 
\draw[gray] 
 (0,-.172)  to (0,-.195); 
\draw[gray] 
 (0,-.21)  to (0,-.225); 
\draw[gray] 
 (0,-.24)  to (0,-.25); 

\end{scope}

}

\newcommand{\level}{k}
\newcommand{\lattice}{K}

\newcounter{FixedFigure}
\newcommand{\figurenumber}{%
    \refstepcounter{FixedFigure}%
    \theFixedFigure%
}

\begin{document}
%%%%%%%%%%%%%%%%%%%%%%%%%%%%%%%%%%%%%%%

%vertical spacing around displayed equations %
\setlength{\abovedisplayskip}{3pt}
\setlength{\belowdisplayskip}{3pt}
\setlength{\abovedisplayshortskip}{-5pt}
\setlength{\belowdisplayshortskip}{3pt}
%%%%%%%%%%%%%%%%%%%%%%%%%%%%%%%%%%%%%%%

\title{
  Engineering of Anyons
  on M5-Probes
  %\\
  via Flux Quantization
}

\date{Feb 2025}

\author{
  \def\arraystretch{.9}
  \begin{tabular}{c}
  Hisham Sati\rlap{${}^{\,\hyperlink{DoS}{a}, \hyperlink{Courant}{b}}$}
  \\
  \footnotesize
  \tt
  {\color{gray}hsati@nyu.edu}
  \end{tabular}
  \;\;\;\;\;\;
  \def\arraystretch{.9}
  \begin{tabular}{c}
  Urs Schreiber\rlap{${}^{\,\hyperlink{DoS}{a}}$}
  \\
  \footnotesize
  \tt
  {\color{gray}us13@nyu.edu}
  \end{tabular}
}

\maketitle

\begin{abstract}
These extended lecture notes survey a novel derivation of anyonic topological order (as seen in fractional quantum Hall systems) on single magnetized M5-branes probing Seifert orbi-singularities (``geometric engineering'' of anyons), which we motivate from fundamental open problems in the field of quantum computing.

\smallskip

The rigorous construction is non-Lagrangian and non-perturbative, based on previously neglected global completion of the M5-brane's tensor field by flux-quantization consistent with its non-linear self-duality and its twisting by the bulk C-field. This exists only in little-studied non-abelian generalized cohomology theories, notably in a twisted equivariant (and ``twistorial'') form of unstable Cohomotopy (``Hypothesis H'').

\smallskip

As a result, topological quantum observables form Pontrjagin homology algebras of mapping spaces from the orbi-fixed worldvolume into a classifying 2-sphere. Remarkably, results from algebraic topology imply from this the quantum observables and modular functor of abelian Chern-Simons theory, as well as braid group actions on defect anyons of the kind envisioned as hardware for topologically protected quantum gates.
\end{abstract}

\begin{center}
\def\arraystretch{1.1}
\begin{tabular}{c}
  prepared for the lecture series
  \\
  \underline{\href{https://ncatlab.org/schreiber/show/Introduction+to+Hypothesis+H}{\bf Introduction to Hypothesis H}}
  \\
  held at 
  \\
  \href{https://conference.math.muni.cz/srni/files/archiv/2025/}{\it 45th Winter School GEOMETRY AND PHYSICS}
  \\
  Srn{\'i}, Czechia (18-25 Jan 2025)
\end{tabular}
\end{center}

\vspace{.2cm}

\begin{center}
\begin{minipage}{8.5cm}
  \tableofcontents
\end{minipage}
\end{center}

\medskip

\vfill

\hrule
\vspace{5pt}

{
\hypertarget{DoS}{}
\footnotesize
\noindent
\def\arraystretch{1}
\tabcolsep=0pt
\begin{tabular}{ll}
${}^a$\,
&
Mathematics, Division of Science; and
\\
&
Center for Quantum and Topological Systems,
\\
&
NYUAD Research Institute,
\\
&
New York University Abu Dhabi, UAE.  
\end{tabular}
\hfill
\adjustbox{raise=-15pt}{
\href{https://ncatlab.org/nlab/show/Center+for+Quantum+and+Topological+Systems}{\includegraphics[width=3cm]{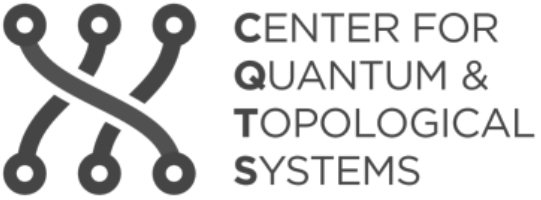}}
}

\vspace{1mm} 
\hypertarget{Courant}{}
\noindent 
\def\arraystretch{1}
\tabcolsep=0pt
\begin{tabular}{ll}
${}^b$\,The Courant Institute for Mathematical Sciences, NYU, NY.
\end{tabular}

\vspace{.2cm}

\noindent
The authors acknowledge the support by {\it Tamkeen} under the 
{\it NYU Abu Dhabi Research Institute grant} {\tt CG008}.
}

\newpage

%%%%%%%%%%%%%%%%%%%%%%%%%%%%%
\section{Motivation: Better Anyon Theory}
%%%%%%%%%%%%%%%%%%%%%%%%%%%%%

While the hopes associated with the idea of {\it quantum computing} 
\cite{NielsenChuang00}\cite{GrumblinHorowitz19}
are hard to over-state \cite{Gill24}\cite{Beterov24}\cite{Preskill24},
there are good arguments that commercial-value quantum computing will ultimately require quantum hardware exhibiting {\it anyonic topological order} \cite{WenEtAl19}\cite{SS24-TopOrd}. 
But microscopic theoretical derivations, from first principles, of such anyonic quantum states in strongly-coupled quantum systems had remained sketchy, which may explain the dearth of experimental realizations to date.

\smallskip 
What we review here (based on \cite{SS24-AbAnyons}\cite{SS25-Seifert}\cite{GSS24-FluxOnM5}) is a rigorous theoretical account via ``geometric engineering on M-branes'' subject to a previously neglected step of ``flux-quantization'' (the latter surveyed in \cite{SS24-Flux}).

\smallskip

First, we expand on the motivation a little further:

\medskip

\noindent
{\bf Ultimate need for Topological Quantum Protection.} Despite the fascinating reality of presently available Noisy Intermediate-Scale Quantum computers (NISQ \cite{Preskill18}) and despite the mid-term prospect of their stabilization at the software-level via Quantum Error Correction (QEC \cite{LidarBrun13}\cite{Preskill23}, at heavy cost of available system scale), serious arguments
\cite{Kak08}\cite{Dyakonov14}\cite{LauLimEtAl22}\cite{Ezratty23a}\cite{Ezratty23b}\cite{HoefletHaenerTroyer23}\cite{Gent23}\cite{Waintal24} and experience \cite{QBI} suggest
that large-scale quantum computation is hardly attainable by incremental optimization of NISQ architectures, but \cite{DasSarma22} 
\footnote{
  \cite{DasSarma22}:
  ``{The qubit systems we have today are a tremendous scientific achievement, but they take us no closer to having a quantum computer that can solve a problem that anybody cares about. [...] What is missing is the breakthrough [...] bypassing quantum error correction by using far-more-stable qubits, in an approach called topological quantum computing.}''
} 
that more fundamental quantum principles will need to be exploited -- notably {\it topological} error {\it protection} already at the hardware-level \cite{Kitaev03}\cite{FKLW03}\cite{Sau17}\cite{SatiValera25} in order to suppress quantum errors occurring in the first place.
\begin{center}
{
  \begin{minipage}{5.4cm}
    \footnotesize
    {\bf Figure \figurenumber: Topological quantum.}
    In order to practically harness the computational
    power of quantum processes, quantum states need to be
    stabilized against decohering environmental noise. 
    Apart from using quantum error correction at the software level, a plausibly necessary way to do so is by 
    using topological quantum processes 
    preventing quatum errors right at the hardware level.
  \end{minipage}
}
\adjustbox{
  raise=.1cm, 
  scale=0.9
}{
\begin{tikzcd}[
  column sep=-20pt,
  row sep=10pt
]
  \adjustbox{
    margin=2pt,
    rndframe=3pt
  }{\def\arraystretch{.9}\def\tabcolsep{2pt} \footnotesize
    \begin{tabular}{c}
      \bf Quantum
      \\
      \rotatebox[origin=c]{-90}{$\Rightarrow$}
      \\
      computational
      \\
      advantage
    \end{tabular}
  }
  \ar[
    dr,
    -Latex
  ]
  &&
  \adjustbox{
    margin=2pt,
    rndframe=3pt
  }{\def\arraystretch{.9}\def\tabcolsep{2pt} \footnotesize
    \begin{tabular}{c}
      \bf Topology
      \\
      \rotatebox[origin=c]{-90}{$\Rightarrow$}
      \\
      hardware-level
      \\
      error-protection
    \end{tabular}
  }
  \ar[
    dl,
    -Latex
  ]
  \\
  & 
  \adjustbox{
    margin=3pt,
    rndframe=3pt
  }{\def\arraystretch{.9}\def\tabcolsep{2pt} \footnotesize
    \begin{tabular}{c}
      \bf Topological Quantum
      \\
      \rotatebox[origin=c]{-90}{$\Rightarrow$}
      \\
      robust large-scale
      \\
      commercial-value
      \\
      quantum computing
    \end{tabular}
  }
\end{tikzcd}
}
\end{center} 
While topological quantum protection is thus possibly indispensable for achieving commercial-value quantum computing, its ambitious development, in theory and practice, is in fact far from mature, is in need of new ideas and of further analysis, and leaves much room for development. 
Since this is not always made clear, to amplify this point:

\begin{itemize}
\item[\bf (i)] {\bf Theoretical challenges:}
While quantum theorists now routinely deal with the algebraic structure (namely: braided fusion categories) commonly {\it expected} \cite{Kitaev06} to describe interaction of anyon species {\it in toto}, 
the {\it microscopic} first-principles  understanding of the formation of anyonic topological order as solitonic states in the many-body (electron) dynamics of quantum materials has remained at most sketchy, even in the best-understood case of the fractional quantum Hall effect \cite{Stormer99}, cf. \cite{Jacek21}.
\footnote{
  \cite[p. 3]{Jacek21}:
  ``Though the Laughlin function very well approximates the true ground state at $\nu = 1/q$, the physical mechanism of related correlations and of the whole hierarchy of the FQHE remained, however, still obscure.
  [...] The so-called HH (Halperin–Haldane) model of consecutive generations of Laughlin states of anyonic quasiparticle excitations from the preceding Laughlin state has been abandoned early because of the rapid growth of the daughter quasiparticle size, which quickly exceeded the sample size. [...] the Halperin multicomponent theory and of the CF model advanced the understanding of correlations in FQHE, however, on  phenomenological level only. CFs were assumed to be hypothetical quasi-particles consisting of electrons and flux quanta of an auxiliary fictitious magnetic field pinned to them. The origin of this field and the manner of attachment of its flux quanta to electrons have been neither explained nor discussed.''
} 

In fact, this is an instance of the general open problem of analytically establishing gapped bound states in any strongly coupled/correlated quantum system: The problem of formulating non-perturbative quantum field theory \cite{BakulevShirkov10}\cite{FerrazEtAl20}. The analogous issue in particle physics (there called the {\it Yang-Mills mass gap} problem \cite{nLabYMMassGap}) has been recognized as being profound enough to be declared one of seven ``Millennium Problems'' \cite{CMI}.

\smallskip 
\item[\bf (ii)] {\bf Practical challenges:}
But without a robust theoretical prediction of anyonic solitons in actual quantum materials, it remains unclear where and how to look for them. As an unfortunate result, experimentalists have turned attention to mere stand-ins, such as ``Majorana zero modes'' at the ends of super/semi-conducting nanowires (\cite{Kitaev01}\cite{LutchyEtAl10} which, even if the doubts about their detection were to be removed \cite{DasSarma23}, are by construction immobile and hence do not serve as hardware-protected quantum braid gates) and quantum-simulation of anyons on NISQ architectures (\cite{IqbalEtAl24}\cite[Fig. 5]{FF24}, which might serve as software-level QEC but again offers no hardware-level protection.
\end{itemize}

\noindent
In short: {\bf Foundation and implementation} of topological quantum computing as a plausible long-term pathway to actual quantum value {\bf deserves and admits thorough re-investigation}.

\smallskip 
Concretely, the intrinsic tension haunting the traditional quantum computing paradigm is (cf. \cite[p 272]{ChenEtAl07}\cite[p 3]{Waintal24}) that:

\hspace{2cm}
\adjustbox{
  margin=3pt,
  bgcolor=lightgray
}{
\def\arraystretch{1.2}
\begin{tabular}{l}
(i) quantum gates are {\it implemented via interaction} of subsystems,
\\
(ii) while quantum coherence requires {\it avoiding all interaction}.
\end{tabular}
}

\medskip

\noindent
{\bf The idea of topological protection} is to cut this Gordian knot 
by {\it quantum gates operating without interaction}. 
The physical principle that allows this to work 
\cite{ASW84}\cite{ASWZ85}\cite[p 6]{FKLW03}\cite[p 50]{Pachos12} 
is the {\it quantum adiabatic theorem} \cite{RigolinOrtiz12}:
Gapped quantum systems frozen at absolute zero in one of several ground states, but dependent on external parameters, will defy interaction with noise quanta below the energy gap and yet have their ground state transformed 
by sufficiently gentle tuning of the parameters: a {\it holonomic quantum gate}. 
This is {\it topological} if it is invariant under local deformations of parameter paths, and thus protected also against classical noise. For an {\it anyonic braid gate} the parameters in question are the positions of defects in a 2-dimensional transverse space within a quantum material.

\begin{center}
\begin{minipage}{5.9cm}
  \footnotesize
  {\bf Figure \figurenumber.
  The idea of topological quantum gates} by anyon braiding: Given an effectively 2-dimensional quantum material with point-like defects and degenerate ground states, the quantum adiabatic theorem implies that sufficiently slow (adiabatic) movement of the defect positions around each other (e.g. by externally tuning the material's properties) entails a unitary transformation on the Hilbert space of ground states. If these unitaries can be made to depend only on the homotopy class of the defect paths relative their endpoints, hence only on the {\it braids} formed by their worldlines, then their operation is topologically protected against noise in the operation of the gate.
\end{minipage}
\adjustbox{
  scale=.8,
  raise=-3cm
}{
\begin{tikzpicture}

  \shade[right color=lightgray, left color=white]
    (3,-3)
      --
      node[above, yshift=-1pt, 
      xshift=14pt,
      sloped]{
        \scalebox{.7}{
          \color{darkblue}
          \bf
          transverse space
        }
      }
    (-1,-1)
      --
    (-1.21,1)
      --
    (2.3,3);

  \draw[]
    (3,-3)
      --
    (-1,-1)
      --
    (-1.21,1)
      --
    (2.3,3)
      --
    (3,-3);

\draw[-Latex]
  ({-1 + (3+1)*.3},{-1+(-3+1)*.3})
    to
  ({-1 + (3+1)*.29},{-1+(-3+1)*.29});

\draw[-Latex]
    ({-1.21 + (2.3+1.21)*.3},{1+(3-1)*.3})
      --
    ({-1.21 + (2.3+1.21)*.29},{1+(3-1)*.29});

\draw[-Latex]
    ({2.3 + (3-2.3)*.5},{3+(-3-3)*.5})
      --
    ({2.3 + (3-2.3)*.49},{3+(-3-3)*.49});

\draw[-latex]
    ({-1 + (-1.21+1)*.53},{-1 + (1+1)*.53})
      --
    ({-1 + (-1.21+1)*.54},{-1 + (1+1)*.54});

  \begin{scope}[rotate=(+8)]
   \draw[dashed]
     (1.5,-1)
     ellipse
     ({.2*1.85} and {.37*1.85});
   \begin{scope}[
     shift={(1.5-.2,{-1+.37*1.85-.1})}
   ]
     \draw[->, -Latex]
       (0,0)
       to
       (180+37:0.01);
   \end{scope}
   \begin{scope}[
     shift={(1.5+.2,{-1-.37*1.85+.1})}
   ]
     \draw[->, -Latex]
       (0,0)
       to
       (+37:0.01);
   \end{scope}
   \begin{scope}[shift={(1.5,-1)}]
     \draw (.43,.65) node
     { \scalebox{.8}{$
     $} };
  \end{scope}
  \draw[fill=white, draw=gray]
    (1.5,-1)
    ellipse
    ({.2*.3} and {.37*.3});
  \draw[line width=3.5, white]
   (1.5,-1)
   to
   (-2.2,-1);
  \draw[line width=1.1]
   (1.5,-1)
   to node[
     above, 
     yshift=-4pt, 
     pos=.85]{
     \;\;\;\;\;\;\;\;\;\;\;\;\;
     \rotatebox[origin=c]{7}
     {
     \scalebox{.7}{
     \color{darkorange}
     \bf
     \colorbox{white}{anyonic defect}
     }
     }
   }
   (-2.2,-1);
  \draw[
    line width=1.1
  ]
   (1.5+1.2,-1)
   to
   (3.5,-1);
  \draw[
    line width=1.1,
    densely dashed
  ]
   (3.5,-1)
   to
   (4,-1);

  \draw[line width=3, white]
   (-2,-1.3)
   to
   (0,-1.3);
  \draw[-latex]
   (-2,-1.3)
   to
   node[
     below, 
     yshift=+3pt,
     xshift=-7pt
    ]{
     \scalebox{.7}{
       \rotatebox{+7}{
       \color{darkblue}
       \bf
       parameter
       }
     }
   }
   (0,-1.3);
  \draw[dashed]
   (-2.7,-1.3)
   to
   (-2,-1.3);

 \draw
   (-3.15,-.8)
   node{
     \scalebox{.7}{
       \rotatebox{+7}{
       \color{darkgreen}
       \bf
       braiding
       }
     }
   };

  \end{scope}

  \begin{scope}[shift={(-.2,1.4)}, scale=(.96)]
  \begin{scope}[rotate=(+8)]
  \draw[dashed]
    (1.5,-1)
    ellipse
    (.2 and .37);
  \draw[fill=white, draw=gray]
    (1.5,-1)
    ellipse
    ({.2*.3} and {.37*.3});
  \draw[line width=3.1, white]
   (1.5,-1)
   to
   (-2.3,-1);
  \draw[line width=1.1]
   (1.5,-1)
   to
   (-2.3,-1);
  \draw[line width=1.1]
   (1.5+1.35,-1)
   to
   (3.6,-1);
  \draw[
    line width=1.1,
    densely dashed
  ]
   (3.6,-1)
   to
   (4.1,-1);
  \end{scope}
  \end{scope}

  \begin{scope}[shift={(-1,.5)}, scale=(.7)]
  \begin{scope}[rotate=(+8)]
  \draw[dashed]
    (1.5,-1)
    ellipse
    (.2 and .32);
  \draw[fill=white, draw=gray]
    (1.5,-1)
    ellipse
    ({.2*.3} and {.32*.3});
  \draw[line width=3.1, white]
   (1.5,-1)
   to
   (-1.8,-1);
\draw
   (1.5,-1)
   to
   (-1.8,-1);
  \draw
    (5.23,-1)
    to
    (6.4-.6,-1);
  \draw[densely dashed]
    (6.4-.6,-1)
    to
    (6.4,-1);
  \end{scope}
  \end{scope}

\draw (1.73,-1.06) node
 {
  \scalebox{.8}{
    $k_{{}_{I}}$
  }
 };

\begin{scope}
[ shift={(-2,-.55)}, rotate=-82.2  ]

 \begin{scope}[shift={(0,-.15)}]

  \draw[]
    (-.2,.4)
    to
    (-.2,-2);

  \draw[
    white,
    line width=1.1+1.9
  ]
    (-.73,0)
    .. controls (-.73,-.5) and (+.73-.4,-.5) ..
    (+.73-.4,-1);
  \draw[
    line width=1.1
  ]
    (-.73+.01,0)
    .. controls (-.73+.01,-.5) and (+.73-.4,-.5) ..
    (+.73-.4,-1);

  \draw[
    white,
    line width=1.1+1.9
  ]
    (+.73-.1,0)
    .. controls (+.73,-.5) and (-.73+.4,-.5) ..
    (-.73+.4,-1);
  \draw[
    line width=1.1
  ]
    (+.73,0+.03)
    .. controls (+.73,-.5) and (-.73+.4,-.5) ..
    (-.73+.4,-1);

  \draw[
    line width=1.1+1.9,
    white
  ]
    (-.73+.4,-1)
    .. controls (-.73+.4,-1.5) and (+.73,-1.5) ..
    (+.73,-2);
  \draw[
    line width=1.1
  ]
    (-.73+.4,-1)
    .. controls (-.73+.4,-1.5) and (+.73,-1.5) ..
    (+.73,-2);

  \draw[
    white,
    line width=1.1+1.9
  ]
    (+.73-.4,-1)
    .. controls (+.73-.4,-1.5) and (-.73,-1.5) ..
    (-.73,-2);
  \draw[
    line width=1.1
  ]
    (+.73-.4,-1)
    .. controls (+.73-.4,-1.5) and (-.73,-1.5) ..
    (-.73,-2);

 \draw
   (-.2,-3.3)
   to
   (-.2,-2);
 \draw[
   line width=1.1,
   densely dashed
 ]
   (-.73,-2)
   to
   (-.73,-2.5);
 \draw[
   line width=1.1,
   densely dashed
 ]
   (+.73,-2)
   to
   (+.73,-2.5);

  \end{scope}
\end{scope}

\begin{scope}[shift={(-5.6,-.75)}]

  \draw[line width=3pt, white]
    (3,-3)
      --
    (-1,-1)
      --
    (-1.21,1)
      --
    (2.3,3)
      --
    (3, -3);

  \shade[right color=lightgray, left color=white, fill opacity=.7]
    (3,-3)
      --
    (-1,-1)
      --
    (-1.21,1)
      --
    (2.3,3);

  \draw[]
    (3,-3)
      --
    (-1,-1)
      --
    (-1.21,1)
      --
    (2.3,3)
      --
    (3, -3);

\draw (1.73,-1.06) node
 {
  \scalebox{.8}{
    $k_{{}_{I}}$
  }
 };

\draw[-Latex]
  ({-1 + (3+1)*.3},{-1+(-3+1)*.3})
    to
  ({-1 + (3+1)*.29},{-1+(-3+1)*.29});

\draw[-Latex]
    ({-1.21 + (2.3+1.21)*.3},{1+(3-1)*.3})
      --
    ({-1.21 + (2.3+1.21)*.29},{1+(3-1)*.29});

\draw[-Latex]
    ({2.3 + (3-2.3)*.5},{3+(-3-3)*.5})
      --
    ({2.3 + (3-2.3)*.49},{3+(-3-3)*.49});

\draw[-latex]
    ({-1 + (-1.21+1)*.53},{-1 + (1+1)*.53})
      --
    ({-1 + (-1.21+1)*.54},{-1 + (1+1)*.54});

  \begin{scope}[rotate=(+8)]
   \draw[dashed]
     (1.5,-1)
     ellipse
     ({.2*1.85} and {.37*1.85});
   \begin{scope}[
     shift={(1.5-.2,{-1+.37*1.85-.1})}
   ]
     \draw[->, -Latex]
       (0,0)
       to
       (180+37:0.01);
   \end{scope}
   \begin{scope}[
     shift={(1.5+.2,{-1-.37*1.85+.1})}
   ]
     \draw[->, -Latex]
       (0,0)
       to
       (+37:0.01);
   \end{scope}
  \draw[fill=white, draw=gray]
    (1.5,-1)
    ellipse
    ({.2*.3} and {.37*.3});
 \end{scope}

   \begin{scope}[shift={(-.2,1.4)}, scale=(.96)]
  \begin{scope}[rotate=(+8)]
  \draw[dashed]
    (1.5,-1)
    ellipse
    (.2 and .37);
  \draw[fill=white, draw=gray]
    (1.5,-1)
    ellipse
    ({.2*.3} and {.37*.3});
\end{scope}
\end{scope}

  \begin{scope}[shift={(-1,.5)}, scale=(.7)]
  \begin{scope}[rotate=(+8)]
  \draw[dashed]
    (1.5,-1)
    ellipse
    (.2 and .32);
  \draw[fill=white, draw=gray]
    (1.5,-1)
    ellipse
    ({.2*.3} and {.37*.3});
\end{scope}
\end{scope}

\begin{scope}
[ shift={(-2,-.55)}, rotate=-82.2  ]

 \begin{scope}[shift={(0,-.15)}]

 \draw[line width=3, white]
   (-.2,-.2)
   to
   (-.2,2.35);
 \draw
   (-.2,.5)
   to
   (-.2,2.35);
 \draw[dashed]
   (-.2,-.2)
   to
   (-.2,.5);

\end{scope}
\end{scope}

\begin{scope}
[ shift={(-2,-.55)}, rotate=-82.2  ]

 \begin{scope}[shift={(0,-.15)}]

 \draw[
   line width=3, white
 ]
   (-.73,-.5)
   to
   (-.73,3.65);
 \draw[
   line width=1.1
 ]
   (-.73,.2)
   to
   (-.73,3.65);
 \draw[
   line width=1.1,
   densely dashed
 ]
   (-.73,.2)
   to
   (-.73,-.5);
 \end{scope}
 \end{scope}

\begin{scope}
[ shift={(-2,-.55)}, rotate=-82.2  ]

 \begin{scope}[shift={(0,-.15)}]

 \draw[
   line width=3.2,
   white]
   (+.73,-.6)
   to
   (+.73,+3.7);
 \draw[
   line width=1.1,
   densely dashed]
   (+.73,-0)
   to
   (+.73,+-.6);
 \draw[
   line width=1.1 ]
   (+.73,-0)
   to
   (+.73,+3.71);
\end{scope}
\end{scope}

\end{scope}

\draw
  (-2.2,-4.2) node
  {
    \scalebox{1.2}{
      $
       \mathllap{
          \raisebox{1pt}{
            \scalebox{.58}{
              \color{darkblue}
              \bf
              \def\arraystretch{.9}
              \begin{tabular}{c}
                some quantum state for
                \\
                fixed defect positions
                \\
                $k_1, k_2, \cdots$
                at time
                {\color{purple}$t_1$}
              \end{tabular}
            }
          }
          \hspace{-5pt}
       }
        \big\vert
          \psi({\color{purple}t_1})
        \big\rangle
      $
    }
  };

\draw[|->]
  (-1.3,-4.1)
  to
  node[
    sloped,
    yshift=5pt
  ]{
    \scalebox{.7}{
      \color{darkgreen}
      \bf
      unitary adiabatic transport
    }
  }
  node[
    sloped,
    yshift=-5pt,
    pos=.4
  ]{
    \scalebox{.7}{
      }
  }
  (+2.4,-3.4);

\draw
  (+3.2,-3.85) node
  {
    \scalebox{1.2}{
      $
        \underset{
          \raisebox{-7pt}{
            \scalebox{.55}{
              \color{darkblue}
              \bf
              \def\arraystretch{.9}
               \begin{tabular}{c}
              another quantum state for
                \\
                fixed defect positions
                \\
                $k_1, k_2, \cdots$
                at time
                {\color{purple}$t_2$}
              \end{tabular}
            }
          }
        }{
        \big\vert
          \psi({\color{purple}t_2})
        \big\rangle
        }
      $
    }
  };

\end{tikzpicture}
}
\end{center}

The remaining problem is to develop a precise mathematical theory describing these {\it anyons}.

\smallskip

\noindent
{\bf Improved Anyon Models via Geometric Engineering on M-branes.}
A remarkable approach to the otherwise elusive microscopic analysis of such strongly-coupled/correlated quantum systems emerges in the guise of ``geometric engineering'' 
\cite{KatzKlemmVafa97}\cite{BourjailyEspahbodi08}
of quantum fields 
on ``M-branes'' probing orbifold singularities, whereby the given dynamics is (partially) mapped onto the fluctuations of Membranes (whence {\it M-theory} \cite{Duff99}), and of higher-dimensional ``M5-branes'' \cite{GSS24-FluxOnM5}, propagating within an auxiliary higher-dimensional gravitating spacetime orbifold \cite{SS20-Orb}.

\vspace{1mm} 
\hspace{.1cm}
\begin{tabular}{p{5.8cm}}
  \footnotesize
  {\bf Figure \figurenumber: Geometric engineering of 
  quantum systems on M-branes}
  provides tools for 
  analyzing otherwise elusive
  strongly coupled/correlated
  quantum phenomena.
\end{tabular}
\quad
\begin{tikzcd}[
  row sep=4pt,
  column sep=20pt
]
  \adjustbox{
    margin=3pt,
    rndframe=3pt
  }{\def\arraystretch{.9}\def\tabcolsep{0pt} \footnotesize
    \begin{tabular}{c}
      Strongly coupled
      \\
      quantum material
    \end{tabular}
  }
  \ar[
    rr,
    -Latex,
    "{
      \scalebox{.7}{
        \color{darkgreen}
        \bf
        match key
      }
    }",
    "{
      \scalebox{.7}{
        \color{darkgreen}
        \bf
        properties
      }
    }"{swap}
  ]
  &\phantom{--}&
  \adjustbox{
    margin=3pt,
    rndframe=3pt
  }{\def\arraystretch{.9}\def\tabcolsep{0pt} \footnotesize
    \begin{tabular}{c}
      M-brane fluctuating in
      \\
      auxiliary gravitational 
      \\
      spacetime orbifold
    \end{tabular}
  }
  \\[-5pt]
  \adjustbox{
    raise=7pt
  }{\def\arraystretch{.9}\def\tabcolsep{0pt} \footnotesize \bf \color{darkblue} 
    \begin{tabular}{c}
      direct analysis
      \\
      unfeasible here
    \end{tabular}
  }  
  \ar[
    u,
    -Latex
  ]
  &&
  \adjustbox{
    raise=3pt
  }{\def\arraystretch{.9}\def\tabcolsep{0pt} \footnotesize  \bf \color{darkblue} 
    \begin{tabular}{c}
      analytical tools
      \\
      exist here
    \end{tabular}
  }  
  \ar[
    u,
    -Latex
  ]
\end{tikzcd}

This procedure is most famous in the (unrealistic) limit of large rank and hence of large numbers $N \to \infty$ of coincident such branes, where it extracts quantum correlators and quantum phase transitions entirely from classical gravitational asymptotics (``holographic duality'' \cite{AGMOOY00}). The application to quantum materials \cite{Zaanen15}\cite{HartnollLucasSachdev18} is now well-studied, notably in the case of quantum critical superconductors engineered in M-theory \cite{HerzogKovtun07}\cite{GauntlettSonnerWiseman10a}\cite{GauntlettSonnerWiseman10b}\cite{Gubser10}\cite{DonosGauntlett13a}\cite{DonosGauntlett13b}\cite{AnLiYang22}.  

\smallskip

But we have established \cite{GSS24-FluxOnM5}\cite{SS24-AbAnyons}\cite{SS25-TQBits}\cite{SS25-Seifert}
that after implementing a previously neglected step of ``flux quantization'' \cite{SS24-Flux} on the M5-brane worldvolume, there provably appear general solitonic and specifically anyonic quantum states already in the more realistic situation of single ($N=1$) coincident branes.
(Similar results for $N = 2$ had previously only been conjectured \cite{ChoGangKim20} by appeal to the expected but notoriously undefined effective quantum field theory on coincident M5-branes.) 

Moreover, in \cite{SS25-ViaAlgTop} we have proven that the resulting topological quantum states and their topological order agrees in fine detail with the expectations for FQH systems as also predicted by abelian Chern-Simons theory, while at the same time predicting that and how {\it defect} anyons in these systems may exhibit the much-desired non-abelian braiding.
It is these results that we survey in the present lecture notes.

\vspace{1.5mm} 
\label{BraneDiagram}
\hypertarget{BraneDiagram}{}
\smallskip 
\hspace{-.62cm}
{\footnotesize
\begin{tabular}{l}
\bf 
Figure \figurenumber: 
\\
Brane diagram for
\\
\bf 
geometric engineering
\\
{\bf of anyons} 
on single M5-
\\
branes wrapping an
orbi-
\\
singularity \cite{SS25-Seifert}.
\\
It is a subtle mechanism
\\
of {\it flux-quantization} 
\cite{SS24-Flux}
\\
of the self-dual tensor-
\\
field on the M5 \cite{GSS24-FluxOnM5} that 
\\
stabilizes \cite{SS24-AbAnyons} its anyonic
\\
soliton configurations.
\end{tabular}
  \adjustbox{
    raise=2pt,
    scale=1.12,
    margin=-2pt,
    fbox
  }{
  \hspace{-12pt}
  \begin{tikzcd}[
    column sep=-5pt,
    row sep=-20pt
  ]
  \scalebox{.7}{
    \color{darkblue}
    \bf
    \def\arraystretch{.85}
    \begin{tabular}{c}
      M5-brane probe
      \\
      worldvolume
    \end{tabular}
  }
  &&
  \scalebox{.7}{
    \color{darkblue}
    \bf
    \def\arraystretch{.85}
    \begin{tabular}{c}
      2-brane worldvolume
      \\
      hosting anyonic solitons
    \end{tabular}
  }
  &&
  \scalebox{.7}{
    \color{darkblue}
    \bf
    \def\arraystretch{.85}
    \begin{tabular}{c}
      M-theory
      \\
      circle
    \end{tabular}
  }
  &&
  \scalebox{.7}{
    \color{darkblue}
    \bf
    \def\arraystretch{.85}
    \begin{tabular}{c}
      cone
      \\
      orbifold
    \end{tabular}
  }
  \\[16pt]
  \Sigma^{1,5}
  &=&
  \mathbb{R}^{1,0}
  \times
  \mathbb{R}^2_{\cpt}
  &\times&
  S^1
  &\times&
  \mathbb{R}^2 \sslash \mathbb{Z}_2
  \\[-12pt]
  &&
  \adjustbox{
    raise=3.8cm,
    scale=.7
  }{
    \begin{tikzpicture}
\begin{scope}[
  scale=.8,
  shift={(.7,-4.9)}
]

  \shade[right color=lightgray, left color=white]
    (3,-3)
      --
    (-1,-1)
      --
        (-1.21,1)
      --
    (2.3,3);

  \draw[dashed]
    (3,-3)
      --
    (-1,-1)
      --
    (-1.21,1)
      --
    (2.3,3)
      --
    (3,-3);

  \node[
    scale=1
  ] at (3.2,-2.1)
  {$\infty$};

  \begin{scope}[rotate=(+8)]
  \draw[dashed]
    (1.5,-1)
    ellipse
    (.2 and .37);
  \draw
   (1.5,-1)
   to 
    node[above, yshift=-1pt]{
     \;\;\;\;\;\;\;\;\;\;\;\;\;
     \rotatebox[origin=c]{7}{
     \scalebox{.7}{
     \color{darkorange}
     \bf
       anyon
     }
     }
   }
    node[below, yshift=+6.3pt]{
     \;\;\;\;\;\;\;\;\;\;\;\;\;\;\;
     \rotatebox[origin=c]{7}{
     \scalebox{.7}{
     \color{darkorange}
     \bf
       worldline
     }
     }
   }
   (-2.2,-1);
  \draw
   (1.5+1.2,-1)
   to
   (4,-1);
  \end{scope}

  \begin{scope}[shift={(-.2,1.4)}, scale=(.96)]
  \begin{scope}[rotate=(+8)]
  \draw[dashed]
    (1.5,-1)
    ellipse
    (.2 and .37);
  \draw
   (1.5,-1)
   to
   (-2.3,-1);
  \draw
   (1.5+1.35,-1)
   to
   (4.1,-1);
  \end{scope}
  \end{scope}
  \begin{scope}[shift={(-1,.5)}, scale=(.7)]
  \begin{scope}[rotate=(+8)]
  \draw[dashed]
    (1.5,-1)
    ellipse
    (.2 and .32);
  \draw
   (1.5,-1)
   to
   (-1.8,-1);
  \end{scope}
  \end{scope}

\end{scope}
    \end{tikzpicture}
    }
    &\times&
    \adjustbox{
      raise=3pt
    }{
    \begin{tikzpicture}
      \draw[
        line width=2.2,
        draw=darkgreen
      ]
        (0,0) circle (1);
    \end{tikzpicture}
    }
    &\times&
    \adjustbox{
      raise=3pt
    }{
\begin{tikzpicture}
\begin{scope}[
  xscale=.7,
  yscale=.5*.7
]
 \shadedraw[draw opacity=0, top color=darkblue, bottom color=cyan]
   (0,0) -- (3,3) .. controls (2,2) and (2,-2) ..  (3,-3) -- (0,0);
 \draw[draw opacity=0, top color=white, bottom color=darkblue]
   (3,3)
     .. controls (2,2) and (2,-2) ..  (3,-3)
     .. controls (4,-3.9) and (4,+3.9) ..  (3,3);
\end{scope}
\end{tikzpicture}
    }
  \end{tikzcd}
  \hspace{-14pt}
  }
}

\vspace{.25cm}

\noindent
Concretely, here we review and explain how this works, aimed at an audience assumed to be familiar with the general mechanism of {\it flux quantization} as surveyed in \cite{SS24-Flux}. 

\smallskip

But first to briefly recall the traditional theory of fractional quantum Hall anyons:

\smallskip 
\noindent
{\bf Quantum Hall effect} (cf.   \cite{PrangeGirvin86}\cite{ChakrabortyPietilainen95}\cite{Stormer99}\cite{PapicBalram24}).
In a very thin (atomic multi-layer) and hence effectively 2-dimensional sheet $\Sigma^2$ of (semi-)conducting material
carrying magnetic flux density $B$,
the energy of electron states is (cf. \cite[(4-12)]{vonKlitzing86}) quantized by
{\it Landau levels} $i \in \mathbb{N}$
as
$$
  E 
    \,=\,
  \hbar \omega_B
  \big(
    i + \tfrac{1}{2}
  \big)
  \,,
$$
where each Landau level comprises of one state per magnetic flux quantum:
$$
  n_{\mathrm{deg}}
  \,=\,
  B/\Phi_0 
  \,,
$$
and the Lorentz force on a longitudinal electron current $J_x$ at filling fraction $\nu$ 
is compensated in equilibrium by an electric {\it Hall field} 
$$
  E_y \,=\, \tfrac{1}{\nu} J_x
  \,.
$$

\medskip

\noindent
\;\;{\bf Integer quantum Hall effect.}
Therefore, Fermi's theory of idealized {\it free} electrons predicts the system to be a conductor away from the energy gaps between a completely filled and the next empty Landau level, 
hence away from the number of electrons being integer multiples 
of the number of flux quanta, where longitudinal conductivity should vanish.
$$n_{\mathrm{el}} = \nu B/\Phi_0
\,,
\;\;
\nu \in \mathbb{N}.$$
This is indeed observed --- in fact, the vanishing conductivity is observed in sizeable neighborhoods of the critical filling fractions (``Hall plateaux'', attributed to subtle disorder effects).

\medskip

\noindent
\;\;{\bf Fractional quantum Hall effect} (FQHE). 
But in reality, the electrons are far from free.
While there is little theory for strongly interacting quantum systems,
experiment shows that 
the Fermi idealization breaks down at low enough temperature, where longitudinal conductivity also decreases in neighborhoods of certain {\it fractional} filling factors $\nu$.
 $$ \nu \,\in\, \mathbb{Q}, 
  \;\;\;\; 
  \scalebox{.8}{prominently for}
  \;\;\;\;\;
  \nu = 1/\lattice
  \,, 
  \;
  \lattice \in 2\mathbb{N} + 1
  \,.
  $$
The traditional heuristic idea is that at these filling fractions the interacting electrons each form a kind of bound state with $\lattice$ flux quanta, making ``composite bosons''
(cf. \cite{Zhang92}) that, as such, condense to produce an insulating mass gap, even inside the Landau level.

\vspace{-.1cm}
\begin{center}
\begin{minipage}{4.4cm}
  \footnotesize
  {\bf Figure \figurenumber: 
  \label{FQHAnyons}
  Anyons in fractional quantum Hall systems} (``quasi-holes'') are (vortices in the electron gas corresponding to) surplus magnetic flux quanta on top of a state of exact rational {\it filling fraction} where each electron is coupled/paired in some subtle way to a fixed number of flux quanta. Compare with Fig. \ref{SolitonicAndDefectAnyons}.
\end{minipage}
\hspace{-.6cm}
\adjustbox{
  raise=-1.5cm
}{
\begin{tikzpicture}

\draw[
  dashed,
  fill=lightgray
]
  (0,0)
  -- (8,0)
  -- (10+.3,2+.3)
  -- (2.8+.3,2+.3)
  -- cycle;

\begin{scope}[
  shift={(2.4,.5)}
]
\shadedraw[
  draw opacity=0,
  inner color=olive,
  outer color=lightolive
]
  (0,0) ellipse (.7 and .3);
\end{scope}

\begin{scope}[
  shift={(4.5,1.5)}
]

\begin{scope}[
 scale=1.8
]
\shadedraw[
  draw opacity=0,
  inner color=olive,
  outer color=lightolive
]
  (0,0) ellipse (.7 and .25);
\end{scope}

\begin{scope}[
 scale=1.45
]
\shadedraw[
  draw opacity=0,
  inner color=olive,
  outer color=lightolive
]
  (0,0) ellipse (.7 and .25);
\end{scope}

\shadedraw[
  draw opacity=0,
  inner color=olive,
  outer color=lightolive
]
  (0,0) ellipse (.7 and .25);

\begin{scope}[
  scale=.2
]
\draw[
  fill=black
]
  (0,0) ellipse (.7 and .25);
\end{scope}

\end{scope}

\begin{scope}[
  shift={(7.2,1)},
  scale=1.7
]
\shadedraw[
  draw opacity=0,
  inner color=white,
  outer color=lightgray
]
  (0,0) ellipse (.7 and .25);
\end{scope}

\node
  at (-.8,.7)
  {
    \adjustbox{
      bgcolor=white,
      scale=.7
    }{
      \color{darkblue}
      \bf
      \def\arraystretch{.9}
      \begin{tabular}{c}
        un-paired
        \\
        flux quantum:
        \\
        \color{purple}
        quasi-hole
      \end{tabular}
    }
  };

\draw[
  white,
  line width=2
]
  (-.4, 1.15)
  .. controls 
  (1,2) and 
  (2,2) ..
  (2.32,.7);
\draw[
  -Latex,
  black
]
  (-.4, 1.15)
  .. controls 
  (1,2) and 
  (2,2) ..
  (2.32,.7);

\node
  at (10,.5)
  {
    \adjustbox{
      scale=.7
    }{
      \color{darkblue}
      \bf
      \def\arraystretch{.9}
      \def\tabcolsep{-5pt}
      \begin{tabular}{c}
        deficit of a
        \\
        flux-quantum:
        \\
        \color{purple}
        quasi-particle
      \end{tabular}
    }
  };

\draw[
  white,
  line width=2
]
  (10,1.0) 
  .. controls 
  (9.5,1.8) and 
  (8.1,2.5) ..
  (7.5,1.18);
\draw[
  -Latex
]
  (10,1.0) 
  .. controls 
  (9.5,1.8) and 
  (8.1,2.5) ..
  (7.5,1.18);

\node
  at (1.3,2.7)
  {
    \adjustbox{
      scale=.7
    }{
      \color{darkblue}
      \bf
      \def\arraystretch{.9}
      \def\tabcolsep{-5pt}
      \begin{tabular}{c}
        $\lattice$ flux-quanta
        \\
        bound to 1 electron:
        \\
        \color{purple}
        composite boson
      \end{tabular}
    }
  };

\draw[
 line width=2.5pt,
  white
]
  (2.4, 3) .. controls 
  (2.8,3.3) and 
  (4,3.5) ..
  (4.3,1.8);

\draw[
  -Latex
]
  (2.4, 3) .. controls 
  (2.8,3.3) and 
  (4,3.5) ..
  (4.3,1.8);

\node at 
  (9,2.6)
  {
   \scalebox{.8}{
     \color{gray}
     (cf. \cite[Fig. 16]{Stormer99})  
   }
  };
\end{tikzpicture}
}
\end{center}

\bigskip 
\noindent
\;\;{\bf Anyonic quasi-particles.}
This heuristic model suggests that in the Hall plateau neighborhood {\it around} such filling fraction, there are {\it unpaired} flux quanta effectively ``bound to'' $1/\lattice$th of a (missing) electron: called ``quasi-particles'' (``quasi-holes''). 
These quasi-particles/holes evidently have fractional charge $\pm e/\mathcolor{purple}{\lattice}$ and are expected to be anyonic with fractional pair exchange phase $e^{\mathrm{i} \pi/\mathcolor{purple}{\lattice}}$. This phase has been experimentally observed \cite{Nakamura20}.

\vspace{.3cm}

\noindent
{\bf Effective abelian Chern-Simons theory.}
The traditional ansatz for an effective field theory description of $\lattice$-fractional quantum Hall systems
postulates that the effective field is a 1-form potential $a$ for the electric current density 2-form $J$, itself minimally coupled to the {\it quasi-hole current} $j$, 
and with effective dynamics encoded by the level = $\level = \lattice/2$ Chern-Simons (CS) Lagrangian  \cite{Zhang92}\cite{Wen95}:

\vspace{.1cm}
$$
  \begin{tikzcd}[
    sep=0pt
  ]
  \scalebox{.7}{
    \bf
    \def\arraystretch{.9}
    \begin{tabular}{c}
      \color{darkblue}
      Electron current
      \\
      \color{gray}
      density 2-form
    \end{tabular}
  }
  &[-10pt]
  J 
  &=&
  \overset{
    \mathclap{
      \adjustbox{
        raise=-2pt,
        scale=.7,
        rotate=+12
      }
      {
        \rlap{
        \color{gray}
        \bf
        current 3-vector
        }
      }
    }
  }{
    \vec J
  }
  \;
  \scalebox{1.2}{$\lrcorner$}
  \;
  \,
  \overset{
    \mathclap{
      \adjustbox{
        raise=-2pt,
        scale=.7,
        rotate=+12
      }
      {
        \rlap{
        \color{gray}
        \bf
        volume form
        }
      }
    }
  }{
    \mathrm{dvol}
  }
  \;\;
  &=:&
  \mathrm{d}\;a
  \hspace{-1pt}
  \adjustbox{
    scale=.7,
    raise=1pt
  }{\;\;
    \color{darkblue}
    \bf
    \def\tabcolsep{0pt}
    \begin{tabular}{c}
      Effective gauge field
    \end{tabular}
  }
  % \;\;\;\;\;
  % \mbox{since}
  % \;\;\;\;
  % \mathrm{div} \,\vec J \,=\, 0
  % \;\;\;
  % \Leftrightarrow
  % \;\;\;
  % \mathrm{d}\, J \,=\, 0
  \\
  \scalebox{.7}{
    \bf
    \def\arraystretch{.9}
    \begin{tabular}{c}
      \color{darkblue}
      Quasi-particle current
      \\
      \color{gray}
      density 2-form
    \end{tabular}
  }
  &
  j 
    &=&
  \vec j 
  \;
  \scalebox{1.2}{$\lrcorner$}
  \;
  \,
  \mathrm{dvol}
  \\
  \scalebox{.7}{
    \bf
    \def\arraystretch{.9}
    \begin{tabular}{c}
      \color{darkblue}
      Background flux
      \\
      \color{gray}
      density 2-form
    \end{tabular}
  }
  &
  F &=&
  \mathrm{d}\, A
  \mathrlap{
    \;
    \adjustbox{
      scale=.7,
      raise=1pt
    }{\;\;
      \color{darkblue}
      \bf
      External gauge field
    }
  }
  \;\;\;\;\;\;\;
  \\
\scalebox{.7}{
  \bf
  \def\arraystretch{.9}
  \begin{tabular}{c}
    \color{darkblue}
    Effective Lagrangian
    \\
    \color{gray}
    density 3-form
  \end{tabular}
}
 &
  L
  &:=&
  \mathrlap{
   \tfrac{\lattice}{2}
  \,
  \grayunderbrace{
    a \, \mathrm{d}a
  }{ 
    \mathclap{
      \mathrm{CS}(a) 
    }
  }
  \;-\;
  \grayunderbrace{
  \mathclap{\phantom{\tfrac{1}{2}}}
  A \, \mathrm{d}a
  }{  A \, J}
  \;+\;
  a \, j
  \;\;\;\;
  \scalebox{.9}{
    \cite[(2.11)]{Wen95}
  }
  }
  \;\;\;\;\;\;\;\;\;\;\;\;
  \end{tikzcd}
$$

\vspace{-.1cm}
\noindent Its Euler-Lagrange equations of motion 
$$
  \frac{
    \delta L
  }{
    \delta a
  }
  \,=\,
  0
  \;\;\;\;\;
  \Leftrightarrow
  \;\;\;\;\;
  \adjustbox{
    margin=3pt,
    bgcolor=lightolive
  }{$
  J \;=\;
  \tfrac{1}{\lattice}\big(
    \,
    F \,-\, j
    \,
  \big)
  $}
$$
in the case of longitudinal electron current
and static quasi-particles
$$
  \begin{tikzcd}[
    sep=-2pt
  ]
  J 
  &\defneq&
  J_0 \, \mathrm{d}x \, \mathrm{d}y
  &-&
  J_x \, \mathrm{d}t \, \mathrm{d}y
  \\
  j 
  &\defneq&
  j_0 \, \mathrm{d}x \, \mathrm{d}y
  \\
  F 
    &\defneq&
  B\, \mathrm{d}x \, \mathrm{d}y
  &-&
  E_y \, \mathrm{d}t \, \mathrm{d}y
  \end{tikzcd}
$$
express just  the hallmark  properties of the FQHE that we saw above, 
at filling fraction $\nu \,=\, 1/\lattice$:
$$
  \Leftrightarrow
  \;
  \left\{\!\!\!
  \def\arraystretch{1.6}
  \begin{array}{ccccl}
    J_x 
      &=&
    \tfrac{1}{\lattice}
    E_y
    &
    \Leftrightarrow
    &
    \mbox{
      Hall conductivity 
      law at $1/\lattice$ filling 
    }
    \\[-3pt]
    J_0 
    &=&
    \tfrac{1}{\lattice}
    \, 
    B
    &
    \Leftrightarrow
    &
    \mbox{
      each electron binds to
      $k$ flux quanta,
      but
    }
    \\[-5pt]
    &&
    \mathllap{-}
    \tfrac{1}{\lattice}
    \,
    j_0
    &&
    \mbox{
      $1/\lattice$th electron missing
      for each quasi-hole\,.
    }
  \end{array}
  \right.
$$
% \;\;\;
% (!)

\medskip

\noindent
{\bf Conceptual problems.} 
However 
this can only be a {\it local} description on a single chart (as is common for Langrangian field theories):
Neither $J$ nor $F$ may admit global coboundaries $a$ and $A$, respectively.
Instead, both must be subjected to some kind of flux-quantization.
For $F$ this must be classical Dirac charge quantization, 
which however is incompatible with integrality of $J$ when $k \neq 1$ (cf. \cite[p. 35]{Witten16}\cite[p 159]{Tong16}).
But without this, the implications break concerning topological order from abelian CS theory (ground state degeneracy, modular functoriality, ...).

Therefore we must ask:

\medskip

\noindent
\adjustbox{
    margin=3pt,
    bgcolor=lightolive
  }{ \rm 
{\bf Question:} {\it Is there a non-Lagrangian theory for quasi-particles of properly flux-quantized FQH systems?}
}

\noindent  \colorbox{lightgray}{{\bf Answer:} Yes!:}

\medskip 
\noindent
{\bf The main result} to be discussed here is that the key features of the anyonic topological order as seen in fractional quantum Hall systems are consistently, rigorously and naturally reflected by the topological light-cone quantization of the self-dual tensor field on M5-brane probes of certain orbi-singularities in 11D supergravity --- once the subtle (non-abelian) flux-quantization of this field is properly taken care of, which is the key step that has not previously received attention. This is what we explain below.

\medskip

\noindent
{\bf Further aspects.}
In fact, fractional quantum Hall systems exhibit further remarkable properties which have not previously been reflected in their effective (Chern-Simons) descriptions, but which are naturally reflected in the M5-brane model, among them {\it hidden supersymmetry}.
We close this introduction by briefly indicating this phenomenon.

\medskip

\noindent
{\bf $N$-Electron ground states of quantum Hall systems.}
While a microscopic derivation of fractional quantum Hall ground states $\Psi$ remains missing, phenomenologically successful Ans{\"a}tze exist:\footnote{
For $N$ electrons in an effectively 2D material, and assumed to be completely spin-polarized by the transverse magnetic field, their wavefunction $\Psi$ is  a skew-symmetric (by Pauli exclusion) $\mathbb{C}$-valued function of $N$ complex numbers $(z^i \in \mathbb{C})_{i = 1}^N$. We omit normaliztion. For the Read-Moore state $N$ must (for $\mathrm{Pf}(-)$ to be defined) be even (which is harmless since $N$ is a macroscopic number of electrons).}

\begin{itemize} 
\item At odd filling fraction $\nu = 1/q$, $q \in 2\mathbb{N} + 1$,
the {\bf Laughlin wavefunction}
$$
\adjustbox{
  margin=2pt,
  bgcolor=lightolive
}{
$
  \Psi_{\mathrm{La}}\big(
    z^1, \cdots, z^N
  \big)
  \,:=\,
  \prod_{i < j}
  \big(z^i - z^j\big)^q
  \, 
  \exp\big(
    - 
    \tfrac{
     1
    }{
      \mathclap{\phantom{\vert^{\vert}}}
      \ell^2_B
    }
    \sum_i 
    \left\vert
      z^i
    \right\vert^2
  \big)
$
}
$$

\newpage 

\item At even filling fraction $\nu = 1/q$, $q \in 2\mathbb{N}$,
the {\bf Read-Moore wavefunction}
 $$
\adjustbox{
  margin=2pt,
  bgcolor=lightolive
}{
$
  \Psi_{\mathrm{RM}}\big(
    z^1, \cdots, z^N
  \big)
  \,:=\,
  \mathrm{Pf}\!
  \left(
    \frac{1}{
      \mathclap{\phantom{\vert^{\vert}}}
      z^{\bullet_{{}_{1}}}
      -
      z^{\bullet_{{}_2}}
    }
  \right)
  \Psi_{\mathrm{La}}(z^1, \cdots z^N)
$
}.
 $$
\end{itemize} 

%\vspace{cm}
\noindent Here the {\it Pfaffian} $\mathrm{Pf}$ of a skew-symmetric $N \times N$ matrix $A$ is the Bererzinian integral
over anti-commuting variables $(\theta^i)_{i = 1}^N$:
\medskip 
$$
  \mathrm{Pf}(A)
  \;:=\;\;
  \underset{
    \smash{
    \mathclap{
      \adjustbox{
        scale=.6,
        raise=-5pt
      }{
        \color{darkgreen}
        \bf
        \def\arraystretch{.9}
        \begin{tabular}{c}
        pick coefficient of
        \\
        top $\theta$-power
        \end{tabular}
      }
    }
    }
  }{
  \textstyle{\int}
  \big(
    \textstyle{\prod_i} \mathrm{d}\theta^i
  \big)
  }
  \, 
  \exp\big(
    \tfrac{1}{2}
    A_{i j}
    \,
    \theta^i  
    \theta^j
  \big)
  \,.
$$

\bigskip

\noindent
{\bf Hidden super-geometry of quantum Hall systems.}
This suggests to promote the plane $\mathbb{C}^1$ to the super-space $\mathbb{C}^{1\vert 1}$ with its {\it super-translation group} structure
$$
  (z,\theta)
  +
  (z',\theta')
  \;=\;
  \big(
    z + z' + \theta \theta'
    ,\,
    \theta + \theta'
  \big)
$$
Here the {\bf super-Laughlin state}
exhibits the Read-Moore state as a super-partner to the Laughlin state (up to normalization)
\cite{Hasebe08}\cite[(13)]{GromovMartinecRyu20}:
\medskip 
$$
\adjustbox{
  margin=2pt,
  bgcolor=lightolive
}{
$
  \Psi_{\mathrm{sLa}}\big(
    (z^1, \theta^1),
    \cdots,
    (z^N, \theta^N)
  \big)
  \;:=\;
  \prod_{i < j}
  \big(
    z^i - z^j - \theta^i \theta^j
  \big)^q
  \, 
  \exp\big(
    - 
    \tfrac{
     1
    }{
      \mathclap{\phantom{\vert^{\vert}}}
      \ell^2_B
    }
    \sum_i 
    \left\vert
      z^i
    \right\vert^2
  \big)
$
}
$$

\bigskip 
$$
  \begin{tikzcd}[
    column sep=100pt,
    row sep=-5pt
  ]
    &
    \Psi_{\mathrm{sLa}}
    \ar[
      d,
      phantom,
      "{
    \scalebox{.7}{
      \color{darkblue}
      \bf
      \def\arraystretch{.9}
      \begin{tabular}{c}
        super-
        \\
        Laughlin state
      \end{tabular}
    }      
      }"{pos=1.8}
    ]
    \ar[
      dl,
      |->,
      "{
        0 \,\mapsfrom\, \theta
      }"{sloped, description},
      "{
        \scalebox{.7}{
          \color{darkgreen}
          \bf
          lowest component
        }      
      }"{sloped, yshift=2}
    ]
    \ar[
      dr,
      |->,
      "{
        \int \prod_i \mathrm{d}\theta^i
      }"{sloped, description},
      "{
        \scalebox{.7}{
          \color{darkgreen}
          \bf
          top component
        }      
      }"{sloped, yshift=2pt}
    ]
    \\
    \smash{
    \mathllap{
      \scalebox{.7}{
        \def\arraystretch{.9}
        \begin{tabular}{c}
          \color{darkblue}
          \bf
          Laughlin state
          \\
          fermionic
          for odd $q$
        \end{tabular}
      }
    }}
    \Psi_{\mathrm{La}}
    &
    {}
    &
    \Psi_{\mathrm{RM}}
    \smash{
    \mathrlap{
      \scalebox{.7}{
        \def\arraystretch{.9}
        \begin{tabular}{c}
          \color{darkblue}
          \bf
          Moore-Read state
          \\
          fermionic
          for even $q$
        \end{tabular}
      }
    }}
  \end{tikzcd}
$$

\vspace{.2cm}

\noindent
{\bf Collective excitations.}
The Moore-Read state is known to have two {\it density-wave excitations} for wave-vectors $k \in \mathbb{C}$:

\begin{itemize} 
\item[{\bf (i)}] The {\bf magneto-roton state}
  $$
  \adjustbox{
    margin=4pt,
    bgcolor=lightolive
  }{
  $
    \Psi_{\mathrm{MR},k}
    (z^1, \cdots, z^N)
    \;:=\;
    \sum_{i}
    \exp\!\big(
      -\mathrm{i} \overline{k}
      \partial_{z^i}
    \big)
    \exp\!\big(
      -
      \tfrac{\mathrm{i}}{2} 
      \overline{k}
      z^i
    \big)
    \,
    \Psi_{\mathrm{MR}}(
      z^1, \cdots, z^N
    )
  $
  }
  $$
  
   \item[{\bf (ii)}] The {\bf neutral fermion} state
  $$
  \adjustbox{
    margin=4pt,
    bgcolor=lightolive
  }{
  $\Psi_{\mathrm{NF}, k}$
  \mbox{
    which originally did not have a closed expression
  }
  }
$$
\end{itemize} 

\noindent
However, lifting the magneto-roton state to super-space, 
for super-wavevector 
$(k, \kappa) \in \mathbb{C}^{1,1}$
$$
\noindent
\adjustbox{
  margin=3pt,
  bgcolor=lightolive
}{
$
  \Psi_{\mathrm{MR}, (k,\kappa)}(
    z^1, \cdots, z^N
  )
  \;:=\;
  \textstyle{\int}
  \big(
    \prod_i \mathrm{d}\theta^i
  \big)
    \sum_{i}
    \exp\!\big(
      -\mathrm{i} \overline{k}
      \partial_{z^i}
    \big)
    \exp\!\big(
      -
      \tfrac{\mathrm{i}}{2} 
      \overline{k}
      z^i
    \big)
    \exp\!\big(
      - \tfrac{\mathrm{i}}{2}
      \overline{\kappa} \theta^i
    \big)
    \,
    \Psi_{\mathrm{sLa}}
    \big(
      (z^1,\theta^1),
      \cdots,
      (z^N,\theta^N)
    \big)
$
}
$$

\vspace{2mm} 
\noindent it reproduces the magneto-roton state 
for even $N$, and the neutral fermion mode when an $(N+1)$st electron is added \cite{GromovMartinecRyu20}:
\vspace{2mm} 
$$
\smallskip 
  \begin{tikzcd}[
    column sep=95pt,
    row sep=0pt
  ]
    & 
    \Psi_{\mathrm{MR}, (k, \kappa)}
    \ar[
      d,
      phantom,
      "{
        \scalebox{.7}{
          \color{darkblue}
          \bf
          \def\arraystretch{.9}
          \begin{tabular}{c}
            super-
            \\
            density excitation
          \end{tabular}
        }
      }"{pos=.5}
    ]
    \ar[
      dl,
      |->,
      "{
        \scalebox{.7}{even $N$}
      }"{sloped, description},
      "{
        \scalebox{.7}{
          \color{darkgreen}
          \bf
          no electron added
        }
      }"{sloped, yshift=2pt}
    ]
    \ar[
      dr,
      |->,
      "{
        \scalebox{.7}{odd $N$}
      }"{sloped, description},
      "{
        \scalebox{.7}{
          \color{darkgreen}
          \bf
          add an electron
        }
      }"{sloped, yshift=2pt}
    ]
    \\
    \mathllap{
      \adjustbox{
        raise=9pt,
        scale=.7
      }{
        \color{darkblue}
        \bf
        \def\arraystretch{.86}
        \begin{tabular}{c}
          magneto-roton 
          \\
          state
        \end{tabular}
     \qquad }
    }
    \hspace{-16pt}
    \Psi_{\mathrm{MR}, k}
    &{}&
    \kappa
    \Psi_{\mathrm{NF}, k}
    \hspace{-16pt}
    \mathrlap{
      \adjustbox{
        scale=.7,
        raise=4pt
      }{\qquad 
        \color{darkblue}
        \bf
        \def\arraystretch{.86}
        \begin{tabular}{c}
          neutral-fermion
          \\
          mode
        \end{tabular}
      }
    }
  \end{tikzcd}
$$

\medskip

\noindent
{\bf Hidden super-symmetry in fractional quantum Hall systems.}
This super-unification predicts hidden supersymmetry in fractional quantum Hall systems --- which is indeed (numerically) observed \cite{PBFGP23}\cite{LZX24} (also \cite[\S 5]{BaeLee21}).

\medskip

This all suggests that an accurate model for fractional quantum Hall systems should in fact itself {\it originate on superspace}, and this is what we start with now.

 \newpage 
%%%%%%%%%%%%%%%%%%%%%%%%%%%
\section{Flux-Quantization on M5-Probes}
\label{FluxQuantizationOnM5Probes}
%%%%%%%%%%%%%%%%%%%%%%%%%%%

The first task now is to understand the flux-quantization on M5-brane probes, according to  \cite{FSS20-H}\cite{FSS22-Twistorial}\cite{SS25-EquTwistorial}.
We will not (need to) explain in full detail the (super-)geometry of probe branes nor of their (super-)gravity backgrounds (full discussion is in \cite{GSS24-SuGra}\cite{GSS24-FluxOnM5}), but do offer the following broad dictionary, for orientation:
\footnote{
  All brane concepts we consider are well-defined and all conclusions have proofs -- at no point do we rely on informal string theory folklore beyond motivation.
}

\medskip

\noindent
{\bf M5-Brane probes} (namely {\it sigma-model} branes, in contrast to {\it black branes}) are 5-dimensional objects propagating in a gravitational target space $X$ (the ``bulk''), along trajectories that are modeled by (super-)immersions of their 6D (and $\mathcal{N} = (2,0)$) worldvolume (super-)manifolds $\Sigma$
\begin{equation}
  \label{M5Immersion}
  \begin{tikzcd}[column sep=large]
    \adjustbox{
      raise=2pt,
      scale=.7
    }{
      \color{darkblue}
      \bf
      \def\arraystretch{.9}
      \begin{tabular}{c}
        probe M5-brane
        \\
        (super-)worldvolume
      \end{tabular}
    }
    \;\;
    \Sigma^{1,5\,\vert\, 2 \cdot \mathbf{8}_+}
    \ar[
      rr,
      "{ \phi_s }",
      "{
        \scalebox{.7}{
          \color{darkgreen}
          \bf
          \def\arraystretch{.9}
          \begin{tabular}{c}
            trajectory
            \\
            (super-)immersion
          \end{tabular}
        }
      }"{swap}
    ]
    &&
    X^{1,10\,\vert\, \mathbf{32}}
    \;\;
    \adjustbox{
      raise=2pt,
      scale=.7
    }{
      \color{darkblue}
      \bf
      \def\arraystretch{.9}
      \begin{tabular}{c}
        target/background
        \\
        (super-)spacetime
      \end{tabular}
    }
  \end{tikzcd}
\end{equation}
Here the admissible (``on-shell'', meaning: satisfying the appropriate equations of motion) immersions $\phi_s$ are controlled by the (super-)geometry of $X$  -- namely the brane's trajectory is subject to the gravitational- and Lorentz-forces exerted by the field content of $X$ -- but $X$ itself remains unaffected by the choice of $\phi_s$ -- meaning that the (gravitational) {\it back-reaction} of the brane on its ambient spacetime is neglected; this is what makes the brane but a {\it probe} of the {\it background} $X$.

\medskip

\noindent
Thereby the probe brane $(\Sigma, \phi_s)$ plays a double role:

\begin{itemize}

\item[(i)] on the one hand it is like a (higher-dimensional) fundamental particle, an  ``observer'' of the bulk $X$ in the sense of mathematical relativity,

\item[(ii)] on the other hand it is itself a (super-)spacetime with its own (quantum) field content:

Remarkably, the magic of super-geometry makes such purely super-geometric immersions $\phi_s$
\eqref{M5Immersion} embody not just the na{\" i}ve (temporal-)spatial worldvolume trajectory, but also a 3-flux density $H_3^s$ {\it on} $\Sigma$ \cite[\S 3.3]{GSS24-FluxOnM5}. 
This is (on-shell) the notorious ``self-dual'' flux density whose accurate quantization (traditionally neglected) is our main concern here.

\end{itemize}

\smallskip 
\noindent
This second aspect is what we are concerned with for the purpose of modeling strongly-coupled quantum systems:
The (1+3)D worldvolume $M^{1,3}$
of a quantum material 
-- or, for the intent of modeling anyons, the effectively  $(1+2)D$-worldvolume $M^{1,2}$ of a sheet-like material (e.g. an atomic mono-layer akin to graphene) --
is to be identified with a sub-quotient of the brane worldvolume, typically with a fixed locus (orbifold singularity) inside the base of a fibration (Kaluza-Klein reduction).
$$
  \begin{tikzcd}[
    row sep=24pt, column sep=large
  ]
    &&
    \mathclap{
      \scalebox{.7}{
        \color{darkblue}
        \bf
        \def\arraystretch{.9}
        \begin{tabular}{c}
          M5-brane
          \\
          worldvolume
        \end{tabular}
      }
    }
    &&
    \mathclap{
      \scalebox{.7}{
        \color{darkblue}
        \bf
        \def\arraystretch{.9}
        \begin{tabular}{c}
          ambient bulk
          \\
          spacetime
        \end{tabular}
      }
    }
    \\[-25pt]
    &&
    \Sigma^{1,5}
    \ar[
      rr,
      "{ \phi }"
    ]
    \ar[
      d,
      ->>
    ]
    &&
    Y^{1,10}
    \ar[
      d,
      ->>,
      "{
        \scalebox{.7}{
          \color{darkgreen}
          \bf
          \def\arraystretch{.9}
          \begin{tabular}{c}
            M/IIA-
            \\
            fibration
          \end{tabular}
        }
      }"
    ]
    \\
    M^{1,2}
    \ar[
      rr,
      hook,
      "{
        \scalebox{.7}{
          \color{darkgreen}
          \bf
          \def\arraystretch{.9}
          \begin{tabular}{c}
            orbi-
            \\
            singularity
          \end{tabular}
        }
      }"
    ]
    &&
    \Sigma^{1,4}
    \ar[rr]
    &&
    X^{1,9}
    \\[-25pt]
    \mathclap{
      \scalebox{.7}{
        \color{purple}
        \bf
        \def\arraystretch{.9}
        \begin{tabular}{c}
          quantum material
          \\
          worldvolume
        \end{tabular}
      }
    }
    &&
    \mathclap{
      \scalebox{.7}{
        \color{darkblue}
        \bf
        \def\arraystretch{.9}
        \begin{tabular}{c}
          D4-brane
          \\
          worldvolume
        \end{tabular}
      }
    }
    &&
  \end{tikzcd}
$$

\noindent
{\bf Their flux quantization} (to recall from \cite{SS24-Flux}) is then encoded in a choice of a fibration $\hotype{A} \xrightarrow{p} \hotype{B}$ of classifying spaces, subject to the constraint that the Bianchi identities for the (duality-symmetric) flux densities on bulk and brane are the closure/flatness condition on $\mathfrak{l} p$-valued differential forms, where $\mathfrak{l}(-)$ forms {\it Whitehead $L_\infty$-algebras} of these classifying fibrations (dual to their minimal relative Sullivan model).

\smallskip 
Given such a choice, the topological sector of the higher gauge fields on bulk and brane are given by maps from the brane-immersion into the classifying fibration:

\vspace{.3cm}

With these comments on perspective out of the way, 
{\bf the plan of this section} are the following topics:

\smallskip 
\begin{itemize}[
  itemsep=-1pt,
  leftmargin=1cm
]

\item[(i)] \hyperlink{BianchiIdentitiesOnM5Probes}{Bianchi identities on magnetized M5-probes}

\item[(ii)] \hyperlink{FluxQuantizationInTwistorialCohomotopy}{Flux quantization in Twistorial Cohomotopy}

\item[(iii)] \hyperlink{ProjectiveSpacesAndHopfFibrations}{Aside: Projective Spaces and their Fibrations}

\item[(iv)] \hyperlink{EquivariantFormOfTwistorialCohomotopy}{Orbi-worldvolumes and Equivariant charges}

\end{itemize}

$$
\adjustbox{
  fbox,
  raise=5pt, scale=0.9
}{
\hspace{-11pt}
\begin{tikzcd}[
  column sep=37pt
]
  \scalebox{.7}{
    \color{darkblue}
    \bf
    brane
  }
  &[-33pt]
  \Sigma
  \ar[
    rr,
    dashed,
    "{
      \scalebox{.7}{
        \color{darkgreen}
        \bf
        \def\arraystretch{.9}
        \begin{tabular}{c}
          densities of
          \\
          brane fluxes
        \end{tabular}
      }
    }"
  ]
  \ar[
    dd,
    "{
      \phi
    }"{swap},
    "{
      \scalebox{.7}{
        \color{olive}
        \bf
        immersion
      }
    }"{sloped, yshift=3pt}
  ]
  &&
  \Omega^1_{\mathrm{dR}}(
    -;
    \mathfrak{l}_{{}_{\hotype{B}}}
    \hotype{A}
  )_{\mathrlap{\mathrm{clsd}}}
  \ar[
    dd,
    "{
      (\mathfrak{l}p)_\ast
    }"
  ]
  &[-8pt]
  \Sigma
  \ar[
    rr,
    dashed,
    "{
      \scalebox{.7}{
        \color{darkgreen}
        \bf
        \def\arraystretch{.9}
        \begin{tabular}{c}
          charges of
          \\
          brane fields
        \end{tabular}
      }
    }",
    "{\ }"{swap, name=s},
  ]
  \ar[
    dd,
    "{ \phi }"{swap},
    "{\ }"{pos=.4, name=t}
  ]
  \ar[
    from=s,
    to=t,
    Rightarrow,
    dashed
  ]
  &&
  \hotype{A}
  \ar[
    dd,
    "{ p }"{swap},
    "{
      \scalebox{.7}{
        \color{olive}
        \bf
        \def\arraystretch{.9}
        \begin{tabular}{c}
          fibration of 
          \\
          classifying spaces
        \end{tabular}
      }
    }"{sloped, yshift=3pt}
  ]
  \\
  \\
  \scalebox{.7}{
    \color{darkblue}
    \bf
    bulk
  }
  &
  X
  \ar[
    rr,
    dashed,
    "{
      \scalebox{.7}{
        \color{darkgreen}
        \bf
        \def\arraystretch{.9}
        \begin{tabular}{c}
          densities of
          \\
          bulk fluxes
        \end{tabular}
      }
    }"{swap}
  ]
  &&
  \Omega^1_{\mathrm{dR}}(-;\mathfrak{l}\hotype{B})_{\mathrlap{\mathrm{clsd}}}
  &
  X
  \ar[
    rr,
    dashed,
    "{
      \scalebox{.7}{
        \color{darkgreen}
        \bf
        \def\arraystretch{.9}
        \begin{tabular}{c}
          charges of
          \\
          bulk fields
        \end{tabular}
      }
    }"{swap},
  ]
  &&
  \hotype{B}
\end{tikzcd}
}
$$

\bigskip 
The first step of flux quantization is to identify the Bianchi identities satisfied by the flux densities:

\medskip 
%\newpage 
\noindent
\hypertarget{BianchiIdentitiesOnM5Probes}{}{\noindent \bf Bianchi identities on M5-Probes of 11D SuGra via super-geometry.}
Consider the 11D super-tangent space
$$
  \begin{tikzcd}
  \underset{
    \mathclap{
      \adjustbox{
        raise=-6pt,
        scale=.7
      }{
        \color{darkblue}
        \bf
        super-Minkowski
      }
    }
  }{
    \mathbb{R}^{1,10\,\vert\,\mathbf{32}}
  }
  \quad 
  \ar[
    r,
    hook
  ]
  &
  \;\;
  \underset{
    \mathclap{
      \adjustbox{
        raise=-1pt,
        scale=.7
      }{
        \color{darkblue}
        \bf
        super-Poincar{\'e}
      }
    }
  }{
  \mathfrak{isom}\big(
  \mathbb{R}^{1,10\,\vert\,\mathbf{32}}  
  \big)
  }
  \;\;
  \ar[
    r,
    ->>
  ]
  &
  \underset{
    \mathclap{
      \adjustbox{
        raise=-1pt,
        scale=.7
      }{
        \color{darkblue}
        \bf
        Lorentz
      }
    }
  }{
    \mathfrak{so}(1,10)
  }
  \end{tikzcd}
$$
with its super-invariant 1-forms
(cf. \cite[\S 2.1]{GSS24-SuGra}):
\vspace{-1mm} 
$$
  \mathrm{CE}\big(
    \mathbb{R}^{1,10\,\vert\,\mathbf{32}}
  \big)
  \;\;
  \simeq
  \;\;
  \underset{
    \mathclap{
      \adjustbox{
        scale=.7,
        raise=-1pt
      }{
        \color{darkblue}
        \bf
        super-transl. invar. forms
      }
    }
  }{\;\;\;
  \Omega^\bullet_{\mathrm{dR}}
  \big(
    \mathbb{R}^{1,10\vert\mathbf{32}}
  \big)^{\mathrm{li}}
  \;\;\;}
  \;\;
  \simeq
  \;\;
  \mathbb{R}_{\mathrm{d}}
  \left[
    \def\arraystretch{1.3}
    \def\arraycolsep{2pt}
    \begin{array}{c}
      (\Psi^\alpha)_{\alpha=1}^{32}
      \\
      (E^a)_{a=0}^{10}
    \end{array}
  \right]
  \Big/
  \left(
    \def\arraystretch{1.1}
    \def\arraycolsep{2pt}
    \begin{array}{rcl}
      \mathrm{d}\, \Psi^\alpha
      &=&
      0
      \\
      \mathrm{d}\, E^a
      &=&
      \big(\hspace{1pt}
        \overline{\Psi}
        \,\Gamma^a\,
        \Psi
      \big)
    \end{array}
  \right)
  \mathrlap{\,.}
$$

\vspace{2mm} 
\noindent Remarkably, the quartic Fierz identities entail that 
\cite{DF82}\cite{NOF86}\cite[Prop. 2.73]{GSS24-SuGra}:
$$
  \hspace{-.6cm}
  \left.
  \def\arraystretch{1.6}
  \begin{array}{ccl}
  G_4^0
  &:=&
  \tfrac{1}{2}
  \big(\hspace{1pt}
    \overline{\Psi}
    \,\Gamma_{a_1 a_2}\,
    \Psi
  \big)
  \, E^{a_1} E^{a_2}
  \\
  G_7^0
  &:=&
  \tfrac{1}{5!}
  \big(\hspace{1pt}
    \overline{\Psi}
    \,\Gamma_{a_1 \cdots a_5}\,
    \Psi
  \big)
  \, E^{a_1} \cdots E^{a_5}
  \end{array}
  \!\!\right\}
  \;
  \in
  \;
  \underset{
    \mathclap{
      \adjustbox{
        scale=.7,
        raise=-3pt
      }{
        \color{darkblue}
        \bf
        fully super-invariant forms
      }
    }
  }{
  \mathrm{CE}\big(
    \mathbb{R}^{1,10\,\vert\,\mathbf{32}}
  \big)^{\mathrm{Spin}(1,10)}
  }
  \;\;\;\;\;\;\;\;\;
  \mbox{satisfy}:
  \;\;\;\;
  \colorbox{lightgray}{$
  \def\arraystretch{1.3}
  \def\arraycolsep{2pt}
  \begin{array}{ccl}
    \mathrm{d}
    \, G_4^0
    &=&
    0
    \\
    \mathrm{d}\, G_7^0
    &=&
    \tfrac{1}{2}
    G_4^0\, G_4^0
  \end{array}
  $}
$$

\medskip

To globalize this situation,
say that an {\bf 11D super-spacetime} $X$ is a super-manifold equipped with a super-Cartan connection, locally on an open cover $\widetilde X \twoheadrightarrow X $ given by
\vspace{2mm} 
$$
  \left.
  \def\arraystretch{1.3}
  \begin{array}{l}
    (\Psi^\alpha)_{\alpha = 1}^{32}
    \\
    (E^a)_{a=0}^{10}
    \\
    \big(
    \Omega^{ab} = 
    -
    \Omega^{ba}
    \big)_{a,b = 0}^{10}
  \end{array}
  \! \right\}
  \;\in\;
  \Omega^1_{\mathrm{dR}}\big(
    \widetilde X
  \big)
  \hspace{.5cm}
  \mbox{\small 
    \begin{tabular}{l}
      such that the
      \\[-3pt]
      super-torsion
      \\[-3pt]
      vanishes
    \end{tabular}
  }
  \hspace{.5cm}
  \mathrm{d}\, E^a 
  -
  \Omega^a{}_b \, E^b
  \;=\;
  \big(\hspace{1pt}
    \overline{\Psi}
    \,\Gamma^a\,
    \Psi
  \big)
  \,,
$$

\vspace{2mm} 
\noindent and say that {\bf C-field super-flux} on such a super-spacetime are super-forms with 
these co-frame components:
\vspace{2mm} 
$$
  \adjustbox{
    margin=-1pt,
    fbox
  }{$
  \def\arraystretch{1.5}
  \begin{array}{lclcl}
    G_4^s
    &:=&
    G_4 \,+\, G_4^0
    &:=&
    \tfrac{1}{4!}
    (G_4)_{a_1 \cdots a_4}
    E^{a_1}\cdots E^{a_4}
    \,+\,
    \tfrac{1}{2}\big(\hspace{1pt}
      \overline{\Psi}
      \,\Gamma_{a_1 a_2}\,
      \Psi
    \big)
    E^{a_1}\, E^{a_2}
    \\
    G_7^s
    &:=&
    G_7 \,+\, G_7^0
    &:=&
    \tfrac{1}{7!}
    (G_4)_{a_1 \cdots a_7}
    E^{a_1}\cdots E^{a_7}
    \,+\,
    \tfrac{1}{5!}\big(\hspace{1pt}
      \overline{\Psi}
      \,\Gamma_{a_1 \cdots a_5}\,
      \Psi
    \big)
    E^{a_1} \cdots E^{a_5}
  \end{array}
  $}
$$

\smallskip 
\medskip

\noindent
{\bf Theorem} \cite[Thm. 3.1]{GSS24-SuGra}:
On an 11D super-spacetime $X$ with C-field super-flux $(G_4^s, G_7^s)$:

\vspace{-.2cm}
\begin{center}
\begin{tikzpicture}
  \node[
    fill=olive,
    fill opacity=.2
  ] 
  at (0,0)
  {\it$
  \mbox{
    \def\arraystretch{1}
    \begin{tabular}{c}
      The duality-symmetric 
      \\
      super-Bianchi identity
    \end{tabular}
  }
  \left\{\!\!
  \def\arraystretch{1.3}
  \begin{array}{ccl}
    \mathrm{d}\, G_4^s
    &=&
    0
    \\
    \mathrm{d}\, G_7^s
    &=&
    \tfrac{1}{2}
    \, G_4^s\, G_4^s
  \end{array}
  \!\! \right\}
  \mbox{
    is equivalent to
  }
  \mbox{
    \def\arraystretch{1}
    \def\tabcolsep{3pt}
    \begin{tabular}{c}
      the full 11D SuGra
      \\
      equations of motion!
    \end{tabular}
  }
  $};
  \node[
  ] 
  at (0,0)
  {\it$
  \mbox{
    \def\arraystretch{1}
    \begin{tabular}{c}
      The duality-symmetric 
      \\
      super-Bianchi identity
    \end{tabular}
  }
  \left\{\!\!
  \def\arraystretch{1.3}
  \begin{array}{ccl}
    \mathrm{d}\, G_4^s
    &=&
    0
    \\
    \mathrm{d}\, G_7^s
    &=&
    \tfrac{1}{2}
    \, G_4^s\, G_4^s
  \end{array}
  \!\! \right\}
  \mbox{
    is equivalent to
  }
  \mbox{
    \def\arraystretch{1}
    \def\tabcolsep{3pt}
    \begin{tabular}{c}
      the full 11D SuGra
      \\
      equations of motion!
    \end{tabular}
  }
  $};
  \end{tikzpicture}
\end{center}
\vspace{-.2cm}

Next, on the super-subspace
$
  \begin{tikzcd}
  \mathbb{R}^{1,5 \,\vert\, 2\cdot \mathbf{8}_+}
  \ar[
    r,
    hook,
    "{ \phi_0 }"
  ]
  &
  \mathbb{R}^{1,10\,\vert\,\mathbf{32}}
  \end{tikzcd}
$
fixed by the involution $\Gamma_{012345} \,\in\, \mathrm{Pin}^+(1,10)$ we have:
$$
  \hspace{-2cm}
  \;\;
  \def\arraycolsep{4pt}
  \begin{array}{ccl}
    H_3^0
    &:=&
    0
  \end{array}
  \;\,
  \in
  \;\,
  \mathrm{CE}\big(
    \mathbb{R}^{1,5\,\vert\,2 \cdot \mathbf{8}_+}
  \big)^{\mathrm{Spin}(1,5)}
  \hspace{.7cm}
  \mbox{satisfies}:
  \hspace{.7cm}
  \colorbox{lightgray}{$
  \def\arraystretch{1.4}
  \begin{array}{c}
  \mathrm{d}\, H_3^0
  \;=\;
  \phi_0^\ast
  \, G_4^0
  \end{array}
  $}
$$

To globalize this situation,
say that a super-immersion $\begin{tikzcd}[column sep=20pt] \Sigma^{1,5\,\vert\,2\cdot\mathbf{8}_+} \ar[r, "{ \phi_s }"] & X^{1,10\,\vert\,\mathbf{32}}\end{tikzcd}$ 
is {\bf $\sfrac{1}{2}$BPS M5} if it is ``locally like'' $\phi_0$,
and say that {\bf B-field super-flux} on such an M5-probe is a super-form with these co-frame components:
\vspace{2mm} 
$$
  \adjustbox{
    margin=1pt,
    fbox,
  }{$
  H_3^s
  \;:=\;
  H_3 
    \,+\,
  H_3^0
  \;:=\;
  \tfrac{1}{3!}
  (H_3)_{a_1 a_2 a_3}
  e^{a_1}\, e^{a_2}\, e^{a_3}
  \,+\,
  0
  $}
  \mathrlap{
    \hspace{1cm}
    \big( 
      e^{a < 6} \,:=\, \phi_s^\ast E^a
    \big)
    \,,
  }
$$
where we are highlighting that with $H_3^0$ vanishing, by the above, the gravitino contribution to the superform vanishes.

\newpage 

\noindent
{\bf Theorem} \cite[\S 3.3]{GSS24-FluxOnM5}: On a super-immersion $\phi_s$ with B-field super-flux $H_3^s$:

\vspace{-.2cm}
\begin{center}
\begin{tikzpicture}
\node[
  fill=olive,
  fill opacity=.2
] at (0,0) {
\it
\;
\begin{tabular}{c}
  The super-Bianchi identity
\end{tabular}
$\Big\{\mathrm{d}\, H_3^s \;=\; 
\phi_s^\ast G_4^s
\Big\}
$
\;
\mbox{is equivalent to}
\;
\begin{tabular}{c}
  the M5's B-field
  \\
  equations of motion.
\end{tabular}
\;
};
\node[
] at (0,0) {
\it
\;
\begin{tabular}{c}
  The 
  super-Bianchi identity
\end{tabular}
$\Big\{\mathrm{d}\, H_3^s \;=\; 
\phi_s^\ast G_4^s
\Big\}
$
\;
\mbox{is equivalent to}
\;
\begin{tabular}{c}
  the M5's B-field
  \\
  equations of motion.
\end{tabular}
\;
};
\end{tikzpicture}
\end{center}
\vspace{-.3cm}

\noindent
In particular, the (non-linear self-)duality conditions on the ordinary fluxes are  {\it implied}: $G_4 \leftrightarrow G_7$ and $H_3 \leftrightarrow H_3$.

\smallskip

Seeing from this that also trivial tangent super-cochains may have non-trivial globalization, observe next that:
$$
  F_2^0
  \;:=\;
  \big(\,
    \overline{\psi}
    \,
    \psi
  \big)
  \;=\;
  0
  \;\in\;
  \mathrm{CE}\big(
    \mathbb{R}^{1,5\,\vert\,2\cdot \mathbf{8}_+}
  \big)^{\mathrm{Spin}(1,5)}
  \hspace{.6cm}
  \mbox{satisfies}:
  \hspace{.6cm}
  \colorbox{lightgray}{\def\arraystretch{1.3}\def\tabcolsep{2pt}
  \begin{tabular}{c}
    $
      \mathrm{d}\, 
      F_2^0
      \;=\;
      0
    $
  \end{tabular}
  }
$$
Globalizing this to $\Sigma^{1,5\,\vert\,2 \cdot \mathbf{8}_+}$ via
\vspace{-2mm} 
\begin{center}
\;
\adjustbox{
  margin=1pt,
  fbox
}
{
$
  F_2^s
  \;:=\;
  F_2 \,+\, F_2^s
  \;:=\;
  \tfrac{1}{2}
  (F_2)_{a_1 a_2}
  \, e^{a_1} e^{a2}
  \,+\,
  0
$
}
\; 
\end{center} 

\vspace{-2mm} 
\noindent we have on top of the above:

\medskip

\noindent
{\bf Theorem} \cite[p 7]{SS25-Seifert}:
\vspace{-.1cm}
\begin{center}
\hspace{2cm} 
\begin{tikzpicture}
\node[
  fill=olive,
  fill opacity=.2
] at (0,0) {
\it
\;
\begin{tabular}{c}
  The 
  super-Bianchi identity
\end{tabular}
$\big\{
  \mathrm{d}\, F_2^s 
    \;=\; 
   0
\big\}
$
\;
\mbox{
  is equivalent to
}
\;
\begin{tabular}{c}
  the Chern-Simons
  \\
  E.O.M.: $F_2 \,=\, 0$.
\end{tabular}
\;
};
\node[
] at (0,0) {
\it
\;
\begin{tabular}{c}
  The 
  super-Bianchi identity
\end{tabular}
$\big\{
  \mathrm{d}\, F_2^s 
    \;=\; 
   0
\big\}
$
\;
\mbox{
  is equivalent to
}
\;
\begin{tabular}{c}
  the Chern-Simons
  \\
  E.O.M.: $F_2 \,=\, 0$.
\end{tabular}
\;
};
\end{tikzpicture}
\end{center}\vspace{-.2cm}

%\newpage

\medskip 
\hypertarget{FluxQuantizationInTwistorialCohomotopy}{}

\noindent{\bf Flux quantization in Twistorial Cohomotopy.}
In summary, a remarkable kind of higher super-Cartan geometry locally modeled on the 11D super-Minkowski spacetime $\mathbb{R}^{1,10\,\vert\, \mathbf{32}}$ entails that on-shell 11D supergravity probed by magnetized $\sfrac{1}{2}$BPS M5-branes  implies and is entirely governed by these Bianchi identities on super-flux densities:

\vspace{2mm} 
\hspace{4.02cm}
\begin{tikzpicture}
  \draw[
    draw opacity=0,
    fill=olive,
    fill opacity=.2
  ]
    (0,0) rectangle (7.8,1.3);
\end{tikzpicture}
\vspace{-1.4cm}
\begin{equation}
  \label{BianchiOnMagnetizedM5}
  \begin{tikzcd}[
    row sep=0pt,
    column sep=-20pt
  ]
    \scalebox{.7}{
      \color{darkblue}
      \bf
      A-field
    }
    &[+18pt]
    \mathrm{d}\, F^s_2 
    &=& 
    \hspace{-48pt}
    0
    &[+45pt]
    \mathrm{d}\, G^s_4 
       &=& 
     0
    &[+20pt]
    \scalebox{.7}{
      \color{darkblue}
      \bf
      C-field
    }
    \\
    \scalebox{.7}{
      \color{darkblue}
      \bf
      \def\arraystretch{.9}
      \def\tabcolsep{-20pt}
      \begin{tabular}{c}
        self-dual
        \\
        B-field
      \end{tabular}
    }
    \;\,
    &
    \mathrm{d}\, H^s_3 
      &=& 
    \phi_s^\ast G^s_4 
    \,+\,
    \mathcolor{purple}\theta \, F^s_2 \,F^s_2
    &
    \mathrm{d}\, G^s_7
    &=& 
    \tfrac{1}{2} G^s_4\, G^s_4    
    &
    \scalebox{.7}{
      \color{darkblue}
      \bf
      \def\arraystretch{.9}
      \def\tabcolsep{-20pt}
      \begin{tabular}{c}
        dual
        \\
        C-field
      \end{tabular}
    }
    \\[+1pt]
    \scalebox{.7}{
      \color{darkblue}
      \bf
      M5 probe
    }
    \hspace{-3pt}
    & 
    &
    \Sigma^{1,5\,\vert\,2\cdot \mathbf{8}_+}
    \ar[
      rrr,
      "{ \phi_s }",
      "{
        \scalebox{.7}{
          \color{darkgreen}
          \bf
          \scalebox{1.24}{$\sfrac{1}{2}$}BPS
          immersion
        }
      }"{swap}
    ]
    &&&
    X^{1,10\,\vert\,\mathbf{32}}
    &&
    \scalebox{.7}{
      \color{darkblue}
      \bf
      SuGra bulk
    }
  \end{tikzcd}
\end{equation}
Here we have observed that the Green-Schwarz term $F_2^s F_2^s$ may equivalently be included for any theta-angle $\mathcolor{purple}{\theta} \in \mathbb{R}$ without affecting the equations of motion (since, recall, the CS e.o.m. $F_2^s = 0$ is already implied by $\mathrm{d}\, F_2^s \,=\, 0$).

\smallskip 
However, non-vanishing theta-angle does affect the admissible flux-quantization laws and hence the global solitonic and torsion charges of the fields. The choice of flux quantization according to {\it Hypothesis H} 
\cite{FSS20-H}\cite{FSS22-Twistorial}
is the following:

\medskip 
% \hspace{-.8cm}
% \begin{tabular}{p{6.2cm}}
% \small
\noindent {\bf
Admissible fibrations of classifying spaces for cohomology theories} with the above character images \eqref{BianchiOnMagnetizedM5}. 
The homotopy quotient of $S^7$ is 

\smallskip 
(i) for $\theta = 0$ by the trivial action and 

(ii) for $\theta \neq 0$ by the principal action of the complex Hopf fibration.
%\end{tabular}

\begin{center} 
\adjustbox{
  raise=+1pt
}{
$
  \begin{tikzcd}[
    row sep=30pt, column sep=large
  ]
    \fbox{$\mathclap{\phantom{\neq}}
    \mathcolor{purple}{\theta} = 0
    $}
    &[-18pt]
    S^7 \!\sslash_{\!\!{}_0}\! \mathrm{U}(1)
    \ar[
      r,
      phantom,
      "{ \simeq }"
    ]
    \ar[
      d,
      ->>,
      gray
    ]
    &[-16pt]
    S^7 \times 
    \mathbb{C}P^\infty
    \ar[
      r,
      ->>
    ]
    &[-5pt]
    S^7
    \ar[
      rr,
      ->>,
      "{
        h_{\mathbb{H}}
      }",
      "{
        \scalebox{.7}{
          \color{darkgreen}
          \bf
          \scalebox{1.4}{$\mathbb{H}$}-Hopf fibration
        }
      }"{swap}
    ]
    \ar[
      d,
      ->>,
      gray,
      "{
        \scalebox{.7}{
        \color{gray}
        \bf
        \scalebox{1.4}{$\mathbb{C}$}-Hopf
        fibration
        }
      }"{description}
    ]
    &\phantom{--}&
    \mathbb{H}P^1
    \ar[
     d,
     equals,
     gray
    ]
    \\
    \fbox{$
    \mathcolor{purple}{\theta} \neq 0$}
    &
    S^7 \!\sslash\! \mathrm{U}(1)
    \ar[
      rr,
      "{ \sim }"
    ]
    &[-16pt]
    &
    \mathbb{C}P^3
    \ar[
      rr,
      ->>,
      "{
        t_{\mathbb{H}}
      }",
      "{
        \scalebox{.7}{
          \color{darkgreen}
          \bf
          Twistor fibration
        }
      }"{swap}
    ]
    &&
    \mathbb{H}P^1
  \end{tikzcd}
$
}
\end{center} 
{\bf Proof.} This may be seen as follows \rlap{\cite[Lem. 2.13]{FSS22-Twistorial}:}

Since the real cohomology of projective space is a truncated polynomial algebra,
\begin{center}
\adjustbox{
  raise=-5pt
}{
$
  \def\arraystretch{1.4}
  \def\arraycolsep{3pt}
  \begin{array}{ccccccc}
  H^\bullet\big(
    \mathbb{C}P^n
    ;\,
    \mathbb{R}
  \big)
  &\simeq&
  \mathbb{R}\big[
    \grayoverbrace{
    c_1}{
      \mathclap{
      \mathrm{deg} = 2
      }
    }
  \big]
  \big/
  (c_1^{n+1})
  &&
  H^\bullet\big(
    \grayoverbrace{
    \mathbb{C}P^\infty
    }{
      \mathclap{
      \simeq\, 
      B \mathrm{U}(1)
      }
    }
    ;\,
    \mathbb{R}
  \big)
  &\simeq&
  \mathbb{R}[c_1]
  \\
  H^\bullet\big(
    \mathbb{H}P^n
    ;\,
    \mathbb{R}
  \big)
  &\simeq&
  \mathbb{R}\big[
    \grayunderbrace{
    \tfrac{1}{2}p_1
    }{
      \mathclap{
        \mathrm{deg} = 4
      }
    }
  \big]
  \big/
  (p_1^{n+1})
  &&
  H^\bullet\big(
    \grayunderbrace{
    \mathbb{H}P^\infty
    }{
      \mathclap{
      \adjustbox{
        scale=.7
      }{$
      \def\arraystretch{1.1}
      \begin{array}{l}
      \,\simeq\, 
      B \mathrm{Sp}(1)
      \,\simeq\,
      B \mathrm{SU}(2)
      \\
      \,\simeq\,
      B \mathrm{Spin}(3)
      \end{array}
      $}
      }
    }
    ;\,
    \mathbb{R}
  \big)
  &\simeq&
  \mathbb{R}[\tfrac{1}{2}p_1]
  \mathrlap{\,,}
  \end{array}
$
}
\end{center}
the minimal dgc-algebra model for $\mathbb{C}P^n$ needs a closed generator $f_2$ to span the cohomology and a generator $h_{2n+1}$ in order to truncate it;
analogously for $\mathbb{H}P^n$.
Since these generators also form a graded linear basis for the rationalized homotopy groups of these spaces, they give the minimal Sullivan models (cf \cite[Prop. 3.7]{SS24-Flux}):
\begin{center}
\adjustbox{
  raise=-16pt
}{
$
  \def\arraystretch{1}
  \begin{array}{ll}
  \mathrm{CE}\big(
    \mathfrak{l}\,
    \mathbb{C}P^n
  \big)
  \;\simeq\;
  \mathbb{R}_{\mathrm{d}}
  \left[
  \def\arraystretch{1.1}
  \def\arraycolsep{0pt}
  \begin{array}{c}
    f_2
    \\
    h_{2n+1}
  \end{array}
  \right]
  \Big/
  \left(
  \def\arraystretch{1.1}
  \def\arraycolsep{2pt}
  \begin{array}{lcl}
    \mathrm{d}\,f_2
    &=&
    0
    \\
    \mathrm{d}\,h_{2n+1}
    &=&
    (f_2)^{n+1}
  \end{array}
  \right)
  \\
  \\
  \mathrm{CE}\big(
    \mathfrak{l}\,
    \mathbb{H}P^n
  \big)
  \;\simeq\;
  \mathbb{R}_{\mathrm{d}}
  \left[
  \def\arraystretch{1.1}
  \def\arraycolsep{0pt}
  \begin{array}{c}
    g_4
    \\
    g_{4n+3}
  \end{array}
  \right]
  \Big/
  \left(
  \def\arraystretch{1.1}
  \def\arraycolsep{2pt}
  \begin{array}{lcl}
    \mathrm{d}\,g_4
    &=&
    0
    \\
    \mathrm{d}\,
    g_{4n+3}
    &=&
    (g_4)^{n+1}
  \end{array}
  \right)
  \mathrlap{\,.}
  \end{array}
$
}
\end{center}
Furthermore, since the second Chern class of a $\mathrm{U}(1) \simeq S\big(\mathrm{U}(1)^2\big) \subset \mathrm{SU}(2)$-bundle 
is minus the cup square of the first Chern class (by the Whitney sum rule), so that (cf. \cite[(216)]{SS23-Mf})
$$
  \begin{tikzcd}[
    row sep=25pt
  ]
    \mathbb{C}P^3
    \ar[
      d,
      "{ t_{\mathbb{H}} }"
    ]
    \ar[
      r,
      hook
    ]
    &
    \mathbb{C}P^\infty
    \ar[
      r,
      phantom,
      "{ \simeq }"
    ]
    \ar[
      d
    ]
    &[-10pt]
    B
    \mathrm{U}(1)
    \ar[
      d,
      "{
        B(
          c \mapsto
          \mathrm{diag}(c,c^\ast)
        )
      }"{description}
    ]
    &[-15pt]
    -(c_1)^2
    \ar[
      d,
      <-|,
      shorten=5pt
    ]
    \\
    \mathbb{H}P^1
    \ar[
      r,
      hook
    ]
    &
    \mathbb{H}P^\infty
    \ar[
      r,
      phantom,
      "{ \simeq }"
    ]
    &
    B \mathrm{SU}(2)
    &
    \tfrac{1}{2}p_1
    =
    c_2
    \mathrlap{\,,}
  \end{tikzcd}
$$
the minimal model of $\mathbb{C}P^3$ {\it relative} to that of $\mathbb{H}P^1 \simeq S^4$
(cf. \cite[Prop. 4.24]{FSS23-Char})
needs to adjoin to the latter not only $f_2$ but also a generator $h_3$ imposing this relation in cohomology, whence it must be
\begin{center}
\adjustbox{
  raise=-15pt
}{
$
  \mathrm{CE}\big(
    \mathfrak{l}_{{}_{\mathbb{H}P^1}}
    \mathbb{C}P^3
  \big)
  \;\simeq\;
  \mathbb{R}_{\mathrm{d}}
  \left[
  \def\arraystretch{1.1}
  \def\arraycolsep{2pt}
  \begin{array}{c}
    f_2
    \\
    h_3
    \\
    g_4
    \\
    g_7
  \end{array}
  \right]
  \Big/
  \left(
  \def\arraystretch{1.1}
  \def\arraycolsep{2pt}
  \begin{array}{ccl}
    \rowcolor{lightolive}
    \mathrm{d}\,f_2
    &=&
    0
    \\
    \rowcolor{lightolive}
    \mathrm{d}\, h_3
    &=&
    g_4 + f_2 f_2
    \\
    \mathrm{d}\,g_4
    &=&
    0
    \\
    \mathrm{d}\, g_7
    &=&
    \tfrac{1}{2} g_4 \, g_4
  \end{array}
  \right)  
  \mathrlap{\,,}
$
}
\end{center}
which is clearly quasi-isomorphic to $\mathrm{CE}(\mathfrak{l}\,\mathbb{C}P^3)$.\hfill $\Box$

\bigskip

The resulting fibration of $L_\infty$-algebras is manifestly just that classifying the desired Bianchi identities \eqref{BianchiOnMagnetizedM5}
\\
(we are showing the case $\theta \neq 0$, which by isomorphic rescaling may be taken to be $\theta = 1$):
$$
  \begin{tikzcd}[
    ampersand replacement=\&,
    row sep=10pt
  ]
    \Sigma^{6}
    \ar[
      r,
      dashed
    ]
    \ar[
      dd,
      "{ \phi }"{swap}
    ]
    \&
    \Omega^1_{\mathrm{dR}}\big(
      -;\,
      \mathfrak{l}_{{}_{\mathbb{H}P^1}}
      \mathbb{C}P^3
    \big)_{\mathrm{clsd}}
    \ar[
      dd,
      "{
        (\mathfrak{l}\,t_{\mathbb{H}})_\ast
      }"
    ]
    \&[-30pt]
    \&[-20pt]
    \Omega^\bullet_{\mathrm{dR}}\big(\Sigma^6\big)
    \ar[
      r,
      <-,
      dashed
    ]
    \ar[
      dd,
      <-,
      "{ \phi^\ast }"{swap}
    ]
    \&
    \mathrm{CE}\big(
      \mathfrak{l}_{_{\mathbb{H}P^1}}
      \mathbb{C}P^3
    \big)
    \ar[
      dd,
      ->,
      "{
        (\mathfrak{l}\,
        t_{\mathbb{H}})^\ast
      }"
    ]
    \&[-22pt]
    \&[-22pt]
    \left.
    \def\arraystretch{1.2}
    \def\arraycolsep{2pt}
    \begin{array}{l}
      F_2
      \\
      H_3
    \end{array}
    \in
    \Omega^\bullet_{\mathrm{dR}}(\Sigma^6)
    \middle\vert
    \def\arraystretch{1.2}
    \def\arraycolsep{2pt}
    \begin{array}{lcl}
      \rowcolor{lightolive}
      \mathrm{d}\, F_2
      &=&
      0
      \\
      \rowcolor{lightolive}
      \mathrm{d}\,H_3
      &=&
      G_4 + F_2 \, F_2
    \end{array}
    \right.
    \\
    \&\&
    \Leftrightarrow
    \&\&\&
    \Leftrightarrow
    \\
    \Sigma^{11}
    \ar[
      r,
      dashed
    ]
    \&
    \Omega^1_{\mathrm{dR}}\big(
      -;\,
      \mathfrak{l}\,
      \mathbb{H}P^1
    \big)_{\mathrm{clsd}}
    \&\&
    \Omega^\bullet_{\mathrm{dR}}(X^{11})
    \ar[
      r,
      <-,
      dashed
    ]
    \&
    \mathrm{CE}(\mathfrak{l}\, \mathbb{H}P^1)
    \&\&
    \left.
    \def\arraystretch{1.2}
    \def\arraycolsep{2pt}
    \begin{array}{l}
      G_4
      \\
      G_7
    \end{array}
    \in
    \Omega^\bullet_{\mathrm{dR}}\big(X^{11}\big)
    \middle\vert
    \def\arraystretch{1.2}
    \def\arraycolsep{2pt}
    \begin{array}{lcl}
      \rowcolor{lightolive}
      \mathrm{d}\, G_4
      &=&
      0
      \\
      \rowcolor{lightolive}
      \mathrm{d}\,G_7
      &=&
      \tfrac{1}{2}
      G_4 \, G_4
    \end{array}
    \right.
  \end{tikzcd}
$$

\medskip\hypertarget{ProjectiveSpacesAndHopfFibrations}{}

\noindent
{\bf Aside: Projective Spaces and their Fibrations} -- Herse we used the following classical facts. 
Consider:

\smallskip

\def\tabcolsep{2pt}
\def\arraystretch{1.4}
\begin{tabular}{ll}
\rowcolor{lightgray}
division algebras 
&
$\mathbb{R} \xhookrightarrow{\;} \mathbb{C} \xhookrightarrow{\;} \mathbb{H}$ 
generically denoted $\mathbb{K} \,\in\, \big\{\mathbb{R},\, \mathbb{C},\, \mathbb{H}\big\}$
\\
groups of units 
&
$
  \mathbb{K}^\times
  \,:=\,
  \mathbb{K} \setminus \{0\}
$ 
understood with the multiplicative group structure
\\
\rowcolor{lightgray}
projective spaces
&
$
  \mathbb{K}P^n
  \;:=\;
  \big(
    \mathbb{K}^{n+1}
    \setminus
    \{0\}
  \big)
  \big/
  \mathbb{K}^\times 
$
\\
higher spheres
&
$
  S^n
  \;\simeq\;
  \big(
    \mathbb{R}^{n+1}
    \setminus 
    \{0\}
  \big)/\mathbb{R}_{{}_{> 0}}
$
\\
\rowcolor{lightgray}
\multicolumn{2}{l}{
$\mathbb{K}$-Hopf fibrations are the quotient co-projections induced by $\iota : \mathbb{R}_{{}_{> 0}} \xhookrightarrow{\;} \mathbb{K}$
}
\end{tabular}

\vspace{2mm}

The classical Hopf fibrations $h_{\mathbb{K}}$ are:
$$
  \begin{tikzcd}[
    column sep=4pt
  ]
    S^0
    \mathrlap{
    \;\simeq\,
    \mathbb{R}^\times
    /\,
    \mathbb{R}_{{}_{>0}}
    }
    \ar[
      d,
      hook,
      "{ \mathrm{ker} }"
    ]
    \\
    S^1 
    \ar[
      r,
      phantom,
      "{ \simeq }"
    ]
    \ar[
      d,
      ->>,
      "{
        h_{\mathbb{R}}
      }"
    ]
    &
    \big(
      \mathbb{R}^2
      \!\setminus\!
      \{0\}
    \big)
    \big/
    \mathbb{R}_{{}_{> 0}}
    \ar[
      d,
      ->>,
      "{
        \iota_\ast
      }"
    ]
    \\
    S^1
    \ar[
      r,
      phantom,
      "{ \simeq }"
    ]
    &
    \grayunderbrace{
    \big(
      \mathbb{R}^2 
      \!\setminus\!
      \{0\}
    \big)
    \big/
    \mathbb{R}^\times
    }{
      \mathbb{R}P^1
    }
  \end{tikzcd}
  \hspace{1.4cm}
  \begin{tikzcd}[
    column sep=4pt
  ]
    S^1
    \mathrlap{
    \;\simeq\,
    \mathbb{C}^\times
    /\,
    \mathbb{R}_{{}_{>0}}
    }
    \ar[
      d,
      hook,
      "{ \mathrm{ker} }"
    ]
    \\
    S^3 
    \ar[
      r,
      phantom,
      "{ \simeq }"
    ]
    \ar[
      d,
      ->>,
      "{
        h_{\mathbb{C}}
      }"
    ]
    &
    \big(
      \mathbb{C}^2
      \!\setminus\!
      \{0\}
    \big)\big/
    \mathbb{R}_{{}_{> 0}}
    \ar[
      d,
      ->>,
      "{
        \iota_\ast
      }"
    ]
    \\
    S^2
    \ar[
      r,
      phantom,
      "{ \simeq }"
    ]
    &
    \grayunderbrace{
    \big(
      \mathbb{C}^2 
      \!\setminus\!
      \{0\}
    \big)
    \big/
    \mathbb{C}^\times
    }{
      \mathbb{C}P^1
    }
  \end{tikzcd}
  \hspace{1.4cm}
  \begin{tikzcd}[
    column sep=4pt
  ]
    S^3
    \mathrlap{
    \;\simeq\,
    \mathbb{H}^\times
    /\,
    \mathbb{R}_{{}_{>0}}
    }
    \ar[
      d,
      hook,
      "{ \mathrm{ker} }"
    ]
    \\
    S^7 
    \ar[
      r,
      phantom,
      "{ \simeq }"
    ]
    \ar[
      d,
      ->>,
      "{
        h_{\mathbb{H}}
      }"
    ]
    &
    \big(
      \mathbb{H}^2
      \!\setminus\!
      \{0\}
    \big)\big/
    \mathbb{R}_{{}_{> 0}}
    \ar[
      d,
      ->>,
      "{
        \iota_\ast
      }"
    ]
    \\
    S^4
    \ar[
      r,
      phantom,
      "{ \simeq }"
    ]
    &
    \grayunderbrace{
    \big(
      \mathbb{H}^2 
      \!\setminus\!
      \{0\}
    \big)
    \big/
    \mathbb{H}^\times
    }{
      \mathbb{H}P^1
    }
  \end{tikzcd}
$$

\begin{tabular}{p{9.2cm}}
\colorbox{lightgray}{The Hopf fibrations in higher dimensions} are the attaching maps exhibiting the topological cell-complex structure of projective spaces \cite{nLabCellStructureOnProjectiveSpaces}, 
from which the (cellular) cohomology follows readily.
\end{tabular}
\qquad \qquad 
\adjustbox{
  raise=4pt
}{
$
  \begin{tikzcd}
    S\big(
      \mathbb{K}^{n+1}
    \big)
    \ar[
      d,
      "{ 
        h_{\mathbb{K}} 
      }"{swap},
      "{\ }"{
        name=t
      }
    ]
    \ar[
      r,
      "{\ }"{
        swap, pos=.4, name=s
      }
    ]
    \ar[
      from=s,
      to=t,
      Rightarrow,
      "{
        \scalebox{.7}{
          \color{gray}
          (po)
        }
      }"
    ]
    &
    \ast
    \ar[
      d
    ]
    \\
    \mathbb{K}P^n
    \ar[r, hook]
    &
    \mathbb{K}P^{n+1}
  \end{tikzcd}
$}

\vspace{.1cm}

\begin{tabular}{p{9.2cm}l}
\colorbox{lightgray}{Further factor-fibrations}
arise by factoring the Hopf fibrations via the stage-wise quotienting along 
$$\mathbb{R}_{{}_{>0}} \xhookrightarrow{\quad} \mathbb{R} \xhookrightarrow{\quad} \mathbb{C} \xhookrightarrow{\quad} \mathbb{H}.$$ 
Notably, the classical quaternionic Hopf fibration 
$h_{\mathbb{H}}$ factors through a higher-dimensional complex Hopf fibration followed by the 
{\bf Calabi-Penrose twistor fibration} $t_{\mathbb{H}}$ \cite[\S 2]{FSS22-Twistorial}.

\vspace{2mm}

\colorbox{lightgray}{Equivariantization:}Since the quotienting is by right actions, these fibrations are equivariant under the left action of 
$$
  \mathrm{Spin}(5)
  \,\simeq\,
  \mathrm{Sp}(2)
  \,:=\,
  \big\{
    g \in \mathrm{GL}_2(\mathbb{H})
    \,\big\vert\,
    g^\dagger \cdot g
    =
    \mathrm{e}
  \big\}
  \,.
  $$
  &
  \qquad 
  $
  \adjustbox{raise=-70pt}{
  \begin{tikzcd}[
    column sep=3pt, row sep=15pt
  ]
    S^1
    \ar[
      r,
      phantom,
      "{ \simeq }"
    ]
    &
    \mathbb{C}^\times
    /
    \mathbb{R}_{{}_{>0}}
    \\
    &&
    S^7
    \ar[
      r,
      phantom,
      "{ \simeq }"
    ]
    \ar[
      dd,
      "{
        h_{\mathbb{C}}
        \mathrlap{
          \scalebox{.7}{
            \color{darkgreen}
            \bf
            \def\arraystretch{.9}
            \begin{tabular}{c}
              complex
              \\
              Hopf fibration
            \end{tabular}
          }
        }
      }"
    ]
    \ar[
      from=ull
    ]
    \ar[
      dddd,
      bend right=30,
      "{
        \mathllap{
          \scalebox{.7}{
            \color{darkgreen}
            \bf
            \def\arraystretch{.9}
            \begin{tabular}{c}
              quaternionic
              \\
              Hopf fibration
            \end{tabular}
          }
        }
        h_{\mathbb{H}}
      }"{swap, pos=.75}
    ]
    &
    \big(
      \mathbb{H}^2
      \!\setminus\!
      \{0\}
    \big)
    \big/\,
    \mathbb{R}_{{}_{>0}}
    \\
    S^2
    \ar[
      r,
      phantom,
      "{ \simeq }"
    ]
    \ar[
      drr,
      crossing over,
      shorten >=-2pt
    ]
    &
    \mathbb{H}^\times
    /\,
    \mathbb{C}^\times
    \\
    &&
    \mathbb{C}P^3
    \ar[
      r,
      phantom,
      "{ \simeq }"
    ]
    \ar[
      dd,
      "{ 
        t_{\mathbb{H}} 
        \mathrlap{
          \scalebox{.7}{
            \color{olive}
            \bf
            \def\arraystretch{.9}
            \begin{tabular}{c}
              Calabi-Penrose
              \\
              twistor fibration
            \end{tabular}
          }
        }
      }"
    ]
    &
    \big(
      \mathbb{H}^2
      \!\setminus\!
      \{0\}
    \big)
    \big/\,
    \mathbb{C}^\times
    \\
    \\
    &&
    \mathbb{H}P^1
    \ar[
      r,
      phantom,
      "{ \simeq }"
    ]
    &
    \big(
      \mathbb{H}^2
      \!\setminus\!
      \{0\}
    \big)
    \big/\,
    \mathbb{H}^\times
  \end{tikzcd}
  }
$
\end{tabular}

\vspace{3mm} 
\begin{tabular}{p{7.5cm}l}
For example, the involution
$
  \sigma 
  \,:=\, 
  \Big[
  \adjustbox{scale=0.8, 
    raise=-1pt
  }{$
  \def\arraycolsep{2pt}
  \def\arraystretch{.9}
  \begin{array}{cc}
    0 & 1 
    \\
    1 & 0
  \end{array}
  $}
  \Big]
  \,\in\,
  \mathrm{Sp}(2)
$ 
swaps the two copies of $\mathbb{H}$:
&
\qquad 
$
  \begin{tikzcd}[
    column sep=-7pt,
    row sep=8pt
  ]
    \mathbb{C}P^3
    \ar[
      rrrrrrrrrrrr,
      "{ t_{\mathbb{H}} }"
    ]
    \ar[
      dddd,
      "{ \sigma }"
    ]
    \ar[
      dr,
      equals
    ]
    &[+8pt]
    &&&&&&&&&&
    &[+8pt]
    \mathbb{H}P^1
    \ar[
      dl,
      equals
    ]
    \ar[
      dddd,
      "{ \sigma }"
    ]
    \\[-6pt]
    &
    \big(
    &
    \mathbb{H}
    \ar[
      ddrr
    ]
    &\times&
    \mathbb{H}
    \ar[
      ddll,
      crossing over
    ]
    &
    \setminus \{0\}
    \big)
    \big/ \mathbb{C}^\times
    \ar[
      rr
    ]
    &\phantom{--}&    
    \big(
    &
    \mathbb{H}
    \ar[
      ddrr
    ]
    &\times&
    \mathbb{H}
    \ar[
      ddll,
      crossing over
    ]
    &
    \setminus \{0\}
    \big)
    \big/ \mathbb{H}^\times
    \\
    \\
    &
    \big(
    &
    \mathbb{H}
    &\oplus&
    \mathbb{H}
    &
    \setminus\{0\}
    \big)
    \big/
    \mathbb{C}^\times
    \ar[
      rr,
    ]
    &&
    \big(
    &
    \mathbb{H}
    &\oplus&
    \mathbb{H}
    &
    \setminus\{0\}
    \big)
    \big/
    \mathbb{H}^\times
    \\[-6pt]
    \mathbb{C}P^3
    \ar[
      rrrrrrrrrrrr,
      "{ t_{\mathbb{H}} }"{swap}
    ]
    \ar[
      ur,
      equals
    ]
    &[+8pt]
    &&&&&&&&&&
    &[+8pt]
    \mathbb{H}P^1
    \ar[
      ul,
      equals
    ]
  \end{tikzcd}
  $
\\
\\
    The resulting 
    $\mathbb{Z}_2$-fixed locus
    is the 2-sphere:
&
\qquad 
$
  \begin{tikzcd}[
    column sep=0pt
  ]
    \big(
      \mathbb{C}P^3
    \big)^{\mathbb{Z}_2 }
    \ar[
      d,
      "{
        (
          t_{\mathbb{H}}
        )^{\mathbb{Z}_2}
      }"
    ]
    &\simeq&
    \big(
      \mathbb{H}
      \!\setminus\!
      \{0\}
    \big)
    \big/\, 
    \mathbb{C}^\times
    \ar[d]
    &\simeq&
    S^2
    \ar[
      d
    ]
    \\
    \big(
      \mathbb{H}P^1
    \big)^{\mathbb{Z}_2}
    &\simeq&
    \big(
      \mathbb{H}
      \!\setminus\!
      \{0\}
    \big)
    \big/\, 
    \mathbb{H}^\times
    &\simeq&
    \ast
  \end{tikzcd}
$
\end{tabular} 
\vspace{.1cm}

\noindent
This is the 2-sphere coefficient that will
end up being responsible for
stabilizing anyons on orbi-worldvolumes!

\noindent
We next discuss how this comes about.

\medskip

\noindent {\bf Aside: Implications of Hypothesis H}, in view of traditional expectations for M-theory.

\smallskip

\noindent
{\bf The plain Hypothesis H} for the bulk theory says that 
the non-perturbative completion of the C-field in 11d supergravity 
is a cocycle in {\it differential Cohomotopy}
$\widehat{\pi}^4$ \cite[\S 4]{FSS15-WZWTerm}\cite[\S 3.1]{GS21}\cite[Ex. 9.3]{FSS23-Char}
and as such involves (exposition in \cite[\S 3.3]{SS24-Flux})
a map $\chi$ from spacetime to the homotopy type of the 4-sphere, with the C-field gauge potentials $(\widehat{C}_3, \widehat{C}_6)$ exhibiting the flux densities $(G_4, G_7)$ as $\mathbb{R}$-rational representatives of $\chi$.

$$
  \adjustbox{
    raise=40pt
  }{
  \begin{tikzcd}[
    column sep=40pt,
    row sep=15pt
  ]
    \underset{
      \mathclap{
        \scalebox{.7}{
          \color{darkblue}
          \bf
          \def\arraystretch{.9}
          \begin{tabular}{c}           
            full nonperturbative
            \\
            11d SuGra C-field 
          \end{tabular}
        }
      }
    }{
      \mbox{
        \color{purple}
        $
        \big(
          \widehat{C}_3
          ,\,
          \widehat{C}_6
        \big)
        $
      }
    }
    \ar[
      r,
      phantom,
      "{ \in }"
    ]
    &[-15pt]
    \overset{
      \mathclap{
        \raisebox{9pt}{
        \scalebox{.7}{
          \color{gray}
          \bf
          \def\arraystretch{.9}
          \begin{tabular}{c}
            canonical differential
            \\
            non-abelian (unstable)
            \\
            4-Cohomotopy
          \end{tabular}
        }
        }
      }
    }{
      \widehat{\pi}^4(X)
    }
    \ar[
      rr,
      "{ 
        \chi 
      }",
      "{
        \scalebox{.7}{
          \color{darkgreen}
          \bf
          topological sector
        }
      }"{swap}
    ]
    \ar[
      dd,
      "{ 
        (G_4, G_7) 
      }"{description},
      "{\;\;
      \scalebox{.7}{
        \color{darkgreen}
        \bf
        \def\arraystretch{.9}
        \begin{tabular}{c}
          flux 
          \\
          densities
        \end{tabular}
    }
      }"{xshift=4pt}
    ]
    &&
    \overset{
      \mathclap{
        \raisebox{9pt}{
        \scalebox{.7}{
          \color{gray}
          \bf
          \def\arraystretch{.9}
          \begin{tabular}{c}
            plain
            \\
            non-abelian (unstable)
            \\
            4-Cohomotopy
          \end{tabular}
        }
        }
      }
    }{
      \pi^4(X)
    }
    \\[-20pt]
    \\
    &
    \underset{
      \mathclap{
        \scalebox{.7}{
          \color{gray}
          \bf
          \def\arraystretch{.9}
          \begin{tabular}{c}
            $\mathfrak{l}S^4$-valued
            \\
            de Rham cohomology
          \end{tabular}
        }
      }
    }{
    H_{\mathrm{dR}}\big(
      X
      ;\,
      \mathfrak{l}S^4
    \big)    
    }
  \end{tikzcd}
  }
  \hspace{-3cm}
  \begin{tikzcd}[
    row sep=-2pt, 
    column sep=10pt
  ]
    &[-5pt]&[-5pt]
    \mathrm{Maps}\big(
      X;\,
      S^4
    \big)
    \ar[
      dd,
      "{ 
        \mathbf{ch} 
      }"
    ]
    &[-5pt]
    \chi
    \mathrlap{
      \hspace{-5pt}
      \scalebox{.7}{
        \color{darkblue}
        \bf
        \def\arraystretch{.9}
        \begin{tabular}{c}
          Cohomotopical
          \\
          charge sector
        \end{tabular}
      }
    }
    \\
    &&& 
    \rotatebox[origin=c]{-90}{$\longmapsto$}
    \\
    \Omega^1_{\mathrm{dR}}\big(
      X
      ;\,
      \mathfrak{l}S^4
    \big)_{\mathrm{clsd}}
    \hspace{-7pt}
    \ar[
      rr,
      shift left=1pt,
      "{
        \eta^{\scalebox{.55}{$\,\shape$}}
      }"
    ]
    &&
    \shape
    \,
    \Omega^1_{\mathrm{dR}}\big(
      X
      ;\,
      \mathfrak{l}S^4
    \big)_{\mathrm{clsd}}
    &
    \mathbf{ch}(\chi)
    \mathrlap{
      \hspace{-5pt}
      \scalebox{.7}{
        \color{darkblue}
        \bf
        \def\arraystretch{.9}
        \begin{tabular}{c}
          character
          \\
          image
        \end{tabular}
      }
    }
    \\
    \underset{
      \mathclap{
        \raisebox{-4pt}{
          \scalebox{.7}{
            \color{darkblue}
            \bf
            \begin{tabular}{c}
              C-field flux densities
            \end{tabular}
          }
        }
      }
    }{
    \big(
      G_4,
      \, 
      G_7
    \big)
    }
    &\longmapsto&
    \eta^{\scalebox{.55}{$\,\shape$}}
    (G_4, G_7)
    \ar[
      ur,
      Rightarrow,
      bend right=20,
      "{ 
        \underset{
          \mathclap{
            \raisebox{-2pt}{
            \scalebox{.7}{
              \color{purple}
              \bf
              gauge potentials
            }
            }
          }
        }{
        (\widehat{C_3}, \widehat{C}_6)
        }
       }"{swap, sloped}
    ]
    \\[+3pt]
  \end{tikzcd}
$$
 
\newpage
As an immediate plausibility check, from the well-known homotopy groups of spheres in low degrees this implies (cf. \cite[(22-3)]{HSS19}\cite[(22)]{SS23-Mf}):

\begin{itemize} 
  \item    
  Integral quantization of charges carried by singular M5-brane branes (cf. the following \eqref{ClassifyingChargeOfM5Brane}):
  \begin{equation}
    \label{ChargesOfFlatM5Brane}
    \def\arraystretch{1.4}
    \begin{array}{l}
    \mbox{
      \color{darkblue}
      $
    \pi^4\big(
      \mathbb{R}^{10,1}
      \setminus
      \mathbb{R}^{5,1}
    \big)
    $}
    \;=\;
    \pi^4\big(
      \mathbb{R}^{5,1}
      \times
      \mathbb{R}_+
      \times
      S^4
    \big)
    \\
    \;=\;
    \pi^4(S^4)
    \,=\,
    \pi_4(S^4)
    \,=\,
    \mbox{\color{darkblue}
    $\mathbb{Z}$\,.
    }
    \end{array}
\end{equation}
\item  
  Integral quantization
  of charges carried by singular M2-branes...
  {\color{gray}plus a torsion-contribution (a first prediction of Hypothesis H: fractional M2-branes)}:
  \begin{equation}
    \label{ChargesOfFlatM2Brane}
    \def\arraystretch{1.4}
    \begin{array}{l}
    {
    \mbox{
    \color{darkblue}
    $
    \pi^4\big(
      \mathbb{R}^{10,1}
      \setminus
      \mathbb{R}^{2,1}
    \big)
    $}
    }
    \;=\;
    \pi^4\big(
      \mathbb{R}^{2,1}
      \times
      \mathbb{R}_+
      \times
      S^7
    \big)
    \\
    \;=\;
    \pi^4(S^7)
    \,=\,
    \pi_7(S^4)
    \,=\,
    \mbox{
      \color{darkblue}
      $\mathbb{Z}$
    }
    \color{gray}
    \oplus
    \mathbb{Z}_{12} \,.
    \end{array}
  \end{equation}
\end{itemize} 

\medskip

\noindent
{\bf On the nature of the spheres.}
In itself, the 4-sphere $S^4$ appearing in Hypothesis H is a  {\it classifying space}, hence an abstract tool of algebraic topology, not a physical space. On the other hand, once it is as such used  for flux quantization of the C-field of 11D supergravity, then every such field configuration relates the two.

In particular, when physical spacetime just so happens to itself be a product of a contractible space with a physical 4-sphere --- notably for spacetimes near M5-branes as in  \eqref{ChargesOfFlatM5Brane} ---
then there arises  an effective identification of the physical spatial sphere with the classifying space 
(cf. \cite[p. 17]{HSS19}\cite[(22)]{SS23-Mf}\cite{SS21-M5Anomaly}):
\begin{equation}
  \label{ClassifyingChargeOfM5Brane}
  \begin{tikzcd}[
    row sep=-4pt,
    column sep=40pt
  ]
    \mathbb{R}^{1,5}
    \times 
    \mathbb{R}^{1}_+
    \times
    S^4
    \ar[
      rr,
      "{ \mathrm{pr}_2 }",
      "{ \sim }"{swap}
    ]
    \ar[
      rrrr,
      bend left=24,
      "{
        \scalebox{.7}{
          \color{darkgreen}
          map classifying brane charge
        }
      }"{description}
    ]
    &&
    S^4
    \ar[
      rr,
      dashed
    ]
    &&
    S^4
    \\
    \mathclap{
      \scalebox{.7}{
        \def\arraystretch{.9}
        \begin{tabular}{c}
          \color{darkblue}
          near-horizont spacetime
          \\
          \color{darkblue}
          of black M5-brane
          \\
          \color{gray}
          (Poincar{\'e} patch of $\mathrm{AdS}_7 \times S^4$
          )
        \end{tabular}
      }
    }
    \ar[
      rr,
      phantom,
      "{
    \mathclap{
      \scalebox{.7}{
        \color{darkgreen}
        \def\arraystretch{.85}
        \begin{tabular}{c}
          homotopy
          \\
          equivalent to
        \end{tabular}
      }
    }      
      }"{pos=.6}
    ]
    &&
    \mathclap{
      \scalebox{.7}{
        \color{darkblue}
        \def\arraystretch{.9}
        \begin{tabular}{c}
          spatial
          \\
          4-sphere
        \end{tabular}
      }
    }
    \ar[
      rr,
      phantom,
      "{
    \mathclap{
      \scalebox{.7}{
        \color{darkgreen}
        \def\arraystretch{.85}
        \begin{tabular}{c}
          effective
          \\
          classifying map
        \end{tabular}
      }
    }      
      }"{pos=.5}
    ]
    &&
    \mathclap{
      \scalebox{.7}{
        \color{darkblue}
        \def\arraystretch{.9}
        \begin{tabular}{c}
          classifying
          \\
          4-sphere
        \end{tabular}
      }
    }
  \end{tikzcd}
\end{equation}
in that for a single M5-brane of {\it unit} charge the effective classifying map (on the right) is in the homotopy class of the {\it identity map} from the spatial to the classifying 4-sphere.
(A vaguely reminiscent kind of identification
may also be recognized in \cite[p. 5-6]{Intriligator00}\cite[p. 5-6]{GanorMotl98}, there thought of as mediated by the scalar fields on the M5-brane.)

Directly analogous comments apply to the role of $S^7$ and the near horizon spacetime geometry of M2-branes, cf. \eqref{ChargesOfFlatM2Brane}.

\medskip

\noindent
{\bf Hypothesis H with curvature corrections.} More generally, curvature corrections from the coupling to the background gravity are postulated to be reflected in {\it tangentially twisted} 4-Cohomotopy \cite{FSS20-H}, analogous to the well-known twisting of the RR-field flux-quantization in K-theory by its background B-field:

$$
\adjustbox{raise=-2cm}{
\def\tabcolsep{4pt}
\def\arraystretch{1.2}
\begin{tabular}{|c|c|}
  \hline
  \color{gray}
  Hypothesis K 
  &
  \color{gray}
  Hypothesis H 
  \\
  \begin{tikzcd}[
    column sep=30pt
  ]
    &[30pt]&[-25pt]
    \mathrm{KU}_0 \!\sslash\! 
    \mathrm{PU}(H)
    \ar[
      dd, 
      gray,
      "{
        \scalebox{.7}{
          \color{darkblue}
          \bf
          \def\arraystretch{.9}
          \begin{tabular}{c}
            twisted
            \\
            K-theory
          \end{tabular}
        }
      }"{description}
    ]
    \\
    \\
    X^9
    \ar[
      rr,
      dashed,
      gray
    ]
    \ar[
      uurr,
      dashed,
      "{
        \scalebox{.7}{
          \color{purple}
          \bf
          RR-field
        }
      }"{sloped}
    ]
    \ar[
      dr,
      gray,
      shorten >=-5pt,
      dashed,
      "{
        \scalebox{.7}{
          \color{darkblue}
          \bf
          twist by
        }
      }"{sloped, yshift=-2pt},
      "{
        \scalebox{.7}{
          \color{darkblue}
          \bf
          \def\arraystretch{.9}
          \begin{tabular}{c}
            background 
            \\
            B-field
          \end{tabular}
        }
      }"{sloped, swap}
    ]
    &&
    \scalebox{.85}{
      \color{gray}
      $B \mathrm{PU}(H)$
    }
    \ar[
      dl,
      gray,
      "{ \sim }"{sloped}
    ]
    \\
    & 
    \scalebox{.85}{
      \color{gray}
      $B^2 \mathrm{U}(1)$
    }
  \end{tikzcd}
  &
  \begin{tikzcd}[
    column sep=35pt
  ]
    &[35pt]&[-30pt]
    S^4 \!\sslash\! \widehat{\mathrm{Sp}(2)}
    \ar[
      dd,
      gray,
      "{
        \scalebox{.7}{
          \color{darkblue}
          \bf
          \def\arraystretch{.9}
          \begin{tabular}{c}
            twisted
            \\
            Cohomotopy
          \end{tabular}
        }
      }"{description}
    ]
    \\
    \\
    T^2 \times X^8
    \ar[
      rr,
      dashed,
      "{
        \scalebox{.7}{
          \color{darkgreen}
          \bf
          Fivebrane structure
        }
      }"
    ]
    \ar[
      uurr,
      dashed,
      "{
        \scalebox{.7}{
          \color{purple}
          \bf
          C-field
        }
      }"{sloped}
    ]
    \ar[
      dr,
      dashed,
      shorten >=-5pt,
      gray,
      "{
        \scalebox{.7}{
          \color{darkblue}
          \bf
          twist by
        }
      }"{sloped, yshift=-1pt},
      "{
        \scalebox{.7}{
          \color{darkblue}
          \bf
          \def\arraystretch{.9}
          \begin{tabular}{c}
            background 
            \\
            gravity
          \end{tabular}
        }
      }"{yshift=+1pt, sloped, swap}
    ]
    &&
    \scalebox{.85}{$
      B \widehat{\mathrm{Sp}(2)}
    $}
    \ar[
      dl,
      dashed,
      gray
    ]
    \\
    &
    \scalebox{.85}{\color{gray}
      $B \mathrm{Spin}(8)$
    }
  \end{tikzcd}
  \\
  \hline
\end{tabular}
}
$$

\medskip

To distinguish M2/M5-charge, the tangential twisting needs to preserve the $\mathbb{H}$-Hopf fibration $\Rightarrow$ tangential $\mathrm{Sp}(2) \hookrightarrow \mathrm{Spin}(8)$-structure \cite[\S 2.3]{FSS20-H}. With this,
integrality of M2's Page charge \& anomaly-cancellation of the M5's Hopf-WZ term follows from trivialization of the Euler 8-class, which means lift to the {\it Fivebrane} 6-group $\widehat{\mathrm{Sp}(2)} \to \mathrm{Sp}(2)$ \cite[\S 4]{FSS19-Hopf}.
 
\medskip

This implies \cite[Prop. 3.13]{FSS20-H}\cite[Thm. 4.8]{FSS19-Hopf}:

\smallskip 
   {\bf (i)} half-integrally shifted
   quantization 
   of M5-brane charge   
   in curved backgrounds, 

\begin{equation}
  [\widetilde{G}_4]
  \;:=\;
  \underbrace{
    [G_4] 
  }_{
    \mathclap{
      \raisebox{-3pt}{
        \scalebox{.7}{
          \color{darkblue}
          \bf
          \def\arraystretch{.9}
          \begin{tabular}{c}
            C-field
            \\
            4-flux
          \end{tabular}
        }
      }
    }
  }
    \;+\; 
  \tfrac{1}{2}\big(
    \underset{
      \mathclap{
        \raisebox{-2pt}{
          \scalebox{.7}{
            \color{darkblue}
            \bf
            \def\arraystretch{.9}
            \begin{tabular}{c}
              integral Spin- 
              \\
              Pontrjagin class
            \end{tabular}
          }
        }
      }
    }{
    \underbrace{
      \tfrac{1}{2}p_1(T X^8)
    }
    }
  \big)
  \,\in\,
  H^4\big(
    X^8;\,
    \mathbb{Z}
  \big)
\end{equation}
  
{\bf (ii)} integral quantization of the Page charge of M2-branes:
   \footnote{
     On the right of \eqref{PageCharge}, the ``hat'' in $\widehat{X}^8$ indicates that this holds locally, namely on suitable fibration over spacetime (cf. \cite[(116)]{FSS20-H}\cite[p. 6]{FSS21-TwistedString}) on which the ``M-theory 3-form'' $H_3$ is globally defined, which it cannot be on $X^8$ unless $\widetilde G_4$ is cohomologically trivial (a basic subtlety that has traditionally been glossed over in discussions of Page charge.)
     But in the present context of flux quantization on M5-branes the analog of this extended spacetime is in fact the worldvolume of the M5-brane, restricted to which $\widetilde G_4$ does trivialize in cohomology, whence we need not further dwell here on the definition of $\widehat{X}^8$.
   }
\begin{equation}
  \label{PageCharge}
     2[\widetilde G_7]
     \;\;
     :=
     \;\;
     2
     \big(
     [G_7]
     +
     \tfrac{1}{2}
     [H_3 \wedge \widetilde G_4]
     \big)
     \;\;
       \in
     \;\;
     H^7(\widehat{X}^8; \mathbb{Z})
\end{equation}

\smallskip

\noindent
Both of these quantization conditions on M-brane charge are thought to be crucial for M-theory to make any sense.
But, previously, item (i) had remained enigmatic and item (ii) had remained wide open. 

\vspace{.2cm}

\noindent
But there is more:

\adjustbox{
  margin=4pt,
  bgcolor=lightolive
}{
\begin{minipage}{6.4cm}
Provable implications from Hypothesis H
\\
of subtle effects expected in M-theory:
\end{minipage}
}
$
\left\{
\adjustbox{
  bgcolor=lightolive
}{
\begin{minipage}{7.5cm}
\def\arraystretch{1.2}
\def\tabcolsep{2pt}
\begin{tabular}{ll}
-- half-integral shift of 4-flux 
&\cite[Prop. 3.13]{FSS20-H}
\\
-- DMW anomaly cancellation
&
\cite[Prop. 3.7]{FSS20-H}
\\
-- the C-field's ``integral EoM'' & \cite[\S3.6]{FSS20-H}
\\
-- M2 Page charge quantization
& \cite[Thm. 4.8]{FSS19-Hopf}
\\
-- integrality of $\tfrac{1}{6}(G_4)^3$ 
&
\cite[Rem. 2.9]{GS21}
\\
-- M5-brane anomaly cancellation & \cite{SS21-M5Anomaly}
\\
-- non-abelian gerbe field on M5 & \cite{FSS21-TwistedString}
\end{tabular}
\end{minipage}
\hspace{5pt}
}
\right.
$

\medskip 
\noindent
It is these and further results that suggest that Hypothesis H goes towards the correct flux-quantization law for the C-field in M-theory.

\medskip

Yet more generally, Hypothesis H applies to orbifold spacetimes, where it postulates flux quantization in (twisted and) {\it equivariant} Cohomotopy \cite{SS20-Tad}\cite{SS21-FracD}. This is what we turn to next.

  \medskip

\hypertarget{EquivariantFormOfTwistorialCohomotopy}{}

\noindent
{\bf Orbi-worldvolumes and Equivariant charges.}
Flux-quantization generalizes to {\it orbifolds} 
\footnote{
  For brevity we consider here only ``very good'' orbifolds, namely global quotients of manifolds by the action of a finite group $G$. This is sufficient for the present purpose and anyways the case understood by default in the string theory literature.
}
by generalizing the cohomology of the charges to {\it equivariant cohomology} \cite{SS20-Orb}. 
In terms of classifying spaces this simply means that all spaces are now equipped with the action of a finite group $G$ and all maps are required to be $G$-equivariant.

\smallskip 
We take $G := \ZTwo$ and the classifying fibration to be the {\bf twistor fibration}
$p := t_{\mathbb{H}}$
equivariant under swapping the $\mathbb{H}$-summands,

$$
  \begin{tikzcd}[
    column sep=70pt
  ]
    \scalebox{.7}{
      \color{darkblue}
      \bf
      \def\arraystretch{.9}
      \begin{tabular}{c}
        orbi-
        \\
        worldvolume
      \end{tabular}
    }
    &[-65pt]
    \Sigma
    \ar[
      in=60,
      out=180-60,
      looseness=3.5,
      shift right=3pt,
      "{
        \!G\!
      }"{description}
    ]
    \ar[
      r,
      dashed,
      "{
        \scalebox{.7}{
          \color{darkgreen}
          \bf
          \def\arraystretch{.9}
          \begin{tabular}{c}
            orbi-brane
            \\
            charges
          \end{tabular}
        }
      }",
      "{\ }"{swap, name=s}
    ]
    \ar[
      d,
      "{ \phi }"{swap},
      "{\ }"{name=t}
    ]
    \ar[
      from=s,
      to=t,
      dashed,
      Rightarrow
    ]
    &[30pt]
    \hotype{A}
    \mathrlap{\,:}
    \ar[
      in=60,
      out=180-60,
      looseness=3.5,
      shift right=3pt,
      "{
        \!G\!
      }"{description}
    ]
    \ar[
      d,
      "{ 
        p 
      }"{swap},
      "{
        \scalebox{.7}{
          \color{olive}
          \bf
          \def\arraystretch{.9}
          \def\tabcolsep{0pt}
          \begin{tabular}{c}
            equivariant
            \\
            classifying
            \\
            fibration...
          \end{tabular}
        }
      }"{xshift=-6pt}
    ]
    \ar[
      r,
      equals
    ]
    &[+20pt]
    \mathbb{C}P^3
    \ar[
      in=60,
      out=180-60,
      looseness=3.5,
      shift right=3pt,
      "{
        \!\ZTwo\!
      }"{description}
    ]
    \ar[
      d,
      "{
        t_{\mathbb{H}}
      }"{swap},
      "{
        \scalebox{.7}{
          \color{olive}
          \bf
          \def\arraystretch{.9}
          \def\tabcolsep{0pt}
          \begin{tabular}{c}
            ...for equivariant
            \\
            twistorial Cohomotopy
          \end{tabular}
        }
      }"{xshift=-6pt}
    ]
    \\
    \scalebox{.7}{
      \color{darkblue}
      \bf
      \def\arraystretch{.9}
      \begin{tabular}{c}
        orbi-
        \\
        spacetime
      \end{tabular}
    }
    &
    X
    \ar[
      in=-55,
      out=180+55,
      looseness=3.8,
      shift left=1pt,
      "{
        \!G\!
      }"{description}
    ]
    \ar[
      r,
      dashed,
      "{
        \scalebox{.7}{
          \color{darkgreen}
          \bf
          \def\arraystretch{.9}
          \begin{tabular}{c}
            orbi-bulk
            \\
            charges
          \end{tabular}
        }
      }"{swap}
    ]
    &
    \hotype{B}
    \mathrlap{\,:}
    \ar[
      in=-55,
      out=180+55,
      looseness=3.8,
      shift left=1pt,
      "{
        \!G\!
      }"{description}
    ]
    \ar[
      r,
      equals
    ]
    &
    S^4
    \ar[
      in=-55,
      out=180+55,
      looseness=3.8,
      shift left=1pt,
      "{
        \!\ZTwo\!
      }"{description}
    ]
  \end{tikzcd}
$$

\vspace{.1cm}

\noindent
and the brane/bulk orbifold we take to be as on \hyperlink{BraneDiagram}{p. 3}:

\vspace{-.1cm}
\begin{tabular}{p{6.7cm}}
  \footnotesize
  {\bf The orbi-brane diagram}
  for a flat M5-brane wrapped on a trivial Seifert-fibered orbi-singularity. 
  Shaded is the $\ZTwo$-fixed locus/orbi-singularity.

  We are adjoining the {\it point at infinity} to the space
  $
    \mathbb{R}^2_{\cpt}
    \underset{\mathrm{homeo}}{\simeq}
    S^2
  $
  which is
  thereby 
  designated as transverse to any worldvolume solitons to be measured in reduced cohomology.
\end{tabular}
\adjustbox{
  raise=-1.8cm
}{
\begin{tikzpicture}
\draw[
  draw opacity=0,
  fill=olive,
  fill opacity=.2
]
  (-2.7,-.9+1.35) 
    rectangle 
  (1.3,-.1+1.35);
\draw[
  draw opacity=0,
  fill=olive,
  fill opacity=.2
]
  (-2.7,-.9) 
    rectangle 
  (1.3,-.1);
\draw[
  draw opacity=0,
  fill=olive,
  fill opacity=.2
]
  (3.5,-.9) 
    rectangle 
  (4.2,-.1);
\node at (0,0)
{$
  \begin{tikzcd}[
    column sep=0pt
  ]
    \Sigma
    \ar[
      in=60,
      out=180-60,
      looseness=3.7,
      shift right=3pt,
      "{
        \!\mathbb{Z}_2\!
      }"{description}
    ]
    \ar[
      d,
      hook,
      "{ 
        \phi 
      }"{swap}
    ]
    &:=&
    \mathbb{R}^{1,0}
    &\times& 
    \mathbb{R}^2_{\cpt}
    &\times& 
    S^1
    &\times&
    \mathbb{R}^2_{\mathrm{sgn}}
    \ar[
      in=60,
      out=180-60,
      looseness=3.5,
      shift right=3pt,
      "{
        \!\mathbb{Z}_2\!
      }"{description}
    ]
    \\
    X
    \ar[
      in=-55,
      out=180+55,
      looseness=3.8,
      shift left=1pt,
      "{
        \!\mathbb{Z}_2\!
      }"{description}
    ]
    &:=&
    \mathbb{R}^{1,0}
    &\times& 
    \mathbb{R}^2_{\cpt}
    &\times& 
    S^1
    &\times&
    \mathbb{R}^2_{\mathrm{sgn}}
    \ar[
      in=60,
      out=180-60,
      looseness=3.5,
      shift right=3pt,
      "{
        \!\mathbb{Z}_2\!
      }"{description}
    ]
    &\times&
    \mathbb{R}^5
    \\[-20pt]
    &&
    \mathclap{
      \scalebox{.7}{
        \color{darkblue}
        \bf
        \def\arraystretch{.9}
        \begin{tabular}{c}
          time
        \end{tabular}
      }
    }
    &&
    \mathclap{
      \scalebox{.7}{
        \color{darkblue}
        \bf
        \def\arraystretch{.9}
        \begin{tabular}{c}
          trnsvrs space
          \\
          to solitons
        \end{tabular}
      }
    }
    &&
    \mathclap{
      \scalebox{.7}{
        \color{darkblue}
        \bf
        \def\arraystretch{.9}
        \begin{tabular}{c}
          M/IIA-
          \\
          circle
        \end{tabular}
      }
    }
    &&
    \mathclap{
      \scalebox{.7}{
        \color{darkblue}
        \bf
        \def\arraystretch{.9}
        \begin{tabular}{c}
          orbi-
          \\
          cone
        \end{tabular}
      }
    }
    &&
    \mathclap{
      \scalebox{.7}{
        \color{darkblue}
        \bf
        \def\arraystretch{.9}
        \begin{tabular}{c}
          trnsvrs space
          \\
          to M5-brane
        \end{tabular}
      }
    }
  \end{tikzcd}
$};
\end{tikzpicture}
}

But since the cone $\ZTwo \acts \; \mathbb{R}^2_{\mathrm{sgn}}$ is equivariantly contractible, 
the inclusion of the $\ZTwo$-fixed loci is actually a homotopy equivalence

$$
  \begin{tikzcd}
    \phantom{\mathrm{pt}}
      \ar[
        out=180-55+90, 
        in=55+90,   
        looseness=4.5, 
      "\scalebox{1}{$
        \;
        \mathclap{
          \mathbb{Z}_2
        }
        \;
      $}"{description},shift right=1]
  \end{tikzcd}
  \hspace{-23pt}
\adjustbox{
 raise=-.92cm
}{
\begin{tikzpicture}
\begin{scope}[yscale=.25, xscale=.6]
 \shadedraw[draw opacity=0, top color=gray, bottom color=cyan]
   (0,0) -- (3,3) .. controls (2,2) and (2,-2) ..  (3,-3) -- (0,0);
 \draw[draw opacity=0, top color=white, bottom color=gray]
   (3,3)
     .. controls (2,2) and (2,-2) ..  (3,-3)
     .. controls (4,-3.9) and (4,+3.9) ..  (3,3);
\node at (0,0) {
  \scalebox{.9}{$\ast$}
};
\end{scope}
\end{tikzpicture}
}
\begin{tikzcd}[
  column sep=20pt
]
{}
\ar[
  rr,
  "{ \sim }"{pos=.1},
  "{ \mathrm{hmtp} }"{swap, pos=.1},
  shorten <=-12pt,
  shorten >=20pt
]
&& 
\ast
      \ar[
        out=180-55+90, 
        in=55+90,   
        looseness=4.5, 
      "\scalebox{1}{$
        \;
        \mathclap{
          \mathbb{Z}_2
        }
        \;
      $}"{description},shift right=1]
\end{tikzcd}
 \;\;\;\;\;\;\;\;\;\;
 \Rightarrow
 \;\;\;\;\;\;\;\;\;\;
  \begin{tikzcd}[
    row sep=15pt, column sep=huge
  ]
    \Sigma^{\ZTwo}
    \ar[
      r,
      "{ \sim }",
      "{ \mathrm{hmtp} }"{swap}
    ]
    \ar[
      d,
      "{ \phi^{\ZTwo} }"{swap}
    ]
    &
    \Sigma
    \ar[
      in=60,
      out=180-60,
      looseness=3.5,
      shift right=3pt,
      "{
        \!\ZTwo\!
      }"{description}
    ]
    \ar[
      d,
      "{ \phi }"
    ]
    \\
    X^{\ZTwo}
    \ar[
      r,
      "{ \sim }",
      "{ \mathrm{hmtp} }"{swap}
    ]
    &
    X
    \ar[
      in=-55,
      out=180+55,
      looseness=3.8,
      shift left=1pt,
      "{
        \!G\!
      }"{description}
    ]
  \end{tikzcd}
$$

\noindent
Therefore, our equivariant classifying maps are determined up to equivariant homotopy by their restriction to the fixed-locus and hence  
the charges are {\it localized on the orbi-singularity} where they take values in 2-Cohomotopy:
$$
  \left\{
  \adjustbox{
    raise=4pt
  }{
  \begin{tikzcd}
    \Sigma
    \ar[
      in=60,
      out=180-60,
      looseness=3.5,
      shift right=3pt,
      "{
        \!\ZTwo\!
      }"{description}
    ]
    \ar[
      rr,
      dashed,
      "{\ }"{swap, name=s}
    ]
    \ar[
      d,
      "{ \phi }"{swap},
      "{\ }"{name=t}
    ]
    \ar[
      from=s,
      to=t,
      Rightarrow,
      dashed
    ]
    &&
    \mathbb{C}P^3
    \ar[
      in=60,
      out=180-60,
      looseness=3.5,
      shift right=3pt,
      "{
        \!\ZTwo\!
      }"{description}
    ]
    \ar[
      d,
      "{ t_{\mathbb{H}} }"
    ]
    \\
    X
    \ar[
      in=-55,
      out=180+55,
      looseness=3.8,
      shift left=1pt,
      "{
        \!\ZTwo\!
      }"{description}
    ]
    \ar[
      rr,
      dashed,
      "{
        \scalebox{.7}{
          \color{darkgreen}
          \bf
          \def\arraystretch{.9}
          \begin{tabular}{c}
            charges on 
            \\
            orbifold
          \end{tabular}
        }
      }"{swap, yshift=-3pt}
    ]
    &&
    S^4
    \ar[
      in=-55,
      out=180+55,
      looseness=3.8,
      shift left=1pt,
      "{
        \!\ZTwo\!
      }"{description}
    ]
  \end{tikzcd}
  }
  \right\}
  \;\;\;\;
  \simeq
  \;\;\;\;
  \left\{
  \adjustbox{
    raise=6pt
  }
  {
  \begin{tikzcd}
    \Sigma^{\mathbb{Z}_2}
    \ar[
      r,
      dashed,
      "{\ }"{swap, name=s2}
    ]
    \ar[
      d,
      "{ 
        \phi^{\mathbb{Z}_2} 
       }"{swap},
       "{\ }"{name=t2}
    ]
    \ar[
      from=s2,
      to=t2,
      dashed,
      Rightarrow
    ]
    &
    (\mathbb{C}P^3)^{\mathbb{Z}_2}
    \ar[
      d,
      "{ 
        t_{\mathbb{H}}^{\mathbb{Z}_2} 
      }"
    ]
    \ar[
      r,
      equals
    ]
    &
    S^2
    \ar[
      d
    ]
    \\
    X^{\mathbb{Z}_2}
    \ar[
      r,
      dashed
    ]
    \ar[
      rr,
      phantom,
      shift right=13pt,
      "{
        \scalebox{.7}{
          \color{darkgreen}
          \bf
          charges localized on
          orbi-singularity
        }
      }"
    ]
    &
    (S^4)^{\mathbb{Z}_2}
    \ar[r,equals]
    &
    \ast
  \end{tikzcd}
  }
  \right\}
  \;\;\;\;
  \simeq
  \;\;\;\;
  \bigg\{\!\!
  \begin{tikzcd}
    \mathbb{R}^2_{\cpt}
    \times S^1
    \ar[
      r,
      dashed,
      "{
        \scalebox{.7}{
          \color{darkgreen}
          \bf
          \def\arraystretch{.9}
          \begin{tabular}{c}
            charges in 2-Cohomotopy
            \\
            of B-field
            solitons
            \\
            on M5 orbi-singularity
          \end{tabular}
        }
      }"{pos=-.3, swap, yshift=-10pt}
    ]
    &
    S^2
  \end{tikzcd}
\!\!\!  \bigg\}
$$

\vspace{.3cm}

\noindent
{\bf Moduli space of worldvolume solitons.}
To be precise, the solitonic charges are to be measured in the {\it reduced} 2-Cohomotopy classified by {\it pointed maps}, enforcing the condition that solitonic fields {\it vanish at infinity} \cite[\S 2.2]{SS24-Flux}.
In the strongly-coupled situation, where the M/IIA circle de-compactifies to $\mathbb{R}^1$, the vanishing-at-infinity must also be applied here, whence (cf. \cite[\S A.2]{SS24-QObs}) the moduli space of topological solitons is the loop space of the reduced 2-Cohomotopy moduli of the transverse space:

\vspace{-.2cm}
\begin{center}
\adjustbox{
  raise=3pt,
  bgcolor=lightolive
}{
$
  \begin{tikzcd}[
    column sep=-20pt
  ]
  \scalebox{.7}{
    \color{darkblue}
    \bf
    \def\arraystretch{.9}
    \begin{tabular}{c}
      Moduli space of solitons
      \\
      on M5 orbi-singularity
    \end{tabular}
  }
  &[20pt]
  \mathrm{Maps}^{\ast}\big(
    \mathbb{R}^2_{\cpt}
    \wedge
    S^1
    ,\,
    S^2
  \big)
  \ar[
    rr,
    phantom,
    "{\simeq}"
  ]
  \ar[
    dr,
    phantom,
    "{\simeq}"{sloped}
  ]
  &&
  \Omega
  \,
  \mathrm{Maps}^{\ast}\big(
    \mathbb{R}^2_{\cpt}
    ,\,
    S^2
  \big)
  &[20pt]
  \scalebox{.7}{
    \color{darkblue}
    \bf
    \def\arraystretch{.9}
    \begin{tabular}{c}
      Loop space of
      \\
      moduli space of solitons
      \\
      on D4 orbi-singularity
    \end{tabular}
  }
  \\
  &
  &
  \mathrm{Maps}^\ast\big(
    \mathbb{R}^2_{\cpt}
    ,\,
    \Omega S^2
  \big)
  \ar[
    ur,
    phantom,
    "{\simeq}"{sloped}
  ]
  \end{tikzcd}
$
}
\end{center}    

(The algebraic topology of maps to $\Omega S^2$ have also found some attention in \cite{Morava23}\cite{Morava24}.)

\medskip

\noindent
{\bf Outlook.}
Strikingly, as we explain next, this moduli space is equivalently a space of {\it worldsheets of strings} in $\mathbb{R}^3$ with unit {\it charged endpoints} forming oriented {\it framed links}! \cite{SS24-AbAnyons}
\begin{center}
\begin{minipage}{6cm}
  \footnotesize
  {\bf Figure \figurenumber: Framed links as stringy worldsheets} are revealed  by careful analysis as being the loops in the moduli space of solitonic cohomotopy charges of the plane. 
\end{minipage}
\;\;\;
\adjustbox{
  raise=-1.2cm
}{
\includegraphics[width=3.6cm]{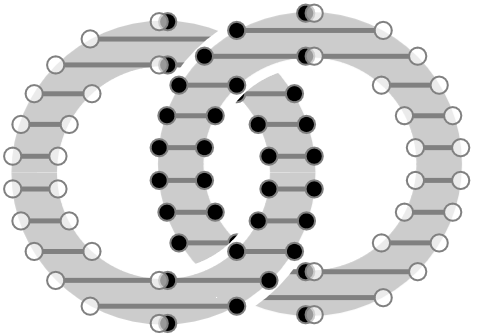}
}
\end{center}
Such link diagrams are just the envisioned topological quantum circuit protocols, and their framing regularizes the anyonic phase observables (``Wilson loop observables'').
 
\begin{center} 
\begin{minipage}{5cm}
 \footnotesize
 {\bf Figure \figurenumber: Traditional protocol for topological quantum computation with anyons,} taken from Rowell (\cite{Rowell22}, following \cite[Fig. 2]{Rowell16}).
\end{minipage}
\;\;
\adjustbox{
  raise=-2cm,
  margin=-2pt,
  fbox
}{
\includegraphics[width=9.5cm]{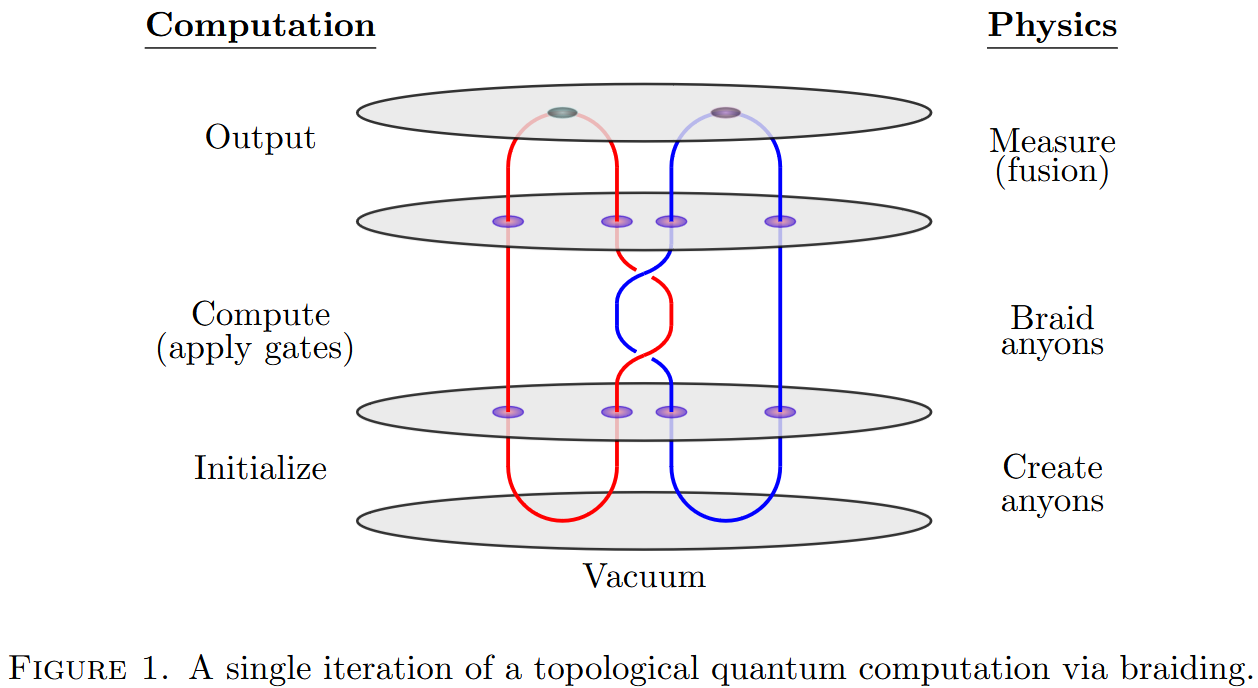}
}
\end{center} 
\newpage

%%%%%%%%%%%%%%%%%%%%%%%%%%%%%%%%%%%%%%%
\section{Cohomotopy Charge of Solitons}
%%%%%%%%%%%%%%%%%%%%%%%%%%%%%%%%%%%%%%%

Remarkably, there is an equivalence between {\it Cohomotopy} of spacetime/worldvolumes and {\it Cobordism} classes of submanifolds behaving like solitonic branes carrying the corresponding Cohomotopy charge \cite[\S 2.2]{SS23-Mf} \cite[\S 2.1]{SS20-Tad}:

\vspace{3pt}
\hspace{-1.1cm}
\def\tabcolsep{3pt}
\def\arraystretch{1}
\begin{tabular}{|ll|}
\hline
&
\\[-7pt]
\begin{minipage}{4cm}
  \small
  The
  {\bf Pontrjagin theorem} \cite{Pontrjagin38}\cite[\S IX]{Kosinski93} 
  identifies the unstable
  $n$-Cohomotopy of a closed manifold with the cobordism classes of its normally framed submanifolds of co-dimension $n$. 
\end{minipage}
&
\hspace{-6pt}
\adjustbox{
  raise=-1cm
}{
\includegraphics[width=14cm
]{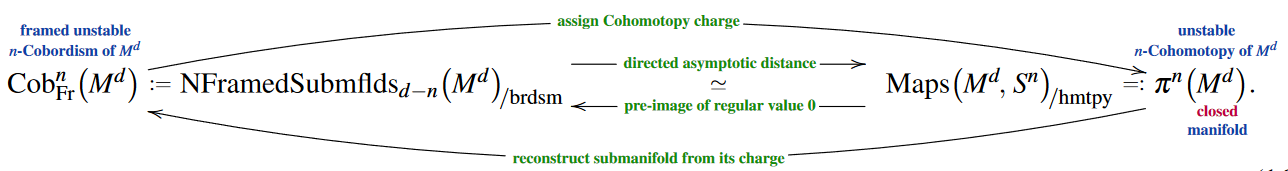}}
\\[-4pt]
&
\\
\hline
\begin{minipage}{4cm}
  \small
  
  The {\bf Cohomotopy charge} of a normally framed submanifold (aka {\it scanning map} or {\it Pontrjagin-Thom collapse}) is represented by mapping points of the ambient space to their directed distance if inside a tubular neighborhood, else to $\infty$.

  \smallskip

  Conversely, every Cohomotopy class is representated by a smooth map with 0 a regular value, whose pre-image is a normally framed submanifold with that Cohomotopy charge.
\end{minipage}
&
\hspace{-6pt}
\adjustbox{raise=-3.3cm}{
\includegraphics[width=14cm]{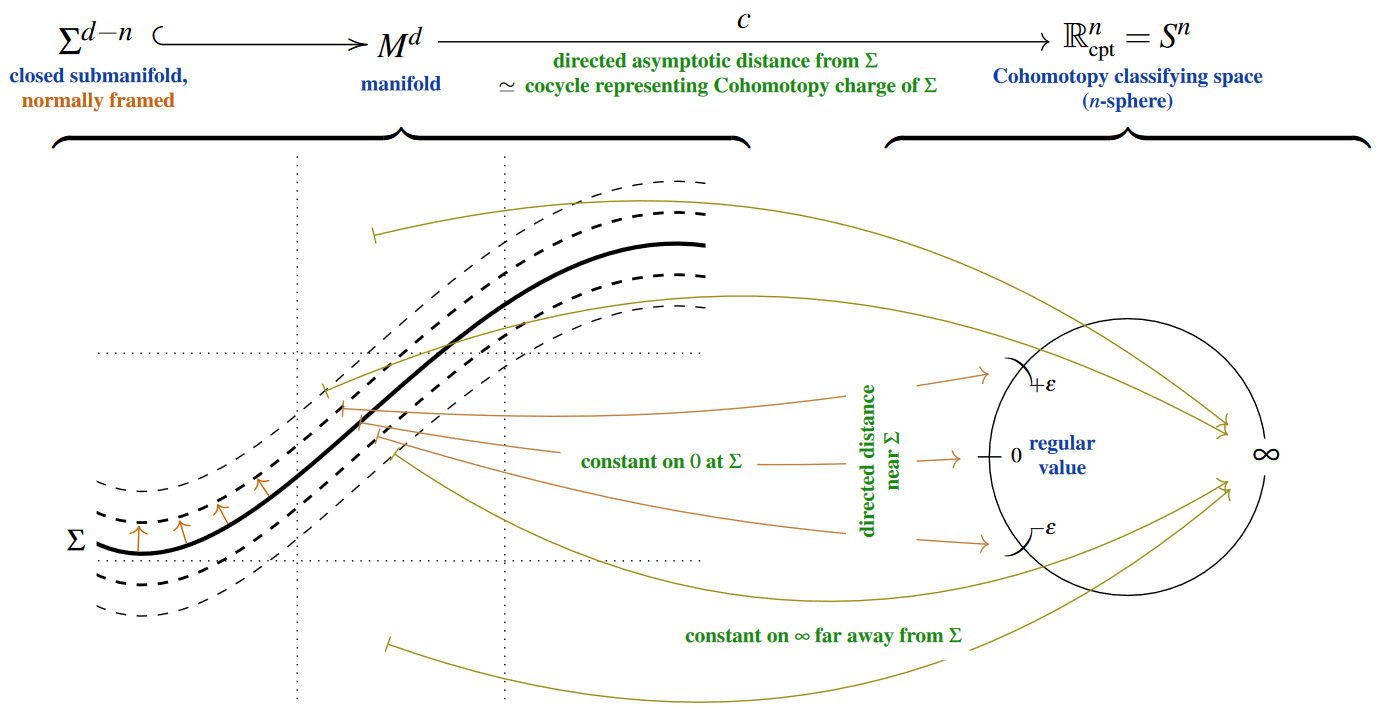}
}
\\[-5pt]
&
\\
\hline
\begin{minipage}{4cm}
\small
Under this relation, 

homotopy of charge maps 

corresponds to nrml. framed

{\bf cobordism} of submnflds.

\smallskip

The cobordism relation exhibits a form of 
pair creation/annihilation of
submanifolds carrying opposite
Cohomotopy charges.

\end{minipage}
&
\adjustbox{
  scale=1.2,
  raise=-2.0cm
}{
\begin{tikzpicture}[
  rotate=90
]

  \draw[fill=lightgray, draw opacity=0, fill opacity=.4]
    (-3,-3) rectangle (0,3);

  \draw[line width=5, lightgray]
    (0,3) to (0,-3);

  \begin{scope}

  \clip (-2.2,-2.2) rectangle (0,2.2);

  \draw[line width=2]
    (0,0) circle (2);

  \end{scope}

  \draw[fill=black]
    (0,2) circle (.1);

  \draw[fill=white]
    (0,-2) circle (.1);

  \draw[->, darkorange]
    (90:2) to  (90:2.5);

  \draw[->, darkorange]
    (90+30:2) to  (90+30:2.5);

  \draw[->, darkorange]
    (90+60:2) to  (90+60:2.5);

  \draw[->, darkorange]
    (90+90:2) to  (90+90:2.5);

  \draw[->, darkorange]
    (-90:2) to  (-90:2.5);

  \draw[->, darkorange]
    (-90-30:2) to  (-90-30:2.5);

  \draw[->, darkorange]
    (-90-60:2) to  (-90-60:2.5);

  \draw
    (-.4, 3.1)
    node
    {
      \rlap{
      \scalebox{.5}{
      \color{darkorange}
      \bf
      \def\arraystretch{.9}
      \begin{tabular}{c}
        normal 
        \\
        framing
        \\
        in space
      \end{tabular}
      }
      }
    };

  \draw
    (+.3, 2.5)
    node
    {
      \rlap{
      \scalebox{.5}{
      \color{darkblue}
      \bf
      \begin{tabular}{c}
        brane
      \end{tabular}
      }
      }
    };

  \draw
    (-.4, -1.85)
    node
    {
      \rlap{
      \scalebox{.5}{
      \color{darkorange}
      \bf
      \def\arraystretch{.9}
      \begin{tabular}{c}
        opposite 
        \\
        normal 
        \\
        framing
      \end{tabular}
      }
      }
    };

  \draw
    (.3, -1.3)
    node
    {
      \rlap{
      \scalebox{.5}{
      \color{darkblue}
      \bf
      \begin{tabular}{c}
        anti-brane
      \end{tabular}
      }
      }
    };

  \draw
    (-1.9, -1.1)
    node
    {
      \rlap{
      \scalebox{.5}{
      \color{darkorange}
      \bf
      \begin{tabular}{c}
        normal framing
        \\
        in spacetime
      \end{tabular}
      }
      }
    };

  \draw
    (-2.7, +1.5)
    node
    {
      \rlap{
      \scalebox{.5}{
      \color{darkblue}
      \bf
      \begin{tabular}{c}
        spacetime
      \end{tabular}
      }
      }
    };

  \node[
    rotate=90,
    xscale=-1
  ] 
  at
    (-1,0)
    {
      $
        {
          \xleftrightharpoons{
          \adjustbox{scale={-1}{1}}{
            \tiny
            \color{darkgreen}
            pair creation
          }
          }
        }
      $
    };

  \node[
    rotate=90
  ]
    at (-1,-.38)
    {
      \raisebox{10pt}{
        \tiny
        \color{darkgreen}
        annihilation
      }
    };

  \node
    at (0,+.9)
    {
      \scalebox{.5}{
      \color{darkblue}
      \bf
        space
      }
    };

\end{tikzpicture}
}
\hspace{-.6cm}
\adjustbox{
  scale=1.1,
  raise=-1.6cm
}{
\begin{tikzpicture}
\def\reduce{.4}
\begin{scope}[shift={(-.8,0)}]
  \draw[fill = black]
    (-.05,1.5-\reduce) rectangle (+.05,-1.5+\reduce);
  \draw[fill=white]
    (0,0) circle (.23);
  \draw[fill=lightgray, fill opacity=.6]
    (0,0) circle (.23);
  \begin{scope}
    \clip
      (0,0) circle (.22);
    \draw (0,0)
      node
      {
        \color{blue}
        $f$
      };
  \end{scope}
  \node 
    at 
    (-.8,0)
    {
      \scalebox{.6}{
        \color{darkorange}
        \bf
        \def\arraystretch{.8}
        \begin{tabular}{c}
          framing
          \\
          charge
        \end{tabular}
      }
    };
\end{scope}

\begin{scope}[shift={(-0,0)}]
  \draw[fill = black]
    (-.05,1.5-\reduce) rectangle (+.05,-1.5+\reduce);
  \draw[fill=white]
    (0,0) circle (.23);
  \draw[fill=lightgray, fill opacity=.6]
    (0,0) circle (.23);
  \begin{scope}
    \clip
      (0,0) circle (.22);
    \draw (0,0)
      node
      {
        \color{blue}
        $w$
      };
  \end{scope}
\end{scope}

\begin{scope}[shift={(+.8,0)}]
  \draw[fill=lightgray, fill opacity=.6]
    (0,0) circle (.23);
  \draw[fill=white, draw opacity=0]
    (-.05,1.5-\reduce) rectangle (+.05,-1.5+\reduce);
  \draw
    (-.05,1.5-\reduce) to (-.05,-1.5+\reduce);
  \draw
    (+.05,1.5-\reduce) to (+.05,-1.5+\reduce);
  \draw[fill=white, draw opacity=0]
    (0,0) circle (.23);
  \draw[fill=lightgray, draw opacity=0, fill opacity=.6]
    (0,0) circle (.23);
  \begin{scope}
    \clip
      (0,0) circle (.22);
    \draw (0,0)
      node
      {
        \color{blue}
        \raisebox{1pt}{
          $\overline{w}$
        }
      };
  \end{scope}
\end{scope}

  \draw
    (2,0)
    node
    {
      \scalebox{2.4}{
        $\rightleftharpoons$
      }
    };

\begin{scope}[shift={(+3.2,0)}]
  \draw[fill = black]
    (-.05,1.5-\reduce) rectangle (+.05,-1.5+\reduce);
  \draw[fill=white]
    (0,0) circle (.23);
  \draw[fill=lightgray, fill opacity=.6]
    (0,0) circle (.23);
  \begin{scope}
    \clip
      (0,0) circle (.22);
    \draw (0,0)
      node
      {
        \color{blue}
        $f$
      };
  \end{scope}
\end{scope}

 \draw[white,line width=1.4]
   (-1, 1.32-\reduce) to (3.4, 1.32-\reduce);
 \draw[white,line width=1.4]
   (-1, 1.1-\reduce) to (3.4, 1.1-\reduce);
 \draw[white,line width=1.4]
   (-1, .89-\reduce) to (3.4, .89-\reduce);

 \draw[white,line width=1.4]
   (-1, -1.32+\reduce) to (3.4, -1.32+\reduce);
 \draw[white,line width=1.4]
   (-1, -1.1+\reduce) to (3.4, -1.1+\reduce);
 \draw[white,line width=1.4]
   (-1, -.89+\reduce) to (3.4, -.89+\reduce);

  \draw
    (2,-.6)
    node
    {
      \tiny
      \color{darkgreen}
      \bf
      \def\arraystretch{.75}
      \begin{tabular}{c}
        creation /
        \\
        annihilation
      \end{tabular}
    };

\draw
  (-.4,-1.8+\reduce)
  node
  {
    \tiny
    \color{darkblue}
    \bf
    branes
  };

\draw
  (+.8,-1.8+\reduce)
  node
  {
    \tiny
    \color{darkblue}
    \bf
    \def\arraystretch{.75}
    \begin{tabular}{c}
      anti-
      \\
      brane
    \end{tabular}
  };

\end{tikzpicture}
}
\\[-6pt]
&
\\
\hline
&
\\[-10pt]
\begin{minipage}{4cm}
  \small
  When making more ambient dimensions available, the cobordism classes eventually (quickly) exhibit {\bf stabilization} on abelian cobordism cohomology groups.
  (This might relate {\it Hypothesis H} to  Vafa's {\it cobordism conjecture}
cf. \cite[\S 4]{SS23-Mf}).
\end{minipage}
&
\hspace{-10pt}
\adjustbox{raise=-1.8cm}{
\includegraphics[width=14.3cm]{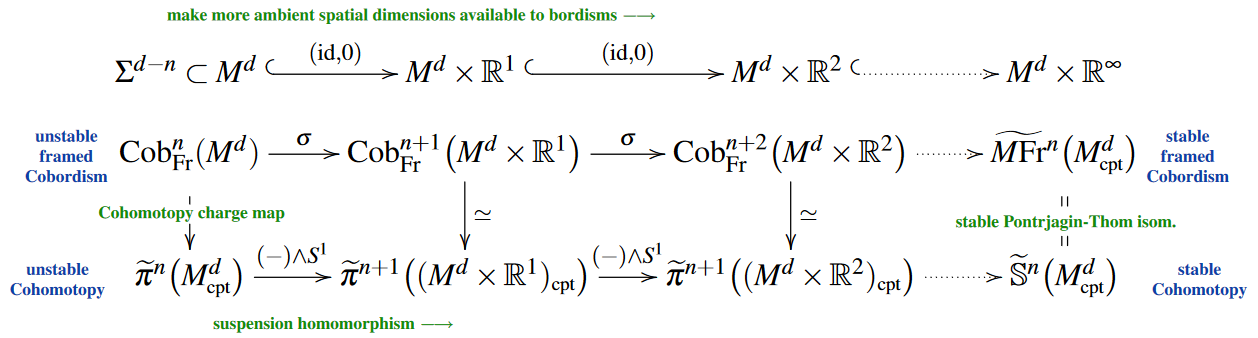}
}\hspace{-5pt}
\\[-10pt]
&
\\
\hline
\begin{minipage}{4cm}
  \small 
  This ``linearized'' Cohomotopy/Cobordism is a {\bf form of K-theory}: algebraic K-theory over the ``absolute base field $\mathbb{F}_1$'' (cf. \cite[Thm. 5.9]{ChuLorscheidSanthanam12}\cite[Cor. 2.25]{BeardleyNakamura24}).
\end{minipage}
&
  \hspace{10pt}
  \begin{tikzcd}[
    column sep=33pt
  ]
    \overset{
      \mathclap{
        \raisebox{4pt}{
          \scalebox{.7}{  
            \color{darkblue}
            \bf
            \def\arraystretch{.9}
            \begin{tabular}{c}
              non-abelian
              \\
              Cohomotopy
            \end{tabular}
          }
        }
      }
    }{
      \pi^\bullet
    }
    \ar[
      rr,
      "{
        \scalebox{.7}{
          \color{darkgreen}
          \bf
          linearize
        }
      }",
      "{
        \scalebox{.7}{
          (i.e.: stabilize)
        }
      }"{swap},
    ]
    &&
    \overset{
      \mathclap{
        \raisebox{4pt}{
          \scalebox{.7}{  
            \color{darkblue}
            \bf
            \def\arraystretch{.9}
            \begin{tabular}{c}
              stable
              \\
              Cohomotopy
            \end{tabular}
          }
        }
      }
    }{
      \mathbb{S}^\bullet
    }
    \ar[
      dr,
      equals,
      "{
        \scalebox{.65}{
          \color{darkgreen}
          \bf
          \def\arraystretch{.9}
          \begin{tabular}{c}
            Barratt-Priddy
            \\
            \& Quillen 
          \end{tabular}
        }
      }"{swap, sloped}
    ]
    \ar[
      rr,
      equals,
      "{
        \scalebox{.65}{
          \color{darkgreen}
          \bf
          Pontrjagin \& Thom
        }
      }"
    ]
    &\phantom{AAAA}&
    \overset{
      \mathclap{
        \raisebox{4pt}{
          \scalebox{.7}{  
            \color{darkblue}
            \bf
            \def\arraystretch{.9}
            \begin{tabular}{c}
              stable framed
              \\
              Cobordism
            \end{tabular}
          }
        }
      }
    }{
      \mathrm{MFr}^\bullet
    }
    \ar[
      dl,
      equals
    ]
    \\
    &&&
    \underset{
      \mathclap{
        \scalebox{.7}{
          \color{darkblue}
          \bf
          \def\arraystretch{.9}
          \begin{tabular}{c}
            algebraic K-theory of
            \\
            ``field with one element''
          \end{tabular}
        }
      }
    }{
      K\mathbb{F}_1^{\,\,\bullet}
    }
  \end{tikzcd}
  \hspace{.3cm}
  \adjustbox{
    scale=.85,
    raise=-3pt
  }{
  \begin{minipage}{5.2cm}
    Thus flux quantization in Cohomotopy lifts to M-theory the same arguments that motivated topological K-theory in type II string theory: its character map reproduces the Bianchi identities \& its equivalence relation models (anti-)brane pair-creation/annihilation.
  \end{minipage}
 \!\!\! }
\\
\hline
\end{tabular}

\newpage\hypertarget{ModuliOfSolitonsFramedLinks}{}

\noindent
{\bf Moduli space of soliton  configurations.}
But the Pontrjagin theorem concerns only the total cohomotopical charge, identifying it with the {\it net} (anti-)brane content. 
Beyond that we have the whole
{\it moduli space} of charges
   (considered now specialized to our 2D transverse space), and {\bf Segal's theorem} \cite{Segal73}
 says that the cohomotopy charge map (scanning map)
identifies this with a moduli space of brane positions, namely with the 
{\it group-completed configuration space of points} \cite{Cohen09}\cite{Williams20}\cite{Kallel25}:

\adjustbox{raise=2pt}{
$
  \begin{tikzcd}[
    row sep=3pt, column sep=60pt
  ]
  \scalebox{.7}{
    \color{darkblue}
    \bf
    \def\arraystretch{.9}
    \begin{tabular}{c}
      Moduli space
      \\
      of solitonic 
      \\
      brane 
      \color{purple} charges
    \end{tabular}
  }
  &[-56pt]
  \mathrm{Maps}^{\ast}\big(
    \mathbb{R}^2_{\cpt}
    ,\,
    S^2
  \big)
  \ar[
    rr,
    "{
      [-]
    }",
    "{
      \scalebox{.7}{
        \color{darkgreen}
        \bf
        net charge
      }
    }"{swap, yshift=-1pt}
  ]
  \ar[
    from=dd,
    shorten=-4pt,
    "{ \sim }"{swap,sloped},
    "{
      \scalebox{.7}{
        \color{olive}
        \bf
        \def\arraystretch{.9}
        \begin{tabular}{c}
          Segal
          \\
          theorem
        \end{tabular}
      }
    }"{xshift=4pt}
  ]
  &&
  \overset{
    \mathclap{
      \adjustbox{
        scale=.7,
        raise=0pt
      }{
        \color{darkblue}
        \bf
        \def\arraystretch{.85}
        \begin{tabular}{c}
          2-Cohomotopy 
        \end{tabular}
      }
    }
  }{
  \pi^2\big(
    \mathbb{R}^2_{\cpt}
  \big)
  }
  \ar[
    dr,
    "{ \sim }"{sloped, swap},
    "{
      \scalebox{.7}{
        \color{olive}
        \bf
        \def\arraystretch{.9}
        \begin{tabular}{c}
          Hopf degree
          \\
          theorem
        \end{tabular}
      }
    }"{sloped}
  ]
  \ar[
    from=dd,
    "{ \sim }"{swap,sloped},
    "{
      \scalebox{.7}{
        \color{olive}
        \bf
        \def\arraystretch{.9}
        \begin{tabular}{c}
          Pontrjagin 
          \\
          theorem
        \end{tabular}
      }
    }"{xshift=4pt}
  ]
  \\
  &
  &&
  &
  \mathbb{Z}
  \\
  \scalebox{.7}{
    \color{darkblue}
    \bf
    \def\arraystretch{.9}
    \begin{tabular}{c}
      Moduli space
      \\
      of solitonic 
      \\
      brane 
      \color{purple} positions
    \end{tabular}
  }
  &
  \underset{
    \mathclap{
      \hspace{0pt}
      \adjustbox{
        scale=.7,
        rotate=-22,
        raise=-3pt,
      }
      {
        \color{darkblue}
        \bf
       \rlap{
         \hspace{-27pt}
           \def\arraystretch{.8}
           \begin{tabular}{c}
           group-
           \\
           completed
           \end{tabular}
       }
      }
   }
  }{
    \mathbb{G}
  }
  \,
  \underset{
    \mathclap{
      \hspace{18pt}
      \adjustbox{
        scale=.7,
        rotate=-22,
        raise=-3pt,
      }
      {
        \color{darkblue}
        \bf
       \rlap{
         \hspace{-30pt}
         config space
       }
      }
   }
  }{
  \mathrm{Conf}(\mathbb{R}^2)
  }
  \ar[
    rr,
    "{ [-] }",
    "{
      \scalebox{.7}{
        \color{darkgreen}
        \bf
        net brane content
      }
    }"{swap, yshift=-1pt}
  ]
  &&
  \underset{
    \mathclap{
      \adjustbox{
        scale=.7,
        raise=-2pt
      }{
        \color{darkblue}
        \bf
        \def\arraystretch{.9}
        \begin{tabular}{c}
          2-Cobordism
          \\
          (unstable)
        \end{tabular}
      }
    }
  }{
  \mathrm{Cob}_{\mathrm{Fr}}^2(
  \mathbb{R}^2
  )
  }
  \ar[
    ur,
    "{ \sim }"{sloped}
  ]
  \end{tikzcd}
$
}

\vspace{.3cm}

\noindent
  where the {\it configuration space of points} is the space of finite subsets of $\mathbb{R}^2$ -- here understood as the space of positions of cores of solitons of unit charge $+1$, 
  \vspace{1mm} 
$$
\mathrm{Conf}(\mathbb{R}^2)
\;\;
=
\;\;
\left\{
\hspace{-1.5cm}
\adjustbox{scale=1.2, raise=-.8cm}{
\begin{tikzpicture}

\vortex{2.1}{.51}

\vortex{2}{.17}

\vortex{4.1}{.53}

\vortex{2}{.17}

\vortex{3}{.25}

\vortex{4.1}{.53}

\draw[white, line width=3]
  (3.2,0) -- (5.2,.7);
\draw[
  draw=lightgray,
  fill=lightgray, 
  fill opacity=.4]
  (.25,0)
    --
  (3.2,0)
    --
  (5.2,.7)
   --
  (2.25,.7)
    --
  (.25,0);

\draw[white,line width=3pt]
  (.25,0)
    --
  (3.2,0);
\draw[gray, line width=1pt]
  (.25,0)
    --
  (3.2,0);

\draw[line width=1pt, gray, dashed]
  (3.2+.6,0)
    --
  (3.2,0);

\node[rotate=19]
  at (.47,.24) {
    \scalebox{.5}{
      \color{gray}
      \bf
      \hspace{60pt} 
      transverse plane
    }
  };

\node
  at (3,.48) {
    \scalebox{.5}{
      \color{darkorange}
      \bf
      \def\arraystretch{.8}
      \begin{tabular}{c}
        positions of
        \\
        soliton cores
      \end{tabular}
    }
  };

\draw (2.1,.51) -- (2.1,.51-.05);  

\draw (2,.17) -- (2,.17-.05);

\draw (4.1,.53) -- (4.1,.53-.05);

\draw (2,.17) -- (2,.17-.05);

\draw (3,.25) -- (3,.25-.05);

\draw (4.1,.53) -- (4.1,.53-.05);

\end{tikzpicture}
}
\right\}
$$

\vspace{.3cm}
\noindent   and its  {\it group completion} $\mathbb{G}(-)$ 
  is the topological completion of the topological partial monoid structure given by disjoint union of soliton configurations. 

  \smallskip 
  Na{\"i}vely this is given by also including {\bf anti-solitons} in the form of configurations of {\it $\pm$-charged points}, topologized such as to allow for their pair annihilation/creation as shown in the left column on the right.

\smallskip 
  Remarkably, closer analysis reveals 
  \cite{Okuyama05}
  that the group completion $\mathbb{G}(-)$ produces configurations of {\bf strings}
  (extending parallel to one axis in $\mathbb{R}^3$) {\bf with charged endpoints} whose pair annihilation/creation is smeared-out to string worldsheets as shown in the right column (\cite[Fig. 2]{SS24-AbAnyons}):

\begin{center}
\begin{minipage}{6cm}
  \footnotesize
  {\bf Figure \figurenumber: The continuous relations in configuration spaces of charged points and strings} exhibit the pair annihiliation/creation of oppositely charged (end-)points.
\end{minipage}
\;\;
\adjustbox{
  scale=.74,
  raise=-.15cm
}
{
\def\arraystretch{1.3}
\begin{tabular}{|c|c|}
\hline
\multicolumn{2}{|c|}{\bf Configurations of charged}
\\
\bf points & \bf strings
\\
\hline
\hline
&
\\[-10pt]
\adjustbox{raise=2cm}
{
\begin{tikzpicture}[decoration=snake]

\draw[line width=1pt,draw=gray,fill=black]
  (+.7,0) circle (.2);
\draw[
  line width=1pt,draw=gray,fill=white,
  fill opacity=.5
]
  (-0.7,0) circle (.2);

\draw[
  decorate,
  ->
] (0,-.3) -- (0,-1);

\draw[line width=1pt,draw=gray,fill=black]
  (.2,-1.4) circle (.2);
\draw[
  line width=1pt,draw=gray,fill=white,
  fill opacity=.5
]
  (0,-1.4) circle (.2);

\begin{scope}[shift={(0,-1.5)}]
\draw[
  decorate,
  ->
] (0,-.3) -- (0,-1);
\end{scope}

\node at (0,-2.8)
  {$
    \varnothing
  $};

\end{tikzpicture}
}

&

\begin{tikzpicture}[decoration=snake]

\begin{scope}[shift={(2,-2)}]

\begin{scope}[shift={(0,0)}]
\draw[line width=2pt, gray]
  (0,0) -- (1,0);
\draw[line width=1pt,draw=gray,fill=white]
  (0,0) circle (.2);
\draw[line width=1pt,draw=gray,fill=white]
  (1,0) circle (.2);
\end{scope}

\begin{scope}[shift={(1.6,0)}]
\draw[line width=2pt, gray]
  (0,0) -- (1,0);
\draw[line width=1pt,draw=gray,fill=black]
  (0,0) circle (.2);
\draw[line width=1pt,draw=gray,fill=black]
  (1,0) circle (.2);
\end{scope}

\begin{scope}[shift={(1.6,-1.3)}]
\draw[line width=2pt, gray]
  (-.05,0) -- (1,0);
\draw[line width=1pt,draw=gray,fill=black]
  (-.2,0) circle (.2);
\draw[line width=1pt,draw=gray,fill=black]
  (1,0) circle (.2);
\end{scope}

\begin{scope}[shift={(0,-1.3)}]
\draw[line width=2pt, gray]
  (0,0) -- (1,0);
\draw[line width=1pt,draw=gray,fill=white]
  (0,0) circle (.2);
\draw[line width=1pt,draw=gray,fill=white, fill opacity=.5]
  (1.2,0) circle (.2);
\end{scope}

\draw[
  decorate,
  ->
] (1.3,-.3) -- (1.3,-1);

\draw[
  decorate,
  ->
] (1.3,-1.7) -- (1.3,-2.4);

\begin{scope}[shift={(0,-1.2)}]
\draw[
  decorate,
  ->
] (1.3,-1.7) -- (1.3,-2.4);
\end{scope}

\begin{scope}[shift={(0,-2.4)}]
\draw[
  decorate,
  ->
] (1.3,-1.7) -- (1.3,-2.4);
\end{scope}

\begin{scope}[shift={(0,-3.9)}]
\draw[
  decorate,
  ->
] (1.3,-1.7) -- (1.3,-2.4);
\end{scope}

\begin{scope}[shift={(0,-2.6)}]
\draw[line width=2pt, gray]
  (0,0) -- (2.6,0);
\draw[line width=1pt,draw=gray,fill=white]
  (0,0) circle (.2);
\draw[line width=1pt,draw=gray,fill=black]
  (2.6,0) circle (.2);
\end{scope}

\begin{scope}[shift={(0,-3.9)}]
\draw[line width=2pt, gray]
  (.6,0) -- (2,0);
\draw[line width=1pt,draw=gray,fill=white]
  (.6,0) circle (.2);
\draw[line width=1pt,draw=gray,fill=black]
  (2,0) circle (.2);
\end{scope}

\begin{scope}[shift={(.3,-5.2)}]
\draw[line width=1pt,draw=gray,fill=black]
  (1.07,0) circle (.2);
\draw[line width=1pt,draw=gray,fill=white, fill opacity=.5]
  (.93,0) circle (.2);
\end{scope}

\draw (1.3, -6.6) node {$\varnothing$};

\end{scope}
  
\end{tikzpicture}
\\
\hline
\multicolumn{2}{|c|}{
  \bf
  \hspace{-1.4cm}
  tracing out
}
\\
\bf
worldlines & \bf worldsheets
\\
\hline
&
\\[-8pt]
\adjustbox{
  raise=1.5cm
}{
\begin{tikzpicture}[decoration=snake]

\draw[line width=1pt,draw=gray,fill=black]
  (+.7,0) circle (.2);
\draw[
  line width=1pt,draw=gray,fill=white,
  fill opacity=.5
]
  (-0.7,0) circle (.2);

\begin{scope}[shift={(0,.5)}]
\draw[line width=1pt,draw=gray,fill=black]
  (.2,-1.4) circle (.2);
\draw[
  line width=1pt,draw=gray,fill=white,
  fill opacity=.5
]
  (0,-1.4) circle (.2);
\end{scope}

\node at (0.1,-2)
  {$
    \varnothing
  $};

\draw[gray]
  (-.7,0) -- (0.1,-1.2); 
\draw[gray]
  (+.71,0) -- (0.1,-1.2); 
\draw[
  gray,
  dashed
]
  (0.1,-1.2) -- (0.1,-1.8); 

\end{tikzpicture}
}
&
\begin{tikzpicture}

\begin{scope}[
  shift={(2,-2)}
]

\begin{scope}[shift={(0,0)}]
\draw[line width=2pt, gray]
  (0,0) -- (1,0);
\draw[line width=1pt,draw=gray,fill=white]
  (0,0) circle (.2);
\draw[line width=1pt,draw=gray,fill=white]
  (1,0) circle (.2);
\end{scope}

\begin{scope}[shift={(2,0)}]
\draw[line width=2pt, gray]
  (0,0) -- (1,0);
\draw[line width=1pt,draw=gray,fill=black]
  (0,0) circle (.2);
\draw[line width=1pt,draw=gray,fill=black]
  (1,0) circle (.2);
\end{scope}

\begin{scope}[shift={(2,-.8)}]
\draw[line width=2pt, gray]
  (-.2,0) -- (1,0);
\draw[line width=1pt,draw=gray,fill=black]
  (-.4,0) circle (.2);
\draw[line width=1pt,draw=gray,fill=black]
  (1,0) circle (.2);
\end{scope}

\begin{scope}[shift={(0,-.8)}]
\draw[line width=2pt, gray]
  (0,0) -- (1.2,0);
\draw[line width=1pt,draw=gray,fill=white]
  (0,0) circle (.2);
\draw[line width=1pt,draw=gray,fill=white, fill opacity=.5]
  (1.4,0) circle (.2);
\end{scope}

\begin{scope}[shift={(0,-1.6)}]
\draw[line width=2pt, gray]
  (0,0) -- (2.8,0);
\draw[line width=1pt,draw=gray,fill=white]
  (0.2,0) circle (.2);
\draw[line width=1pt,draw=gray,fill=black]
  (2.8,0) circle (.2);
\end{scope}

\begin{scope}[shift={(.15,-2.4)}]
\draw[line width=2pt, gray]
  (.5,0) -- (2,0);
\draw[line width=1pt,draw=gray,fill=white]
  (.5,0) circle (.2);
\draw[line width=1pt,draw=gray,fill=black]
  (2.2,0) circle (.2);
\end{scope}

\begin{scope}[shift={(.5,-3.2)}]
\draw[line width=1pt,draw=gray,fill=black]
  (1.07,0) circle (.2);
\draw[line width=1pt,draw=gray,fill=white, fill opacity=.5]
  (.93,0) circle (.2);
\end{scope}

\draw (1.5, -4) node {$\varnothing$};

\end{scope}

\draw[
  gray,
  smooth,
  fill=gray,
  fill opacity=.3,
  draw opacity=.3
]
  plot 
  coordinates{
    (2,-1.06) 
    (2,-2.8)
    (2.2,-3.6)
    (2.65,-4.4)
    (3.5,-5.3)
    (4.35,-4.4)
    (4.8,-3.6)
    (5,-2.8)
    (5,-1.06)
  }
  -- (4.05,-1.06)
  plot 
  coordinates {
    (4.05,-1.06)
    (4,-2)
    (3.6, -2.8)
    (3.4, -2.8)
    (3,-2.1)
    (2.95,-1.06)
  }
  --(2,-1.06);

\draw[
  line width=2pt,
  white
]
  (1.9,-1.5) -- (5.1,-1.5);
\draw[
  line width=2pt,
  white
]
  (1.9,-1.34) -- (5.1,-1.34);
\draw[
  line width=2pt,
  white
]
  (1.9,-1.18) -- (5.1,-1.18);
  
\end{tikzpicture}
\\
\hline
\end{tabular}
}
\end{center}

  \medskip

  This means (cf. \cite[Prop. 3.14]{SS24-AbAnyons})
  that the {\bf vacuum-to-vacuum soliton scattering processes},  
  forming the loop space $\Omega \, \mathbb{G}\, \mathrm{Conf}(\mathbb{R}^2)$,
  are identified with {\it framed links}
  (\cite[p 15]{Ohtsuki01}); for instance:
  
  \vspace{-.2cm}
  \begin{center}
  \begin{minipage}{6cm}
    \footnotesize
    {\bf Figure \figurenumber: 
    Loops in the configuration space of charged strings} in the plane may be identified with (diagrams for) framed oriented links. 
  \end{minipage}
  \;\;\;
  \adjustbox{
    raise=-3cm
  }{
  \includegraphics[width=10cm]{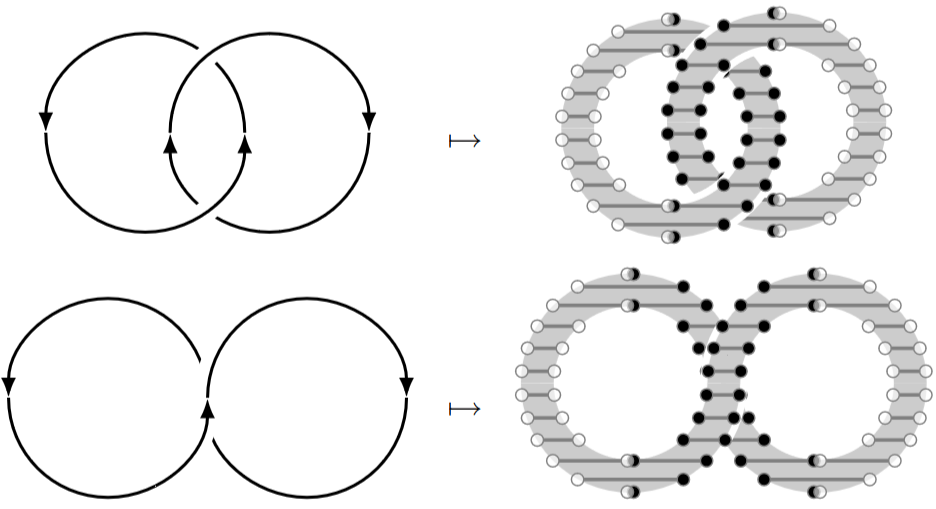}
  }
  \end{center}
  \vspace{-.3cm}
  subject to {\it link cobordism} (cf. \cite{Lobb24}):

\begin{center}
\adjustbox{
  raise=-2.5cm,
  margin=-6pt,
  fbox
}{
\hspace{-4pt}\includegraphics[width=9.2cm]{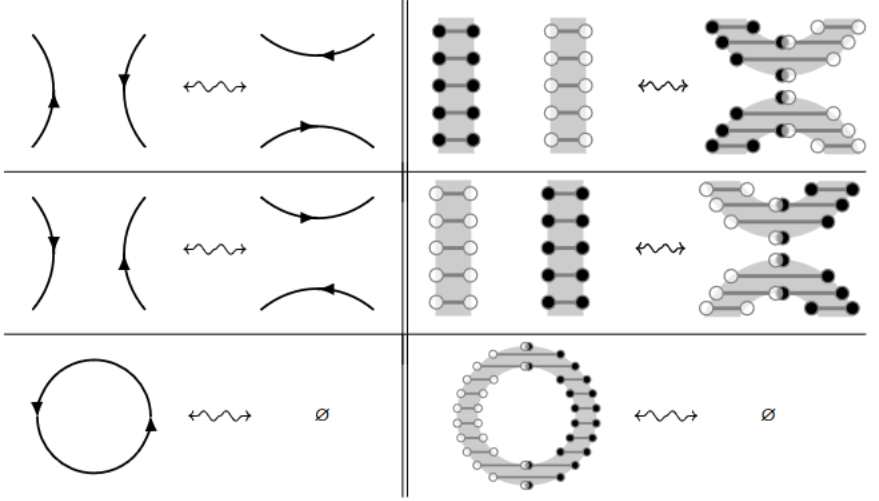}\hspace{-4pt}
}
\hspace{.3cm}
\begin{minipage}{5cm}
  \footnotesize 
  {\bf Figure \figurenumber:
  Link cobordism from deformations of charged string moduli.}
  Shown on the right are evident continuous deformations of paths in the above configuration space of charged strings, hence in the group-completed configuration space of points. Shown on the left are the local deformations of corresponding framed links, generating the relation of {\it link cobordism}.
\end{minipage}
\end{center}

It follows \cite[Thm 3.17]{SS24-AbAnyons}
that the 
charge of a soliton scattering process $L$ is the sum over crossings of the {\it crossing number}
$$
\adjustbox{
  scale=.8
}{
$
\#\left(\!
\adjustbox{raise=-.63cm, scale=.5}{
\begin{tikzpicture}

\draw[
  line width=1.2,
  -Latex
]
  (-.7,-.7) -- (.7,.7);
\draw[
  line width=7,
  white
]
  (+.7,-.7) -- (-.7,.7);
\draw[
  line width=1.2,
  -Latex
]
  (+.7,-.7) -- (-.7,.7);
 
\end{tikzpicture}
}
\!\right)
\,=\,
+1
\,,
\hspace{.2cm}
\#\left(\!
\adjustbox{raise=-.63cm, scale=0.5}{
\begin{tikzpicture}[xscale=-1]

\draw[
  line width=1.2,
  -Latex
]
  (-.7,-.7) -- (.7,.7);
\draw[
  line width=7,
  white
]
  (+.7,-.7) -- (-.7,.7);
\draw[
  line width=1.2,
  -Latex
]
  (+.7,-.7) -- (-.7,.7);
 
\end{tikzpicture}
}
\!\right)
\,=\,
-1\,,
$
}
$$
which equals the linking+framing number: 

\begin{center}
\begin{tabular}{c}
\rowcolor{lightolive}
$
  \begin{tikzcd}[
    column sep=35pt,
    row sep=0pt,
  ]
    \Omega
    \mathbb{G}
    \mathrm{Conf}(\mathbb{R}^2)
    \ar[
      r,
      shorten=-2pt,
      "{ \sim }"{yshift=2pt}
    ]
    &
    \Omega\mathrm{Maps}^{\!\ast/}\!\big(
      \mathbb{R}^2_{\cpt}
      S^2
    \big)
    \ar[
     r,
      shorten=-2pt,
     "{ [-] }"
    ]
    &
    \pi_3(S^2)
    \simeq
    \mathbb{Z}
    \\
    L 
    \ar[
      rr,
      |->,
      "{
        \scalebox{.7}{
          \color{darkgreen}
          \bf
          total crossing number =
        }
      }",
      "{
        \scalebox{.7}{
          \color{darkgreen}
          \bf
          linking + framing number
        }
      }"{swap}
    ]
    &&
    \# L
  \end{tikzcd}
$
\end{tabular}
\end{center} 
\vspace{.1cm}

\noindent But this turns out to be  precisely the  {\bf Wilson loop observable} of $L$ {\bf in} (abelian) 
{\bf Chern-Simons theory}!
This is due to \cite[\S 3]{SS24-AbAnyons}, and we proceed to review how this comes about, culminating in \eqref{KnotTheoreticWilsonLoopObservables} below.

\medskip 
\noindent
{\bf The $k$-Soliton sector.}
More generally, we may consider loops based in the $\mathcolor{purple}{k}$th connected component of the moduli space (cf. \cite[Rem. 3.20]{SS24-AbAnyons}), corresponding to scattering process from $k$ to $k$ net number of solitons.
\vspace{-1mm} 
$$
  \begin{tikzcd}[
    row sep=10pt
  ]
    \mathclap{
      \scalebox{.7}{
        \color{darkblue}
        \bf
        net charge $k$
      }
    }
    &
    \mathclap{
      \scalebox{.7}{
        \color{darkblue}
        \bf
        Hopf degree $k$
      }
    }
    \\[-13pt]
    \mathbb{G}
    \mathrm{Conf}_{
      \mathcolor{purple}{k}
    }
    (\mathbb{R}^2)
    \ar[
      r,
      "{ \sim }"
    ]
    \ar[
      d,
      hook
    ]
    &
    \mathrm{Maps}^\ast_{
      \mathcolor{purple}{k}
    }
    \big(
      \mathbb{R}^2_{\cpt}
      ,\,
      S^2
    \big)
    \ar[
      d,
      shorten=-1pt,
      hook
    ]
    \\
    \mathbb{G}
    \mathrm{Conf}(\mathbb{R}^2)
    \ar[
      r,
      "{ \sim }"
    ]
    &
    \mathrm{Maps}^\ast
    \big(
      \mathbb{R}^2_{\cpt}
      ,\,
      S^2
    \big)
  \end{tikzcd}
$$
Since the double loop space $\mathrm{Maps}^{\ast}\big(\mathbb{R}^2_{\cpt}, S^2\big)$ admits the structure of a topological group,
all these connected components have the same homotopy type, and hence these scattering processes $L$ are again classified by the integer total crossing number $\# L$ which is the abelian Chern-Simons Wilson-loop observable.
$$
\adjustbox{
  raise=4pt
}{
$
  \begin{tikzcd}[
    row sep=10pt, column sep=large
  ]
    \Omega_{\mathcolor{purple}{k}}
    \,
    \mathbb{G}\mathrm{Conf}(\mathbb{R}^2)
    \ar[
      d,
      ->>
    ]
    \ar[
      drr,
      end anchor={
        [yshift=3pt]
      },
      "{
        L 
        \;\;\mapsto\;\;
        \# L
      }"{sloped}
    ]
    \\
    \pi_0
    \Omega_{\mathcolor{purple}{k}}
    \,
    \mathbb{G}\mathrm{Conf}(\mathbb{R}^2)
    \ar[
      rr,
      "{ \sim }"{swap}
    ]
    &&
    \mathbb{Z}\;.
  \end{tikzcd}
$
}
$$

\noindent
For instance, a generic $\mathcolor{purple}{k=3}$ process looks like this:

$$
\adjustbox{
      raise=-1.5cm, 
      scale=0.75
    }{
\begin{tikzpicture}

\begin{scope}[shift={(-2.36,0)}]
\draw[dashed, line width=2]
  (0,0) ellipse (.6 and 1.8);
\draw[white,fill=white]
  (.2,   1.44) rectangle 
  (.2+1,-1.5);
\end{scope}

\begin{scope}[shift={(2-2.36,0)}]
\draw[dashed, line width=2]
  (0,0) ellipse (.6 and 1.8);
\draw[white,fill=white]
  (.2,   1.44) rectangle 
  (.2+1,-1.5);
\end{scope}

\begin{scope}[shift={(4-2.36,0)}]
\draw[dashed, line width=2]
  (0,0) ellipse (.6 and 1.8);
\draw[white,fill=white]
  (.2,   1.44) rectangle 
  (.2+1,-1.5);
\end{scope}

\draw[
  line width=7,
  white
]
  (0,-1.5)
  .. controls (0,0) and (2,0) ..
  (2,+1.5);
\draw[
  line width=1.8,
  -Latex
]
  (0,-1.5)
  .. controls (0,0) and (2,0) ..
  (2,+1.5);

\begin{scope}[shift={(0,-.03)}]
\draw[
  line width=8,
  white
]
  (2,-1.5) 
    .. controls (2,0) and (-2,0) ..
  (-2,+1.5);
\end{scope}
\begin{scope}
\clip
  (-1.5,.8) rectangle (0,-.8);
\begin{scope}[shift={(0,-.07)}]
\draw[
  line width=8,
  white
]
  (2,-1.5) 
    .. controls (2,0) and (-2,0) ..
  (-2,+1.5);
\end{scope}
\end{scope}

\draw[
  line width=1.8,
  -Latex
]
  (2,-1.5) 
    .. controls (2,0) and (-2,0) ..
  (-2,+1.5);

\begin{scope}[shift={(.1,0)}]
\draw[
  line width=8,
  white
]
  (-2,-1.5)
    .. controls (-2,0) and (0,0) ..
  (0,+1.5);
\end{scope}
\draw[
  line width=1.8,
  -Latex
]
  (-2,-1.5)
    .. controls (-2,0) and (0,0) ..
  (0,+1.5);

\begin{scope}[shift={(.8,.65)}]
\begin{scope}[shift={(0,-.04)}]
\draw[line width=8, white]
  (403:1.3 and .3) arc 
  (403:472:1.4 and .3);
\end{scope}
\draw[line width=1.8]
  (403:1.3 and .3) arc 
  (403:472:1.4 and .3);
\draw[line width=7, white]
  (141:1.3 and .3) arc 
  (141:390:1.4 and .3);
\draw[line width=1.8]
  (141:1.3 and .3) arc 
  (141:390:1.4 and .3);
\draw[line width=1.8,-Latex]
  (250:1.3 and .3) arc 
  (250:264:1.4 and .3);
\end{scope}

\begin{scope}[shift={(-2,-.5)}, rotate=-40]

\begin{scope}[shift={(.23,0)}]
\draw[line width=5, white]
  (180:.35) arc 
  (180:274:.35); 
\draw[line width=1.8]
  (180:.35) arc 
  (180:274:.35); 
\draw[line width=1.8]
  (315:.35) arc 
  (315:360:.35); 
\end{scope}

\begin{scope}[shift={(-.23,0)}]
\draw[line width=5,white]
  (0:.35) arc 
  (0:180:.35); 
\draw[line width=1.8]
  (0:.35) arc 
  (0:180:.35); 
\end{scope}

\begin{scope}[shift={(.23,0)}]
\draw[line width=5, white]
  (0:.35) arc 
  (0:180:.35); 
\draw[line width=1.8]
  (0:.35) arc 
  (0:180:.35); 
\end{scope}

\begin{scope}[shift={(-.23,0)}]
\draw[line width=5, white]
  (180:.35) arc 
  (180:360:.35); 
\draw[line width=1.8]
  (180:.35) arc 
  (180:360:.35); 
\end{scope}

\end{scope}

\end{tikzpicture}      
    }
$$

\vspace{-.1cm}

\noindent
and via the framed cobordism moves

\vspace{-2mm}

\begin{center}
\adjustbox{scale=0.7}{
\def\tabcolsep{9pt}
\begin{tabular}{ccccccccc}
&&&&
\\[-8pt]
\adjustbox
{
  scale=.7,
  raise=-1.4cm
}{
\hspace{-16pt}
\begin{tikzpicture}

\draw[
  line width=1.8,
  -Latex
]
  (-1.6,-1.6) -- (+1.6,+1.6);

\draw[
  line width=15,
  white
]
  (+1.6,-1.6) -- (-1.6,+1.6);
\draw[
  line width=1.8,
  -Latex
]
  (+1.6,-1.6) -- (-1.6,+1.6);
  
\end{tikzpicture}}
&
\hspace{-7pt}
\adjustbox{
  raise=-5pt,
  scale=.8
}{
\begin{tikzpicture}[decoration=snake]
  \draw[decorate, ->]
    (0,0) -- (+.55,0);
  \draw[decorate, ->]
    (0,0) -- (-.55,0);
\end{tikzpicture}
}
\hspace{-7pt}
&
\adjustbox{
  scale=.7,
  raise=-1.3cm
}
{
\begin{tikzpicture}

\draw[
  line width=1.8,
  -Latex
]
  (-1.7,-1.7) -- (0,0);

\draw[
  -Latex,
  line width=1.8,
]
  (+1.7,-1.7) -- (-1.7,1.7);

\draw[
  white,fill=white
]
  (-1.5,-1.5) rectangle (1.5,1.5);

\draw[line width=1.8]
  (-1.5,-1.5)
   .. controls
     (-.8,-.8) and (-.8,+.8) ..
  (-1.5,+1.5);

\begin{scope}[shift={(-.42,0)}]
\draw[
  line width=1.8
]
  (45:.3) arc (45:360-45:.3);
\end{scope}

\draw[line width=1.8]
  (+1.6,-1.6) -- (-.2,.2);
\draw[line width=1.8, -Latex]
  (+.2,+.2) -- (1.6,1.6);

\draw[white,fill=white]
  (-1.1,-.2) rectangle (-.5,+.2);
\clip
  (-1.1,-.2) rectangle (-.5,+.2);

\draw[line width=1.8]
  (-.82,.24) circle (.17);

\draw[line width=1.8]
  (-.81,-.24) circle (.18);
\end{tikzpicture}
}
&
\hspace{-7pt}
\adjustbox{
  raise=-5pt,
  scale=.8
}{
\begin{tikzpicture}[decoration=snake]
  \draw[decorate, ->]
    (0,0) -- (+.55,0);
  \draw[decorate, ->]
    (0,0) -- (-.55,0);
\end{tikzpicture}
}
\hspace{-7pt}
&
\adjustbox{
  scale=.7,
  raise=-1.3cm
}
{
\begin{tikzpicture}

\draw[
  line width=1.8,
  -Latex
]
  (-1.7,-1.7) -- (0,0);

\draw[
  line width=1.8,
  -Latex, 
]
  (+1.7,-1.7) -- (-1.7,+1.7);

\draw[
  white,fill=white
]
  (-1.5,-1.5) rectangle (1.5,1.5);

\draw[line width=1.8]
  (-1.5,-1.5)
   .. controls
     (-.8,-.8) and (-.8,+.8) ..
  (-1.5,+1.5);

\begin{scope}[shift={(-.42,0)}]
\draw[
  line width=1.85
]
  (45:.3) arc (45:360-45:.3);
\end{scope}

\draw[line width=1.8]
  (+1.7,-1.7) -- (-.2,.2);
\draw[line width=1.8, -Latex]
  (+.2,+.2) -- (1.7,1.7);
 
\end{tikzpicture}
}
&
\hspace{-7pt}
\adjustbox{
  raise=-5pt,
  scale=.8
}{
\begin{tikzpicture}[decoration=snake]
  \draw[decorate, ->]
    (0,0) -- (+.55,0);
  \draw[decorate, ->]
    (0,0) -- (-.55,0);
\end{tikzpicture}
}
\hspace{-7pt}
&
\adjustbox{
  scale=.7,
  raise=-1.3cm
}
{
\begin{tikzpicture}

\draw[
  line width=1.8,
]
  (-1.7,-1.7) -- (-1.5,-1.5);
\draw[
  line width=1.8,
  -Latex
]
  (-1.5,+1.5) -- (-1.7,+1.7);

\draw[
  line width=1.8,
]
  (+1.7,-1.7) -- (+1.5,-1.5);
\draw[line width=1.8, -Latex]
  (+1.5,+1.5) -- (1.7,1.7);

\draw[
  line width=1.8,
  -Latex
]
  (+.25,-.25) -- (-.25,+.25);

\draw[line width=1.8]
  (-1.5,-1.5)
   .. controls
     (-.8,-.8) and (-.8,+.8) ..
  (-1.5,+1.5);

\begin{scope}[shift={(-.42,0)}]
\draw[
  line width=1.85
]
  (45:.3) arc (45:360-45:.3);
\end{scope}

\begin{scope}[
  shift={(+.42,0)},
  xscale=-1
]
\draw[
  line width=1.85
]
  (45:.3) arc (45:360-45:.3);
\end{scope}

\begin{scope}[
  xscale=-1
]

\draw[line width=1.8]
  (-1.5,-1.5)
   .. controls
     (-.8,-.8) and (-.8,+.8) ..
  (-1.5,+1.5);

\draw[white,fill=white]
  (-1.1,-.2) rectangle (-.5,+.2);
\clip
  (-1.1,-.2) rectangle (-.5,+.2);

\draw[line width=1.8]
  (-.82,.24) circle (.17);

\draw[line width=1.8]
  (-.81,-.24) circle (.18);

\end{scope}
 
\end{tikzpicture}
}
&
\hspace{-7pt}
\adjustbox{
  raise=-5pt,
  scale=.8
}{
\begin{tikzpicture}[decoration=snake]
  \draw[decorate, ->]
    (0,0) -- (+.55,0);
  \draw[decorate, ->]
    (0,0) -- (-.55,0);
\end{tikzpicture}
}
\hspace{-7pt}
&
\adjustbox{
  scale=.7,
  raise=-1.3cm
}
{
\begin{tikzpicture}

\draw[
  line width=1.8,
]
  (-1.7,-1.7) -- (-1.5,-1.5);
\draw[
  line width=1.8,
  -Latex
]
  (-1.5,+1.5) -- (-1.7,+1.7);

\draw[
  line width=1.8,
]
  (+1.7,-1.7) -- (+1.5,-1.5);
\draw[line width=1.8, -Latex]
  (+1.5,+1.5) -- (1.7,1.7);

\draw[
  line width=1.8,
  -Latex
]
  (+.25,-.25) -- (-.25,+.25);

\draw[line width=1.8]
  (-1.5,-1.5)
   .. controls
     (-.8,-.8) and (-.8,+.8) ..
  (-1.5,+1.5);

\begin{scope}[shift={(-.42,0)}]
\draw[
  line width=1.85
]
  (45:.3) arc (45:360-45:.3);
\end{scope}

\begin{scope}[
  shift={(+.42,0)},
  xscale=-1
]
\draw[
  line width=1.85
]
  (45:.3) arc (45:360-45:.3);
\end{scope}

\begin{scope}[
  xscale=-1
]

\draw[line width=1.8]
  (-1.5,-1.5)
   .. controls
     (-.8,-.8) and (-.8,+.8) ..
  (-1.5,+1.5);

\end{scope}
 
\end{tikzpicture}
}
\\[-8pt]
&&&&
\end{tabular}
}

\end{center}

\vspace{-2mm} 
\noindent
it computes to the trivial scattering process accompanied by $\# L$ vacuum pair braiding processes:

\vspace{-2mm}

\begin{center}
\adjustbox{
  raise=-1.7cm, scale=0.7
}{
\begin{tikzpicture}

\begin{scope}[shift={(-2.36,0)}]
\draw[dashed, line width=2]
  (0,0) ellipse (.6 and 1.8);
\draw[white,fill=white]
  (.2,   1.44) rectangle 
  (.2+1,-1.5);
\end{scope}

\begin{scope}[shift={(2-2.36,0)}]
\draw[dashed, line width=2]
  (0,0) ellipse (.6 and 1.8);
\draw[white,fill=white]
  (.2,   1.44) rectangle 
  (.2+1,-1.5);
\end{scope}

\begin{scope}[shift={(4-2.36,0)}]
\draw[dashed, line width=2]
  (0,0) ellipse (.6 and 1.8);
\draw[white,fill=white]
  (.2,   1.44) rectangle 
  (.2+1,-1.5);
\end{scope}

\draw[
  line width=7,
  white
]
  (0,-1.5)
  .. controls (0,0) and (2,0) ..
  (2,+1.5);
\draw[
  line width=1.8,
  -Latex
]
  (0,-1.5)
  .. controls (0,0) and (2,0) ..
  (2,+1.5);

\begin{scope}[shift={(0,-.03)}]
\draw[
  line width=8,
  white
]
  (2,-1.5) 
    .. controls (2,0) and (-2,0) ..
  (-2,+1.5);
\end{scope}
\begin{scope}
\clip
  (-1.5,.8) rectangle (0,-.8);
\begin{scope}[shift={(0,-.07)}]
\draw[
  line width=8,
  white
]
  (2,-1.5) 
    .. controls (2,0) and (-2,0) ..
  (-2,+1.5);
\end{scope}
\end{scope}

\draw[
  line width=1.8,
  -Latex
]
  (2,-1.5) 
    .. controls (2,0) and (-2,0) ..
  (-2,+1.5);

\begin{scope}[shift={(.1,0)}]
\draw[
  line width=8,
  white
]
  (-2,-1.5)
    .. controls (-2,0) and (0,0) ..
  (0,+1.5);
\end{scope}
\draw[
  line width=1.8,
  -Latex
]
  (-2,-1.5)
    .. controls (-2,0) and (0,0) ..
  (0,+1.5);

\end{tikzpicture}
\hspace{+.4cm}
\adjustbox{raise=1.5cm}{
\begin{tikzpicture}[decoration=snake]
  \draw[decorate, ->]
    (0,0) -- (+.55,0);
  \draw[decorate, ->]
    (0,0) -- (-.55,0);
\end{tikzpicture}
}
\hspace{.9cm}
\begin{tikzpicture}

\begin{scope}[shift={(-2.36,0)}]
\draw[dashed, line width=2]
  (0,0) ellipse (.6 and 1.8);
\draw[white,fill=white]
  (.2,   1.44) rectangle 
  (.2+1,-1.5);
\end{scope}

\begin{scope}[shift={(2-2.36,0)}]
\draw[dashed, line width=2]
  (0,0) ellipse (.6 and 1.8);
\draw[white,fill=white]
  (.2,   1.44) rectangle 
  (.2+1,-1.5);
\end{scope}

\begin{scope}[shift={(4-2.36,0)}]
\draw[dashed, line width=2]
  (0,0) ellipse (.6 and 1.8);
\draw[white,fill=white]
  (.2,   1.44) rectangle 
  (.2+1,-1.5);
\end{scope}

\draw[
  line width=1.8,
  -Latex
]
  (-2,-1.5) -- (-2,+1.5);
\draw[
  line width=1.8,
  -Latex
]
  (0,-1.5) -- (0,+1.5);
\draw[
  line width=1.8,
  -Latex
]
  (+2,-1.5) -- (+2,+1.5);

\begin{scope}[
  shift={(3.7,-.7)}
]
\begin{scope}[shift={(-.42,0)}]
\draw[
  line width=1.8
]
  (45:.3) arc (45:360-45:.3);
\end{scope}

\begin{scope}[
  shift={(+.42,0)},
  xscale=-1
]
\draw[
  line width=1.8
]
  (45:.3) arc (45:360-45:.3);
\end{scope}

\draw[line width=1.8, -Latex]
  (+.22,-.22) -- (-.27,+.27);

\end{scope}

\begin{scope}[
  shift={(3.7,+.5)},
  xscale=-1
]
\begin{scope}[shift={(-.42,0)}]
\draw[
  line width=1.8
]
  (45:.3) arc (45:360-45:.3);
\end{scope}

\begin{scope}[
  shift={(+.42,0)},
  xscale=-1
]
\draw[
  line width=1.8
]
  (45:.3) arc (45:360-45:.3);
\end{scope}

\draw[line width=1.8, -Latex]
  (+.22,-.22) -- (-.27,+.27);

\end{scope}

\end{tikzpicture}
}

\end{center}

\medskip

\noindent
{\bf Chern-Simons level.}
We will see below further meanings of the number $lattice$:
\begin{center}
  \adjustbox{
    margin=4pt,
    bgcolor=lightolive
  }{
  This integer $\mathcolor{purple}{\lattice}$
  is equivalently
  $
  \left\{\!\!
  \adjustbox{}{
  \def\tabcolsep{-3pt}
  \begin{tabular}{l}
  the {\it number} of fractional quasi-hole vortices
  in a quantum Hall system,
  \\
  twice the 
  {\it level}
  of their effective abelian Chern-Simons theory,
  \\
  the {\it maximal denominator} for filling fractions 
  of their quantum states.
  \end{tabular}
  }
  \right.
  $
  }
\end{center}

\noindent
Generally, we will recover in a novel {\it non-Lagrangian} way
the features of quantum Chern-Simons theory that are traditionally argued starting with the $k$th multiple of the local Lagrangian density $a \wedge \mathrm{d}a$ for a gauge potential 1-form $a$.

\medskip

\noindent
{\bf The situation on the 2-Sphere.}
Furthermore, consider $\lattice$ solitons on the actual 2-sphere $S^2$. Here,
the 2-Cohomotopy moduli space satisfies (cf. \cite{Hansen74}):

$$
  \pi_0
  \Omega_{\mathcolor{purple}{\lattice}}
  \,
  \mathrm{Maps}\big(
    S^2
    ,\,
    S^2
  \big)
  \;\simeq\;
  \mathbb{Z}_{2\vert 
    \mathcolor{purple}{\lattice} 
  \vert}
  \,,
$$

\vspace{1pt}

\noindent
and the long homotopy fiber sequence induced by point evaluation shows that the generator of this cyclic group is again identified with the basic half-braiding operation:
\vspace{-3mm} 
$$
  \begin{tikzcd}[
    row sep=2pt
  ]
  &
  \mathrm{Maps}^\ast\big(
    \mathbb{R}^2_{\cpt},
    S^2
  \big)
  \ar[
    r,
    "{   
      \scalebox{.7}{
        \color{darkgreen}
        \bf
        fiber of...
      }
    }"
  ]
  &
  \mathrm{Maps}\big(
    S^2,
    S^2
  \big)  
  \ar[
    r,
    "{ 
      \scalebox{.7}{
        \color{darkgreen}
        \bf
        \def\arraystretch{.9}
        \begin{tabular}{c}
          point-
          \\
          evaluation
        \end{tabular}
      }
    }"
  ]
  &
  S^2
  \\
  \grayunderbrace{
    \pi_2\big(S^2\big)
  }{
    \mathcolor{black}{\mathbb{Z}}
  }
  \ar[
    r,
    "{
      2 
      \mathcolor{purple}{\lattice}
    }"
  ]
  &
  \grayunderbrace{
  \pi_0
  \Omega_{\mathcolor{purple}{\lattice}}
  \mathrm{Maps}^\ast\big(
    \mathbb{R}^2_{\cpt},
    S^2
  \big)
  }{
    \mathcolor{black}{
      \mathbb{Z}
    }
  }
  \ar[
    r,
  ]
  &
  \grayunderbrace{
  \pi_0
  \Omega_{\mathcolor{purple}{\lattice}}
  \mathrm{Maps}\big(
    S^2,
    S^2
  \big)  
  }{
   \mathbb{Z}_{2\vert 
     \mathcolor{purple}{\lattice} 
   \vert}
  }
  \ar[
    r,
  ]
  &
  \grayunderbrace{
  \pi_1\big(S^2\big)
  }{
    \mathcolor{black}{1}
  }
  \\[-10pt]
  & 
\adjustbox{
  raise=.2cm, 
  scale=.5}{
\begin{tikzpicture}

\draw[
  line width=1.2,
  -Latex
]
  (-.7,-.7) -- (.7,.7);
\draw[
  line width=7,
  white
]
  (+.7,-.7) -- (-.7,.7);
\draw[
  line width=1.2,
  -Latex
]
  (+.7,-.7) -- (-.7,.7);
 
\end{tikzpicture}
}
 \ar[
   r,
   |->,
   shorten=20pt
 ]
  &
\left[
\adjustbox{
  raise=.2cm, 
  scale=.5}{
\begin{tikzpicture}

\draw[
  line width=1.2,
  -Latex
]
  (-.7,-.7) -- (.7,.7);
\draw[
  line width=7,
  white
]
  (+.7,-.7) -- (-.7,.7);
\draw[
  line width=1.2,
  -Latex
]
  (+.7,-.7) -- (-.7,.7);
 
\end{tikzpicture}
}
\right]
\end{tikzcd}
$$

With flux-quantized fields being equipped with a classifying space $\hotype{A}$, there is a neat way to directly obtain the topological quantum observables -- via the following observation:

\vspace{.1cm}

\noindent
{\bf Topological flux observables in Yang-Mills theory -- \bf Theorem} \cite[\S 1]{SS24-QObs}. For $G$-Yang-Mills theory on $\mathbb{R}^{1,1} \times \Sigma^2$, 
with a choice of $\mathrm{Ad}$-invariant lattice $\Lambda \subset \mathfrak{g}$:

\vspace{-1mm} 
\begin{itemize}
\item[{\bf (i)}] Non-perturbative quantization of the algebra of flux observables through the closed surface $\Sigma^2$ is given by the group $C^\ast$-algebra $\mathbb{C}[-]$ of the Fr{\'e}chet-Lie group of smooth maps $\Sigma^2 \to G \ltimes (\mathfrak{g}/\Lambda)$. 

\vspace{-1mm} 
\noindent
\item[{\bf (ii)}] The corresponding group algebra of topological observables (observing only the connected components of flux) coincides with the Pontrjagin homology algebra of pointed maps $(\mathbb{R}^1 \times \Sigma^2)_{\cpt} \xrightarrow{\quad} B\big( G \ltimes (\mathfrak{g}/\Lambda) \big)$:
\end{itemize}  
\noindent
\adjustbox{
  bgcolor=lightolive
}{
$
  \hspace{-.15cm}
  \begin{tikzcd}[
    column sep=12pt,
    decoration=snake
  ]
  \underset{
    \adjustbox{
      raise=-3pt,
      scale=.7
    }{
      \color{darkblue}
      \bf
      \def\arraystretch{.9}
      \begin{tabular}{c}
        Non-perturbative quantum 
        algebra of
        \\
        observables on flux through \scalebox{1.1}{$\Sigma^2$}
      \end{tabular}
    }  
  }{
  \mathbb{C}\Big[
  C^\infty\big(
    \Sigma^2
    ,\,
    G
  \big)
  \ltimes
  C^\infty\big(
    \Sigma^2
    ,\,
    (\mathfrak{g}/\Lambda)
  \big)
  \Big]
  }
  \ar[
    r,
    decorate,
    shorten <=-2pt,
    shorten >=-2pt,
    "{
      \pi_0
    }"{swap, pos=.3}
  ]
  &[-2pt]
  \underset{
    \adjustbox{
      raise=-3pt,
      scale=.7
    }{
      \color{darkblue}
      \bf
      \def\arraystretch{.9}
      \begin{tabular}{c}
        corresponding algebra of
        \\
        topological observables
      \end{tabular}
    }  
  }{
  \mathbb{C}\Big[
    H^0\big(
      \Sigma^2
      ;\,
      G
    \big)
    \ltimes
    H^1\big(
      \Sigma^2
      ;\,
      \Lambda
    \big)
  \Big]
  }
  \ar[
    r,
    phantom,
    "{ \simeq }"
  ]
  &[-12pt]
  \underset{
    \adjustbox{
      raise=-3pt,
      scale=.7
    }{
      \color{darkblue}
      \bf
      \def\arraystretch{.9}
      \begin{tabular}{c}
        Pontrjagin homology algebra
        of 
        \\moduli space
        of soliton charges 
      \end{tabular}
    }  
  }{
  H_0
  \Big(
  \mathrm{Maps}^\ast
  \scalebox{1.25}{$($}
    (\mathbb{R}^1
    \times
    \Sigma^2)_{\cpt}
    ,\,
    B(
      G \ltimes
      (\mathfrak{g}/\Lambda)
  \scalebox{1.25}{$)$}
  ;\,
  \mathbb{C}
  \Big)
  }
  \end{tikzcd}
$\hspace{-4pt}}

\vspace{2mm}

For example, in electromagnetism,
with $G = \mathrm{U}(1)$ 
and $\Lambda := \mathbb{Z} \xhookrightarrow{\;} \mathbb{R}$ this gives \cite[\S 2]{SS24-QObs}:

$$
\adjustbox{raise=2pt}{
$
  \mathbb{C}
  \Big[
  \grayunderbrace{
    H^1(\Sigma^2; \mathbb{Z})
  }{
    \scalebox{.7}{
      electric
    }
  }
  \times
  \grayunderbrace{
    H^1(\Sigma^2; \mathbb{Z})
  }{
    \scalebox{.7}{
      magnetic
    }
  }
  \Big]
  \simeq
  H_0
  \Big(
  \mathrm{Maps}^\ast
  \scalebox{1.25}{$($}
    (\mathbb{R}^1
    \times
    \Sigma^2)_{\cpt}
    ,\,
    \grayunderbrace{
    B \mathrm{U}(1)
    \times 
    B \mathrm{U}(1)
    }{
      \mathclap{
      \scalebox{.7}{
        \def\arraystretch{.9}
        \begin{tabular}{c}
          classifying space for
          \\
          Dirac flux quantization
        \end{tabular}
      }
      }
    }
  \scalebox{1.25}{$)$}
  ;\,
  \mathbb{C}
  \Big).
$}
$$

\noindent
This allows us to generalize \cite[\S 3,4]{SS24-QObs}:

\smallskip 
\noindent
{\bf Topological flux observables of any higher gauge theory.}
For a higher gauge theory flux-quantized in $\hotype{A}$-cohomology,
the quantum algebra of topological flux observables 
on a spacetime of the form
$\mathbb{R}^{1,1}\times \Sigma^{D-2}$
is the Pontrjagin homology algebra of the soliton moduli,
hence in $\mathrm{deg}=0$ is
the group algebra of
vacuum soliton processes 
``{\bf on the light-cone}'':
$$
\adjustbox{
  bgcolor=lightolive
}{
$
  \def\arraystretch{2}
  \begin{array}{ccc}
  \mathrm{Obs}_\bullet
  &:=&
  H_\bullet\Big(
    \mathrm{Maps}^\ast
    \scalebox{1.2}{$($}
      (\mathbb{R}^1
      \times 
      \Sigma^{D-2})_{\cpt}
      ,\,
      \hotype{A}
    \scalebox{1.2}{$)$}
    ;\,
    \mathbb{C}
  \Big)
  \\[-2pt]
  &\simeq&
  H_\bullet\Big(
    \Omega
    \,
    \mathrm{Maps}
    \scalebox{1.2}{$($}
      \Sigma^{D-2}
      ,\,
      \hotype{A}
    \scalebox{1.2}{$)$}
    ;\,
    \mathbb{C}
  \Big)
  \\[+3pt]
  \mathrm{Obs}_0
  &=&
  \mathbb{C}\Big[
    \pi_0
    \Omega
    \,
    \mathrm{Maps}\big(
      \Sigma^{D-2}
      ,\,
      \hotype{A}
    \big)
  \Big]
  \end{array}
$
}
$$

\vspace{1mm}
\noindent For this, note that the star-involution
is given by the {\it combination} of
complex conjugation (time reversal) and loop reversal (hence $x$-reversal), where
$
  \mathbb{R}^{1,1}
  \,\simeq\,
  \mathbb{R}\langle t, x \rangle
  \,,
$
and the  operator product is given 
by loop  concatenation:

\medskip 
\hspace{-.6cm}
\begin{tikzpicture}

\node at (0,0){
$
\left[
\adjustbox{
  raise=-2.8cm
}{
\begin{tikzpicture}

\begin{scope}[
  shift={(-2,-2)}
]

\draw[
  line width=1,
  -Latex,
  gray
]
  (0,-.5) --
  (0,5);

\node[
  color=gray
]
  at (-.25,4.5) 
 {$t$};

\draw[
  line width=1,
  -Latex,
  gray
]
  (-.5,0) --
  (6,0);

\node[gray] at (5.5,-.25) 
 {$x$};

\draw[
  line width=1,
  -Latex,
  gray
]
  (180+46:.5) --
  (46:6);

\node[gray] at (46:.6) 
 {\colorbox{white}{$\Sigma$}};

\end{scope}

\begin{scope}[
  yscale=.6,
  rotate=-60,
  scale=.6
]

\draw[
  line width=27,
  olive!10
]
  (41+90:2) arc 
  (41+90:360-41+90:2);

\begin{scope}[
  shift={(0,5.65)},
  yscale=-1,
]

\draw[
  line width=27,
  olive!10
]
  (45+90:2) arc 
  (45+90:360-45+90:2);
\end{scope}

\begin{scope}[
  shift={(-.4,1.4)}
]

\draw[
  line width=27,
  olive!10
]
 (+1.8,0) -- (+1.8-.9,.9);

\draw[
  line width=27,
  olive!10
]
 (1.8-1.9,1.9) -- (-1.1,2.9);

\draw[
  line width=27,
  olive!10
]
 (-1,0) -- (+1.9,2.9);
\end{scope}

%%%%%%%%%%%%%%%%%%%%%

\draw[
  line width=22,
  olive!20
]
  (41+90:2) arc 
  (41+90:360-41+90:2);

\begin{scope}[
  shift={(0,5.65)},
  yscale=-1,
]

\draw[
  line width=22,
  olive!20
]
  (45+90:2) arc 
  (45+90:360-45+90:2);
\end{scope}

\begin{scope}[
  shift={(-.4,1.4)}
]

\draw[
  line width=22,
  olive!20
]
 (+1.8,0) -- (+1.8-.9,.9);

\draw[
  line width=22,
  olive!20
]
 (1.8-1.9,1.9) -- (-1.1,2.9);

\draw[
  line width=22,
  olive!20
]
 (-1,0) -- (+1.9,2.9);
\end{scope}

%%%%%%%%%%%%%%%%%%%%%%%%%%%%

\draw[
  line width=17,
  olive!30
]
  (41+90:2) arc 
  (41+90:360-41+90:2);

\begin{scope}[
  shift={(0,5.65)},
  yscale=-1,
]

\draw[
  line width=17,
  olive!30
]
  (45+90:2) arc 
  (45+90:360-45+90:2);
\end{scope}

\begin{scope}[
  shift={(-.4,1.4)}
]

\draw[
  line width=17,
  olive!30
]
 (+1.8,0) -- (+1.8-.9,.9);

\draw[
  line width=17,
  olive!30
]
 (1.8-1.9,1.9) -- (-1.1,2.9);

\draw[
  line width=17,
  olive!30
]
 (-1,0) -- (+1.9,2.9);
\end{scope}

%%%%%%%%%%%%%%%%%%%%%%%%%%%%%%%%%%%%

\draw[
  line width=12,
  olive!40
]
  (41+90:2) arc 
  (41+90:360-41+90:2);

\begin{scope}[
  shift={(0,5.65)},
  yscale=-1,
]

\draw[
  line width=12,
  olive!40
]
  (45+90:2) arc 
  (45+90:360-45+90:2);
\end{scope}

\begin{scope}[
  shift={(-.4,1.4)}
]

\draw[
  line width=12,
  olive!40
]
 (+1.8,0) -- (+1.8-.9,.9);

\draw[
  line width=12,
  olive!40
]
 (1.8-1.9,1.9) -- (-1.1,2.9);

\draw[
  line width=12,
  olive!40
]
 (-1,0) -- (+1.9,2.9);
\end{scope}

%%%%%%%%%%%%%%%%%%%%%%%%%%%%%%%%%%

\draw[
  line width=7,
  olive!50
]
  (41+90:2) arc 
  (41+90:360-41+90:2);

\begin{scope}[
  shift={(0,5.65)},
  yscale=-1,
]

\draw[
  line width=7,
  olive!50
]
  (45+90:2) arc 
  (45+90:360-45+90:2);
\end{scope}

\begin{scope}[
  shift={(-.4,1.4)}
]

\draw[
  line width=7,
  olive!50
]
 (+1.8,0) -- (+1.8-.9,.9);

\draw[
  line width=7,
  olive!50
]
 (1.8-1.9,1.9) -- (-1.1,2.9);

\draw[
  line width=7,
  olive!50
]
 (-1,0) -- (+1.9,2.9);
\end{scope}

\end{scope}

\end{tikzpicture}
}
\right]
\;\;\;
\scalebox{3}{$\cdot$}
\;\;\;
\left[
\adjustbox{
  raise=-2.8cm
}{
\begin{tikzpicture}

\begin{scope}[
  shift={(-2,-2)}
]

\draw[
  line width=1,
  -Latex,
  gray
]
  (0,-.5) --
  (0,5);

\node[gray] at (-.25,4.5) 
 {$t$};

\draw[
  line width=1,
  -Latex,
  gray
]
  (-.5,0) --
  (6,0);

\node[gray] at (5.5,-.25) 
 {$x$};

\draw[
  line width=1,
  -Latex,
  gray
]
  (180+46:.5) --
  (46:6);

\node[gray] at (46:.6) 
 {\colorbox{white}{$\Sigma$}};

\end{scope}

\begin{scope}[
  rotate=+20,
  shift={(2,0)},
  yscale=.6,
  xscale=1,
]

\draw[
  line width=27,
  olive!10
]
  (0,0)
  circle
  (1.3);
\draw[
  line width=22,
  olive!20
]
  (0,0)
  circle
  (1.3);
\draw[
  line width=17,
  olive!30
]
  (0,0)
  circle
  (1.3);
\draw[
  line width=12,
  olive!40
]
  (0,0)
  circle
  (1.3);
\draw[
  line width=7,
  olive!50
]
  (0,0)
  circle
  (1.3);
  
\end{scope}

\begin{scope}[
  rotate=-30,
  shift={(.7,+.4)},
  yscale=.8,
  xscale=1.2,
]

\draw[
  line width=32,
  olive!0
]
  (0,0)
  circle
  (1.3);
\draw[
  line width=27,
  olive!10
]
  (0,0)
  circle
  (1.3);
\draw[
  line width=22,
  olive!20
]
  (0,0)
  circle
  (1.3);
\draw[
  line width=17,
  olive!30
]
  (0,0)
  circle
  (1.3);
\draw[
  line width=12,
  olive!40
]
  (0,0)
  circle
  (1.3);
\draw[
  line width=7,
  olive!50
]
  (0,0)
  circle
  (1.3);
  
\end{scope}

\end{tikzpicture}
}
\right]
$
};

\node at (0,-6.7)
{
$
\scalebox{1.8}{$=$}
\;\;\;
\left[
\adjustbox{
  raise=-2.9cm
}{
\begin{tikzpicture}

\begin{scope}[
  shift={(4,1.8)}
]

\begin{scope}[
  rotate=+20,
  shift={(2,0)},
  yscale=.6,
  xscale=1,
]

\draw[
  line width=27,
  olive!10
]
  (0,0)
  circle
  (1.3);
\draw[
  line width=22,
  olive!20
]
  (0,0)
  circle
  (1.3);
\draw[
  line width=17,
  olive!30
]
  (0,0)
  circle
  (1.3);
\draw[
  line width=12,
  olive!40
]
  (0,0)
  circle
  (1.3);
\draw[
  line width=7,
  olive!50
]
  (0,0)
  circle
  (1.3);
  
\end{scope}

\begin{scope}[
  rotate=-30,
  shift={(.7,+.4)},
  yscale=.8,
  xscale=1.2,
]

\draw[
  line width=32,
  olive!0
]
  (0,0)
  circle
  (1.3);
\draw[
  line width=27,
  olive!10
]
  (0,0)
  circle
  (1.3);
\draw[
  line width=22,
  olive!20
]
  (0,0)
  circle
  (1.3);
\draw[
  line width=17,
  olive!30
]
  (0,0)
  circle
  (1.3);
\draw[
  line width=12,
  olive!40
]
  (0,0)
  circle
  (1.3);
\draw[
  line width=7,
  olive!50
]
  (0,0)
  circle
  (1.3);
  
\end{scope}

\end{scope}

\begin{scope}[
  shift={(-2,-1.8)}
]

\draw[
  line width=1,
  -Latex,
  gray
]
  (0,-.5) --
  (0,5);

\node[
  color=gray
]
  at (-.25,4.5) 
 {$t$};

\draw[
  line width=1,
  -Latex,
  gray
]
  (-.5,0) --
  (6,0);

\node[gray] at (5.5,-.25) 
 {$x$};

\draw[
  line width=1,
  -Latex,
  gray
]
  (180+46:.5) --
  (46:6);

\node[gray] at (46:.6) 
 {\colorbox{white}{$\Sigma$}};

\end{scope}

\begin{scope}[
  yscale=.6,
  rotate=-60,
  scale=.6
]

\begin{scope}[
  shift={(0,5.65)},
  yscale=-1,
]

\draw[
  line width=32,
  white
]
  (45+90:2) arc 
  (45+90:360-45+90:2);

\end{scope}

\draw[
  line width=27,
  olive!10
]
  (41+90:2) arc 
  (41+90:360-41+90:2);

\begin{scope}[
  shift={(0,5.65)},
  yscale=-1,
]

\draw[
  line width=27,
  olive!10
]
  (45+90:2) arc 
  (45+90:360-45+90:2);
\end{scope}

\begin{scope}[
  shift={(-.4,1.4)}
]

\draw[
  line width=27,
  olive!10
]
 (+1.8,0) -- (+1.8-.9,.9);

\draw[
  line width=27,
  olive!10
]
 (1.8-1.9,1.9) -- (-1.1,2.9);

\draw[
  line width=27,
  olive!10
]
 (-1,0) -- (+1.9,2.9);
\end{scope}

%%%%%%%%%%%%%%%%%%%%%

\draw[
  line width=22,
  olive!20
]
  (41+90:2) arc 
  (41+90:360-41+90:2);

\begin{scope}[
  shift={(0,5.65)},
  yscale=-1,
]

\draw[
  line width=22,
  olive!20
]
  (45+90:2) arc 
  (45+90:360-45+90:2);
\end{scope}

\begin{scope}[
  shift={(-.4,1.4)}
]

\draw[
  line width=22,
  olive!20
]
 (+1.8,0) -- (+1.8-.9,.9);

\draw[
  line width=22,
  olive!20
]
 (1.8-1.9,1.9) -- (-1.1,2.9);

\draw[
  line width=22,
  olive!20
]
 (-1,0) -- (+1.9,2.9);
\end{scope}

%%%%%%%%%%%%%%%%%%%%%%%%%%%%

\draw[
  line width=17,
  olive!30
]
  (41+90:2) arc 
  (41+90:360-41+90:2);

\begin{scope}[
  shift={(0,5.65)},
  yscale=-1,
]

\draw[
  line width=17,
  olive!30
]
  (45+90:2) arc 
  (45+90:360-45+90:2);
\end{scope}

\begin{scope}[
  shift={(-.4,1.4)}
]

\draw[
  line width=17,
  olive!30
]
 (+1.8,0) -- (+1.8-.9,.9);

\draw[
  line width=17,
  olive!30
]
 (1.8-1.9,1.9) -- (-1.1,2.9);

\draw[
  line width=17,
  olive!30
]
 (-1,0) -- (+1.9,2.9);
\end{scope}

%%%%%%%%%%%%%%%%%%%%%%%%%%%%%%%%%%%%

\draw[
  line width=12,
  olive!40
]
  (41+90:2) arc 
  (41+90:360-41+90:2);

\begin{scope}[
  shift={(0,5.65)},
  yscale=-1,
]

\draw[
  line width=12,
  olive!40
]
  (45+90:2) arc 
  (45+90:360-45+90:2);
\end{scope}

\begin{scope}[
  shift={(-.4,1.4)}
]

\draw[
  line width=12,
  olive!40
]
 (+1.8,0) -- (+1.8-.9,.9);

\draw[
  line width=12,
  olive!40
]
 (1.8-1.9,1.9) -- (-1.1,2.9);

\draw[
  line width=12,
  olive!40
]
 (-1,0) -- (+1.9,2.9);
\end{scope}

%%%%%%%%%%%%%%%%%%%%%%%%%%%%%%%%%%

\draw[
  line width=7,
  olive!50
]
  (41+90:2) arc 
  (41+90:360-41+90:2);

\begin{scope}[
  shift={(0,5.65)},
  yscale=-1,
]

\draw[
  line width=7,
  olive!50
]
  (45+90:2) arc 
  (45+90:360-45+90:2);
\end{scope}

\begin{scope}[
  shift={(-.4,1.4)}
]

\draw[
  line width=7,
  olive!50
]
 (+1.8,0) -- (+1.8-.9,.9);

\draw[
  line width=7,
  olive!50
]
 (1.8-1.9,1.9) -- (-1.1,2.9);

\draw[
  line width=7,
  olive!50
]
 (-1,0) -- (+1.9,2.9);
\end{scope}

\end{scope}

\end{tikzpicture}
}
\right]
$
};

\node[
  scale=.8
] at (-7,-4.6) {
  \clap{
  \color{darkblue}
  \bf
  \def\arraystretch{.9}
  \begin{tabular}{c}
    topological classes 
    \\
    of
    \\[+5pt]
    vacuum-to-vacuum 
    \\[-2pt]
    processes of
    \\
    quantized flux
    \\
    along $t-x$
    \\[+8pt]
    and their 
    \\
    concatenation
  \end{tabular}
  }
};

\draw[
  -Latex,
  gray
]
  (-8.35,-3.4)
  .. controls 
  (-9.5,-3.5) and 
  (-9.5,-1.8) ..
  (-8.1,-1.3);

\draw[
  white,
  line width=3
]
 (-5.9,-4.4) .. controls
 (-3.8,-4) and
 (-3.8,-3) ..
 (-4.5,-1.1);
\draw[
  -Latex,
  gray
]
 (-5.9,-4.4) .. controls
 (-3.8,-4) and
 (-3.8,-3) ..
 (-4.5,-1.1);

\draw[
  white,
  line width=2pt
]
  (-5.8,-5.7) .. controls
  (-4,-5) and
  (-2,-5) .. 
  (1,-5.6);

\draw[
  -Latex,
  gray
]
  (-5.8,-5.7) .. controls
  (-4,-5) and
  (-2,-5) .. 
  (1,-5.6);

\end{tikzpicture}
\vspace{-,3cm}
\begin{center}
  \footnotesize
  {\bf Figure \figurenumber: Vacuum-to-vacuum processes of flux} and their consecutive evolution in light-cone time.
\end{center}

\newpage

%%%%%%%%%%%%%%%%%%%%%%%%%%
\section{The topological Quantum States}
\label{TheTopologicalQuantumStates}
%%%%%%%%%%%%%%%%%%%%%%%%%%

With the theory thus set up, we here turn to analyzing its predictions for topological quantum states and their topological order to be observed on closed surfaces. Remarkably, we find close agreement with the fine-detail of predictions of abelian Chern-Simons theory, even though the approach here is completely different (non-Lagrangian but properly flux-quantized). Details of the following discussion are spelled out in \cite[\S 3.1-4]{SS25-ViaAlgTop}.

\medskip

To summarize so far,
we have seen that the topological sector of the flux-quantized phase space of solitons on magnetized M5-probes $\Sigma$ wrapping Seifert orbi-singularities is

\vspace{-1mm} 
$$
  \begin{tikzcd}[
    row sep=0pt,
    column sep=8pt
  ]
  \mathrm{Maps}\!\left(
  \def\arraycolsep{0pt}
  \def\arraystretch{.9}
  \begin{array}{c}
    \Sigma
    \\
    \downarrow
    \\
    X
  \end{array}
  ,\,
  \def\arraycolsep{0pt}
  \def\arraystretch{.9}
  \begin{array}{c}
    \mathbb{C}P^3
    \\
    \downarrow
    \\
    S^4
  \end{array}
  \right)^{\!\!\ZTwo}
  \ar[
   r,
   phantom,
   "{ \simeq }"
  ]
  &
  \mathrm{Maps}^{\!\ast/}\!\big(
    \mathbb{R}^2_{\cpt}
    \wedge 
    S^1
    ,\,
    S^2
  \big)
  \ar[
    r,
    phantom,
    "{ \simeq }"
  ]
  &
  \Omega_0
  \,
  \mathbb{G}\mathrm{Conf}(\mathbb{R}^2)
  \ar[
    rr,
    "{ [-] }"
  ]  
  &&
  \pi_0\, 
  \Omega_0
  \,
  \mathbb{G}\mathrm{Conf}(\mathbb{R}^2)
  \ar[
    r,
    phantom,
    "{ \simeq }"
  ]
  &
  \mathbb{Z}
  \\[-16pt]
  &&
  L
  \ar[
    rrr,
    |->
  ]
  &&&
  \#L
  \\[-3pt]
  \mathclap{
    \scalebox{.7}{
      \color{darkblue}
      \bf
      \def\arraystretch{.9}
      \begin{tabular}{c}
        topological sector
        \\
        of flux-quantized
        \\
        phase space 
      \end{tabular}
    }
  }
  &
  \mathclap{
    \scalebox{.7}{
      \color{darkblue}
      \bf
      \def\arraystretch{.9}
      \begin{tabular}{c}
        2-Cohomotopy
        \\
        cocycle space
      \end{tabular}
    }
  }
  &
  \mathclap{
    \scalebox{.7}{
      \color{darkblue}
      \bf
      \def\arraystretch{.9}
      \begin{tabular}{c}
        loop space of
        \\
        group-completed
        \\
        configuration space
      \end{tabular}
    }
  }
  &
  &&
  \mathclap{
    \scalebox{.7}{
      \color{darkblue}
      \bf
      \def\arraystretch{.9}
      \begin{tabular}{c}
        net
        \\
        charge
      \end{tabular}
    }
  }
  \end{tikzcd}
$$

\noindent
{\bf The topological quantum states of this system} now follow 
\cite{SS24-QObs}\cite[\S 3]{SS24-AbAnyons}
by general algebraic quantum theory:

\vspace{.1cm}

\noindent
The gauge-invariant topological 
{\bf observables}
form the (higher) homology of this space

$$
  \def\arraystretch{.9}
  \begin{array}{l}
    \mathrm{Obs}_\bullet
    \;:=\;
    H_\bullet\big(
      \Omega_0
      \,
      \mathbb{G}
      \mathrm{Conf}(\mathbb{R}^2)
      ;\,
      \mathbb{C}
    \big)
  \end{array}
$$
 making a (star-)algebra under concatenation (reversion) of loops --- the {\it Pontrjagin algebra}.

\vspace{.1cm}

$$
\begin{tikzcd}[
  row sep=8pt,
  column sep=65pt
]
  \Omega_0
  \, \mathbb{G}\mathrm{Conf}(\mathbb{R}^2)
  \ar[
    rr,
    "{
      \mathrm{rev}
    }"{swap},
    "{
      \scalebox{.7}{
        \color{darkgreen}
        \bf
        loop reversal
      }
    }"
  ]
  &&
  \Omega_0
  \, \mathbb{G}\mathrm{Conf}(\mathbb{R}^2)  
  \\[+1pt]
  H_\bullet\big(
  \Omega_0
  \, \mathbb{G}\mathrm{Conf}(\mathbb{R}^2)    
  ;\,
  \mathbb{C}
  \big)
  \ar[
    drr,
    "{
      \scalebox{.7}{
        \color{darkgreen}
        \bf
        \def\arraystretch{.9}
        \begin{tabular}{c}
          Hermitian conjugation
          \\
          of quantum observables
        \end{tabular}
      }
    }"{swap, sloped, pos=.5}
  ]
  \ar[
    rr,
    "{
      \mathrm{rev}_\ast
    }"{swap},
    "{
      \scalebox{.7}{
        \color{darkgreen}
        \bf
        \def\arraystretch{.85}
        \begin{tabular}{c}
          Pontr.
          antipode
        \end{tabular}
      }
    }"
  ]
  &&
  H_\bullet\big(
  \Omega_0
  \, \mathbb{G}\mathrm{Conf}(\mathbb{R}^2)    
  ;\,
  \mathbb{C}
  \big)
  \ar[
    d,
    shorten=-3pt,
    "{
      \overline{(\mbox{-})}
    }"{swap},
    "{
      \scalebox{.7}{
        \color{darkgreen}
        \bf
        \def\arraystretch{.85}
        \begin{tabular}{c}
          cmplx
          \\
          cnjgtn
        \end{tabular}
      }
    }"{xshift=-3pt}
  ]
  \\[+10pt]
  &&
  H_\bullet\big(
  \Omega_0
  \, \mathbb{G}\mathrm{Conf}(\mathbb{R}^2)    
  ;\,
  \mathbb{C}
  \big)  
\end{tikzcd}
$$

\vspace{0.5mm}
\noindent  This means that time-reversal goes along with the 
reversal of looping around the M/IIA-circle, whence we are dealing with
  a version of 
  {\it discrete light-cone quantization}
  in their topological sectors.
\vspace{-2mm} 
\begin{center}
\begin{minipage}{6cm}
  \footnotesize
  {\bf Figure \figurenumber: Worldline of a particle travelling around a compact dimension.} In the limit of high momentum/boost the particle travels on a compactified light-cone.
\end{minipage}
\;\;
\adjustbox{
  raise=-1.2cm
}{
\includegraphics[width=2.5cm]{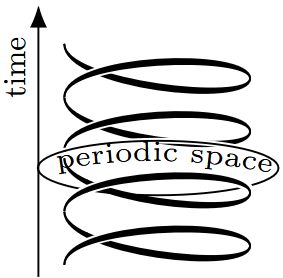}
}
\end{center}

The basic ordinary (degree=0) observables detect the deformation class of a framed link $L$. 
\begin{equation}
\label{BasicBraidObservables}
\adjustbox{
  raise=4pt,
  bgcolor=lightolive
}{
\hspace{-8pt}
$
  \begin{tikzcd}[
    row sep=-3pt,
    column sep=-6pt
  ]
    \mathrm{Obs}_0
    \ar[
      rr,
      "{ \sim }"
    ]
    &&
    \mathbb{C}\Big[
      \pi_0 
      \big(
        \Omega_0 
        \mathbb{G}
        \mathrm{Conf}(\mathbb{R}^2)
      \big)
    \Big]
    \ar[
      rr,
      "{ \sim }"
    ]
    &&
    \mathbb{C}[\mathbb{Z}]
    \\
    \mathcal{O}_L
    &:=&
    \delta_{[L]}
    &=&
    \delta_{\#\!L}
    \\
    \mathcal{O}_L
    \cdot
    \mathcal{O}_{L'}
    &=&
    \delta_{L \sqcup L'}
    &=&
    \delta_{\#\!L + \#\!L'}
  \end{tikzcd}
$
}
\end{equation}

Since these observables commute among each other, their {\it pure} topological quantum states are their 
(real \& positive) algebra homomorphisms:
$$
  \begin{tikzcd}[
    column sep=4pt,
    row sep=-2pt,
    /tikz/column 3/.append style={anchor=base west}
  ]
  \mathrm{PureQStates}_0
  \ar[
    d,
    hook
  ]
  &
  \underset{
    \mathclap{
      \hspace{-14pt}
      \adjustbox{
        scale=.7,
        raise=-8pt,
      }{
        \color{gray}
        \bf
        \def\arraystretch{.9}
        \begin{tabular}{c}
          on commuting
          \\
          observables
        \end{tabular}
      }
    }
  }{
    \simeq
  }
  &
  \Big\{
    \rho
    :
    \mathrm{Obs}_0
    \xrightarrow{
      \scalebox{.7}{
        \color{darkgreen}
        \bf
        homo
      }
    }
    \mathbb{C}
    \;\Big\vert\;
    \rho \in \mathrm{MixedQStates}_0
  \Big\}
  \\
  \mathrm{MixedQStates}_0
  &:=&
  \bigg\{
    \rho
    :
    \mathrm{Obs}_0
    \xrightarrow{
      \scalebox{.7}{
        \color{darkgreen}
        \bf
        linear
      }
    }
    \mathbb{C}
  \;\Big\vert\;
  \underset{
    \mathcal{O}
    \in \mathrm{Obs}_\bullet
  }{\forall}
  \Big(
    \underset{
      \mathclap{
        \raisebox{-1pt}{
          \scalebox{.7}{
            \color{darkblue}
            \bf
            reality
          }
        }
      }
    }{
    \rho\big(
      \mathcal{O}^\ast
    \big)
    \,=\,
    \rho(\mathcal{O})^\ast
    }
    ,\;
  \underset{
      \mathclap{
        \raisebox{-1pt}{
          \scalebox{.7}{
            \color{darkblue}
            \bf
            (semi-)positivity
          }
        }
      }  
  }{
  \rho(\mathcal{O}^\ast 
    \!\cdot\! 
  \mathcal{O})
  \;\geq 0\;
  \in
  \mathbb{R}
  \hookrightarrow
  \mathbb{C}
  }
  \Big)
  \,,
  \underset{
      \mathclap{
        \raisebox{-1pt}{
          \scalebox{.7}{
            \color{darkblue}
            \bf
            normalization
          }
        }
      }  
  }{
  \;\;
  \rho(1) = 1
  }\;\,
  \bigg\}
  \,.
  \end{tikzcd}
$$

\noindent
Therefore pure topological states $\vert \filling \rangle$ are determined by an 
{\bf anyonic phase} $\exp(\pi \mathrm{i}/\filling)$ 
assigned to any crossing,
$$
\adjustbox{
  scale=.55
}{
\adjustbox
{
  scale=.8,
  raise=-1.4cm
}{
\begin{tikzpicture}

\draw[
  line width=1.2,
  -Latex
]
  (-1.6,-1.6) -- (+1.6,+1.6);

\draw[
  line width=15,
  white
]
  (+1.6,-1.6) -- (-1.6,+1.6);
\draw[
  line width=1.2,
  -Latex
]
  (+1.6,-1.6) -- (-1.6,+1.6);
  
\end{tikzpicture}}
\hspace{-4pt}
\begin{tikzpicture}[decoration=snake]
  \draw[decorate,->]
    (0,0) -- (+.55,0);
  \draw[decorate,->]
    (0,0) -- (-.55,0);
\end{tikzpicture}
\hspace{-4pt}
\adjustbox{
  scale=.8,
  raise=-1.3cm
}
{
\begin{tikzpicture}

\draw[
  line width=1.2,
  -Latex
]
  (-1.7,-1.7) -- (0,0);

\draw[
  line width=1.2,
  -Latex, 
]
  (+1.7,-1.7) -- (-1.7,+1.7);

\draw[
  white,fill=white
]
  (-1.5,-1.5) rectangle (1.5,1.5);

\draw[line width=1.2]
  (-1.5,-1.5)
   .. controls
     (-.8,-.8) and (-.8,+.8) ..
  (-1.5,+1.5);

\begin{scope}[shift={(-.42,0)}]
\draw[
  line width=1.25
]
  (45:.3) arc (45:360-45:.3);
\end{scope}

\draw[line width=1.2]
  (+1.7,-1.7) -- (-.2,.2);
\draw[line width=1.2, -Latex]
  (+.2,+.2) -- (1.7,1.7);
 
\end{tikzpicture}
}
}
$$
accumulating to 
the exponentiated 
crossing number
$$
  \begin{tikzcd}[row sep=-3pt, column sep=small]
    \mathrm{Obs}_0
    \ar[
      rr,
      "{
        \langle \filling\vert
        - 
        \vert \filling \rangle
      }"
    ]
    &&
    \mathbb{C}
    \\
    \mathcal{O}_L
    &\longmapsto&
    e^{
      \tfrac{
        \pi \mathrm{i}
      }{
        \filling
      }
       \# L
    }
  \end{tikzcd}
$$

\noindent
{\bf The resulting expectation values}
\begin{equation}
  \label{KnotTheoreticWilsonLoopObservables}
  \adjustbox{
    margin=10pt,
    bgcolor=lightolive
  }{$
  \langle \filling \vert
  \mathcal{O}_L
  \vert \filling \rangle
  \;=\;
  \exp\big(
    \tfrac{
      \pi \mathrm{i}
    }{
      \filling
    }
    \,
    \# L
  \big)
  \;=\;
  \exp\bigg(
    \tfrac{
      \pi \mathrm{i}
    }{
      \filling
    }
    \Big(
      \displaystyle{
        \sum_{i \neq j \in \pi_0(L)}
      }
      \underset{
        \mathclap{
          \adjustbox{
            raise=-1pt,
            scale=.7
          }{
            \color{gray}
            \bf
            \def\arraystretch{.85}
            \begin{tabular}{c}
              linking
              \\
              numbers
            \end{tabular}
          }
        }
      }{
        \mathrm{lnk}(L_i, L_j)
      }
      \,+\,
      \displaystyle{
        \sum_{i \in \pi_0(L)}
      }
      \underset{
        \mathclap{
          \adjustbox{
            raise=-1pt,
            scale=.7
          }{
            \color{gray}
            \bf
            \def\arraystretch{.85}
            \begin{tabular}{c}
              framing
              \\
              numbers
            \end{tabular}
          }
        }
      }{
        \mathrm{frm}(L_i)
      }
    \Big)
  \bigg)
  $}
\end{equation}
are \cite[\S 3]{SS24-AbAnyons} just those of {\it Wilson loop observables} in ``spin'' {\it Chern-Simons theory}, as {\bf expected for abelian anyons}!
For example: 
\vspace{-1mm} 
$$
\Bigg\langle \filling\, \Bigg\vert
\adjustbox{
  raise=-2.6cm,
  scale=.4
}{
\begin{tikzpicture}
\foreach \n in {0,1,2} {
\begin{scope}[
  rotate=\n*120-4
]
\draw[
  line width=2,
  -Latex
]
 (0,-1)
   .. controls
   (-1,.2) and (-2,2) ..
 (0,2)
   .. controls
   (1,2) and (1,1) ..
  (.9,.7);
\end{scope}

\node[darkgreen]
  at (\n*120+31:.7) {
    \scalebox{1}{$+$}
  };
};
\end{tikzpicture}
}
\hspace{-7pt}
\Bigg\vert
\, \filling
\Bigg\rangle
\;\;
=
\;\;
\Bigg\langle \filling\, \Bigg\vert
\adjustbox{
  raise=-2.6cm,
  scale=.4
}{
\begin{tikzpicture}
\foreach \n in {0,1,2} {
\begin{scope}[
  rotate=\n*120-4
]
\draw[
  line width=2,
  -Latex
]
 (0,-1)
   .. controls
   (-1,.2) and (-2,2) ..
 (0,2)
   .. controls
   (1,2) and (1,1) ..
  (.9,.7);
\end{scope}
};

\draw[white,fill=white]
  (-.9,-.7)
  rectangle
  (.9,.2);

\draw[line width=2]
  (-.79,.21) 
    .. controls
    (-.7,-.25) and (+.7,-.25) ..
  (.79,.21);

\draw[line width=2]
  (-.27,-.7) 
  .. controls
    (-.4,-.2) and 
    (+.4,-.2) ..
  (+.27,-.7);

\draw[white,fill=white]
 (-.5,.8) 
 rectangle (+.5,-.2);

\draw[line width=2]
  (-.5,.58)
  .. controls
    (-.0,.7) and (-.0,-.26)
    ..
  (-.5,-.06);

\draw[line width=2]
  (+.5,.58)
  .. controls
    (+.0,.7) and (+.0,-.26)
    ..
  (+.5,-.06);

\end{tikzpicture}
}
\hspace{-7pt}
\Bigg\vert
\, \filling
\Bigg\rangle
\;\;
=
\;\;
\exp\big(
  \pi \mathrm{i}
  \,
  \tfrac{3}{\filling}
\big).
$$

\vspace{-.2cm}
\noindent 
Applying the GNS-construction to such state produces a 1-dimensional Hilbert space 
\begin{equation}
\label{1dHilbertSpaceOnSphere}
\grayoverbrace{\mathbb{C}[\theta,\theta^{-1}]}{\mathbb{C}[\mathbb{Z}]}
  \big/
\big( 
  e^{\pi \mathrm{i}/\filling} - \theta 
\big) \,\simeq\, \mathbb{C}
\,,
\end{equation}

\vspace{1mm} 
\noindent which is as expected for the quantum states of abelian Chern-Simons theory on $\mathbb{R}^2_{\cpt}$. (More on this on p \pageref{AnyonicTopologicalOrder}.)

\medskip

\noindent
{\bf Remark.} At this point, $m \in \mathbb{R} \neq 0$ may be irrational, but its rationality will be enforced by requiring compatibility with states on more general domain surfaces, see p. \pageref{AnyonicTopologicalOrder} and p. \pageref{Surfaces}.

\smallskip

\noindent
{\bf Remark.} These {\it solitonic} anyons are {\it not} yet the controllable/parameterized defect anyons that could be used for topological braid quantum gates operating by adiabatic movement of anyonic {\it defects} or (quasi-){\it holes}. But the latter arise as defect points among the former, we come to this on \hyperlink{DefectAnyons}{p. \pageref{DefectAnyons}}.

\smallskip

\noindent
{\bf Remark.} The appearance of framed links along just the above lines is known in the condensed matter theory of anyonic defect lines in the 3D ``8-band model'' (\cite[pp 15]{FreedmanHastingsNayakQiWalkerWang11}, following \cite{TeoKane10}): From this perspective, the Cohomotopy classifying space $S^2$ plays the role of the classifying space for electron  band Hamiltonians on a crystal lattice.

\medskip

\medskip\label{AnyonicTopologicalOrder}
\noindent
{\bf Anyonic topological order on Flux-quantized M5-probes.} We now identify the promised topological order on M5-probes flux-quantized in equivariant twistorial Cohomotopy, by considering M5s wrapping closed surfaces:

\medskip

\noindent
{\bf Anyonic quantum observables on closed surfaces.}
Consider now a {\it closed orientable surface} $\mathcolor{darkblue}{\Sigma^2_g}$ of genus $g \in \mathbb{N}$ to replace the previous factor $\mathbb{R}^2_{\cpt}$ in the brane diagram:

$$
\adjustbox{
  bgcolor=lightgray
}{
$
  \begin{tikzcd}[
    column sep=0pt,
    background color=lightgray
  ]
    \Sigma^{\mathrlap{1,5}}
    \ar[
      in=60,
      out=180-60,
      looseness=3.7,
      shift right=3pt,
      "{
        \!\mathbb{Z}_2\!
      }"{description}
    ]
    \;
    &:=&
    \mathbb{R}^{1,0}
    &\times& 
    \mathcolor{darkblue}{\Sigma^2_g}
    &\times& 
    S^1
    &\times&
    \mathbb{R}^2_{\mathrm{sgn}}
    \ar[
      in=60,
      out=180-60,
      looseness=3.5,
      shift right=3pt,
      "{
        \!\mathbb{Z}_2\!
      }"{description}
    ]
  \end{tikzcd}
$
}
$$

\noindent
Directly analogous analysis as before gives that the topological quantum observables on the flux-quantized self-dual tensor field form the group algebra of the fundamental group of the 2-cohomotopy moduli space in the $\lattice$th connected component
\begin{equation}
  \adjustbox{
    margin=5pt,
    bgcolor=lightolive
  }{$
  \mathrm{Obs}_0\big(\Sigma^2_g\big)
  \;:=\;
  H_0\Big(
    \Omega_{\lattice}
    \, \mathrm{Maps}\big(
      \Sigma^2_g
      ,\,
      S^2
    \big)
    ;\,
    \mathbb{C}
  \Big)
  \;\simeq\;
  \mathbb{C}
  \Big[
    \pi_0
    \Omega_{\lattice}
    \, \mathrm{Maps}\big(
      \Sigma^2_g
      ,\,
      S^2
    \big)
  \Big]
  $}
  \,,
\end{equation}
where $\lattice \in \mathbb{N}$ is the degree of the classifying maps,
corresponding under the Pontrjagin theorem to a net number of $\lattice$ (anti-)solitons on $\Sigma^2_g$.

\vspace{.1cm}

\noindent
{\bf Theorem} (using \cite[Thm 1]{Hansen74}\cite[Thm 1]{LarmoreThomas80}\cite[Cor 7.6]{Kallel01}). 
This group of 2-cohomotopy charge sectors is identified as
{\it twice} the integer Heisenberg group extension (cf. \cite{LeePacker96}) of $\mathbb{Z}^{2g}$ by $\mathbb{Z}_{2\vert \lattice\vert}$:
\footnote{Here $\mathbb{Z}_n := \mathbb{Z}/(n)$ (with $\mathbb{Z}_0 = \mathbb{Z}$) are the (in-)finite cyclic groups.}
\begin{center}
\adjustbox{
  margin=5pt,
  bgcolor=lightolive
}{
$
  \pi_0 \Omega_\lattice
  \mathrm{Maps}\big(
    \Sigma^2_g
    ,\,
    S^2
  \big)
  \;\simeq\;
  \left\{\!
    \big(
      \vec a
      ,\,
      \vec b
      ,\,
      [n]
    \big)
    \,\in\,
    \mathbb{Z}^g 
    \times
    \mathbb{Z}^g
    \times
    \mathbb{Z}_{2\vert \lattice \vert}
    \,,
    \def\arraystretch{1.4}
    \begin{array}{l}
      \big(
        \vec a
        ,\,
        \vec b
        ,\,
        [n]
      \big)
      \cdot
      \big(
        \vec a'
        ,\,
        \vec b'
        ,\,
        [n']
      \big)
      \;=\;
      \\
      \big(
        \vec a + \vec a'
        ,\,
        \vec b + \vec b'
        ,\,
        [
          n + n' 
            + 
          \mathcolor{purple}{
          \vec a \cdot \vec b'
          -
          \vec a' \cdot \vec b
          }
          \,
        ]
      \big)      
    \end{array}
  \!\! \right\}
  \;=:\;
  \widehat{\mathbb{Z}^{2g}}
$
}
\end{center}

%\medskip

\noindent {\bf Ground state degeneracy.}
Hence the observable group-algebra $\mathrm{Obs}_0$ 
for $g = 1$,  $\Sigma^2_1 = T^2$, 
has generators 
$$
  \left\{
  \def\arraystretch{1.2}
  \def\arraycolsep{2pt}
  \begin{array}{ccl}
    W_a &:=& (1,0,[0]) 
    \\
    W_b &:=& (0,1,[0]) 
    \\
    \zeta &:=& (0,0, [1]) 
  \end{array}
  \right\}
$$
subject to the relations
\vspace{1mm} 
$$
  \left\{\!
  \def\arraystretch{1.2}
  \begin{array}{l}
  W_a \cdot W_b
  \;=\;
  \zeta^{\mathcolor{purple}{2}} \,
  W_b \cdot W_a
  \\
  \zeta^{\mathcolor{purple}{2}\lattice} = 1
  \\
  {[\zeta, -] = 0}
  \end{array}
\!  \right\}
  \,.
$$

\vspace{.1cm}

\noindent
This algebra is just 
the observable algebra expected \cite[(5.28)]{Tong16} for anyonic topological order on the torus as described by abelian ``spin'' Chern-Simons theory at {\it lattice norm} $\lattice$ and {\it level} $\level = \lattice/2$. 

\begin{equation}
\adjustbox{}{
\def\arraystretch{1.6}
\def\tabcolsep{10pt}
\begin{tabular}{ccccc}
  \rowcolor{lightgray}
  & {\bf symbol} & {\bf in ordinary CS} & {\bf in ``spin'' CS}
  &
  $
    \exp\big(
      \tfrac{\mathrm{i}}{\hbar}
      S_{CS}
    \big)
    =
  $
  \\
  CS level 
  &
  $k$ & $\in \phantom{2}\mathbb{N}_{>0}$ & $\in \tfrac{1}{2}\mathbb{N}_{>0}$
  & 
  $e^{
    2 \pi \mathrm{i} 
    \, k \,
    \int\!\! A \, \mathrm{d}A
  }$
  \\
  CS lattice
  &
  $K \defneq 2k$ 
  & 
  $\in 2\mathbb{N}_{> 0}$
  &
  $\in \phantom{\tfrac{1}{2}}\mathbb{N}_{> 0}$
  &
  $e^{
    \pi \mathrm{i} 
    \, K \,
    \int\!\! A \, \mathrm{d}A
  }$
\end{tabular}
}
\end{equation}

The non-trivial irreps have:

-- dimension $\lattice$, this being the expected {\it ground state degeneracy} on the torus,

-- are labeled by 
\colorbox{lightolive}{$\nu := p/\lattice$}, $p \in \{1,2,\cdots,\lattice\}$, as expected for fractional 
{\it filling factors}.

\begin{center}
  \scalebox{.7}{
    \color{darkblue}
    \bf
    \def\arraystretch{.8}
    \begin{tabular}{c}
      Hilbert space of
      \\
      quantum states
      \\
      on the torus
    \end{tabular}
  }
\adjustbox{
  margin=5pt,
  bgcolor=lightolive
}{
$
  \HilbertSpace{H}_{T^2}
  \coloneqq
  \mathrm{Span}\Big(
    \big\vert [n] \big\rangle
    , [n] \in \mathbb{Z}_{\vert \lattice \vert}
  \Big)
  \;\in\;
  \mathrm{Obs}_0(T^2)
  \mathrm{Modules}
  \,,\;
  \mathrm{dim}\big(
    \HilbertSpace{H}_{T^2}
  \big)
  \;=\;
  \lattice
  \,,
$ 
}

\vspace{2mm}
\adjustbox{scale=1}{
$
  \def\arraystretch{1.5}
  \def\arraycolsep{2pt}
  \begin{array}{rcr}
  W_a \big\vert [n] \big\rangle
  &:=&
  e^{
    2 \pi \mathrm{i} 
    n \nu
  }
  \big\vert [n] \big\rangle
  \\
  W_b \big\vert [n] \big\rangle
  &:=&
  \big\vert [n+1] \big\rangle
  \\
  \zeta \big\vert [n] \big\rangle
  &:=&
  e^{
    \pi \mathrm{i}
    \nu
  }
  \big\vert [n] \big\rangle \;.
  \end{array}
$
}
\end{center}

\vspace{-.2cm}

\noindent

{\bf Modular equivariance.} Strikingly, in this construction modular symmetry {\bf is  manifest}, since the looped mapping space  is canonically acted on by the mapping class group $\mathrm{MCG}$ of $\Sigma^2_g$ (cf. \cite[\S 2.1]{FarbMargalit12}), simply by precomposition of maps!
Inspection of the above theorem (cf. \cite[bottom of p 153]{Hansen74}) shows that this MCG-action action identifies indeed as the canonical action of $\mathrm{Sp}_{2g}(\mathbb{Z})$ on $\widehat{\mathbb{Z}^{2g}}$.
$$
\adjustbox{
  raise=5pt
}{
$
  \begin{tikzcd}[
    row sep=12pt,
    column sep=0pt
  ]
  \grayoverbrace{
  \pi_0 \mathrm{Homeos}_{\mathrm{or}}\big(\Sigma^2_g\big)
  }{
    \mathrm{MCG}(\Sigma^2_g)
  }
  \ar[
    d,
    shorten=1pt,
    "{
      \scalebox{.7}{
        \color{gray}
        \cite[\S 6.3]{FarbMargalit12}
      }
    }"
  ]
  &\acts&
  \pi_0 \Omega_{\lattice} \, \mathrm{Maps}\big(\Sigma^2_g,\, S^2 \big)
  \ar[
    d,
    phantom,
    "{ \simeq }"{rotate=90}
  ]
  \\
  \mathrm{Sp}_{2g}(\mathbb{Z})
  &\acts&
  \widehat{ \mathbb{Z}^{2g} }
  \end{tikzcd}
$
}
$$

Hence, we may ask for a lift of the $\widehat{\mathbb{Z}^{2g}}$ action on quantum states to an action of the semidirect product $\widehat{\mathbb{Z}^{2g}} \rtimes \mathrm{Sp}_{2g}(\mathbb{Z})$.
For $g = 1$ and even $\lattice$
one readily checks
that this gives the modular
transformations 
of states known \cite[p 65]{Manoliu98} from abelian Chern-Simons theory:

\bigskip
$$
  \overset{
    \mathclap{
      \adjustbox{
        rotate=14,
        scale=.7
      }{
        \color{darkblue}
        \bf
        \rlap{
          \hspace{-10pt}
          \def\arraystretch{.9}
          \def\tabcolsep{-5pt}
          \begin{tabular}{l}
            modular
            \\
            action on observables
          \end{tabular}
        }
      }
    }
  }{
    m(W)
  }
  \,\cdot\,
  \overset{
    \mathclap{
      \adjustbox{
        rotate=14,
        scale=.7,
      }{
        \color{darkblue}
        \bf
        \rlap{
          \hspace{-15pt}
          and on states
        }
      }
    }
  }{
  m
  \scalebox{1.25}{$($}
    \big\vert
      [n]
    \big\rangle
  \scalebox{1.25}{$)$}
  }
  \;=\;
  m
  \Big(
    W
    \big\vert
      [n]
    \big\rangle
  \Big)
  \,,
  \;\;\;\;\;
  \forall
  \left\{
    \def\arraycolsep{2pt}
    \def\arraystretch{1.3}
    \begin{array}{ccc}
      m &\in& \mathrm{Sp}_{2g}(\mathbb{Z})
      \\[+3pt]
      W &\in& 
      \widehat{\mathbb{Z}^{2g}}
      \\
      \big\vert [n] \big\rangle 
        &\in& 
      \HilbertSpace{H}_{g}
    \end{array}
  \right.  
$$

\vspace{4pt}

$$
  \hspace{2cm}
  S\Big(
    \big\vert
      [n]
    \big\rangle
  \Big)
  \;=\;
  \tfrac{1}{\sqrt{\vert \lattice \vert }}
  \sum_{[\widehat{n}]}
  e^{
    2 \pi \mathrm{i} 
    \tfrac{
      n
      \, 
      \widehat{n}
    }{ \lattice }
  }
  \big\vert [\widehat n] \big\rangle
  \,,\;\;\;\;\;\;
  T\Big(
    \big\vert
      [n]
    \big\rangle
  \Big)
  \;=\;
  e^{
    \big(
      - \pi \mathrm{i} / 12
      +
      \mathrm{i} \pi
      \tfrac{ n^2 }{ \lattice }
    \big)
  }
  \big\vert
    [n]
  \big\rangle
  \,.
$$

\smallskip

Generally, writing $(\vec e_i \in \mathbb{Z}^g)_{1 = 1}^g$ for the canonical basis vectors,
the observable group-algebra $\mathrm{Obs}_0$ 
for general $g$ 
has generators 
$$
  \left\{
  \def\arraystretch{1.4}
  \def\arraycolsep{2pt}
  \begin{array}{ccl}
    W^i_a &:=& (\vec e_i,0,[0]) 
    \\
    W^i_b &:=& (0,\vec e_j,[0]) 
    \\
    \zeta &:=& (0,0, [1]) 
  \end{array}
  ,
  1 \leq i \leq g
  \right\}
$$
subject to  the relations
\vspace{3mm} 
$$
  \left\{
  \def\arraycolsep{1pt}
  \def\arraystretch{1.1}
  \begin{array}{l}
  W^i_a \cdot W^j_b
  \;=\;
  \delta^{i j} \zeta^{\mathcolor{purple}{2}} \,
  W^j_b \cdot W^i_a
  \\
  \zeta^{\mathcolor{purple}{2}\lattice} = 1
  \\
  \scalebox{.93}{
    all other commutators vanish
  }
  \end{array}
  \hspace{-4pt}
  \right\}.
$$

\vspace{.1cm}

\noindent
Requiring the reps $\HilbertSpace{H}_g$ of this algebra to analogously support modular equivariance requires them to  have dimension $\vert \lattice\vert^g$ --- which is the result expected \cite[p 40]{Manoliu98}
for abelian topological order on $\Sigma^2_g$:

\vspace{1mm}

\begin{center}
  \hspace{-1.3cm}
  \scalebox{.7}{
    \color{darkblue}
    \bf
    \def\arraystretch{.8}
    \begin{tabular}{c}
      Hilbert space of
      \\
      quantum states
      \\
      on genus=$g$ surface
    \end{tabular}
  }
\adjustbox{
  margin=5pt,
  bgcolor=lightolive
}{
$
  \HilbertSpace{H}_{\Sigma^2_g}
  \;\in\;
  \mathrm{Obs}_0(\Sigma^2_g)
  \mathrm{Modules}
  \,,
  \;\;\;\;\;
  \mathrm{dim}\big(
    \HilbertSpace{H}_{\Sigma^2_g}
  \big)
  \;=\;
  \vert \lattice \vert^g
  \,,
$ 
}
\end{center}

\label{Surfaces}
Here, the generators $W^i_{a,b}$ correspond to the classical generators of the surface's fundamental group.
{\bf Oriented closed surfaces}
are all obtained (cf. \cite[p 100]{GallierXu13})
by identifying in 
the regular $4g$-gon,
for {\it genus} $g \in \mathbb{N}$:

\vspace{-.1cm}

\noindent
\begin{minipage}{9cm}
   \begin{itemize}[
    leftmargin=.85cm,
    itemsep=.5pt]
  \item[(i)] all boundary vertices with a single point;  \newline 
  and, going clockwise for $r \in \{0, \cdots, g-1\}$,
  \item[(iia)] the $(4r + 1)$st boundary edge with the reverse of the $(4r + 3)$rd,
  \item[(iib)] the $(4r + 2)$nd boundary edge with the reverse of the $(4r + 4)$th.
  \end{itemize}
\end{minipage}
\quad
\adjustbox{
  raise=-.3cm
}{
\adjustbox{
  scale=.8,
  raise=-.6cm
}{
\begin{tikzpicture}
  \node  at (1,2.8)
  {
    \scalebox{.9}{
      \color{darkblue}
      \bf
      sphere
    }
  };
  \draw[
    line width=3,
    fill=lightgray
  ]
    (0,0)
    rectangle 
    (2,2);

\draw[fill=black]
  (0,0) circle (.08);

\draw[fill=black]
  (2,0) circle (.08);

\draw[fill=black]
  (0,2) circle (.08);

\draw[fill=black]
  (2,2) circle (.08);

\node
  at (1,.9) {$
    \Sigma^2_0 \simeq S^2
  $};

\end{tikzpicture}
}
%%%%%%%%%%%%%%%%
\;\;\;
\adjustbox{
  scale=.8,
  raise=-.64cm
}{
\begin{tikzpicture}
  \node  at (1,2.8)
  {
    \scalebox{.9}{
      \color{darkblue}
      \bf
      torus
    }
  };
  \draw[
    draw opacity=0,
    fill=lightgray
  ]
    (0,0)
    rectangle 
    (2,2);

\draw[
  dashed,
  color=darkgreen,
  line width=1.6,
]
  (0,2) -- 
  node[yshift=8pt]{
    \scalebox{1}{
      \color{black}
      $a$
    }
  }
  (2,2);
\draw[
  -Latex,
  darkgreen,
  line width=1.6
]
  (1.3-.01,2) -- 
  (1.3,2);

\draw[
  dashed,
  color=darkgreen,
  line width=1.6,
]
  (0,0) -- (2,0);
\draw[
  -Latex,
  darkgreen,
  line width=1.6
]
  (1.3-.01,0) -- 
  (1.3,0);

\draw[
  dashed,
  color=olive,
  line width=1.6,
]
  (0,0) -- (0,2);
\draw[
  -Latex,
  olive,
  line width=1.6
]
  (0, .8+.01) -- 
  (0,.8);

\draw[
  dashed,
  color=olive,
  line width=1.6,
]
  (2,0) -- 
  node[xshift=8pt]{
    \scalebox{1}{
      \color{black}
      $b$
    }
  }
  (2,2);
\draw[
  -Latex,
  olive,
  line width=1.6
]
  (2, .8+.01) -- 
  (2,.8);

\draw[fill=black]
  (0,0) circle (.08);

\draw[fill=black]
  (2,0) circle (.08);

\draw[fill=black]
  (0,2) circle (.08);

\draw[fill=black]
  (2,2) circle (.08);

\node
  at (1,.9)
  {
    $\Sigma^2_1 \simeq \mathbb{T}^2$
  };

\end{tikzpicture}
}
\hspace{-23pt}
%%%%%%%%%%%%%%%%%%%%
\adjustbox{
  raise=-1.4cm
}{
  \begin{tikzpicture}[
    scale=.65
  ]

  \node[
    rotate=39
  ] at (-1.8,1.8) {
    \scalebox{.64}{
      \color{darkblue}
      \bf
      \def\arraystretch{.7}
      \begin{tabular}{c}
        2-holed
        \\
        torus
      \end{tabular}
    }
  };

  \draw[
    draw opacity=0,
    fill=lightgray
  ]
    (90-22.5:2) --
    (45+90-22.5:2) --
    (2*45+90-22.5:2) --
    (3*45+90-22.5:2) --
    (4*45+90-22.5:2) --
    (5*45+90-22.5:2) --
    (6*45+90-22.5:2) --
    (7*45+90-22.5:2) --
    cycle;

    \draw[
      line width=1.6,
      dashed,
      darkgreen
    ]
      (90-22.5:2)
      --
      (90+22.5:2);

    \draw[
      line width=1.6,
      dashed,
      olive
    ]
      (-45+90-22.5:2)
      --
      (-45+90+22.5:2);

    \draw[
      line width=1.6,
      dashed,
      darkgreen
    ]
      (-90+90-22.5:2)
      --
      (-90+90+22.5:2);

    \draw[
      line width=1.6,
      dashed,
      olive
    ]
      (-135+90-22.5:2)
      --
      (-135+90+22.5:2);

    \draw[
      line width=1.6,
      dashed,
      darkblue
    ]
      (-180+90-22.5:2)
      --
      (-180+90+22.5:2);

    \draw[
      line width=1.6,
      dashed,
      darkblue
    ]
      (-180+90-22.5:2)
      --
      (-180+90+22.5:2);

    \draw[
      line width=1.6,
      dashed,
      purple
    ]
      (-225+90-22.5:2)
      --
      (-225+90+22.5:2);

    \draw[
      line width=1.6,
      dashed,
      darkblue
    ]
      (-270+90-22.5:2)
      --
      (-270+90+22.5:2);

    \draw[
      line width=1.6,
      dashed,
      purple
    ]
      (-315+90-22.5:2)
      --
      (-315+90+22.5:2);

\draw[
  line width=1.3,
  -Latex,
  darkgreen
]
  (0.3,1.83) -- (.35,1.83);

\begin{scope}[
  rotate=-45
]
\draw[
  line width=1.3,
  -Latex,
  olive
]
  (0.24,1.83) -- 
  (.25,1.83);
\end{scope}
\begin{scope}[
  xscale=-1,
  rotate=+90
]
\draw[
  line width=1.3,
  -Latex,
  darkgreen
]
  (0.25,1.83) -- 
  (.3,1.83);
\end{scope}
\begin{scope}[
  xscale=-1,
  rotate=+135
]
\draw[
  line width=1.3,
  -Latex,
  olive
]
  (0.25,1.83) -- 
  (.3,1.83);
\end{scope}

\begin{scope}[
  rotate=+180
]
\draw[
  line width=1.3,
  -Latex,
  darkblue
]
  (.23,1.83) -- 
  (.33,1.83);
\end{scope}

\begin{scope}[
  rotate=-225
]
\draw[
  line width=1.3,
  -Latex,
  purple
]
  (.2,1.83) -- 
  (.3,1.83);
\end{scope}

\begin{scope}[
  rotate=-270
]
\draw[
  line width=1.3,
  -Latex,
  darkblue
]
  (-.20,1.83) -- 
  (-.25,1.83);
\end{scope}
\begin{scope}[
  rotate=-315
]
\draw[
  line width=1.3,
  -Latex,
  purple
]
  (-.20,1.83) -- 
  (-.25,1.83);
\end{scope}

\foreach \n in {1,...,8} {
  \draw[
    fill=black
  ]
    (22.5+\n*45:2)
    circle
    (.11);
};

\node[
  scale=.9
] at
  (0,2.3) {
    $a_1$
  };

\begin{scope}[
  rotate=-45
]
  \node[scale=.9] at
    (0,2.3) {
      $b_1$
    };
\end{scope}

\begin{scope}[
  rotate=-180
]
  \node[scale=.9] at
    (0,2.3) {
      $a_2$
    };
\end{scope}

\begin{scope}[
  rotate=-180-45
]
  \node[scale=.9] at
    (0,2.3) {
      $b_2$
    };
\end{scope}

\node
  at (0,0) {
    $\Sigma^2_2$
  };
  
\end{tikzpicture}
}
}

\noindent
  In other words, the homotopy type of the surface sits in a 
  (pointed) homotopy co-fiber sequence of this form:

$$
\adjustbox{
  margin=3pt,
  bgcolor=lightolive
}{
$
  \begin{tikzcd}
    S^1
    \ar[
      rr,
      "{
        \prod_i
        [ a_i, b_i ]
      }"
    ]
    &&
    \bigvee_g
    \big(
      S^1_a \vee S^1_b
    \big)
    \ar[
      r
    ]
    &
    \mathcolor{blue}{\Sigma^2_g}
    \ar[
      r,
      "{
        \delta
      }"
    ]
    &
    S^2
  \end{tikzcd}
$
}
$$
\vspace{.1cm}

\noindent 
whence its fundamental group  
 is the quotient of the free group on $2g$ generators $(a_i, b_i)_{i = 1}^g$ by the normal subgroup generated by that polygon's boundary:
  $$
    \pi_1\big(
      \Sigma^2_{g}
    \big)
    \;\simeq\;
    \big\langle
      a_1, b_1, \cdots,
      a_g, b_g
    \big\rangle
    \big/
    \prod_i [a_i, b_i]
  $$

\vspace{.2cm}

\noindent
 {\bf 2-Cohomotopy moduli of oriented closed surfaces.}
  Mapping this co-fiber sequence into $S^2$ and applying $\pi_0 \Omega_{\lattice}$, it collapses 
  \cite[Prop. 2]{Hansen74}
  to twice 
  \cite[Thm 1]{LarmoreThomas80}
  the integer Heisenberg central extension
  of $\mathbb{Z}^{2g}$ by
  $\mathbb{Z}_{2 \vert g \vert}$:
$$
  \begin{tikzcd}[
    column sep=11pt,
    row sep=10pt
  ]
    1 
    \ar[r]
    &
    \;
    \grayunderbrace{
    \pi_0
    \Omega_{\lattice}
    \,
    \mathrm{Maps}\big(
      S^2
      ,\,
      S^2
    \big)
    }{
      \mathcolor{black}{
        \mathbb{Z}_{
          2 \vert \lattice \vert
        }
      }
    }
    \ar[
      rr,
      "{ \delta^\ast }"
    ]
    &&
    \grayunderbrace{
    \pi_0
    \Omega_{\lattice}
    \,
    \mathrm{Maps}\big(
      \Sigma^2_g
      ,\,
      S^2
    \big)
    }{
      \scalebox{.7}{
        \color{darkblue}
        \bf
        integer Heisenberg group
      }
    }
    \ar[
      rr
    ]
    &&
    \grayunderbrace{
    \pi_0
    \Omega_\ast
    \,
    \mathrm{Maps}^\ast\big(
      \,
      \textstyle{\bigvee_g}
      (S^1_a \vee S^1_b)
      ,\,
      S^2
    \big)
    }{
      \mathcolor{black}{
        \mathbb{Z}^{
          \mathcolor{purple}{2g}
        }
      }
    }
    \;
    \ar[r]
    & 
    1
    \mathrlap{\,.}
  \end{tikzcd}
$$

\vspace{0cm}

\noindent
{\bf The phase generators.}
Hence these integer Heisenberg groups inject into each other as the surfaces are surjected onto each other by collapsing pairs of 1-cycles:
$$
  \begin{tikzcd}[
    row sep=0pt
  ]
    \Sigma^2_{g}
    \ar[
      ->>,
      from=rr,
      "{ p }"{swap}
    ]
    &&
    \Sigma^2_{g + 1}
    \\
    \pi_0 
    \Omega_{\lattice}\,
    \mathrm{Maps}\big(
      \Sigma^2_{g}
      ,\,
      S^2
    \big)
    \ar[
      d,
      phantom,
      "{ \simeq }"{sloped}
    ]
    \ar[
      rr,
      "{ 
        \pi_1(p^\ast, \lattice)
      }"
    ]
    &&
    \pi_0 
    \Omega_{\lattice}\,
    \mathrm{Maps}\big(
      \Sigma^2_{g+1}
      ,\,
      S^2
    \big)
    \ar[
      d,
      phantom,
      "{ \simeq }"{sloped}
    ]
    \\[10pt]
    \widehat{
      \mathbb{Z}^{2g}
    }
    \ar[
      rr,
      hook
    ]
    &&
    \widehat{
      \mathbb{Z}^{2(g+1)}
    }
    \mathrlap{\,.}
  \end{tikzcd}
$$

Thereby, their central generator $\zeta$ represents the previously identified half-braiding operation of solitons on these surfaces.
This is the ``reason'' for the central extension being by $\mathbb{Z}_{\mathcolor{purple}{2} \vert \lattice \vert}$ instead of just $\mathbb{Z}_{\vert \lattice \vert}$: 

The phase generator $\zeta$ does not correspond to full rotations (such as around the square on the right) but to ``{\it particle exchange}'' by half-braiding ---
as expected for anyons.
\vspace{-2mm} 
$$
\begin{tikzcd}[
  row sep=-5pt,
  column sep=30pt
]
\adjustbox{
  scale=.8,
  raise=-.6cm
}{
\begin{tikzpicture}
  \draw[
    line width=3,
    fill=lightgray
  ]
    (0,0)
    rectangle 
    (2,2);

\draw[fill=black]
  (0,0) circle (.08);

\draw[fill=black]
  (2,0) circle (.08);

\draw[fill=black]
  (0,2) circle (.08);

\draw[fill=black]
  (2,2) circle (.08);

\node
  at (1,.8) {$
    \Sigma^2_0 \simeq S^2
  $};

\end{tikzpicture}
}
\ar[
  from=rr,
  ->>,
  "{ p }"{swap}
]
&&
\adjustbox{
  scale=.8,
  raise=-.7cm
}{
\begin{tikzpicture}
  \draw[
    draw opacity=0,
    fill=lightgray
  ]
    (0,0)
    rectangle 
    (2,2);

\draw[
  dashed,
  color=darkgreen,
  line width=1.5,
]
  (0,2) -- 
  node[yshift=7pt]{
    \scalebox{.9}{
      \color{black}
      $W_a$
    }
  }
  (2,2);
\draw[
  -Latex,
  darkgreen,
  line width=1.6
]
  (1.3-.01,2) -- 
  (1.3,2);

\draw[
  dashed,
  color=darkgreen,
  line width=1.5,
]
  (0,0) -- (2,0);
\draw[
  -Latex,
  darkgreen,
  line width=1.6
]
  (1.3-.01,0) -- 
  (1.3,0);

\draw[
  dashed,
  color=olive,
  line width=1.5,
]
  (0,0) -- (0,2);
\draw[
  -Latex,
  olive,
  line width=1.6
]
  (0, .8+.01) -- 
  (0,.8);

\draw[
  dashed,
  color=olive,
  line width=1.5,
]
  (2,0) -- 
  node[xshift=12pt, yshift=-2pt]{
    \scalebox{.9}{
      \color{black}
      $W_b$
    }
  }
  (2,2);
\draw[
  -Latex,
  olive,
  line width=1.6
]
  (2, .8+.01) -- 
  (2,.8);

\draw[fill=black]
  (0,0) circle (.08);

\draw[fill=black]
  (2,0) circle (.08);

\draw[fill=black]
  (0,2) circle (.08);

\draw[fill=black]
  (2,2) circle (.08);

\node
  at (1,.9)
  {
    $\Sigma^2_1 \simeq \mathbb{T}^2$
  };

\end{tikzpicture}
}
\hspace{-10pt}
\\
\mathbb{Z}_{2\vert k \vert}
\ar[
  rr,
  hook
]
&&
\widehat{
  \mathbb{Z}^2
}
\\[-5pt]
\Bigg[
\adjustbox{
  raise=.2cm, 
  scale=.5}{
\begin{tikzpicture}

\draw[
  line width=1.2,
  -Latex
]
  (-.7,-.7) -- (.7,.7);
\draw[
  line width=7,
  white
]
  (+.7,-.7) -- (-.7,.7);
\draw[
  line width=1.2,
  -Latex
]
  (+.7,-.7) -- (-.7,.7);

\node[
  rotate=-90
] at 
  (1.6,0)
  {
  \scalebox{1.3}{
  \color{darkblue}
  \bf
  \def\arraystretch{.9}
  \begin{tabular}{c}
    particle
    \\
    exchange
  \end{tabular}
}
  };
 
\end{tikzpicture}
}
\hspace{-23pt}
\Bigg]
&\longmapsto&
\hspace{-2pt}
\mathllap{
\adjustbox{
  rotate=90,
  scale=.7,
  raise=-20pt
}{
  \color{darkblue}
  \bf
  \def\arraystretch{.85}
  \begin{tabular}{c}
    phase
    \\
    generator
  \end{tabular}
}}\;\,
\zeta
\mathrlap{
  \;\;
  \mapsto\,
  e^{
    \pi \mathrm{i}
    \nu
  }
}
\end{tikzcd}
$$

%\vspace{m}

\noindent
\begin{minipage}{9.3cm}
{\bf Non-orientable closed surfaces}
are all obtained by identifying in the regular $2h$-gon, for {\it crosscap number} $h \in \mathbb{N}_{\geq 1}$:
  
   \vspace{1mm} 
   \begin{itemize}[
    leftmargin=.7cm,
    itemsep=-1pt
  ]
  \item[(i)] all boundary vertices with a single point
  and, going clockwise for $r \in \{0, \cdots, h-1\}$,
  \item[(ii)] the $(2r + 1)$st boundary edge with the reverse of the $(2r + 2)$nd.
  \end{itemize}
\end{minipage}
\hspace{8pt}
\adjustbox{
  raise=-.6cm
}{
\adjustbox{
  raise=-.8cm
}{
\begin{tikzpicture}[
  scale=.95
]

\node[
  rotate=48
] at 
  (-1.4,.8) {
    \scalebox{.8}{
      \color{darkblue}
      \bf
      \def\arraystretch{.85}
      \begin{tabular}{c}
        projective 
        \\
        plane
      \end{tabular}
    }
  };

\draw[
  line width=1.3,
  color=darkgreen,
  dashed,
  fill=lightgray
]
  (0,0)
  circle
  (1.15);
  
  \draw[
    fill=black
  ]
    (-1.15,0)
    circle 
    (.08);

  \draw[
    fill=black
  ]
    (1.15,0)
    circle 
    (.08);

\draw[
  line width=1.3,
  color=darkgreen,
  -Latex
]
  (0+.12,1.15) --
  (0+.2,1.15);

\node
  at (0,1.4)
  {
    \scalebox{.85}{$a$}
  };

\draw[
  line width=1.3,
  color=darkgreen,
  -Latex
]
  (0-.12,-1.15) --
  (0-.2,-1.15);

\draw[
  dashed,
  gray,
  Latex-Latex
] 
  (45:1.15) -- 
  (180+45:1.15);

\draw[
  dashed,
  gray,
  Latex-Latex
] 
  (-45:1.15) -- 
  (180-45:1.15);

\node[
] at
  (-.5,0) {
    \scalebox{.9}{
      $\Sigma^2
        _{\overline{1}}=$
    }
  };

\node[
] at
  (.5,0) {
    \scalebox{.9}{
      $\mathbb{R}P^2$
    }
  };
\end{tikzpicture}
}
\;\;\;\;\;
\adjustbox{
  scale=.95,
  raise=-1cm
}{
\begin{tikzpicture}

  \node[
    rotate=90
  ]  at (-.4,1)
  {
    \scalebox{.8}{
      \color{darkblue}
      \bf
      Klein bottle
    }
  };
  \draw[
    draw opacity=0,
    fill=lightgray
  ]
    (0,0)
    rectangle 
    (2,2);

\draw[
  dashed,
  color=darkgreen,
  line width=1.6,
]
  (0,2) -- 
  node[yshift=8pt]{
    \scalebox{1}{
      \color{black}
      $a$
    }
  }
  (2,2);
\draw[
  -Latex,
  darkgreen,
  line width=1.6
]
  (1.3-.01,2) -- 
  (1.3,2);

\draw[
  dashed,
  color=darkorange,
  line width=1.6,
]
  (0,0) -- (2,0);
\draw[
  -Latex,
  olive,
  line width=1.6
]
  (.85+.01,0) -- 
  node[yshift=-10pt, xshift=5pt]{
    \scalebox{1}{
      \color{black}
      $b$
    }
  }
  (.85,0);

\draw[
  dashed,
  color=olive,
  line width=1.6,
]
  (0,0) -- (0,2);
\draw[
  -Latex,
  olive,
  line width=1.6
]
  (0,1.2-.01) -- 
  (0,1.2);

\draw[
  dashed,
  color=darkgreen,
  line width=1.6,
]
  (2,0) -- 
  (2,2);

\draw[
  -Latex,
  darkgreen,
  line width=1.6
]
  (2, .8+.01) -- 
  (2,.8);

\draw[fill=black]
  (0,0) circle (.08);

\draw[fill=black]
  (2,0) circle (.08);

\draw[fill=black]
  (0,2) circle (.08);

\draw[fill=black]
  (2,2) circle (.08);

\node
  at (1,.9)
  {
    $\Sigma^2_{\overline{2}}$
  };

\end{tikzpicture}
}
}

\noindent
  In other words, the homotopy type of the surface sits in a 
  (pointed) homotopy co-fiber sequence of this form:
$$
\adjustbox{
  margin=3pt,
  bgcolor=lightolive
}{
$
  \begin{tikzcd}
    S^1
    \ar[
      rr,
      "{
        \prod_i
        a_i^2
      }"
    ]
    &&
    \bigvee_h
      S^1
    \ar[
      r
    ]
    &
    \mathcolor{blue}{
      \Sigma
        ^2
        _{\overline{h}}
    }
    \ar[
      r,
      "{
        \delta
      }"
    ]
    &
    S^2
  \end{tikzcd}
$
}
$$

\vspace{.1cm}

\noindent
 {\bf 2-Cohomotopy moduli of non-orientable closed surfaces.}
  Mapping this co-fiber sequence into $S^2$ and applying $\pi_0 \Omega_k$, it induces 
  \cite[Prop. 3]{Hansen74}
  an extension
  of $\mathbb{Z}^{\mathcolor{purple}{h-1}}$ by
  $\mathbb{Z}_{2}$ which as such is trivial \cite[Thm. 2]{LarmoreThomas80}:
$$
  \begin{tikzcd}[
    column sep=11pt,
    row sep=10pt
  ]
    1 
    \ar[r] 
    &
    \;
    \grayunderbrace{
      \mathrm{coker}
      \Big(
        \big(
        \Sigma
        \,
        \textstyle{\prod_i}
        a_i^2
        \big)^\ast
      \Big)
    }{
      \mathcolor{black}{
        \mathbb{Z}_{
          2 
        }
      }
    }
    \ar[
      rr,
      "{ \delta^\ast }"
    ]
    &&
    \grayunderbrace{
    \pi_0
    \Omega_k
    \,
    \mathrm{Maps}^\ast\big(
      \Sigma
        ^2
        _{\overline{h}}
      ,\,
      S^2
    \big)
    }{
      \mathcolor{black}{
      \mathbb{Z}_2
      \,\times\,
      \mathbb{Z}
        ^{
          \mathcolor{purple}{h-1}
        }
      }
    }
    \ar[
      rr
    ]
    &&
    \grayunderbrace{
      \mathrm{ker}\Big(
        \big(
        \textstyle{\prod_i}
        a_i^2
        \big)^\ast
      \Big)
    }{
      \mathcolor{black}{
        \mathbb{Z}^{ 
          \mathcolor{purple}{h-1}
        }
      }
    }
    \;
    \ar[r]
    & 
    1
    \mathrlap{\,.}
  \end{tikzcd}
$$
Again, the exponent appearing, $h-1$, is just that expected for abelian Chern-Simons ground state degeneracy, where
(cf. \cite[(73)]{CTY16}):
$$
  \mathrm{dim}\big(
    \HilbertSpace{H}_{\Sigma^2_{\overline{\mathcolor{purple}{h}}}}
  \big)
  \;=\;
  \vert k \vert^{
    \mathcolor{purple}{h-1}}.
$$

\medskip

%%%%%%%%%%%%%%%%%%%%%%%%%%%%%%%%%%%%%%%%%%
\section{The topological Quantum Gates}
\label{DefectAnyons}\hypertarget{DefectAnyons}{}
%%%%%%%%%%%%%%%%%%%%%%%%%%%%%%%%%%%%%%%%%%

Where the results of the previous \S\ref{TheTopologicalQuantumStates} establish that on closed (non-punctured) surfaces the predictions of our theory on solitonic anyons and topological order agree with the fine detail of those of abelian Chern-Simons theory, we now analyze the corresponding predictions for punctured surfaces, and find that the punctures behave like possibly non-abelian {\it defect anyons}. 
Details for the following material are spelled out in \cite[\S 3.5-7]{SS25-ViaAlgTop}.

\medskip

\noindent
{\bf Defects via punctured worldvolumes.}
It is now immediate to bring {\it adiabatically movable defect anyons} into the picture, missing in traditional discussion but crucially needed for topological quantum gates (cf. \cite[\S 3]{MyersSatiSchreiber23-TQG}).
Namely, we may simply further generalize the surfaces $\Sigma^2_g$ to their $n$-{\it punctured} versions, 
obtained by deleting the positions of a subset of points -- thus literally creating defects!
\vspace{2mm} 
$$
  \def\arraystretch{1.5}
  \begin{array}{l}
  \rowcolor{lightolive}
  \Sigma^2_{g,n} 
  \;:=\;
  \Sigma^2_{g,n}
  \setminus \{ s_1, \cdots s_n \}
  \\
  \mbox{for}\; 
  \{ s_1, \cdots s_n \} \subset \Sigma^2_g \end{array}
\hspace{30pt}
\adjustbox{raise=-.5cm}{
\begin{tikzpicture}
  \draw[
    dashed,
    fill=lightgray
  ]
    (0,0) 
    -- (2+.4,0) 
    -- (3.9+.4,1)
    -- (1.9,1)
    -- cycle;

\begin{scope}[
  shift={(1.2,.3)}
]
\draw[
  draw=gray,
  fill=white,
  rotate=5pt
] 
  (0,0) 
    ellipse 
  (3pt and 1.5pt);

\node[
  rotate=-20,
  scale=.7
]
at (.25,-.08)
{$s_1$};

\end{scope}

\begin{scope}[
  shift={(2.2,.4)}
]
\draw[
  draw=gray,
  fill=white,
  rotate=5pt,
  scale=.9
] 
  (0,0) 
    ellipse 
  (3pt and 1.5pt);

\node[
  rotate=-20,
  scale=.7
]
at (.25,-.08)
{$s_2$};
\end{scope}

\begin{scope}[
  shift={(2.1,.8)},
  scale=.8
]
\draw[
  draw=gray,
  fill=white,
  rotate=5pt
] 
  (0,0) 
    ellipse 
  (3pt and 1.5pt);

\node[
  rotate=-20,
  scale=.7
]
at (.25,-.08)
{$s_3$};
\end{scope}

\end{tikzpicture}
}
$$
\vspace{1pt}

\noindent
 
That these defects are void of the dynamical solitons is elegantly enforced by identifying all their positions with the point-at-infinity (where, recall, the soliton's very nature is to not be present):
$$
  \scalebox{.7}{
    \color{darkblue}
    \bf
    \def\arraystretch{.9}
    \begin{tabular}{c}
      domain for solitons in the presence of $n$ defects =
      \\
      one-point compactification of 
      $n$-punctured surface
    \end{tabular}
  }
  \def\arraystretch{1.6}
  \begin{array}{l}
  \big(
    \Sigma^2_{g,n}
  \big)_{\cpt}
  \;\;\;\;\;\;\;\;
  \mbox{e.g.:}
  \;\;\;\;\;
  (\Sigma^2_{0,1})_{\cpt}
  \;\simeq\;
  \mathbb{R}^2_{\cpt}
  \mathrlap{\,.}
  \end{array}
$$

\vspace{2pt}

\noindent
In this generality, our previous brane diagram now is the following,
with algebra of 
soliton quantum observables
as shown, by the same kind of argument as before:
$$
\adjustbox{
  bgcolor=lightgray,
  raise=-7pt
}{
$
  \begin{tikzcd}[
    column sep=0pt,
    background color=lightgray
  ]
    \Sigma^{\mathrlap{1,6}}
    \ar[
      in=60,
      out=180-60,
      looseness=3.7,
      shift right=3pt,
      "{
        \!\mathbb{Z}_2\!
      }"{description}
    ]
    \;
    &:=&
    \mathbb{R}^{1,0}
    &\times& 
    \mathcolor{darkblue}{
      \big(
        \Sigma^2_{g,n}
      \big)_{\cpt}
    }
    &\times& 
    S^1
    &\times&
    \mathbb{R}^2_{\mathrm{sgn}}
    \mathrlap{\,,}
    \ar[
      in=60,
      out=180-60,
      looseness=3.5,
      shift right=3pt,
      "{
        \!\mathbb{Z}_2\!
      }"{description}
    ]
  \end{tikzcd}
$
}
\;\;
 \rightsquigarrow
\;\;
\adjustbox{
}{
$
  \mathrm{Obs}_0\big(
    \Sigma^2_{g,n}
  \big)
  \;:=\;
  H_0\Big(
    \Omega_{\lattice}
    \,
    \mathrm{Maps}^{\ast}
    \big(
      (\Sigma^2_{g,n})_{\cpt}
      ,\,
      S^2
    \big)
    ;\,
    \mathbb{C}
  \Big)
  \,.
$
}
$$

\medskip

\noindent
\begin{minipage}{12.5cm}
{\bf Braid group action.}
This algebra of observables is faithfully acted on by the mapping class group 
of the punctured surface
-- again simply by precomposition of maps. 

But, with punctures, that group is now an extension 
(cf. \cite[Thm. 3.13]{Massuyeau21})
of the plain mapping class group by the  {\it surface braid group} that acts by (``adiabatically'') moving the defects around each other! 
\end{minipage}
\hspace{1.5cm}
\adjustbox{
  raise=-2.4cm
}{
  \includegraphics[width=3cm]{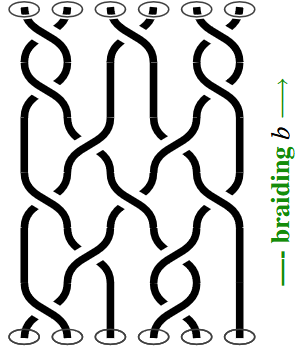}
}

\vspace{-.8cm}

$$
\hspace{-2cm} 
\begin{tikzcd}[
    row sep=-2pt,
    column sep=30pt
  ]
    \mathllap{
      1 \xrightarrow{\;}\;
    }
    \mathrm{Br}_n(\Sigma^2_g)
    \ar[r, hook]
    &
    \pi_0 \mathrm{Homeos}^{\ast}_{\mathrm{or}}\big(
      (\Sigma^2_{g,n})_{\cpt}
    \big)
    \ar[r,->>]
    &[+10pt]
    \mathrm{MCG}(\Sigma^2_g)
    \mathrlap{
      \; \xrightarrow{\;} 1
    }
    \\
    \mathclap{
      \scalebox{.7}{
        \color{darkgreen}
        \bf
        surface braid group
      }
    }
    &
    \mathclap{
      \scalebox{.7}{
        \color{purple}
        \bf
        \def\arraystretch{.9}
        \begin{tabular}{c}
          mapping class group
          \\
          of punctured surface
        \end{tabular}
      }
    }
    &
    \mathclap{
      \scalebox{.7}{
        \color{darkblue}
        \bf
        \def\arraystretch{.9}
        \begin{tabular}{c}
          mapping class group
          \\
          of plain surface
        \end{tabular}
      }
    }
  \end{tikzcd}
$$
In deducing this, we used that
$
  \mathrm{Homeos}^{\ast}
  \big(
    (\Sigma^2_{g,n})_{\cpt}
  \big)
  \;\simeq\;
  \mathrm{Homeos}
  \big(
    \Sigma^2_{g,n}
  \big)
$,
since $(-)_{\cpt}$ is functorial on homeos.

Concretely, observe that
the homotopy type of the one-point compactification of a punctured closed surface is the wedge sum of the original surface with $n-1$ circles (cf. \cite[p 11]{Hatcher02}, whose graphics we are adapting):
\begin{equation}
\label{PinchingOfSphere}
\big(\Sigma^2_{g,n}\big)_{\cpt}
  \;\simeq\;
  \Sigma^2_g 
  \,\vee\, 
  \textstyle{
    \underset{\mathclap{n-1}}{\bigvee}
  } 
  \, S^1
  \hspace{1cm}
  \adjustbox{
    raise=-2cm
  }{
\begin{tikzpicture}
  \node at (0,0) {
  \includegraphics[width=5cm]{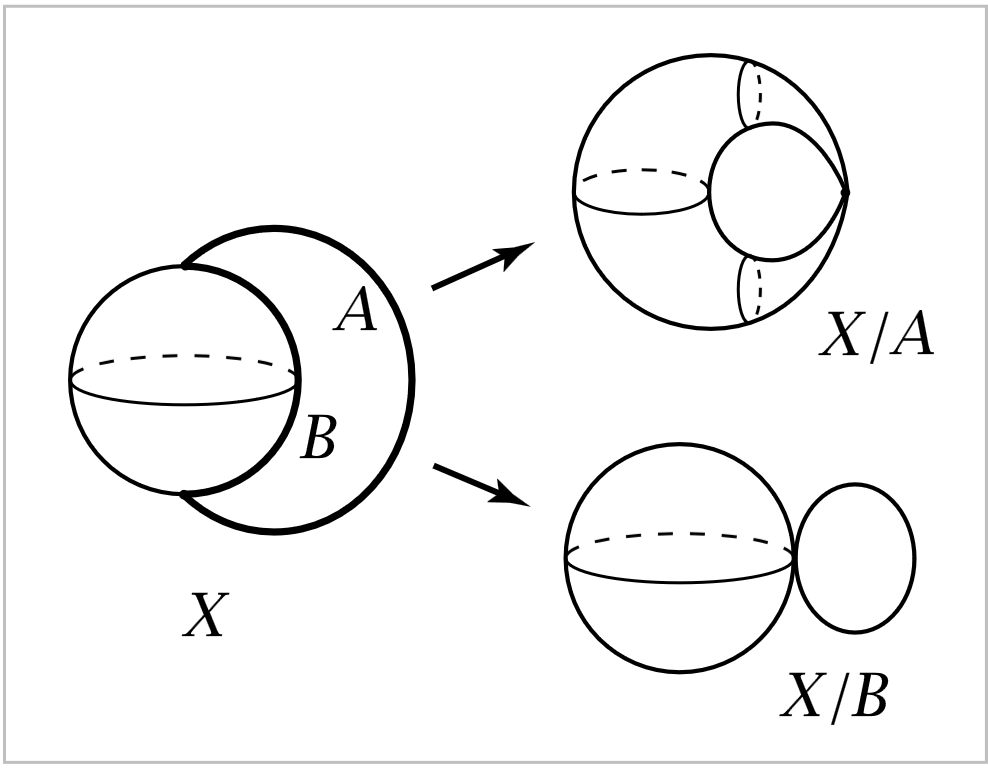}
  };
  \node[
    scale=.5,
    rotate=-40
  ] at (-1.87,.86)
  {
    \color{purple}
    puncture
  };
  \begin{scope}[
    yscale=-1
  ]
  \node[
    scale=.5,
    rotate=+40
  ] at (-1.85,.85)
  { \color{purple}
    puncture
  };
  \end{scope}
  \node[
    scale=.8
  ] 
    at (-2.15,-.53)
  {$\Sigma^2_0$};
  \node[
    scale=.8
  ] 
    at (.8,.14)
  {$(\Sigma^2_{0,2})_{\cpt}\,
  \rotatebox[origin=c]{10}{$=$}\,$};
  \node[scale=.8] at (.7,-1.7)
  {
    $\Sigma^2_0 \vee S^1
   \;
  \adjustbox{rotate=10,raise=2pt}{$=$}\,$
  };
  \node[
    rotate=-20,
    scale=.8
  ]
    at (-.2,-.6)
    {$\sim$};
  \node[
    rotate=+20,
    scale=.8
  ]
    at (-.18,+.66)
    {$\sim$};
  \node[
    scale=.7
  ]
  at (1.98,.97) 
  {$\infty$};
  \begin{scope}[
    shift={(-1.47,.57)}
  ]
  \draw[purple]
    (180-5:.2 and .1) arc
    (180-5:360+5:.11 and .05);
  \end{scope}
  \begin{scope}[
    yscale=-1,
    shift={(-1.47,.52)},
  ]
  \draw[purple]
    (180-5:.2 and .1) arc
    (180-5:360+5:.11 and .05);
  \end{scope}
\end{tikzpicture}
  }
\end{equation}
This means that the punctures are effectively topology changing defects (as such reminiscent of the {\it genon}-ic anyon defects considered in \cite{BarkeshliJianQi13}) and it implies that their bare quantum observables are (the group algebra) of:
\begin{equation}
  \label{QuantumObservablesInPresenceOfDefects}
  \pi_1
  \mathrm{Maps}^{\ast}\Big(
    \big(\Sigma^2_{g,n}\big)_{\cpt}
    ,\,
    S^2
  \Big)
  \; \simeq \;
  \pi_1
  \mathrm{Maps}^{\ast}\big(
    \Sigma^2_g
    ,\,
    S^2
  \big)
  \times
  \mathbb{Z}^{n-1}
  \;
  % \\
  % &
  \underset{
    \mathclap{g = 0}
  }{\simeq} \;\;
  \mathbb{Z}^n
  \,,
\end{equation}
where on the right we recognize that associated with each of the $n$ punctures is one copy of the braid phase observable algebra \eqref{BasicBraidObservables}
for the nearby solitonic anyons.
This observable algebra of defects  gets further enhanced by the corresponding mapping class group:

\smallskip

\noindent
{\bf Side-remark: Defect-braiding on M5s as a quantum-gravitational effect.}
Noting that the mapping class group is equivalently the group of large {\it diffeomorphisms} of the punctured surface (cf. \cite[p 45]{FarbMargalit12}),
\begin{equation}
  \label{HomeosAsDiffeos}
  \pi_0
  \mathrm{Homeos}^{\ast}_{\mathrm{or}}
  \big(
    (\Sigma^2_{g,n})_{\cpt}
  \big)
  \;\simeq\;
  \pi_0 \, \mathrm{Diffeos}_{\mathrm{or}}(\Sigma^2_{g,n})
  \,,
\end{equation}
we see that braiding of anyonic defects is reflected in equipping the moduli spaces of cohomotopical charges on the brane worldvolume with the action by diffeomorphisms, hence by passing to the action {\it groupoid} of moduli quotiented by diffeos:
$$
  \mathrm{GnrlCovariantModuli}(\Sigma)
  \;\simeq\;
  \mathrm{Moduli}(\Sigma)
  \sslash
  \mathrm{Diffeos}(\Sigma)
  \,.
$$
But this is the hallmark of {\it generally covariant} systems (cf. \cite{Dul23}), such as are our probe M5-branes.

Ultimately we are to consider surfaces $\Sigma^2_{g,n,\mathcolor{purple}{b}}$ that may feature $\mathcolor{purple}{b} \in \mathbb{N}$ boundary components, and then determine the normal subgroup {\it pure gauge} diffeomorphisms inside \eqref{HomeosAsDiffeos} which are trivial on the boundary. The resulting quotient group will be the (ADM/BSM-like) group of diffeos that serve in practice as experimentally observable boundary charges.

\smallskip
\noindent
{\bf Observables on soliton + defect anyons.}
So the covariantized quantum observables on the disk in the presence of $n$ defects is the 
group algebra of the 
subgroup of vanishing total framing of the 
spherical {\it framed braid group} \cite{KoSmolinsky92}, namely of the 
{\it wreath product} 
$\mathbb{Z} \wr \mathrm{Br}_n$
(cf. \cite[\S 8]{BhattacharjeeMacphersonMollerNeumann06})
of the soliton monodromy group $\mathbb{Z}$ 
with the actual braid group $\mathrm{Br}_n(\Sigma^2_0)$ of the defect anyons,

\vspace{.2cm}
\begin{equation}
  \label{WreathProductWuantumObservables}
  \mathbb{C}
  \big[
  \overset{
    \mathclap{
      \adjustbox{
        scale=.7,
        rotate=12
      }
      {
        \rlap{
          \color{darkblue}
          \bf solitonic anyons
        }
      }
    }
  }{
    \mathbb{Z}^{n-1}
  }
  \rtimes
  \overset{
    \mathclap{
      \adjustbox{
        scale=.7,
        rotate=12
      }
      {
        \rlap{
          \color{darkblue}
          \bf defect anyons
        }
      }
    }
  }{
    \mathrm{Br}_n
  }
  \big]
  \;\;
  \subset
  \;\;
  \mathbb{C}
  \big[
    \mathbb{Z}^n
  \rtimes
    \mathrm{Br}_n
  \big]
  \;\;
  =
  \;\;  
  \mathbb{C}\big[
    \mathbb{Z} 
    \wr 
    \mathrm{Br}_n
  \big]
  \;\;
  =
  \;\;
  \mathbb{C}\big[
    \mathrm{FBr}_n
  \big]
  \,,
\end{equation}
where the braid group acts on $\mathbb{Z}^n$ through permutation of the factors
$\mathrm{Sym}_n$ of the defect anyons:
\begin{equation}
  \label{BraidWreathSurjectingOntoSymmetricWreath}
  \begin{tikzcd}
  \mathbb{Z}^n
  \rtimes 
  \mathrm{Br}_n(\Sigma^2_g)
  \ar[r, ->>]
  &
  \mathbb{Z}^n
  \rtimes 
  \mathrm{Sym}_n
  \;\;
  \simeq
  \;\;
  \Big\{
    \big(
      (n_i)_{i =1}^n
      ,\,
      \sigma 
      \big)
      \;\big\vert\;
      \big(
        (n_\bullet), \sigma
       \big)
        \cdot
      \big(
        (n'_\bullet), \sigma'
       \big)
       \;=\;
       \big(
         (n_\bullet + n'_{\sigma(\bullet)})
         ,\,
         \sigma \sigma'
       \big)
  \Big\}.
  \end{tikzcd}
\end{equation}
Just such {\it para-statistical}
(cf. \cite{WangHazzard25})
wreath-group statistics of defect anyons is seen in condensed matter \cite{FreedmanHastingsNayakQiWalkerWang11}.

\begin{center}
\begin{minipage}{4.5cm}
  \footnotesize
  {\bf Figure \figurenumber:
  \label{SolitonicAndDefectAnyons}
  Solitonic and defect anyons from flux quantized in cohomotopy.}  
  Compare the similarity to 
  the situation of FQH anyons in 
  Figure \ref{FQHAnyons}.
\end{minipage}
\;\;
\begin{tabular}{c}
\begin{tabular}{r|ccc}
  \adjustbox{}{\bf\def\arraystretch{1}\def\tabcolsep{1pt}\begin{tabular}{c}
      Anyons as seen
      \\
      in Cohomotopy
    \end{tabular}
  }
  &
  \bf
  Nature
  &
  \bf
  Number
  &
  \bf Braiding
  \\
  \hline
  \rowcolor{lightgray}
  \bf
  Solitonic anyons 
  &
  \def\arraystretch{.9}
  \def\tabcolsep{1pt}
  \begin{tabular}{c}
    concentrations
    \\
    of flux density
  \end{tabular}
  &
  \def\arraystretch{.9}
  \def\tabcolsep{-4pt}
  \begin{tabular}{c}
    net charge,
    \\
    CS-level: $k$
  \end{tabular}
  &
  \def\arraystretch{.9}
  \def\tabcolsep{-4pt}
  \begin{tabular}{c}
    by (LC-)time
    \\
    evolution
  \end{tabular}
  \\
  \bf
  Defect anyons
  &
  \def\arraystretch{.9}
  \def\tabcolsep{-4pt}
  \begin{tabular}{c}
    punctures in
    \\
    worldvolume
  \end{tabular}
  &
  \def\arraystretch{.9}
  \def\tabcolsep{-4pt}
  \begin{tabular}{c}
    $n$ in $\Sigma^2_{g,n}$
  \end{tabular}
  &
  \def\arraystretch{.9}
  \def\tabcolsep{-4pt}
  \begin{tabular}{c}
    by worldvolume
    \\
    diffeomorphisms
  \end{tabular}
\end{tabular}
\end{tabular}
\end{center}

\medskip 
\begin{center}
\begin{tikzpicture}

\draw[
  dashed,
  fill=lightgray
]
  (0,0)
  -- (8,0)
  -- (10,2)
  -- (2.8,2)
  -- cycle;

\begin{scope}[
  shift={(2.4,.5)}
]
\shadedraw[
  draw opacity=0,
  inner color=olive,
  outer color=lightolive
]
  (0,0) ellipse (.7 and .3);
\end{scope}

\begin{scope}[
  shift={(6.3,.6)}
]
\shadedraw[
  draw opacity=0,
  inner color=olive,
  outer color=lightolive
]
  (0,0) ellipse (.7 and .3);
\end{scope}

\begin{scope}[
  shift={(4.5,1.5)}
]
\shadedraw[
  draw opacity=0,
  inner color=olive,
  outer color=lightolive
]
  (0,0) ellipse (.7 and .25);
\end{scope}

\begin{scope}[shift={(4.8,.7)}]
\draw[
  fill=white,
  draw=gray
]
  (0,0) ellipse 
  (.17 and 
  0.5*0.17);
\end{scope}

\begin{scope}[shift={(7.5,1.2)}]
\draw[
  fill=white,
  draw=gray
]
  (0,0) ellipse 
  (.17 and 
  0.5*0.17);
\end{scope}

\draw[
  white,
  line width=2
]
  (-1.4, .7)
  .. controls (1,2) and (2,2) ..
  (2.32,.7);
\draw[
  -Latex,
  black
]
  (-1.4, .7)
  .. controls (1,2) and (2,2) ..
  (2.32,.7);

\node
  at (-1.1,.7)
  {
    \adjustbox{
      bgcolor=white,
      scale=.7
    }{
      \color{darkblue}
      \bf
      \def\arraystretch{.9}
      \def\tabcolsep{-5pt}
      \begin{tabular}{c}
        field \color{purple}solitons/
        \\
        quasi-particles/ 
        \\
        -holes/vortices:
        \\
        frmd submanifolds
      \end{tabular}
    }
  };

\draw[
  white,
  line width=2
]
  (10,1.1) 
  .. controls 
  (9.5,1.8) and 
  (8.5,1.8) ..
  (7.7,1.3);
\draw[
  -Latex
]
  (10,1.1) 
  .. controls 
  (9.5,1.8) and 
  (8.5,1.8) ..
  (7.7,1.3);

\node
  at (10.2,.5)
  {
    \adjustbox{
      scale=.7
    }{
      \color{darkblue}
      \bf
      \def\arraystretch{.9}
      \def\tabcolsep{-5pt}
      \begin{tabular}{c}
        flux-expelling 
        {\color{purple}defects}:
        \\
        punctures in the surface
      \end{tabular}
    }
  };
  
\end{tikzpicture}
\end{center}

\vspace{3mm}

\noindent

\noindent
{\bf Topologically protected rotation gates.}
The above action of $\mathrm{Sym}_n$ on $\mathbb{Z}^{n-1} \subset \mathbb{C}^{n-1}$ is the  
``{\it standard representation}'' of the symmetric group (the complement of the trivial 1d representation inside the defining permuation representation)
$$
  \HilbertSpace{H}_{\Sigma^2_{0,1,n}}
  \;\simeq\;
  \adjustbox{
    raise=-.2cm
  }{
  \begin{tikzpicture}
    \draw (0,0) rectangle (.3,.3);
    \draw (0+.3,0) rectangle (.3+.3,.3);    
    \draw (0+2*.3,0) rectangle (.3+2*.3,.3);    
    \draw[densely dotted] (0+3*.3,0) rectangle (.3+3*.3,.3);    
    \draw(0+4*.3,0) rectangle (.3+4*.3,.3);    
    \draw (0,-.3) rectangle (.3,.3-.3);
  \end{tikzpicture}
  }
  \,.
$$

Via the cyclic subgroup $\mathbb{Z}_n \subset \mathrm{Sym}_n$ of cyclic permutations, this standard representation contains what in quantum computing are known as q\textit{d}it-based {\it rotation gates} \cite{YeEtAl11}. For example, for $n = 3$ inspection readily shows that the unitarized standard representation is generated from a Pauli {\it Z-gate} and a qbit-rotation around the (conventional) $y$-axis, like this:
$$
  \adjustbox{
    raise=-.2cm
  }{
  \begin{tikzpicture}
    \draw (0,0) rectangle (.3,.3);
    \draw (0+.3,0) rectangle (.3+.3,.3);    
    \draw (0,-.3) rectangle (.3,.3-.3);
  \end{tikzpicture}
  }
  \;\simeq\;
  \Bigg\{
  (213) \mapsto
  \grayunderbrace{
  \left[
  \def\arraystretch{1}
  \def\arraycolsep{2pt}
  \begin{array}{cc}
    1 & 0
    \\
    0 & -1
  \end{array}
  \right]
  }{ Z }
  \,,\;\;\;
  (231) \mapsto
  \grayunderbrace{
  \left[
  \def\arraystretch{1}
  \def\arraycolsep{2pt}
  \begin{array}{cc}
    \mathrm{cos}(\alpha) & -\mathrm{sin}(\alpha)
    \\
    \mathrm{sin}(\alpha) & \;\mathrm{cos}(\alpha)
  \end{array}
  \right]  
  }{
    R_y(2\alpha)
  }
  \mbox{where $\alpha = 4\pi/3$}
  \Bigg\}
  \,,
$$
which implies at once that in the standard rep 
$  \adjustbox{
    raise=-.2cm
  }{
  \begin{tikzpicture}
    \draw (0,0) rectangle (.3,.3);
    \draw (0+.3,0) rectangle (.3+.3,.3);    
    \draw (2*.3,0) rectangle 
          (2*.3+.3,.3);    
    \draw (3*.3,0) rectangle 
          (3*.3+.3,.3);    
    \draw (4*.3,0) rectangle 
          (4*.3+.3,.3);    
    \draw (0,-.3) rectangle (.3,.3-.3);
  \end{tikzpicture}
  }
$
of $\mathrm{Sym}_6$ we find also the corresponding qbit-\textit{controlled} rotation, and so on.

Together with the global phase rotations of solitonic anyons given by the first wreath factor in \eqref{WreathProductWuantumObservables}, (controlled), such rotation gates are the workhorse in the quantum Fourier transform \cite[\S 5]{NielsenChuang00}\cite[\S 3.2.1]{WangHuEtAl20} (hence notably in Shor's algorithm for prime faxctorization) and their precision and error protection is a major bottleneck in the implementation of useful quantum algorithms (cf. \cite[\S III]{FowlerHollenberg07}).

Here we see that our geometric engineering predicts the relevant gates to have topologically error-protected implementation by braiding of defects in FQH systems.
\begin{center}
\begin{minipage}{7.6cm}
  \small
  {\bf Figure \figurenumber: Topological rotation gates}, obtained by cyclic braiding of defect anyons, combined with the global phase rotations given by braiding of  solitonic anyons, would provide intrinsically exact and topologically protected gates of the kind that make up the quantum Fourier transform (in qdit-bases), and with it many other quantum algorithms. 
\end{minipage}
\;\;\;\;
\adjustbox{
  raise=-.9cm
}{
\begin{tikzpicture}

  \draw[line width=1.5]
    (.5,0) .. controls
    (.5,1) and
    (0,1) ..
    (0,2);
  \draw[line width=1.5]
    (1,0) .. controls
    (1,1) and
    (.5,1) ..
    (.5,2);

  \node at 
    (1.4,.8) {\bf\dots};

  \draw[line width=1.5]
    (2,0) .. controls
    (2,1) and
    (1.5,1) ..
    (1.5,2);
    (1.3,1) {\bf\dots};

  \draw[
    line width=6,
    white
  ]
    (0,0) .. controls
    (0,1) and
    (2,1) ..
    (2,2);
  \draw[line width=1.5]
    (0,0) .. controls
    (0,1) and
    (2,1) ..
    (2,2);

\begin{scope}
  \draw
    (115:.18 and .08) arc 
    (110:418:.18 and .08);
\end{scope}
\begin{scope}[shift={(.5,0)}]
  \draw
    (115:.18 and .08) arc 
    (110:418:.18 and .08);
\end{scope}
\begin{scope}[shift={(1,0)}]
  \draw
    (115:.18 and .08) arc 
    (110:418:.18 and .08);
\end{scope}
\begin{scope}[shift={(2,0)}]
  \draw
    (115:.18 and .08) arc 
    (110:418:.18 and .08);
\end{scope}
\begin{scope}[shift={(0,2)}]
  \draw[line width=1.7, white]
    (0:.18 and .08) arc 
    (0:360:.18 and .08);
  \draw
    (0:.18 and .08) arc 
    (0:360:.18 and .08);
\end{scope}
\begin{scope}[shift={(.5,2)}]
  \draw[line width=1.7, white]
    (0:.18 and .08) arc 
    (0:360:.18 and .08);
  \draw
    (0:.18 and .08) arc 
    (0:360:.18 and .08);
\end{scope}
\begin{scope}[shift={(1.5,2)}]
  \draw[line width=1.7, white]
    (0:.18 and .08) arc 
    (0:360:.18 and .08);
  \draw
    (0:.18 and .08) arc 
    (0:360:.18 and .08);
\end{scope}
\begin{scope}[shift={(2,2)}]
  \draw[line width=1.7, white]
    (0:.18 and .08) arc 
    (0:360:.18 and .08);
  \draw
    (0:.18 and .08) arc 
    (0:360:.18 and .08);
\end{scope}

\end{tikzpicture}
}
\end{center}

\medskip

\noindent
{\bf Side-remark: ``Parastatistics'' as the most stable anyon braid gates.}
Such braid representations on irreps of the symmetric group have traditionally received little to no attention in topological quantum computing (popular are instead solutions to the Knizhnik-Zamolodchikov equation, cf. \cite{SS24-TopOrd},  and of the Yang-Baxter equation). Elsewhere they are discussed as speculative {\it parastatistics} \cite{HartleTaylor69}\cite{Polychronakos96} of  fundamental particles instead of as adiabatic Berry-transformations of defect anyons. Therefore Jordan 2010 \cite{Jordan10a}, who is the first to propose symmetric irreps as a model for quantum computation -- aka {\it permutational quantum computing} \cite{Jordan10b} --, admitted ``not [to] worry too much about the physical justification for the model'' \cite[p 109]{Jordan10a}. 

This seems to be a blind spot in the literature:
Irreps of $\mathrm{Sym}_n$ are in particular surface braid representations via the surjection $\mathrm{Br}_n(\Sigma^2_g) \twoheadrightarrow \mathrm{Sym}_n$ --- regarded as anyon braid gates they are in fact the {\it most stabilized} such, in that the gate operation is independent not just of isotopy but even of homotopy of the adiabatic transformation.

\medskip 
\noindent

%%%%%%%%%%%%%%%%%%%%%%%%%%%%%%%%%%%%%%%%%%%%%%%%%%%%%
\section{Conclusion: Better Anyon Theory}
%%%%%%%%%%%%%%%%%%%%%%%%%%%%%%%%%%%%%%%%%%%%%%%%%%%%%

\noindent
{\bf New theory of anyonic topological order, engineered on flux-quantized M5s.}
In summary, we have seen that global completion by flux-quantization of 11D supergravity with M5-probes (here: in equivariant twistorial cohomotopy -- ``Hypothesis H''), makes the quantized topological sector of the self-dual tensor field on M5-probes (wrapping Seifert orbi-singularities) reproduce key phenomena of abelian Chern-Simons theory thought of as an effective field theory for abelian anyons in fractional quantum Hall (FQH) systems:

\medskip

\noindent {\bf (i) Flux tubes bound to anyons.} The central assumption in the traditional heuristic understanding of the FQHE is that the anyonic solitons have flux quanta ``attached'' to them \cite[p 883]{Stormer99}. It is crucially this assumption that motivates and justifies abelian Chern-Simons theory as an effective field theory for FQH anyons, since variation of the sum of the abelian Chern-Simons term with the standard source term predicts that the gauge field flux is localized at the source particles  (cf. \cite[(5.25)]{Tong16}\cite[(3.6)]{Witten16}).

\vspace{.1cm}

\noindent
\begin{minipage}{6.5cm}
In contrast, in the present approach {\bf this effect is a consequence of cohomotopical flux-quantization}, via the Pontrjagin theorem: The classifying map of the 2-Cohomotopy charge identifies an open neighborhood of each anyon with the 2-sphere minus its point at infinity, and the flux density $F_2$ is the pullback of the sphere's volume form along this map (cf. p \pageref{WhiteheadIntegralFormula}), hence supported on just these open neighborhoods.
\end{minipage}
\hspace{-3pt}
 \adjustbox{
    raise=-2.3cm,
    scale=1
  }{
   \begin{tikzpicture}
\begin{scope}[
  scale=.8,
  shift={(.7,-4.9)}
]

\draw[
  line width=.8,
  ->,
  darkgreen
]
  (1.5,1) 
  .. controls (2,1.6) and (3,2.6) .. 
  (6,1);

\node[
  scale=.7,
  rotate=-18
] at (4.7,1.8) {
  \color{darkblue}
  \bf
  classifying map
};

\node[
  scale=.7,
  rotate=-18
] at (4.5,1.4) {
  \color{darkblue}
  \bf
  $n$
};

  \shade[
    right color=gray, left color=lightgray,
    fill opacity=.9
  ]
    (3,-3)
      --
    (-1,-1)
      --
        (-1.21,1)
      --
    (2.3,3);

  \draw[dashed]
    (3,-3)
      --
    (-1,-1)
      --
    (-1.21,1)
      --
    (2.3,3)
      --
    (3,-3);

  \node[
    scale=1
  ] at (3.2,-2.1)
  {$\infty$};

  \begin{scope}[rotate=(+8)]
  \shadedraw[
    dashed,
    inner color=olive,
    outer color=lightolive,
  ]
    (1.5,-1)
    ellipse
    (.2 and .37);
  \draw
   (1.5,-1)
   to 
    node[above, yshift=-1pt]{
     \;\;\;\;\;\;\;\;\;\;\;
     \rotatebox[origin=c]{7}{
     \scalebox{.7}{
     \color{darkorange}
     \bf
       anyon
     }
     }
   }
    node[below, yshift=+6.3pt]{
     \;\;\;\;\;\;\;\;\;\;\;\;
     \rotatebox[origin=c]{7}{
     \scalebox{.7}{
     \color{darkorange}
     \bf
       worldline
     }
     }
   }
   (-2.2,-1);
  \draw
   (1.5+1.2,-1)
   to
   (4,-1);
  \end{scope}

  \begin{scope}[shift={(-.2,1.4)}, scale=(.96)]
  \begin{scope}[rotate=(+8)]
  \shadedraw[
    dashed,
    inner color=olive,
    outer color=lightolive,
  ]
    (1.5,-1)
    ellipse
    (.2 and .37);
  \draw
   (1.5,-1)
   to
   (-2.3,-1);
  \draw
   (1.5+1.35,-1)
   to
   (4.1,-1);
  \end{scope}
  \end{scope}
  \begin{scope}[shift={(-1,.5)}, scale=(.7)]
  \begin{scope}[rotate=(+8)]
  \shadedraw[
    dashed,
    inner color=olive,
    outer color=lightolive,
  ]
    (1.5,-1)
    ellipse
    (.2 and .32);
  \draw
   (1.5,-1)
   to
   (-1.8,-1);
  \end{scope}
  \end{scope}
  
\end{scope}

\node[
  scale=.73,
  rotate=-27
] at (2.21,-5.21) {
  \color{darkblue}
  \bf
  \def\arraystretch{.9}
  \begin{tabular}{l}
    flux
    \\
    $F_2 =$ 
    \\
    $\;\;n^\ast(\mathrm{dvol}_{S^2})$
  \end{tabular}
};

\node[
  scale=.73,
] at (2.05,-5) {
  \color{darkblue}
  \bf
};

\node[
  rotate=-140
] at (6,-4) {
  \includegraphics[width=2cm]{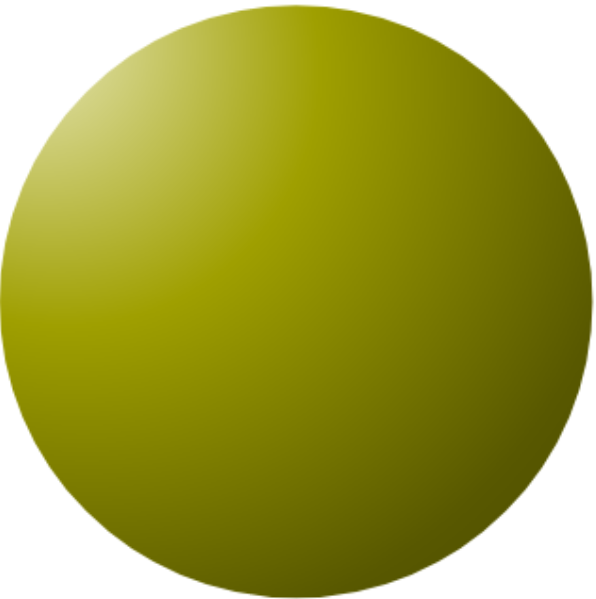}
};

\node[
  scale=.7
] at (6,-5.2) {
  \color{darkblue}
  \bf
  2-sphere $S^2$
};

\node[
  scale=.8,
  rotate=-22
] at (.05,-4.55)
{$\Sigma^2$};

\end{tikzpicture}
}
\hspace{-.5cm}
\begin{minipage}{2cm}
  \footnotesize%
  \raggedright%
  {\bf Figure \figurenumber:}
  Anyon flux quantized in
  2-cohomotopy via Pontrjagin's theorem.
\end{minipage}
\vspace{2pt}

\noindent {\bf (ii) Anyons subject to each other's Aharonov-Bohm phases.}
Traditional discussion furthermore assumes from these attached flux tubes that the anyons must pick up Aharonov-Bohm quantum phases when circling around each other. While this is plausible, rigorous quantum field-theoretic derivation of this statement may not have found much attention.

\vspace{2pt}

\noindent

In contrast, in the approach discussed here, this effect is again a direct consequence of cohomotopical flux-quantization, now via algebro-topological theorems of Segal and others, which serve to identify the cohomotopy charge moduli space with configuration spaces of soliton cores, whose fundamental group reflects the anyon braid phases (and thereby also the ground state degeneracy / topological order).

$$
  \begin{tikzcd}[
    row sep=3pt,
    column sep=90pt
  ]
  \pi_0
  \mathrm{Maps}^{\ast}\big(
    \mathbb{R}^2_{\cpt}
    ,\,
    \colorbox{lightolive}{$S^2$}
  \big)
  \ar[
    r,
    shorten=-2pt,
    "{ \sim }"
  ]
  \ar[
    d,
    phantom,
    "{ \simeq }"{sloped}
  ]
  &
  \pi_0
  \mathrm{Maps}^{\ast}\big(
    \mathbb{R}^2_{\cpt}
    ,\,
    B^2 \mathbb{Z}
  \big)
  \ar[
    d,
    phantom,
    "{ \simeq }"{sloped}
  ]
  \\
  \mathbb{Z}
  \ar[
    r,
    phantom,
    "{
      \scalebox{.7}{
        \color{darkgreen}
        \bf
        same net charges...
      }
    }"
  ]
  &
  \mathbb{Z}
  \\
  \pi_1
  \mathrm{Maps}^{\ast}\big(
    \mathbb{R}^2_{\cpt}
    ,\,
    \colorbox{lightolive}{$S^2$}
  \big)
  \ar[
    r,
    shorten=-2pt,
  ]
  \ar[
    d,
    phantom,
    "{ \simeq }"{sloped}
  ]
  &
  \pi_1
  \mathrm{Maps}^{\ast}\big(
    \mathbb{R}^2_{\cpt}
    ,\,
    B^2 \mathbb{Z}
  \big)
  \ar[
    d,
    phantom,
    "{ \simeq }"{sloped}
  ]
  \\
  \pi_1 \mathbb{G}
  \underset{
    \mathclap{
      \adjustbox{
        scale=.7,
        raise=+1pt
      }{
        \color{gray}
        \bf
        config space
      }
    }
  }{
  \mathrm{Conf}(\mathbb{R}^2)
  }
  \ar[
    r,
    phantom,
    "{
      \scalebox{.7}{
        \color{darkgreen}
        \bf
        ...but different moduli
      }
    }"
  ]
  &
  \underset{
    \mathclap{
      \adjustbox{
        scale=.7,
        raise=-2pt
      }{
        \color{gray}
        \bf
        no structure
      }
    }
  }{
  1
  }
  \end{tikzcd}
$$

\noindent
Note how both these effects come about by changing the traditional flux-quantization of the Chern-Simons field from the classifying space for complex line bundles
to just its first ``cell''.
This preserves the quantization of charges but makes their moduli exhibit anyonic effects.
$$
\adjustbox{
  raise=3pt
}{
$
  \begin{tikzcd}[
    row sep=0pt,
    column sep=90pt
  ]
    \colorbox{lightolive}{$
     S^2
     $}
     \,\simeq\,
     \mathbb{C}P^1
     \quad 
    \ar[
      r, 
      hook,
      "{
        \scalebox{.7}{
          \color{darkgreen}
          \bf
          1st cell inclusion
        }
      }"{swap, yshift=-1pt}
    ]
    &
    \quad 
    \mathbb{C}P^\infty
    \,\simeq\,
    B^2 \mathbb{Z}
    \\
    \mathclap{
    \scalebox{.7}{
      \color{darkblue}
      \bf
      \def\arraystretch{.9}
      \begin{tabular}{c}  
        classifying space
        \\
        for 
        {\color{olive}2-Cohomotopy}
      \end{tabular}
    }
    }
    &
    \mathclap{
    \scalebox{.7}{
      \color{darkblue}
      \bf
      \def\arraystretch{.9}
      \begin{tabular}{c}  
        classifying space for
        \\
        ordinary 2-cohomology
      \end{tabular}
    }
    }
  \end{tikzcd}
$
}
$$

\noindent {\bf (iii) Topological order.} The traditional way of establishing topological order is by applying geometric quantization to Wilson line observables, with respect to some effective action, which
is a somewhat convoluted process   
involving ad-hoc choices and regularizations.
In contrast, in the approach discussed here, the quantum observables obtain immediately, without further choices, from the topological light-cone quantization of the flux-quantized moduli space (as its Pontrjagin homology algebra).
$$
\adjustbox{
  raise=-4pt
}{
$
  \begin{tikzcd}[
    decoration=snake,
    row sep=8pt, column sep=huge
  ]
    & 
    \adjustbox{
      scale=.7,
      margin=1pt,
      bgcolor=lightgray
    }{\def\arraystretch{.9}\def\tabcolsep{0pt}
      \begin{tabular}{c}
        phase
        \\
        space
      \end{tabular}
    }   
    \ar[
      rr,
      decorate,
      "{
        \scalebox{.7}{
          \def\arraystretch{.9}
          \begin{tabular}{c}
            choose prequantum line
            \\
            bundle \& polarization
          \end{tabular}
        }
      }"{yshift=5pt},
      "{
        \scalebox{.7}{
          \color{darkgreen}
          \bf
          \def\arraystretch{.9}
          \begin{tabular}{c}
            traditional
            quantization
          \end{tabular}
        }
      }"{swap, yshift=-11pt}
    ]
    &&[+2pt]
    \adjustbox{
      scale=.7,
      margin=1pt,
      bgcolor=lightgray
    }{\def\arraystretch{.9}\def\tabcolsep{0pt}
      \begin{tabular}{c}
        Hilbert
        \\
        space
      \end{tabular}
    }   
    \ar[
      dr,
      decorate,
      "{
        \scalebox{.7}{
          \def\arraystretch{.9}
          \begin{tabular}{c}
            choose \& regularize 
            \\
            operators
            \\
            to represent
          \end{tabular}
        }
      }"{sloped, yshift=2pt}
    ]
    &[+3pt]
    \\
    \adjustbox{
      scale=.7,
      margin=3pt,
      bgcolor=lightgray
    }{\def\arraystretch{.9}\def\tabcolsep{0pt}\color{darkblue}\bf
      \begin{tabular}{c}
        flux-quantized
        \\
        gauge fields
      \end{tabular}
    }
    \ar[
      rrrr,
      "{
        \scalebox{.7}{
          \color{darkgreen}
          \bf
          topological 
          light-cone
          quantization
        }
      }"{swap, yshift=-3pt}
    ]
    \ar[
      ur,
      decorate,
      "{
        \scalebox{.7}{
          \def\arraystretch{.85}
          \begin{tabular}{c}
            choose
            \\
            effective action
          \end{tabular}
        }
      }"{sloped, yshift=3pt}
    ]
    &&&&
    \adjustbox{
      scale=.7,
      margin=3pt,
      bgcolor=lightgray
    }{\def\arraystretch{.9}\def\tabcolsep{0pt}\color{darkblue}\bf
      \begin{tabular}{c}
        topological
        \\
        observable
        \\
        algebra
      \end{tabular}
    }   
  \end{tikzcd}
$
}
$$
Here the looping $\Omega_k$ that drives this quantum dynamics reflects dependence of moduli on the M/IIA circle.(!)

\medskip

\noindent {\bf (iv) Defect anyons} --- as opposed to the solitonic anyons tracing out ``Wilson lines'' --- seem to have previously found little to no attention in  quantum Hall theory in general and its effective abelian Chern-Simons theories in particular. And yet, it is only such classically parameterized and hence, in principle, externally controllable defect anyons which may support braid quantum gates as envisioned in topological quantum computation. 

In our approach, defect braiding emerges just as readily as the solitonic anyons, as a mild kind of quantum gravitational effect on M5-worldvolumes having a punctured surface factor space. This may be seen as a theoretical prediction of defect anyons in quantum Hall systems which might inform future search for experimental realization.

\newpage

\hrule
\vspace{-8pt}
\noindent
\colorbox{white}{\bf Summary of results:}

\vspace{-8pt}

\begin{center}
On super-space, the equations of motion 

of {\bf 11D supergravity} with magnetic $\sfrac{1}{2}$BPS {\bf M5-brane} probes

are equivalent to these Bianchi identities on the super-flux densities:

\vspace{4pt}

\hspace{-3pt}
\begin{tikzpicture}
  \draw[
    draw opacity=0,
    fill=olive,
    fill opacity=.2
  ]
    (0,0) rectangle (7.8,1.3);
\end{tikzpicture}
\vspace{-1.4cm}
$$
  \begin{tikzcd}[
    row sep=0pt,
    column sep=-20pt
  ]
    \scalebox{.7}{
      \color{darkblue}
      \bf
      A-field
    }
    &[+18pt]
    \mathrm{d}\, F^s_2 
    &=& 
    \hspace{-48pt}
    0
    &[+45pt]
    \mathrm{d}\, G^s_4 
       &=& 
     0
    &[+12pt]
    \scalebox{.7}{
      \color{darkblue}
      \bf
      C-field
    }
    \\
    \scalebox{.7}{
      \color{darkblue}
      \bf
      \def\arraystretch{.9}
      \def\tabcolsep{-20pt}
      \begin{tabular}{c}
        self-dual
        \\
        B-field
      \end{tabular}
    }
    \;\,
    &
    \mathrm{d}\, H^s_3 
      &=& 
    \phi_s^\ast G^s_4 
    \,+\,
    \mathcolor{purple}\theta \, F^s_2 \,F^s_2
    &
    \mathrm{d}\, G^s_7
    &=& 
    \tfrac{1}{2} G^s_4\, G^s_4    
    &
    \scalebox{.7}{
      \color{darkblue}
      \bf
      \def\arraystretch{.9}
      \def\tabcolsep{-20pt}
      \begin{tabular}{c}
        dual
        \\
        C-field
      \end{tabular}
    }
    \\[+1pt]
    \scalebox{.7}{
      \color{darkblue}
      \bf
      M5 probe
    }
    \hspace{-3pt}
    & 
    &
    \Sigma^{1,5\,\vert\,2\cdot \mathbf{8}_+}
    \ar[
      rrr,
      "{ \phi_s }",
      "{
        \scalebox{.7}{
          \color{darkgreen}
          \bf
          \scalebox{1.24}{$\sfrac{1}{2}$}BPS
          immersion
        }
      }"{swap}
    ]
    &&&
    X^{1,10\,\vert\,\mathbf{32}}
    &&
    \scalebox{.7}{
      \color{darkblue}
      \bf
      SuGra bulk
    }
  \end{tikzcd}
$$
\end{center}

\hrule

\begin{center}
One admissible choice of {\bf flux-quantization} law (the simplest in number of CW cells)

is {\bf twistorial Cohomotopy}, 
where the charges are classified by dashed maps like this:

\vspace{2pt}

$
  \begin{tikzcd}[
    column sep=15pt,
    row sep=15pt
  ]
    \scalebox{.7}{
      \color{darkblue}
      \bf
      \def\arraystretch{.9}
      \def\tabcolsep{-5pt}
      \begin{tabular}{c}
        M5-brane
        \\
        worldvolume
      \end{tabular}
    }
    &
    \Sigma^{1,5}
    \ar[
      rr,
      dashed,
      "{
        (a_1, b_2)
      }"{description},
      "{
        \scalebox{.7}{
          \color{darkorange}
          \bf
          A- \& B-field charge
        }
      }"{yshift=3pt}
    ]
    \ar[
      dd,
      "{
        \phi
      }"{description},
      "{
        \scalebox{.7}{
          \color{darkgreen}
          \bf
          immersion
        }
      }"{swap, sloped, yshift=-5pt}
    ]
    &\phantom{---------}&
    \mathbb{C}P^3
    \ar[
      dd,
      "{
        t_{\mathbb{H}}
      }"{description},
    ]
    \\
    \\
    \scalebox{.7}{
      \color{darkblue}
      \bf
      \def\arraystretch{.9}
      \def\tabcolsep{-5pt}
      \begin{tabular}{c}
        bulk
        \\
        spacetime
      \end{tabular}
    }
    &
    X^{1,10}
    \ar[
      rr,
      dashed,
      "{
        (c_3, c_6)
      }"{description},
      "{
        \scalebox{.7}{
          \color{darkorange}
          \bf
          C-field charge
        }
      }"{yshift=+5pt}
    ]
    &&
    S^4
  \end{tikzcd}
$

For  (very good) $G \subset \mathrm{Sp}(2)$-orbifold domains,
these maps are to be $G$-equivariant. 

\end{center}

\hrule

\begin{center}

This flux-quantization implies a list of topological effects
expected in M-theory.

$\Rightarrow$ {\bf Hypothesis H}: This is the right choice of flux-quantization for M-theory.

\end{center}

\hrule

\vspace{-.1cm}
\begin{center}
Choosing (``engineering'') the M5-probe to be:

\vspace{.1cm}

 \adjustbox{
    raise=2pt,
    scale=.8,
    margin=-2pt,
    fbox
  }{
  \hspace{-12pt}
  \begin{tikzcd}[
    column sep=-5pt,
    row sep=-20pt
  ]
  \scalebox{.7}{
    \color{darkblue}
    \bf
    \def\arraystretch{.85}
    \begin{tabular}{c}
      M5-brane probe
      \\
      worldvolume
    \end{tabular}
  }
  &&
  \scalebox{.7}{
    \color{darkblue}
    \bf
    \def\arraystretch{.85}
    \begin{tabular}{c}
      2-brane worldvolume
      \\
      hosting anyonic solitons
    \end{tabular}
  }
  &&
  \scalebox{.7}{
    \color{darkblue}
    \bf
    \def\arraystretch{.85}
    \begin{tabular}{c}
      M-theory
      \\
      circle
    \end{tabular}
  }
  &&
  \scalebox{.7}{
    \color{darkblue}
    \bf
    \def\arraystretch{.85}
    \begin{tabular}{c}
      cone
      \\
      orbifold
    \end{tabular}
  }
  \\[16pt]
  \Sigma^{1,5}
  &=&
  \mathbb{R}^{1,0}
  \times
  \Sigma^2_{g,n}
  &\times&
  S^1
  &\times&
  \mathbb{R}^2_{\mathrm{sgn}} \sslash \mathbb{Z}_2
  \\[-12pt]
  &&
  \adjustbox{
    raise=3.8cm,
    scale=.7
  }{
    \begin{tikzpicture}
\begin{scope}[
  scale=.8,
  shift={(.7,-4.9)}
]

  \shade[right color=lightgray, left color=white]
    (3,-3)
      --
    (-1,-1)
      --
        (-1.21,1)
      --
    (2.3,3);

  \draw[dashed]
    (3,-3)
      --
    (-1,-1)
      --
    (-1.21,1)
      --
    (2.3,3)
      --
    (3,-3);

  \node[
    scale=1
  ] at (3.2,-2.1)
  {$\infty$};

  \begin{scope}[rotate=(+8)]
  \draw[dashed]
    (1.5,-1)
    ellipse
    (.2 and .37);
  \draw
   (1.5,-1)
   to 
    node[above, yshift=-1pt]{
     \;\;\;\;\;\;\;\;\;\;\;\;\;
     \rotatebox[origin=c]{7}{
     \scalebox{.7}{
     \color{darkorange}
     \bf
       anyon
     }
     }
   }
    node[below, yshift=+6.3pt]{
     \;\;\;\;\;\;\;\;\;\;\;\;\;\;\;
     \rotatebox[origin=c]{7}{
     \scalebox{.7}{
     \color{darkorange}
     \bf
       worldline
     }
     }
   }
   (-2.2,-1);
  \draw
   (1.5+1.2,-1)
   to
   (4,-1);
  \end{scope}

  \begin{scope}[shift={(-.2,1.4)}, scale=(.96)]
  \begin{scope}[rotate=(+8)]
  \draw[dashed]
    (1.5,-1)
    ellipse
    (.2 and .37);
  \draw
   (1.5,-1)
   to
   (-2.3,-1);
  \draw
   (1.5+1.35,-1)
   to
   (4.1,-1);
  \end{scope}
  \end{scope}
  \begin{scope}[shift={(-1,.5)}, scale=(.7)]
  \begin{scope}[rotate=(+8)]
  \draw[dashed]
    (1.5,-1)
    ellipse
    (.2 and .32);
  \draw
   (1.5,-1)
   to
   (-1.8,-1);
  \end{scope}
  \end{scope}

\end{scope}
    \end{tikzpicture}
    }
    &\times&
    \adjustbox{
      raise=3pt
    }{
    \begin{tikzpicture}
      \draw[
        line width=2.2,
        draw=darkgreen
      ]
        (0,0) circle (1);
    \end{tikzpicture}
    }
    &\times&
    \adjustbox{
      raise=3pt
    }{
\begin{tikzpicture}
\begin{scope}[
  xscale=.7,
  yscale=.5*.7
]
 \shadedraw[draw opacity=0, top color=darkblue, bottom color=cyan]
   (0,0) -- (3,3) .. controls (2,2) and (2,-2) ..  (3,-3) -- (0,0);
 \draw[draw opacity=0, top color=white, bottom color=darkblue]
   (3,3)
     .. controls (2,2) and (2,-2) ..  (3,-3)
     .. controls (4,-3.9) and (4,+3.9) ..  (3,3);
\end{scope}
\end{tikzpicture}
    }
  \end{tikzcd}
  \hspace{-14pt}
  }

\vspace{.15cm}

the {\bf moduli space of solitons}
becomes:

$
  \mathrm{Moduli}
  \;\simeq\;
  \mathrm{Maps}^\ast\Big(
    \big(
      \mathbb{R}^1 \times \Sigma^2_{g,n}
    \big)_{\cpt}
    ,\,
    S^2
  \Big)
$

\end{center}

\hrule

\begin{center}

The algebra of {\bf topological quantum observables} on theses solitons is:

$
  \begin{tikzcd}[
    column sep=0pt,
    row sep=-10pt
  ]
  \mathrm{Obs}_0
  &:=&
  H_0\bigg(
  \mathrm{Maps}^\ast\Big(
    \big(
      \mathbb{R}^1 \times \Sigma^2_{g,n}
    \big)_{\cpt}
    ,\,
    S^2
  \Big)
  ;\,
  \mathbb{C}
  \bigg)
  &\simeq&
  \mathbb{C}\Big[
    \pi_1
    \mathrm{Maps}^\ast\big(
      \Sigma^2_{g,n}
      ,\,
      S^2
    \big)
  \Big]
  \,,
  \\
  \mathclap{
    \scalebox{.7}{
      \color{darkblue}
      \bf
      \def\arraystretch{.9}
      \begin{tabular}{c}
        topological 
        \\
        quantum observables
      \end{tabular}
    }
  }
  &&
  \mathclap{
    \scalebox{.7}{
      \color{darkblue}
      \bf
      \def\arraystretch{.9}
      \begin{tabular}{c}
        Pontrjagin homology algebra
      \end{tabular}
    }
  }
  &&
  \mathclap{
    \scalebox{.7}{
      \color{darkblue}
      \bf
      \def\arraystretch{.9}
      \begin{tabular}{c}
        group algebra
      \end{tabular}
    }
  }
  \end{tikzcd}
$

acted on by large diffeomorphisms ({\bf general covariance} on the brane):

$
  \begin{tikzcd}[
    row sep=-2pt,
    column sep=20pt
  ]
    1
    \ar[r]
    &
    \mathrm{Br}_n(\Sigma^2_g)
    \ar[rr, hook]
    &&
    \pi_0 \mathrm{Homeos}^{\ast}_{\mathrm{or}}\big(
      (\Sigma^2_{g,n})_{\cpt}
    \big)
    \ar[rr,->>]
    &&
    \mathrm{MCG}(\Sigma^2_g)
    \ar[r]
    &
    1
    \\
    &
    \mathclap{
      \scalebox{.7}{
        \color{darkblue}
        \bf
        braid group
      }
    }
    &&
    \mathclap{
      \scalebox{.7}{
        \color{darkblue}
        \bf
        large diffeomorphism group
      }
    }
    &&
    \mathclap{
      \scalebox{.7}{
        \color{darkblue}
        \bf
        mapping class group
      }
    }
  \end{tikzcd}
$

\end{center}

\hrule

\begin{center}

The corresponding {\bf topological quantum states}:

\vspace{.1cm}

\def\arraystretch{1.3}
\begin{tabular}{ll}
on $\Sigma^2_{0,0} = S^2$ 
&
reflect abelian braiding of {\bf \color{darkblue}solitonic anyons}
\\
on $\Sigma^2_{g,0} = \Sigma^2_{1,0} \# \cdots \# \Sigma^2_{1,0}$ 
&
have $k^g$-fold degeneracy:
{\bf \color{darkblue}topological order}
\\
on $\Sigma^2_{1,0} = \mathbb{T}^2$ 
&
exhibit irred $\mathrm{SL}_2(\mathbb{Z})$-{\bf \color{darkblue}modular equivariance}
\\
on $\Sigma^2_{0,n} = S^2 \setminus \{z^1, \cdots, z^n\}$ 
&
reflect abelian braiding of 
$
\underset{
  \mathclap{
  \adjustbox{
    scale=.85
  }{
  \color{purple}
  \def\arraystretch{.9}
  \begin{tabular}{c}
    new \& needed for
    \\
    topological
    quantum gates!
  \end{tabular}
  }
  }
}{
  \mbox{\textit{\textbf{\color{darkblue}defect anyons}}}
}
$
\end{tabular}
\hspace{-.6cm}
\adjustbox{
  scale=.75,
  rotate=-90
}{
\clap{
\hspace{-62pt}
\def\arraystretch{.9}
\begin{tabular}{c}
  as for abelian
  \\
  {\bf Chern-Simons QFT}
  \\
  $\overbrace{\phantom{------}}$
\end{tabular}
}
}

\end{center}

\hrule

\hspace{-.8cm}
\adjustbox{
  rotate=90
}{
  \clap{
    \colorbox{white}{
    \Huge
    \color{gray}
    Engineering of Anyons 
    \hspace{6pt} on M5-Branes 
    \hspace{6pt} via Flux-Quantization\hspace{-3pt}}
    \hspace{-23.6cm}
  }
}

\newpage

A broad lesson following immediately from our successful geometric engineering of topological qbits is the plausible existence of more exotic anyonic states than traditionally envisioned: Namely the ``duality symmetry'' \cite{Polchinski17}\cite[\S 6]{Duff99} of M-theory predicts that any geometrically engineered quantum system has ``dual'' incarnations with isomorphic quantum observables but entirely different geometric realization, where ordinary space is replaced by more abstract parameter spaces.
Notably ``T-duality'' 
\cite{Waldorf24}\cite{FSS18-TDual}\cite{GSS24-TDual}
applied to topological quantum materials
has been argued 
\cite{MathaiThiang15}\cite{MathaiThiang16}\cite{Hannabuss18}
to exchange the roles of ordinary space with that of reciprocal ``momentum space''.

\medskip

\noindent
{\bf (2) Novel experimental pathways towards anyons.}
Indeed, while anyonic solitons are traditionally envisioned as being localized in ``position space'' (meaning that the anyon cores are points in the plane of the crystal lattice) the physical principle behind topological quantum gates --- namely 
\cite{ASW84}\cite{ASWZ85}\cite[p 6]{FKLW03}\cite[p 50]{Pachos12} 
the {\it quantum adiabatic theorem} \cite{RigolinOrtiz12} --- is unspecific to position space and only requires the material's Hamiltonian to {\it depend on any continuous parameters} (such as external voltage or strain) varying in any abstract parameter space.

\medskip 
\hspace{.1cm}
\begin{tabular}{p{6cm}}
  \footnotesize
  {\bf The general physical conditions for topological quantum gates} given by the {\it quantum adiabatic theorem}, listed (a) - (e) on the right, are much more general than traditionally considered for anyon braid gates --- the latter are only the special case where the parameters are configurations of points in the plane of the 2D crystal lattice.

\end{tabular}
\quad 
\adjustbox{
  raise=3pt,
  scale=.85,
  fbox
}{
\hspace{-.1cm}
\begin{minipage}{11cm}
\begin{itemize}[
  itemsep=-3pt
]
\item[(a)] {\bf Ground state degeneracy}
{\small
(when frozen at absolute zero, the system still has
more than one state to be in, even up to phase).
}

\item[(b)] {\bf Spectral gap}
{\small
(quanta of energy smaller than a given
gap $\epsilon > 0$ cannot excite these
ground states).
}

\item[(c)] {\bf Control parameters}
{\small
(the above properties hold for a range
of continuously tunable external parameters).
}

\item[(d)] {\bf Parameter topology}
{\small 
(there exist 
closed parameter paths that cannot be continuously contracted).
}

\item[(e)] {\bf Local invariance}
{\small
(continuously deformed parameter paths induce the same transformation on ground states).
}

\end{itemize}
\end{minipage}
}

\medskip

\hspace{-.7cm}
\begin{tabular}{p{6.4cm}}
This means that, in principle, the possibilities in which anyonic quantum states could arise in the laboratory are far more general than what has been explored to date.

Concretely, a key example of alternative parameters for ground states of a quantum material are points in their reciprocal  {\it momentum space}: This is the space of (quasi-)momenta, hence of wave-vectors for plane quasi-particle waves going through the crystalline material.
\end{tabular}
\qquad 
\adjustbox{
  raise=-2.7cm,
  scale=.9
}{
\begin{tikzpicture}

\begin{scope}[
  shift={(3,0)}
]

\begin{scope}
\clip
  (-2,-2) rectangle (2,2);

\node at (0,-3.51){
\clap{
\smash{
\adjustbox{
  rotate=23
}{
\includegraphics[width=5.33cm]{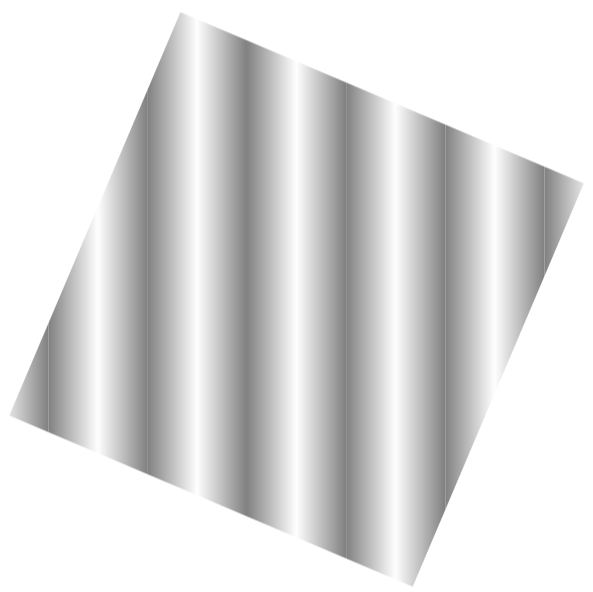}
}
}
}
};
\end{scope}

\draw[
 line width=2,
 dashed
]
  (-2,-2) rectangle (2,2);

\draw[
  -Latex,
  line width=1.6,
  purple
]
  (-2,-2) to
  (-2+1.19,-2+.56);
\end{scope}

\begin{scope}[
  shift={(-3,0)}
]
\draw[
 line width=2,
 dashed
]
  (-2,-2) rectangle (2,2);

\draw[
  -Latex,
  line width=1.6,
  purple
]
  (-2,-2) to
  (-2+1.19,-2+.56);

\draw[
  draw opacity=0,
  fill=gray
]
  (-2+1.19,-2+.56)
  circle (.15);

\end{scope}

\draw[
  <->,
  line width=1.2
]
  (-.6,0) 
  -- 
  node[
    yshift=9pt
  ] {\scalebox{.7}{
    \color{darkgreen}
    \bf
    duality
  }}  
  (.6,0);

\node
 at (-3,-2.65) {\footnotesize
   \def\arraystretch{.9}
   \begin{tabular}{c}
     {\bf Position space:}
     \\
     a {\color{purple}point} is a position
     \\
     on the crystal lattice
   \end{tabular}
 };

\node
 at (+3,-2.65) {\footnotesize
   \def\arraystretch{.9}
   \begin{tabular}{c}
     {\bf Momentum space:}
     \\
     a {\color{purple}point} is a plane wave
     \\
     on the crystal lattice
   \end{tabular}
 };
 
\end{tikzpicture}
}

\medskip 
We have observed before that candidate anyon-like solitons localized (not in position space but) in momentum space are plausible both theoretically \cite{SS24-TopOrd} as well as experimentally \cite{WuSoluyanovBzdusek19}\cite{TiwariBzdusek20}\cite{JiangEtAl21} and may have been hiding in plain sight: as band nodes of (interacting) topological semimetals.

\vspace{-0cm}
\hspace{-.64cm}
\begin{tabular}{p{11.2cm}}
Indeed, momentum space naturally features key properties that are typically assumed for anyon braid gates but remain elusive in position space:

\vspace{1mm} 
{\bf (i) toroidal base topology}
is routinely assumed
\cite{Wen89}\cite{WenDagottoFradkin90}\cite{Liu24} in order to achieve the required ground-state degeneracy, but is quite unrealistic in position space, even more so when meant to be punctured by defect anyons  --- while
the momentum space of a crystal is automatically a torus (the {\it Brillouin torus}).

\vspace{1mm} 
{\bf (ii) stable defect points}  
need special engineering 
in
position space but arise automatically in momentum space in the guise of {\it band nodes}
of topological semi-metals
\cite[Fig. 6]{SS24-TopOrd}

\vspace{1mm} 
{\bf (iii) defect point movement}
in a  {\it controlled} way is necessary for braid gates but remains elusive in position space, while band nodes in momentum space have already shown to be movable in a variety of systems, by tuning of external parameters (e.g., strain).
\end{tabular}
\hspace{.5cm}
\adjustbox{
  raise=-3.2cm
}{
\includegraphics[width=4.2cm]{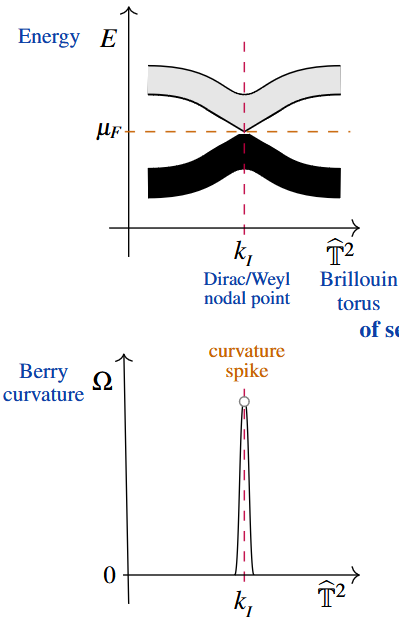}}

\smallskip

The geometric engineering of anyons discussed here goes towards providing also fundamental theoretical underpinning of the possibility of more ``exotic'' anyon realizations than have traditionally been envisioned.

\newpage

%%%%%%%%%%%%%%%%%%%%%%%%%%%
\section{Digest for Algebraic Topologists}
%%%%%%%%%%%%%%%%%%%%%%%%%%%

We are concerned with 
\adjustbox{
  margin=1pt,
  bgcolor=lightolive
}
{algebro-topological phenomena} arising
when 
\adjustbox{
  margin=1pt,
  bgcolor=lightolive
}{magnetic flux} penetrates
a semi-conducting surface $\Sigma^2$.
The ``gauge group'' of the electromagnetic field is 
$G \defneq \mathrm{U}(1)$
and {\it ordinarily} such flux is classified by maps to
$B \mathrm{U}(1) \,\simeq\, \mathbb{C}P^\infty$.
The following theorem turns the analysis of this situation first into a problem of differential topology, and then into a problem of algebraic topology.

\vspace{-2mm} 
\begin{center}
\hspace{0cm}
\clap{
\adjustbox{
  raise=-1.9cm
}{
\begin{tikzpicture}
\node at (0,0)
{
\includegraphics[width=4.1cm]{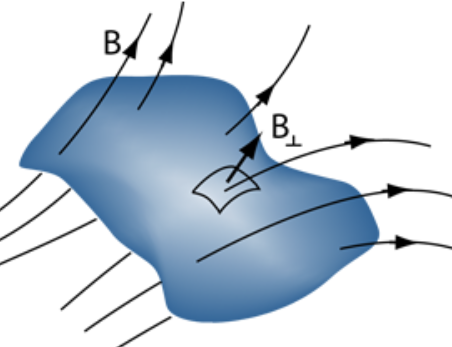}
};

\node[
  scale=.7,
  rotate=-81
] at (2.2,-.2) {
  \color{darkblue}
  \bf
  magnetic flux 
};

\node[
  scale=.7,
  rotate=-48
] at (-.74,-.24) {
  \color{darkorange}
  \bf
  \begin{tabular}{c}
    surface 
    \\
    \;\;\;\;\;\;\;$\Sigma^2$
  \end{tabular}
};
\end{tikzpicture}
}
\begin{minipage}{6.5cm}
\def\tabcolsep{6pt}
\def\arraystretch{1.2}
\begin{tabular}{cl}
  $G$ & Lie group (``gauge group'')
  \\
  $\mathfrak{g}$ & its Lie algebra
  \\
  $C^\infty(\mbox{-},\mbox{-})$
  &
  manifold of smooth functions
  \\
  $(\mbox{-})\ltimes(\mbox{-})$
  &
  semidirect product via adjoint
  \\
  $\mathbb{C}[-]$
  &
  group convolution
  $C^\ast$-algebra
  \\
  $\pi_0(-)$
  &
  path-connected components
\end{tabular}
\end{minipage}
\;\;
\begin{minipage}{6cm}
  \def\tabcolsep{3pt}
  \begin{tabular}{l}
    soliton on $X$
  \\
  \begin{tabular}{ll}
  =
  &
  topological field configuration 
  \\
  &
  that vanishes at the ends of $X$
  \end{tabular}
  \\[10pt]
  \begin{tabular}{ll}
   $\Rightarrow$
   &
   classified by {\it pointed} map
   \\
   &
  $
    X_{\cpt} \xrightarrow{\;} B G
  $
  \\
  &
  from one-point compactification
  \end{tabular}
  \end{tabular}
\end{minipage}
}
\end{center}

\smallskip

\noindent
{\bf Theorem \cite{SS24-QObs} (Yang-Mills flux quantum observables):} 
\textit{For ordinary gauge fields on a spacetime $\simeq \mathbb{R}^{1,1} \times \Sigma^2$}, 
and for $\Lambda \subset \mathfrak{g}$ an $\mathrm{Ad}$-invariant lattice.
\textit{the \textbf{quantum observables of field flux}
through $\Sigma^2$}
\textit{form the group-convolution $C^\ast$-algebra}
\vspace{-1mm} 
$$
\adjustbox{
  margin=-1pt -.5pt -2pt 2pt,
  bgcolor=lightolive
}{
$
  \underset{
    \mathclap{
      \adjustbox{
        raise=-0pt,
        scale=.7
      }
      {
        \color{darkblue}
        \bf
        quantum flux observables
      }
    }
  }{
  \mathbb{C}\Big[
  C^\infty\big(
    \Sigma^2
    ,\,
    G \ltimes
    (\mathfrak{g}/\Lambda)
  \big)
  \Big]
  }
$
}
$$

\smallskip 
\noindent
\textbf{Commercial-value quantum computing} 
will require
\textbf{robust} quantum observables, insensitive to local fluctuations, 
only depending on {\bf topological sectors} of field configurations.

\vspace{1pt}
$$
  \begin{tikzcd}[column sep=large]
  \underset{
    \mathclap{
      \adjustbox{
        raise=-3pt,
        scale=.7
      }{
        \color{darkblue}
        \bf
        \def\arraystretch{.9}
        \begin{tabular}{c}
          all quantum flux observables
        \end{tabular}
      }
    }
  }{
  C^\infty\big(
    \Sigma^2
    ,\,
    G \ltimes
    (\mathfrak{g}/\Lambda)
  \big)
  }
  \ar[
    r,
    "{
      [-]
    }"
  ]
  &
  \adjustbox{
    margin=1pt 0pt 1pt 2pt,
    bgcolor=lightolive
  }{$
  \underset{
    {
      \adjustbox{
        raise=-3pt,
        scale=.7
      }{
        \color{darkblue}
        \bf
        \def\arraystretch{.9}
        \begin{tabular}{c}
          robust topological observables
        \end{tabular}
      }
    }
  }{
  \mathcolor{purple}{\pi_0}\,
  C^\infty\big(
    \Sigma^2
    ,\,
    G \ltimes
    (\mathfrak{g}/\Lambda)
  \big)
  }
  $}
  \end{tikzcd}
$$

\noindent
{\bf Proposition \cite{SS24-QObs} (topological sector observables):}
The topological flux quantum observables  form the homology
Pontrjagin  algebra of 
maps from space to classifying space.
(shown now assuming $\Lambda = 0$, for simplicity):
\vspace{1mm} 
$$
  \def\arraystretch{1.8}
  \begin{array}{l}
    \overset{
      \adjustbox{
        raise=3pt,
        scale=.7
      }{
        \color{darkblue}
        \bf
        Topological flux 
        quantum observables
      }
    }{
    \mathbb{C}\Big[
    \pi_0 
    \, 
    C^\infty\big(
      \Sigma^2
      ,\,
      G 
    \big)
    \Big]
    \;\simeq\;
    \mathbb{C}\Big[
    \pi_0 
    \,  
    \mathrm{Maps}\big(
      \Sigma^2
      ,\,
      G 
    \big)
    \Big]
    }
%    \\
    \adjustbox{
      margin=1pt 0pt 1pt 2pt,
      bgcolor=lightolive
    }{$
    \;\simeq\;\;\;\;
    \underset{
      \mathclap{
        \adjustbox{
          raise=-3pt,
          scale=.7
        }{
          \color{darkblue}
          \bf
          \def\arraystretch{.9}
          \begin{tabular}{c}
            group algebra of
            fundamental group
            \\
            of maps to classifying space
          \end{tabular}
        }
      }
    }{
    \mathbb{C}\Big[
    \pi_1 
    \,  
    \mathrm{Maps}\big(
      \Sigma^2
      ,\,
      B G 
    \big)
    \Big]
    }
    \;\;\;\;\simeq\;
    \underset{
      \mathclap{
        \adjustbox{
          raise=-3pt,
          scale=.7
        }{
          \color{darkblue}
          \bf
          \def\arraystretch{.9}
          \begin{tabular}{c}
            homology Pontrjagin algebra of
            \\
            \color{darkorange}
            soliton moduli space
          \end{tabular}
        }
      }
    }{
    H_0\Big(
    \mathrm{Maps}^\ast
    \scalebox{1.3}{$($}
      \big(
        \mathbb{R}^1 \times
        \Sigma^2
      \big)_{\!\cpt}
      ,\,
      B G 
    \scalebox{1.3}{$)$}
    ;\,
    \mathbb{C}
    \Big)
    }
    $}
  \end{array}
$$

%\end{minipage}

\vspace{5pt}

\noindent
{\bf Example:} 
$
  \mathbb{C}\big[
    \pi_0
    \,
    \mathrm{Maps}\big(
      \Sigma^2_g
      ,\,
      \mathrm{U}(1)
    \big)
  \big]
  \;\simeq\;
  \mathbb{C}\big[
   H^1(\Sigma^2_g;\, \mathbb{Z})
  \big]
  \;\simeq\;
  \adjustbox{
    margin=1pt 2pt 1pt 2pt,
    bgcolor=lightolive
  }{$
  \mathbb{C}\big[
    \mathbb{Z}^{2g}
  \big]
  $}
$
\;\;
for 
 $\Sigma^2_g$ 
an
 orientable surface of genus=$g$.

\vspace{3mm}

\noindent
{\bf Effective flux of ``fractional quantum Hall systems''}(FQH).
{\color{purple}However}, at very low temperature, experiment
suggests 
instead of $\mathbb{Z}^{2g}$
its 2nd 
\adjustbox{
  margin=0pt 1pt 2pt 1pt,
  bgcolor=lightolive
}{
integer Heisenberg extension
$\widehat{\mathbb{Z}^{2g}}$}
\vspace{-2mm} 
$$
\begin{tikzpicture}
\node at (0,0) {
$
  \widehat{\mathbb{Z}^{2g}}
  \;:=\;
  \left\{
  \def\arraycolsep{-2pt}
  \def\arraystretch{1.3}
  \begin{array}{c}
    \big(
      \vec a, 
      \vec b,
      n
    \big)
    \,\in\,
    \mathbb{Z}^{g}
    \times
    \mathbb{Z}^g
    \times
    \mathbb{Z}
  %  \\[+10pt]
    \;,
    \big(
      \vec a, 
      \vec b,
      n
    \big)
    \cdot
    \big(
      \vec a', 
      \vec b',
      n'
    \big)
    \,=\,
  %  \\
    \big(
      \vec a + \vec a'
      ,\,
      \vec b + \vec b'
      ,\,
      n + n' + 
      \mathcolor{purple}{
        \vec a 
        \cdot 
        \vec b'
        -
        \vec a' \cdot \vec b
      }
  \,  \big)
  \end{array}
  \right\}
$
};
\node[scale=.7] 
at (6,.5) {
  \color{darkblue}
  \bf
  \def\arraystretch{.9}
  \begin{tabular}{c}
    twice the unit
    \\
    central extension
  \end{tabular}
};
\end{tikzpicture}
$$
being the observables of an 
``\textbf{effective Chern-Simons field}'',
where the center
$\mathbb{Z} \hookrightarrow \widehat{\mathbb{Z}^{2g}}$
observes an \textbf{anyon braiding phase}.

\vspace{-7mm}
\begin{center} 
\begin{minipage}{5cm}
  \footnotesize
  {\bf Figure \figurenumber:
  Fundamental polygons for closed surfaces.}
\end{minipage}
\;\;\;\;
\begin{minipage}{9cm}
\adjustbox{
  scale=.8,
  raise=-.65cm
}{
\begin{tikzpicture}
  \node  at (1,2.8)
  {
    \scalebox{.9}{
      \color{darkblue}
      \bf
      sphere
    }
  };
  \draw[
    line width=3,
    fill=lightgray
  ]
    (0,0)
    rectangle 
    (2,2);

\draw[fill=black]
  (0,0) circle (.08);

\draw[fill=black]
  (2,0) circle (.08);

\draw[fill=black]
  (0,2) circle (.08);

\draw[fill=black]
  (2,2) circle (.08);

\node
  at (1,.9) {$
    \Sigma^2_0 \simeq S^2
  $};

\end{tikzpicture}
}
%%%%%%%%%%%%%%%%
\;\;\;
\adjustbox{
  scale=.8,
  raise=-.64cm
}{
\begin{tikzpicture}
  \node  at (1,2.8)
  {
    \scalebox{.9}{
      \color{darkblue}
      \bf
      torus
    }
  };
  \draw[
    draw opacity=0,
    fill=lightgray
  ]
    (0,0)
    rectangle 
    (2,2);

\draw[
  dashed,
  color=darkgreen,
  line width=1.6,
]
  (0,2) -- 
  node[yshift=8pt]{
    \scalebox{1}{
      \color{black}
      $a$
    }
  }
  (2,2);
\draw[
  -Latex,
  darkgreen,
  line width=1.6
]
  (1.3-.01,2) -- 
  (1.3,2);

\draw[
  dashed,
  color=darkgreen,
  line width=1.6,
]
  (0,0) -- (2,0);
\draw[
  -Latex,
  darkgreen,
  line width=1.6
]
  (1.3-.01,0) -- 
  (1.3,0);

\draw[
  dashed,
  color=olive,
  line width=1.6,
]
  (0,0) -- (0,2);
\draw[
  -Latex,
  olive,
  line width=1.6
]
  (0, .8+.01) -- 
  (0,.8);

\draw[
  dashed,
  color=olive,
  line width=1.6,
]
  (2,0) -- 
  node[xshift=8pt]{
    \scalebox{1}{
      \color{black}
      $b$
    }
  }
  (2,2);
\draw[
  -Latex,
  olive,
  line width=1.6
]
  (2, .8+.01) -- 
  (2,.8);

\draw[fill=black]
  (0,0) circle (.08);

\draw[fill=black]
  (2,0) circle (.08);

\draw[fill=black]
  (0,2) circle (.08);

\draw[fill=black]
  (2,2) circle (.08);

\node
  at (1,.9)
  {
    $\Sigma^2_1 \simeq \mathbb{T}^2$
  };

\end{tikzpicture}
}
\hspace{3pt}
%%%%%%%%%%%%%%%%%%%%
\adjustbox{
  raise=-1.4cm
}{
  \begin{tikzpicture}[
    scale=.65
  ]

  \node[
    rotate=39
  ] at (-1.8,1.8) {
    \scalebox{.64}{
      \color{darkblue}
      \bf
      \def\arraystretch{.7}
      \begin{tabular}{c}
        2-holed
        \\
        torus
      \end{tabular}
    }
  };

  \draw[
    draw opacity=0,
    fill=lightgray
  ]
    (90-22.5:2) --
    (45+90-22.5:2) --
    (2*45+90-22.5:2) --
    (3*45+90-22.5:2) --
    (4*45+90-22.5:2) --
    (5*45+90-22.5:2) --
    (6*45+90-22.5:2) --
    (7*45+90-22.5:2) --
    cycle;

    \draw[
      line width=1.6,
      dashed,
      darkgreen
    ]
      (90-22.5:2)
      --
      (90+22.5:2);

    \draw[
      line width=1.6,
      dashed,
      olive
    ]
      (-45+90-22.5:2)
      --
      (-45+90+22.5:2);

    \draw[
      line width=1.6,
      dashed,
      darkgreen
    ]
      (-90+90-22.5:2)
      --
      (-90+90+22.5:2);

    \draw[
      line width=1.6,
      dashed,
      olive
    ]
      (-135+90-22.5:2)
      --
      (-135+90+22.5:2);

    \draw[
      line width=1.6,
      dashed,
      darkblue
    ]
      (-180+90-22.5:2)
      --
      (-180+90+22.5:2);

    \draw[
      line width=1.6,
      dashed,
      darkblue
    ]
      (-180+90-22.5:2)
      --
      (-180+90+22.5:2);

    \draw[
      line width=1.6,
      dashed,
      purple
    ]
      (-225+90-22.5:2)
      --
      (-225+90+22.5:2);

    \draw[
      line width=1.6,
      dashed,
      darkblue
    ]
      (-270+90-22.5:2)
      --
      (-270+90+22.5:2);

    \draw[
      line width=1.6,
      dashed,
      purple
    ]
      (-315+90-22.5:2)
      --
      (-315+90+22.5:2);

\draw[
  line width=1.3,
  -Latex,
  darkgreen
]
  (0.3,1.83) -- (.35,1.83);

\begin{scope}[
  rotate=-45
]
\draw[
  line width=1.3,
  -Latex,
  olive
]
  (0.24,1.83) -- 
  (.25,1.83);
\end{scope}
\begin{scope}[
  xscale=-1,
  rotate=+90
]
\draw[
  line width=1.3,
  -Latex,
  darkgreen
]
  (0.25,1.83) -- 
  (.3,1.83);
\end{scope}
\begin{scope}[
  xscale=-1,
  rotate=+135
]
\draw[
  line width=1.3,
  -Latex,
  olive
]
  (0.25,1.83) -- 
  (.3,1.83);
\end{scope}

\begin{scope}[
  rotate=+180
]
\draw[
  line width=1.3,
  -Latex,
  darkblue
]
  (.23,1.83) -- 
  (.33,1.83);
\end{scope}

\begin{scope}[
  rotate=-225
]
\draw[
  line width=1.3,
  -Latex,
  purple
]
  (.2,1.83) -- 
  (.3,1.83);
\end{scope}

\begin{scope}[
  rotate=-270
]
\draw[
  line width=1.3,
  -Latex,
  darkblue
]
  (-.20,1.83) -- 
  (-.25,1.83);
\end{scope}
\begin{scope}[
  rotate=-315
]
\draw[
  line width=1.3,
  -Latex,
  purple
]
  (-.20,1.83) -- 
  (-.25,1.83);
\end{scope}

\foreach \n in {1,...,8} {
  \draw[
    fill=black
  ]
    (22.5+\n*45:2)
    circle
    (.11);
};

\node[
  scale=.9
] at
  (0,2.3) {
    $a_1$
  };

\begin{scope}[
  rotate=-45
]
  \node[scale=.9] at
    (0,2.3) {
      $b_1$
    };
\end{scope}

\begin{scope}[
  rotate=-180
]
  \node[scale=.9] at
    (0,2.3) {
      $a_2$
    };
\end{scope}

\begin{scope}[
  rotate=-180-45
]
  \node[scale=.9] at
    (0,2.3) {
      $b_2$
    };
\end{scope}

\node
  at (0,0) {
    $\Sigma^2_2$
  };
  
\end{tikzpicture}
}

\end{minipage}
\end{center}

\noindent
{\bf Question:} 
Is there classifying space 
$\hotype{A}$
for this effective CS field?

\medskip 
\noindent {\bf Answer:}
Yes! The 2-sphere
\adjustbox{
  margin=0pt .5pt 0pt 2pt,
  bgcolor=lightolive
}{
$ 
  \mathcolor{purple}{S^2} \,\simeq\, 
  \mathbb{C}P^1
  \hookrightarrow
  \mathbb{C}P^\infty
  \simeq
  B \mathrm{U}(1)
$
}

\newpage 
\noindent {\bf Theorem \cite{Hansen74}\cite{LarmoreThomas80}:} 
The cofiber presentation of the surface
$$
  \begin{tikzcd}
    S^1
    \ar[
      r,
      "{
        \prod_i [a_i, b_i]
      }"
    ]
    &[+5pt]
    \bigvee_{g}(
      S^1_a \vee S^1_b
    )
    \ar[
      rr
    ]
    &&
    \mathcolor{darkblue}{\Sigma^2_g}
    \ar[rr]
    &&
    S^2
  \end{tikzcd}
$$
induces  short exact sequence exhibiting the Heisenberg extension:
$$
  \begin{tikzcd}[
    column sep=20pt
  ]
    1 
    \ar[
      r
      %,shorten=-4pt
    ]
    &
    \grayunderbrace{
    \pi_1
    \mathrm{Maps}\big(
      S^2
      ,\,
      \mathcolor{purple}{S^2}
    \big)
    }{
      \mathcolor{black}{\mathbb{Z}}
    }
    \ar[
      rr
      %, shorten=-4pt
    ]
    &&
    \grayunderbrace{
    \pi_1
    \mathrm{Maps}\big(
      \Sigma^2_g
      ,\,
      \mathcolor{purple}{S^2}
    \big)
    }{
      \mathcolor{black}{
        \widehat{\mathbb{Z}^{2g}}
      }
    }
    \ar[
      rr
      %,       shorten=-4pt
    ]
    &&
    \grayunderbrace{
    \pi_1
    \mathrm{Maps}^\ast
    \big(
      \textstyle{\bigvee_{\!2g}}
      S^1
      ,\,
      \mathcolor{purple}{S^2}
    \big)
    }{
      \mathcolor{black}{
      \mathbb{Z}^{2g}
      }
    }
    \ar[
      r
      %,      shorten=-4pt
    ]
    &
    1\;.
  \end{tikzcd}
$$

\noindent
\begin{minipage}{10.5cm}

{\bf Question:} Can we identify the center $\mathbb{Z}$ as arising from braiding?

{\bf Answer:} Yes! 

\end{minipage} 

\medskip 
\noindent {\bf Theorem \cite{SS24-AbAnyons}:}
$\mathrm{Maps}^\ast(S^2, S^2)$ is configurations of charged strings
such that $\Omega \mathrm{Maps}^\ast(S^2, S^2)$ is framed links subject to cobordism,
$\pi_1 \mathrm{Maps}^\ast(S^2, S^2)$ generated from framed unknot with 1 braiding
$$
\begin{tikzcd}[
  row sep=-4pt
]
  \Omega
  \mathrm{Maps}^\ast\big(
    S^2, \, S^2
  \big)
  \ar[
    rr,
    "{
      [-]
    }"
  ]
  &&
  \pi_3(S^2)
  \,\simeq\,
  \mathbb{Z}
  \\
  L
  \ar[
    rr,
    |->,
    shorten=8pt
  ]
  &&
  \# L
  \\
  \mathclap{
    \adjustbox{
      scale=.7
    }{
      \color{darkblue}
      \bf
      framed link
    }
  }
  &&
  \mathclap{
    \adjustbox{
      scale=.7
    }{
      \color{darkblue}
      \bf
      \begin{tabular}{c}
        linking + framing
        \\
        number
      \end{tabular}
    }
  }
\end{tikzcd}
$$

\vspace{-2mm} 
\noindent is CS 
observable
(``Wilson loop'').

\begin{center} 
\begin{minipage}{9cm}

\includegraphics[width=7cm]{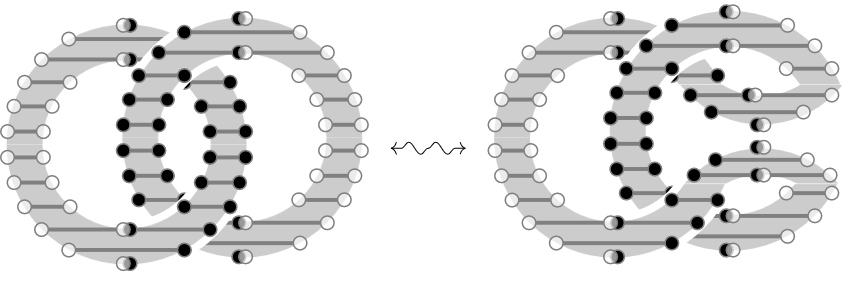}

\vspace{4pt}

\adjustbox{
  scale=1.1
}{
$
\#\left(\!
\adjustbox{raise=-.63cm, scale=.5}{
\begin{tikzpicture}

\draw[
  line width=1.2,
  -Latex
]
  (-.7,-.7) -- (.7,.7);
\draw[
  line width=7,
  white
]
  (+.7,-.7) -- (-.7,.7);
\draw[
  line width=1.2,
  -Latex
]
  (+.7,-.7) -- (-.7,.7);
 
\end{tikzpicture}
}
\!\right)
\,=\,
+1
\,,
\hspace{.4cm}
\#\left(\!
\adjustbox{raise=-.63cm, scale=0.5}{
\begin{tikzpicture}[xscale=-1]

\draw[
  line width=1.2,
  -Latex
]
  (-.7,-.7) -- (.7,.7);
\draw[
  line width=7,
  white
]
  (+.7,-.7) -- (-.7,.7);
\draw[
  line width=1.2,
  -Latex
]
  (+.7,-.7) -- (-.7,.7);
 
\end{tikzpicture}
}
\!\right)
\,=\,
-1\,,
$
}
\end{minipage}
\end{center}

\noindent

\noindent
{\bf Ergo:} 
Remarkably, topological quantum observables of effective flux
in quantum Hall systems is algebro-topologically described by
replacing the classifying space 
$B \mathrm{U}(1) \,\simeq\, \mathbb{C}P^\infty$
with its 2-skeleton 
$S^2 \,\simeq\, \mathbb{C}P^1$.

\vspace{3pt}

\noindent 
{\bf Question 1:} Is there a deeper rationale for such replacement?

\noindent 
{\bf Answer:} Yes \cite{SS24-Flux}\cite{SS25-Seifert}: {\it Hypothesis H}.

\vspace{5pt}

\noindent 
{\bf Question 2:} Does this new model make novel predictions?

\noindent 
{\bf Answer:} Yes --  
{\it defect anyons} in FQH-systems:

\vspace{-3mm} 

\noindent
\begin{minipage}{10cm}
With the classifying space identified for known situations, 
\\
we find its implications for previously inaccessible cases:

\vspace{2pt}

Namely generalize now to 
\adjustbox{
  margin=2pt,
  bgcolor=lightolive
}{{\it $\mathcolor{purple}{n}$-punctured} surfaces $\Sigma^2_{g,\mathcolor{purple}{n}}$}, 

reflecting $\mathcolor{purple}{n}$ {\it defect points} in the semiconductor 

where the magnetic field is {\it expelled} 

(type-I superconducting spots).
\end{minipage}
\hspace{-1.8cm}
\adjustbox{
  raise=-1cm
}{
\begin{tikzpicture}

\draw[
  dashed,
  fill=lightgray
]
  (0,0)
  -- (8-.9,0)
  -- (10-.9,2)
  -- (2.8,2)
  -- cycle;

\begin{scope}[
  shift={(2.4,.5)}
]
\shadedraw[
  draw opacity=0,
  inner color=olive,
  outer color=lightolive
]
  (0,0) ellipse (.7 and .3);
\end{scope}

\begin{scope}[
  shift={(4.3,1.3)}
]
\shadedraw[
  draw opacity=0,
  inner color=olive,
  outer color=lightolive
]
  (0,0) ellipse (.7 and .25);
\end{scope}

\begin{scope}[shift={(5.5,.7)}]
\draw[
  fill=white,
  draw=purple
]
  (0,0) ellipse 
  (.17 and 
  0.5*0.17);
\end{scope}

\begin{scope}[shift={(7.2,1.2)}]
\draw[
  fill=white,
  draw=purple
]
  (0,0) ellipse 
  (.17 and 
  0.5*0.17);
\end{scope}

\node
  at (2,-.5)
  {
    \adjustbox{
      bgcolor=white,
      scale=.7
    }{
      \color{darkblue}
      \bf
      \def\arraystretch{.9}
      \def\tabcolsep{-5pt}
      \begin{tabular}{c}
        field solitons:
        \\
        Pontrjagin submanifolds
      \end{tabular}
    }
  };

\node
  at (5.8,-.5)
  {
    \adjustbox{
      scale=.7
    }{
      \color{darkblue}
      \bf
      \def\arraystretch{.9}
      \def\tabcolsep{-5pt}
      \begin{tabular}{c}
        flux-expelling 
        {\color{purple} defects}:
        \\
        punctures
      \end{tabular}
    }
  };
  
\end{tikzpicture}
}

\bigskip 

\noindent
{\bf Proposition.}
 The observables are, in this generality:
\vspace{-2mm} 
$$
  \def\arraystretch{2}
  \begin{array}{rcl}
  \mathrm{Obs}_0
  &\simeq&
  \mathbb{C}
  \Big[
  \pi_1
  \mathrm{Maps}^{\ast}
  \scalebox{1.4}{$($}
    \mathcolor{darkblue}{
    \big(\Sigma^2_{g,b,\mathcolor{purple}{n}}\big)_{\cpt}
    }
    ,\,
    S^2
  \scalebox{1.4}{$)$}
  \Big]
  \\
  &\simeq&
  \mathbb{C}
  \Big[
  \pi_1 
  \mathrm{Maps}^\ast
  \scalebox{1.2}{$($}
    \mathcolor{darkgreen}{
    \Sigma^2_{g,b}
    \vee
    \underset{\mathclap{n-1}}{
      \bigvee
    }
    S^1
    }
    ,\,
    S^2
  \scalebox{1.2}{$)$}
  \Big]
  \\
  &\simeq&
  \mathbb{C}\Big[
  \pi_1
  \mathrm{Maps}^{\ast}\big(
    \Sigma^2_{g,b}
    ,\,
    S^2
  \big)
  \times
  \mathbb{Z}^{n-1}
  \Big]
  \\
  &
  \underset{
    \mathclap{
      \substack{
        {g = 0} 
        \\
        {b=1}
      }
    }
  }{\simeq} 
  &
  \mathbb{C}\big[
    \mathbb{Z}^{n-1}
  \big]
  \end{array}
\qquad 
\adjustbox{
  raise=-2cm
}{
\begin{tikzpicture}
  \node at (0,0) {
  \includegraphics[width=5cm]{PinchingSphereAtTwoPoints.png}
  };
  \node[
    scale=.5,
    rotate=-40
  ] at (-1.87,.86)
  {
    \color{purple}
    puncture
  };
  \begin{scope}[
    yscale=-1
  ]
  \node[
    scale=.5,
    rotate=+40
  ] at (-1.85,.85)
  { \color{purple}
    puncture
  };
  \end{scope}
  \node[
    scale=.8
  ] 
    at (-2.15,-.53)
  {$\Sigma^2_0$};
  \node[
    scale=.8
  ] 
    at (.8,.14)
  {$\mathcolor{darkblue}{(\Sigma^2_{0,2})_{\cpt}}\,
  \rotatebox[origin=c]{10}{$=$}\,$};
  \node[scale=.8] at (.7,-1.7)
  {
    $\mathcolor{darkgreen}{\Sigma^2_0 \vee S^1}
   \;
  \adjustbox{rotate=10,raise=2pt}{$=$}\,$
  };
  \node[
    rotate=-20,
    scale=.8
  ]
    at (-.2,-.6)
    {$\sim$};
  \node[
    rotate=+20,
    scale=.8
  ]
    at (-.18,+.66)
    {$\sim$};
  \node[
    scale=.7
  ]
  at (1.98,.97) 
  {$\infty$};
  \begin{scope}[
    shift={(-1.47,.57)}
  ]
  \draw[purple]
    (180-5:.2 and .1) arc
    (180-5:360+5:.11 and .05);
  \end{scope}
  \begin{scope}[
    yscale=-1,
    shift={(-1.47,.52)},
  ]
  \draw[purple]
    (180-5:.2 and .1) arc
    (180-5:360+5:.11 and .05);
  \end{scope}
\end{tikzpicture}
}
\hspace{-5pt}
\adjustbox{
  rotate=-90,
  scale=.7
}{
  \clap{
  \def\arraystretch{1}
  \begin{tabular}{l}
    Topology change due to defects!
    \\
    (cf. \cite[p 11]{Hatcher02}) Different to but  
    \\
    not unlike the {\it genon}-proposal
    \cite{BarkeshliJianQi13}.
  \end{tabular}
  }
}
$$
%\hspace{2.2cm}
subject to the 
{\bf diffeomorphism} 
{\bf action} by: 
$$
\begin{tikzcd}[
    row sep=-2pt,
    column sep=33pt
  ]
   1 \ar[r]
   &
    \mathrm{Br}_n(\Sigma^2_g)
    \ar[r, hook]
    &
    \pi_0 \mathrm{Homeos}^{\ast}_{\mathrm{or}}\big(
      (\Sigma^2_{g,n})_{\cpt}
    \big)
    \ar[r,->>]
    &[+10pt]
    \mathrm{MCG}(\Sigma^2_g)
    \ar[r] 
    &
     1 \;.
    \\[-1pt]
    &
    \mathclap{
      \scalebox{.7}{
        \color{darkgreen}
        \bf
        surface braid group
      }
    }
    &
    \mathclap{
      \scalebox{.7}{
        \color{purple}
        \bf
        \def\arraystretch{.9}
        \begin{tabular}{c}
          mapping class group
          \\
          of punctured surface
        \end{tabular}
      }
    }
    &
    \mathclap{
      \scalebox{.7}{
        \color{darkblue}
        \bf
        \def\arraystretch{.9}
        \begin{tabular}{c}
          mapping class group
          \\
          of plain surface
        \end{tabular}
      }
    }
  \end{tikzcd}
$$
\\
\begin{minipage}{13cm}
Therefore the 
\adjustbox{
  margin=2pt,
  bgcolor=lightolive
}{equivariant quantum states (jargon: ``generally covariant'')}
on 
$\Sigma^2_{0,\mathcolor{purple}{n}}$ 

are representations of the 
{\it wreath product of solitonic and defect phases}:

\vspace{13pt}

$$
    \mathbb{Z} 
    \wr 
    \mathrm{Br}_{\mathcolor{purple}{n}}(\Sigma^2_0)
  \;\;
  =
  \;\;
  \overset{
    \mathclap{
      \adjustbox{
        scale=.7,
        rotate=16
      }
      {
        \rlap{
          \color{darkblue}
          \bf solitonic anyons
        }
      }
    }
  }{
    \mathbb{Z}^{\mathcolor{purple}{n}-1}
  }
  \rtimes
  \overset{
    \mathclap{
      \adjustbox{
        scale=.7,
        rotate=16
      }
      {
        \rlap{
          \color{purple}
          \bf defect anyons
        }
      }
    }
  }{
    \mathrm{Br}_{\mathcolor{purple}{n}}(\Sigma^2_0)
  }
  \;\twoheadrightarrow\;
  \mathbb{Z}^{\mathcolor{purple}{n}-1}
  \rtimes
  \mathrm{Sym}_{\mathcolor{purple}{n}}
$$
\end{minipage}
\hspace{25pt}
\adjustbox{
  raise=-1.7cm
}{
  \includegraphics[width=3cm]{braiding.png}
}

\medskip 
Such {\it braid representations for defects} have not previously been derived for FQH systems --
but are just what is needed for the grand goal of 
{\it topological quantum gates}:
programmable unitary transformations of quantum systems,
insensitive to continuous deformations
(hence to noise!)
\begin{center}
\adjustbox{
  scale=.8,
  raise=0cm
}{
\begin{tikzpicture}

\clip
  (-8.4, 1.3) rectangle
  (4.9,-4.7);

  \shade[right color=lightgray, left color=white]
    (3,-3)
      --
    (-1,-1)
      --
    (-1.21,1)
      --
    (2.3,3);

  \draw[]
    (3,-3)
      --
    (-1,-1)
      --
    (-1.21,1)
      --
    (2.3,3)
      --
    (3,-3);

\draw[-Latex]
  ({-1 + (3+1)*.3},{-1+(-3+1)*.3})
    to
  ({-1 + (3+1)*.29},{-1+(-3+1)*.29});

\draw[-Latex]
    ({-1.21 + (2.3+1.21)*.3},{1+(3-1)*.3})
      --
    ({-1.21 + (2.3+1.21)*.29},{1+(3-1)*.29});

\draw[-Latex]
    ({2.3 + (3-2.3)*.5},{3+(-3-3)*.5})
      --
    ({2.3 + (3-2.3)*.49},{3+(-3-3)*.49});

\draw[-latex]
    ({-1 + (-1.21+1)*.53},{-1 + (1+1)*.53})
      --
    ({-1 + (-1.21+1)*.54},{-1 + (1+1)*.54});

  \begin{scope}[rotate=(+8)]
   \draw[dashed]
     (1.5,-1)
     ellipse
     ({.2*1.85} and {.37*1.85});
   \begin{scope}[
     shift={(1.5-.2,{-1+.37*1.85-.1})}
   ]
     \draw[->, -Latex]
       (0,0)
       to
       (180+37:0.01);
   \end{scope}
   \begin{scope}[
     shift={(1.5+.2,{-1-.37*1.85+.1})}
   ]
     \draw[->, -Latex]
       (0,0)
       to
       (+37:0.01);
   \end{scope}
   \begin{scope}[shift={(1.5,-1)}]
     \draw (.43,.65) node
     { \scalebox{.8}{$
     $} };
  \end{scope}
  \draw[fill=white, draw=gray]
    (1.5,-1)
    ellipse
    ({.2*.3} and {.37*.3});
  \draw[line width=3.5, white]
   (1.5,-1)
   to
   (-2.2,-1);
  \draw[line width=1.1]
   (1.5,-1)
   to node[
     above, 
     yshift=-4pt, 
     pos=.85]{
     \;\;\;\;\;\;\;\;\;\;\;\;\;
     \rotatebox[origin=c]{7}
     {
     \scalebox{.7}{
     \color{darkorange}
     \bf
     \colorbox{white}{anyonic defect}
     }
     }
   }
   (-2.2,-1);
  \draw[
    line width=1.1
  ]
   (1.5+1.2,-1)
   to
   (3.5,-1);
  \draw[
    line width=1.1,
    densely dashed
  ]
   (3.5,-1)
   to
   (4,-1);

  \draw[line width=3, white]
   (-2,-1.3)
   to
   (0,-1.3);
  \draw[-latex]
   (-2,-1.3)
   to
   node[
     below, 
     yshift=+3pt,
     xshift=-7pt
    ]{
     \scalebox{.7}{
       \rotatebox{+7}{
       \color{darkblue}
       \bf
       parameter
       }
     }
   }
   (0,-1.3);
  \draw[dashed]
   (-2.7,-1.3)
   to
   (-2,-1.3);

 \draw
   (-3.15,-.8)
   node{
     \scalebox{.7}{
       \rotatebox{+7}{
       \color{darkgreen}
       \bf
       braiding
       }
     }
   };

  \end{scope}

  \begin{scope}[shift={(-.2,1.4)}, scale=(.96)]
  \begin{scope}[rotate=(+8)]
  \draw[dashed]
    (1.5,-1)
    ellipse
    (.2 and .37);
  \draw[fill=white, draw=gray]
    (1.5,-1)
    ellipse
    ({.2*.3} and {.37*.3});
  \draw[line width=3.1, white]
   (1.5,-1)
   to
   (-2.3,-1);
  \draw[line width=1.1]
   (1.5,-1)
   to
   (-2.3,-1);
  \draw[line width=1.1]
   (1.5+1.35,-1)
   to
   (3.6,-1);
  \draw[
    line width=1.1,
    densely dashed
  ]
   (3.6,-1)
   to
   (4.1,-1);
  \end{scope}
  \end{scope}

  \begin{scope}[shift={(-1,.5)}, scale=(.7)]
  \begin{scope}[rotate=(+8)]
  \draw[dashed]
    (1.5,-1)
    ellipse
    (.2 and .32);
  \draw[fill=white, draw=gray]
    (1.5,-1)
    ellipse
    ({.2*.3} and {.32*.3});
  \draw[line width=3.1, white]
   (1.5,-1)
   to
   (-1.8,-1);
\draw
   (1.5,-1)
   to
   (-1.8,-1);
  \draw
    (5.23,-1)
    to
    (6.4-.6,-1);
  \draw[densely dashed]
    (6.4-.6,-1)
    to
    (6.4,-1);
  \end{scope}
  \end{scope}

\draw (1.73,-1.06) node
 {
  \scalebox{.8}{
    $k_{{}_{I}}$
  }
 };

\begin{scope}
[ shift={(-2,-.55)}, rotate=-82.2  ]

 \begin{scope}[shift={(0,-.15)}]

  \draw[]
    (-.2,.4)
    to
    (-.2,-2);

  \draw[
    white,
    line width=1.1+1.9
  ]
    (-.73,0)
    .. controls (-.73,-.5) and (+.73-.4,-.5) ..
    (+.73-.4,-1);
  \draw[
    line width=1.1
  ]
    (-.73+.01,0)
    .. controls (-.73+.01,-.5) and (+.73-.4,-.5) ..
    (+.73-.4,-1);

  \draw[
    white,
    line width=1.1+1.9
  ]
    (+.73-.1,0)
    .. controls (+.73,-.5) and (-.73+.4,-.5) ..
    (-.73+.4,-1);
  \draw[
    line width=1.1
  ]
    (+.73,0+.03)
    .. controls (+.73,-.5) and (-.73+.4,-.5) ..
    (-.73+.4,-1);

  \draw[
    line width=1.1+1.9,
    white
  ]
    (-.73+.4,-1)
    .. controls (-.73+.4,-1.5) and (+.73,-1.5) ..
    (+.73,-2);
  \draw[
    line width=1.1
  ]
    (-.73+.4,-1)
    .. controls (-.73+.4,-1.5) and (+.73,-1.5) ..
    (+.73,-2);

  \draw[
    white,
    line width=1.1+1.9
  ]
    (+.73-.4,-1)
    .. controls (+.73-.4,-1.5) and (-.73,-1.5) ..
    (-.73,-2);
  \draw[
    line width=1.1
  ]
    (+.73-.4,-1)
    .. controls (+.73-.4,-1.5) and (-.73,-1.5) ..
    (-.73,-2);

 \draw
   (-.2,-3.3)
   to
   (-.2,-2);
 \draw[
   line width=1.1,
   densely dashed
 ]
   (-.73,-2)
   to
   (-.73,-2.5);
 \draw[
   line width=1.1,
   densely dashed
 ]
   (+.73,-2)
   to
   (+.73,-2.5);

  \end{scope}
\end{scope}

\begin{scope}[shift={(-5.6,-.75)}]

  \draw[line width=3pt, white]
    (3,-3)
      --
    (-1,-1)
      --
    (-1.21,1)
      --
    (2.3,3)
      --
    (3, -3);

  \shade[right color=lightgray, left color=white, fill opacity=.7]
    (3,-3)
      --
    (-1,-1)
      --
    (-1.21,1)
      --
    (2.3,3);

  \draw[]
    (3,-3)
      --
    (-1,-1)
      --
    (-1.21,1)
      --
    (2.3,3)
      --
    (3, -3);

\draw (1.73,-1.06) node
 {
  \scalebox{.8}{
    $k_{{}_{I}}$
  }
 };

\draw[-Latex]
  ({-1 + (3+1)*.3},{-1+(-3+1)*.3})
    to
  ({-1 + (3+1)*.29},{-1+(-3+1)*.29});

\draw[-Latex]
    ({-1.21 + (2.3+1.21)*.3},{1+(3-1)*.3})
      --
    ({-1.21 + (2.3+1.21)*.29},{1+(3-1)*.29});

\draw[-Latex]
    ({2.3 + (3-2.3)*.5},{3+(-3-3)*.5})
      --
    ({2.3 + (3-2.3)*.49},{3+(-3-3)*.49});

\draw[-latex]
    ({-1 + (-1.21+1)*.53},{-1 + (1+1)*.53})
      --
    ({-1 + (-1.21+1)*.54},{-1 + (1+1)*.54});

  \begin{scope}[rotate=(+8)]
   \draw[dashed]
     (1.5,-1)
     ellipse
     ({.2*1.85} and {.37*1.85});
   \begin{scope}[
     shift={(1.5-.2,{-1+.37*1.85-.1})}
   ]
     \draw[->, -Latex]
       (0,0)
       to
       (180+37:0.01);
   \end{scope}
   \begin{scope}[
     shift={(1.5+.2,{-1-.37*1.85+.1})}
   ]
     \draw[->, -Latex]
       (0,0)
       to
       (+37:0.01);
   \end{scope}
  \draw[fill=white, draw=gray]
    (1.5,-1)
    ellipse
    ({.2*.3} and {.37*.3});
 \end{scope}

   \begin{scope}[shift={(-.2,1.4)}, scale=(.96)]
  \begin{scope}[rotate=(+8)]
  \draw[dashed]
    (1.5,-1)
    ellipse
    (.2 and .37);
  \draw[fill=white, draw=gray]
    (1.5,-1)
    ellipse
    ({.2*.3} and {.37*.3});
\end{scope}
\end{scope}

  \begin{scope}[shift={(-1,.5)}, scale=(.7)]
  \begin{scope}[rotate=(+8)]
  \draw[dashed]
    (1.5,-1)
    ellipse
    (.2 and .32);
  \draw[fill=white, draw=gray]
    (1.5,-1)
    ellipse
    ({.2*.3} and {.37*.3});
\end{scope}
\end{scope}

\begin{scope}
[ shift={(-2,-.55)}, rotate=-82.2  ]

 \begin{scope}[shift={(0,-.15)}]

 \draw[line width=3, white]
   (-.2,-.2)
   to
   (-.2,2.35);
 \draw
   (-.2,.5)
   to
   (-.2,2.35);
 \draw[dashed]
   (-.2,-.2)
   to
   (-.2,.5);

\end{scope}
\end{scope}

\begin{scope}
[ shift={(-2,-.55)}, rotate=-82.2  ]

 \begin{scope}[shift={(0,-.15)}]

 \draw[
   line width=3, white
 ]
   (-.73,-.5)
   to
   (-.73,3.65);
 \draw[
   line width=1.1
 ]
   (-.73,.2)
   to
   (-.73,3.65);
 \draw[
   line width=1.1,
   densely dashed
 ]
   (-.73,.2)
   to
   (-.73,-.5);
 \end{scope}
 \end{scope}

\begin{scope}
[ shift={(-2,-.55)}, rotate=-82.2  ]

 \begin{scope}[shift={(0,-.15)}]

 \draw[
   line width=3.2,
   white]
   (+.73,-.6)
   to
   (+.73,+3.7);
 \draw[
   line width=1.1,
   densely dashed]
   (+.73,-0)
   to
   (+.73,+-.6);
 \draw[
   line width=1.1 ]
   (+.73,-0)
   to
   (+.73,+3.71);
\end{scope}
\end{scope}

\end{scope}

\draw[
  draw=white,
  fill=white
]
  (-8,-2.4) rectangle
  (3,-3.8);

\draw[
  draw=white,
  fill=white
]
  (-1,-1.7) rectangle
  (3.4,-2.5);

\begin{scope}[
  shift={(0,1.3)}
]
\draw
  (-2.2,-4.2) node
  {
    \adjustbox{
      bgcolor=white,
      scale=1.2
    }{
      $
       \mathllap{
          \raisebox{1pt}{
            \scalebox{.58}{
              \color{darkblue}
              \bf
              \def\arraystretch{.9}
              \begin{tabular}{c}
                some quantum state for
                \\
                fixed defect positions
                \\
                $k_1, k_2, \cdots$
                at time
                {\color{purple}$t_1$}
              \end{tabular}
            }
          }
          \hspace{-5pt}
       }
        \big\vert
          \psi({\color{purple}t_1})
        \big\rangle
      $
    }
  };

\draw
  (+3.2,-3.85) node
  {
    \adjustbox{
      bgcolor=white,
      scale=1.2
    }{
      $
        \underset{
          \raisebox{-7pt}{
            \scalebox{.55}{
              \color{darkblue}
              \bf
              \def\arraystretch{.9}
               \begin{tabular}{c}
              another quantum state for
                \\
                fixed defect positions
                \\
                $k_1, k_2, \cdots$
                at time
                {\color{purple}$t_2$}
              \end{tabular}
            }
          }
        }{
        \big\vert
          \psi({\color{purple}t_2})
        \big\rangle
        }
      $
    }
  };

\draw[|->]
  (-1.3,-4.1)
  to
  node[
    sloped,
    yshift=5pt
  ]{
    \scalebox{.7}{
      \color{darkgreen}
      \bf
      unitary adiabatic transport
    }
  }
  node[
    sloped,
    yshift=-5pt,
    pos=.4
  ]{
    \scalebox{.7}{
      }
  }
  (+2.4,-3.4);
\end{scope}

\end{tikzpicture}
}
\end{center}

\vspace{-1cm}
Concretely, the action on $\mathbb{Z}^{-1} \subset \mathbb{C}^{n-1}$ is that of the {\it standard representation}, the complement of the trivial 1d rep inside the defining permutation representation of $\mathrm{Sym}_n$.
This yields what are known as 
\adjustbox{
  margin=2pt,
  bgcolor=lightolive
}
{\it controlled qdit-rotation gates},
the workhorse of quantum algorithms
\& the bottleneck for noise-protection,
here topologically protected
as cylic defect braidings:

e.g.:
$
  \adjustbox{
    raise=-.2cm
  }{
  \begin{tikzpicture}
    \draw (0,0) rectangle (.3,.3);
    \draw (0+.3,0) rectangle (.3+.3,.3);    
    \draw (0,-.3) rectangle (.3,.3-.3);
  \end{tikzpicture}
  }
  \;\simeq\;
  \Bigg\{
  (213) \mapsto
  \grayoverbrace{
  \left[
  \def\arraystretch{1}
  \def\arraycolsep{2pt}
  \begin{array}{cc}
    1 & 0
    \\
    0 & -1
  \end{array}
  \right]
  }{ Z }
  \,,\;\;\;
  (231) \mapsto
  \grayoverbrace{
  \left[
  \def\arraystretch{1}
  \def\arraycolsep{2pt}
  \begin{array}{cc}
    \mathrm{cos}(\alpha) & -\mathrm{sin}(\alpha)
    \\
    \mathrm{sin}(\alpha) & \;\mathrm{cos}(\alpha)
  \end{array}
  \right]  
  }{
    R_y(2\alpha)
  }
  \mbox{
    \def\tabcolsep{-5pt}
    \begin{tabular}{l}
      where 
      \\
      $\alpha = 4\pi/3$
    \end{tabular}
    }
  \;
  \Bigg\}
$
\qquad \quad 
\adjustbox{
  raise=-.9cm
}{
\begin{tikzpicture}[
  yscale=.8
]

  \draw[line width=1.5]
    (.5,0) .. controls
    (.5,1) and
    (0,1) ..
    (0,2);
  \draw[line width=1.5]
    (1,0) .. controls
    (1,1) and
    (.5,1) ..
    (.5,2);

  \node at 
    (1.4,.8) {\bf\dots};

  \draw[line width=1.5]
    (2,0) .. controls
    (2,1) and
    (1.5,1) ..
    (1.5,2);
    (1.3,1) {\bf\dots};

  \draw[
    line width=6,
    white
  ]
    (0,0) .. controls
    (0,1) and
    (2,1) ..
    (2,2);
  \draw[line width=1.5]
    (0,0) .. controls
    (0,1) and
    (2,1) ..
    (2,2);

\begin{scope}
  \draw
    (115:.18 and .08) arc 
    (110:418:.18 and .08);
\end{scope}
\begin{scope}[shift={(.5,0)}]
  \draw
    (115:.18 and .08) arc 
    (110:418:.18 and .08);
\end{scope}
\begin{scope}[shift={(1,0)}]
  \draw
    (115:.18 and .08) arc 
    (110:418:.18 and .08);
\end{scope}
\begin{scope}[shift={(2,0)}]
  \draw
    (115:.18 and .08) arc 
    (110:418:.18 and .08);
\end{scope}
\begin{scope}[shift={(0,2)}]
  \draw[line width=1.7, white]
    (0:.18 and .08) arc 
    (0:360:.18 and .08);
  \draw
    (0:.18 and .08) arc 
    (0:360:.18 and .08);
\end{scope}
\begin{scope}[shift={(.5,2)}]
  \draw[line width=1.7, white]
    (0:.18 and .08) arc 
    (0:360:.18 and .08);
  \draw
    (0:.18 and .08) arc 
    (0:360:.18 and .08);
\end{scope}
\begin{scope}[shift={(1.5,2)}]
  \draw[line width=1.7, white]
    (0:.18 and .08) arc 
    (0:360:.18 and .08);
  \draw
    (0:.18 and .08) arc 
    (0:360:.18 and .08);
\end{scope}
\begin{scope}[shift={(2,2)}]
  \draw[line width=1.7, white]
    (0:.18 and .08) arc 
    (0:360:.18 and .08);
  \draw
    (0:.18 and .08) arc 
    (0:360:.18 and .08);
\end{scope}

\end{tikzpicture}
}

\vspace{.3cm}

\noindent
{\bf Conclusion \& Outlook.} 
  With non-linear flux-quantization laws taken into account in physics,
  substantial algebraic topology reveals previously unrecognized
  phenomena
  potentially visible in experiment 
  and relevant for quantum technology
  (potentially a more fruitful commercial AlgTop-application than topological data analysis).
  This opens the opportunity to make AlgTop research inform quantum technology. 

\smallskip

\noindent
{\bf Symmetry-protection and Equivariant homotopy theory.}
A noteworthy class of open problems in this regard is the generalization of all of the phenomena discussed here from spaces to {\it $G$-spaces} equipped with continuous actions of a finite group, with $G$-equivariant maps between them.
\begin{itemize}
\item On the physics side this corresponds to the generic situation of {\it $G$-symmetry protected} topological materials (see pointers in \cite[\S 2.3]{SS24-TopOrd}) particularly important for crystalline symmetry in ``anomalous'' quantum Hall systems \cite{SS25-FQAH}.

\item On the math side this corresponds to enhancing the flux quantization laws to exotic {\it equivariant cohomology}, specifically to equivariant Cohomotopy (cf. \cite[\S 4.5]{SS21-EquBund}\cite[\S 6]{SS20-Orb}).
\end{itemize}

Concretely for $G\acts \Sigma^2_{g,n}$ a surface equipped with (crystalline) $G$-symmetry action for $G \subset \mathrm{Pin}(2)$ a finite subgroup, and understanding the canonical $\mathrm{Pin}(2) \hookrightarrow \mathrm{Spin}(3) \twoheadrightarrow \mathrm{SO}(3)$-action on $S^2 \,\simeq\, S(\mathbb{R}^3)$, the $G$-symmetry protected enhancement of the above algebra of quantum observables will be formed with the subspace $\mathrm{Map}(-,-)^{\!G} \subset \mathrm{Map}(-,-)$ of $G$-equivariant maps
$$
  \scalebox{.7}{
    \color{darkblue}
    \bf
    \def\arraystretch{.9}
    \begin{tabular}{@{}c@{}}
      topological quantum observables
      \\
      in \scalebox{1.1}{$G$}-symmetry protected material
    \end{tabular}
  }
  \;\;\;\;
  \mathcolor{purple}{G}\mathrm{Obs}_0
  \;\;
  :=
  \;\;
  \mathbb{C}
  \Big[
    \pi_1 
    \, 
    \mathrm{Map}\big(
      \Sigma^2_{g,n}
      ,\,
      S^2
    \big)^{\!\mathcolor{purple}{G}}
  \Big]
  \scalebox{.7}{
    \color{darkblue}
    \bf
    \def\arraystretch{.9}
    \begin{tabular}{@{}c@{}}
      fundamental group algebra of 
      \\
      $G$-equivariant mapping space
    \end{tabular}
  }
$$
While the analog of the above theorem for non-trivial such $G$ actions remains open, this formula reduces a great deal of subtle physics of topologically ordered quantum materials to a precise question in pure algebraic topology.

\vspace{4mm}

\newpage

\noindent
{\bf Vista: Homotopy Quantum Logic.}
Let us shift gears. We have seen that:

\vspace{2pt}

\begin{center} 
\adjustbox{
  rndfbox=4pt
}{
\begin{minipage}{5.3cm}
Topological quantum states 
$\HilbertSpace{H}_\Sigma$
\\
of solitonic field fluxes 
\\
with classifying space $\hotype{A}$
\\
on spacetime domain $\mathbb{R}^{1,1} \times \Sigma$
\end{minipage}
\begin{minipage}{6cm}
form representations of 
$\pi_1$
\\
of the soliton moduli space
\\
\adjustbox{
  margin=2pt 5pt,
  bgcolor=lightolive
}{$\mathrm{Fields}_\Sigma
\,:=\,
\mathrm{Maps}^\ast\big(\Sigma,\, \hotype{A}\big) \!\sslash\! \mathrm{Aut}^\ast(\Sigma)$}
\\
{}
\end{minipage}
}
\end{center}

\vspace{.3cm}

\noindent
\begin{minipage}{12cm}
This is remarkable
because
such representations are equivalent to 
{\it vector bundles} $\HilbertSpace{H}_\Sigma$ on $\mathrm{Fields}_\Sigma$
{\it with flat connections} $\nabla$, that is {\it local systems} on moduli
with the homotopy type of $\mathrm{Fields}_\Sigma$
understood as an $\infty$-groupoid,
(physics newspeak: generalized symmetry)
flat vector bundles are equivalently
functors $\vdash\HilbertSpace{H}_\Sigma$
to the groupoid $\mathrm{Mod}_{\mathbb{C}}$:
\end{minipage}
\quad
\adjustbox{
  raise=6pt
}{
$
  \begin{tikzcd}[column sep=50pt]
    \big(
      \overset{
        \mathclap{
          \adjustbox{
            rotate=21,
            scale=.65
          }{
            \rlap{
            \hspace{-3pt}
            \color{darkblue}
            \bf
            quantum states
            }
          }
        }
      }{
        \HilbertSpace{H}_{\Sigma}
      }
      ,\,
      \overset{
        \mathclap{
          \adjustbox{
            rotate=21,
            scale=.65
          }{
            \rlap{
            \hspace{-3pt}
            \color{darkblue}
            \bf
            \& observables
            }
          }
        }
      }{
        \nabla
      }
    \big)
    \ar[
      d,
      ->>
    ]
    \ar[
      dr,
      phantom,
      "{
        \scalebox{.7}{\color{gray}(pb)}
      }"{pos=.2}
    ]
    \ar[
      r
    ]
    &
    \mathrm{Mod}^{\mathbb{C}/}_{\mathbb{C}}
    \ar[
      d,
      ->>
    ]
    \\
    \mathrm{Fields}_\Sigma
    \ar[
      r,
      "{
        \vdash \HilbertSpace{H}_\Sigma
      }"
    ]
    &
    \mathrm{Mod}_{\mathbb{C}}
  \end{tikzcd}
$
}
\begin{minipage}{8cm}

\end{minipage}

\medskip 
\noindent 
\begin{minipage}{11cm}
This is the special case
of {\it $\infty$-local systems} \cite{SS23-Entanglement}:
chain complex-bundles 
with flat $\infty$-connection.
These are equivalently $\mathrm{Fields}_\Sigma$-
{\it parameterized module spectra}
for the $E_\infty$-ring $H\mathbb{C}$
hence 
\adjustbox{
  margin=2pt 3pt,
  bgcolor=lightolive
}{$H\mathbb{C}[\Omega \mathrm{Fields}_\Sigma]$-modules}
detecting higher structure
in the moduli space:
\end{minipage}
\quad 
\adjustbox{
  raise=2pt
}{
$
  \begin{tikzcd}[
    column sep=32pt
  ]
    \adjustbox{
      scale=.7,     
  rndfbox=4pt
    }{
      \color{darkblue}
      \bf
      \def\arraystretch{.9}
      \def\tabcolsep{-5pt}
      \begin{tabular}{c}
        Higher quantum states
        \\
        \& higher observables
      \end{tabular}
    }
    \ar[
      d,
      ->>
    ]
    \ar[
      dr,
      phantom,
      "{
        \scalebox{.7}{\color{gray}(pb)}
      }"{pos=.2}
    ]
    \ar[
      r
    ]
    &
    \mathrm{Mod}^{H\mathbb{C}/}_{H\mathbb{C}}
    \ar[
      d,
      ->>
    ]
    \\
    \mathrm{Fields}_\Sigma
    \ar[
      r,
      "{
        \vdash 
         \HilbertSpace{H}
          ^\infty
          _\Sigma
      }"
    ]
    &
    \mathrm{Mod}_{H\mathbb{C}}
  \end{tikzcd}
$
}

Here 
\begin{itemize} 
\item $H \mathbb{C}$ denotes the 
{\it homotopy complex numbers}:
the EM-ring spectrum of $\mathbb{C}$;
\item  $H \mathbb{C}[\Omega \mathrm{Fields}_\Sigma]$ is the
 {\it homotopy Pontrjagin algebra}
 whose $\pi_\bullet$ is 
 $\mathrm{Obs}_\bullet$.
\end{itemize} 

\vspace{.3cm}

\noindent
These objects form the {\bf tangent $\infty$-topos} $T \mathrm{Grpd}_\infty$ (over $H \mathbb{C}$), which is \cite{SS23-Monadology}\cite{SS23-Entanglement}:

{\bf (i)} the arena of parameterized stable homotopy theory,

{\bf (ii) }categorial semantics of a 
novel quantum programming language.

\medskip

\noindent
Remarkably, this provides an AlgTop angle on 
an ill-understood but central physics aspect:

\smallskip 
\adjustbox{
  margin=2pt,
  bgcolor=lightolive
}{
\begin{minipage}{11cm}
\it What exactly is \textbf{quantum measurement}
of anyonic topological order?
\end{minipage}
}

\bigskip

\noindent {\bf Fact.} \cite{SS23-Monadology}
Given quantum states $\HilbertSpace{H} \in  \mathrm{Mod}_{\mathbb{C}}^{\mathrm{Fields}}$,
\begin{itemize}
\item  a {\it quantum measurement basis} is

-- a choice of space $W$ (of ``{\it possible worlds}''), 

 \vspace{-3mm} 
-- a map $W \xrightarrow{i} \mathrm{Fields}$ whose 
 base change is {\it ambidextrous}:
%\hspace{2pt}
\begin{tikzcd}
  \mathrm{Mod}_{\mathbb{C}}^{W}
  \ar[
    rr,
    shift left=5pt,
    "{
      i_! \,\simeq\, i_\ast
    }",
  ]
  \ar[
    rr,
    phantom,
    "{
      \bot \;\, \top
    }"{scale=.7}
  ]
  \ar[
    from=rr,
    shift left=5pt,
    "{
      i^\ast
    }"
  ]
  &&
  \mathrm{Mod}_{\mathbb{C}}^{\mathrm{Fields}}, 
\end{tikzcd}

 \vspace{-3mm} 
-- a $V \in \mathrm{Mod}_{\mathbb{C}}^{W}$
which (co)induces $\HilbertSpace{H} \simeq i_\ast V$;

\vspace{3pt}

\item  the {\it measurement \& collapse operation} is
is the counit
$
  \begin{tikzcd}
    i^\ast \HilbertSpace{H}
    \simeq
    i^\ast i_\ast V
    \ar[
      r,
      "{
        \unit{i}{V}
      }"
    ]
    &
    V
    \,.
  \end{tikzcd}
$
\end{itemize}

\medskip 
\noindent {\bf Example.} 
Focusing on 
$\mathrm{Fields} := \ast \sslash \pi_0\mathrm{Homeo}(\Sigma^2_{g,n,b})$, 
such measurement bases are given by 
finite index subgroups of $\pi_0\mathrm{Homeo}(\Sigma^2_{g,n,b})$.
There is a rich theory of these,
potentially of direct relevance for
realizing topological quantum computing...

\vspace{.4cm}

More on these 
quantum-information theoretic
aspects can be found in 
\cite{Schreiber25-ICMAT}\cite{SS25-ViaAlgTop}.

\newpage

\appendix

%%%%%%%%%%%%%%%%%%%%%%%%%%%%
\section{Background on Homotopy Theory}
%%%%%%%%%%%%%%%%%%%%%%%%%%%%

We collect some notions used in the main text to establish notation and give basic pointers to the literature.

\vspace{.1cm}

\noindent
{\bf Homotopy theory} (cf. \cite{Strom11}).
For $f_0, f_1 \,:\, X \xrightarrow{\;} Y$ a pair of continuous maps between (topological) 
spaces
a {\it homotopy} $\eta : f_0 \Rightarrow f_1$ is a continuous deformation between them: a continuous map
$ \eta : [0,1] \times X \xrightarrow{\;} Y$ such that 
$$
  \begin{array}{l}
    \eta(0,x) = f_0(x),
    \\[1pt]
    \eta(1,x) = f_1(x),
  \end{array}
  \qquad 
\mathrm{denoted}
  \qquad 
  \begin{tikzcd}
    X
    \ar[
      r,
      bend left=30,
      "{ f_0 }",
      "{\ }"{swap, name=s}
    ]
    \ar[
      r,
      bend right=30,
      "{ f_1 }"{swap},
      "{\ }"{name=t}
    ]
    \ar[
      from=s,
      to=t,
      Rightarrow,
      shift right=3pt,
      "{ \eta }"
    ]
    &
    Y
    \mathrlap{.}
  \end{tikzcd}
$$
For example, a 
square ``homotopy-
commutative diagram''  
$$
  \begin{tikzcd}[column sep=large]
    \Sigma 
    \ar[
      d,
      "{ \phi }"{swap},
      "{\ }"{name=t}
    ]
    \ar[
      r,
      dashed,
      "{ b }",
      "{\ }"{swap, name=s}
    ]
    & 
    \hotype{A}
    \ar[
      d,
      "{ p }"
    ]
    \ar[
      from=s,
      to=t,
      Rightarrow,
      dashed,
      "{ \eta }"
    ]
    \\
    X 
    \ar[
      r,
      dashed,
      "{ c }"{swap}
    ]
      & 
    \hotype{B}
  \end{tikzcd}
  \qquad 
\mathrm{means \; that }
\qquad 
  \def\arraystretch{1.3}
  \begin{array}{l}
  \eta : [0,1] \times \Sigma \xrightarrow{\;}
  \hotype{B}
  \\
  \eta(0,s) = p\big(b(s)\big)
  \mathrlap{,}
  \\
  \eta(1,s) = c\big(\phi(s)\big)
  \mathrlap{.}
  \end{array}
$$

\smallskip 
If one declares -- and we do -- to work in a ``convenient'' full sub-category of all topological spaces (such as that of {\it compactly generated} or of {\it Delta-generated} topological spaces, cf. \cite[p 21, 131]{SS21-EquBund}) then the topological space $\mathrm{Maps}(X,Y)$ of all continuous maps $X \xrightarrow{\;} Y$ satisfies the adjointness relation
$$
  \big\{
  P 
  \xrightarrow{\;}
  \mathrm{Maps}(X,Y)
  \big\}
  \;\simeq\;
  \big\{
  P \times X
  \xrightarrow{\;}
  Y
  \big\}.
$$
For $P \defneq [0,1]$, this says that homotopies are equivalently paths in mapping spaces, and that homotopy-classes of maps are the mapping spaces' path-connected components:
$$
  \pi_0 \, \mathrm{Maps}(X,Y)
  \;\simeq\;
  \mathrm{Maps}(X,Y)_{\!/\mathrm{hmtp}}
  \,.
$$

\vspace{.1cm}

\noindent
Since homotopies are maps themselves, there are homotopies-between-homotopies and ever higher-homotopies.

\vspace{.1cm}

\noindent
Thereby, topological spaces constitute a model for {\bf higher categorical symmetry} namely for higher groupoids. As such, they represent both cohomology as well as higher gauge fields in the topological sector.
\footnote{
  Beyond the topological sector, full higher gauge fields are still represented by maps $X \xrightarrow{\;} \hotype{B}$ etc., only that now $\hotype{B}$ is no longer just a topological space but a ``smooth $\infty$-stack'', cf. \cite{FSS14-Stacky}\cite[pp 41]{FSS23-Char}.
}

\begin{center} 
\adjustbox{
  bgcolor=lightolive
}{
\hspace{-10pt}
\def\arraystretch{.9}
\begin{tabular}{c||c|c|c|c}
  \rowcolor{lightgray}$\mathclap{\phantom{\vert^{\vert^{\vert}}}}$
  {\bf Cohomology} 
  & cocycle & coboundary
  &
  higher coboundary
  &
  ...
  \\
  homotopy & 
  \begin{tikzcd}[
    column sep=15pt
  ]
    X 
    \ar[
      r,
      "{ f }"
    ]
    &
    \hotype{B}
  \end{tikzcd}
  & 
  \begin{tikzcd}
    X 
    \ar[
      r,
      bend left=35,
      "{ f }",
      "{\ }"{swap, name=s}
    ]
    \ar[
      r,
      bend right=35,
      "{ f' }"{swap},
      "{\ }"{name=t}
    ]
    \ar[
      from=s,
      to=t,
      shift right=1pt,
      Rightarrow,
      "{ \eta }"
    ]
    &
    \hotype{B}
  \end{tikzcd}
  &
  \begin{tikzcd}[
    column sep=50pt,
    background color=lightolive
  ]
    X 
    \ar[
      r,
      bend left=45,
      "{ f }"{description},
      "{\ }"{swap, name=s}
    ]
    \ar[
      r,
      bend right=45,
      "{ f' }"{description},
      "{\ }"{name=t}
    ]
    \ar[
      from=s,
      to=t,
      Rightarrow,
      bend left=30,
      shift left=4pt,
      "{ \eta' }",
      "{\ }"{swap, name=t2}
    ]
    \ar[
      from=s,
      to=t,
      Rightarrow,
      shift right=6pt,
      bend right=30,
      "{ \eta }"{swap},
      "{\ }"{name=s2}
    ]
    \ar[
      from=s2,
      to=t2,
      Rightarrow,
      shorten=-3pt
    ]
    &
    \hotype{B}
  \end{tikzcd}
  &
  ...
  \\
  \rowcolor{lightgray}$\mathclap{\phantom{\vert^{\vert^{\vert}}}}$
  {\bf Physics} & field & gauge transf.
  &
  higher gauge transf.
  &
  ...
\end{tabular}
\hspace{-4pt}}

\end{center}

\noindent
In this vein, spaces are homotopy-{\it equivalent} $\hotype{B} \simeq \hotype{B}'$ 
if they are gauge equivalent namely if we have maps

$$
  \begin{tikzcd}
    \hotype{B}
    \ar[
      r,
      shift left=4pt,
      "{ f }"
    ]
    \ar[
      r,
      <-,
      shift right=4pt,
      "{ g }"{swap}
    ]
    &
    \hotype{B}'
  \end{tikzcd}
\qquad 
\mathrm{with}
\qquad 
  \def\arraystretch{1.1}
  \begin{array}{l}
  g \circ f \Rightarrow \mathrm{id}_{\hotype{B}}
  \\
  f \circ g \Rightarrow \mathrm{id}_{\hotype{B}'}
  \end{array}
$$
 
For example $\mathbb{R}^n \simeq \ast$ in homotopy theory, 
reflecting the fact that there is no non-trivial topological sector for fields on $\mathbb{R}^n$.

\vspace{2mm}

For actually computing homotopy classes of maps --- hence cohomology, hence gauge-equivalence classes of fields in the topological sector --- tools from {\it model category theory} are indispensable, which largely say how to ``absorb homotopies into spaces'' (cf. \cite[\S 1]{FSS23-Char}).

\noindent
For example, if $p : \hotype{A} \xrightarrow{\;} \hotype{B}$ is a {\it Serre fibration}, such as a fiber bundle,
and $\Sigma$ is a {\it cell complex}, such as a manifold, then sections-up-to-homotopy of $p$ pulled back to $\Sigma$ are homotopy equivalent to plain sections:
$$
  \left\{
  \adjustbox{
    raise=3pt
  }{
  \begin{tikzcd}[column sep=large]
    \Sigma 
    \ar[
      d,
      "{ \phi }"{swap},
      "{\ }"{name=t}
    ]
    \ar[
      r,
      dashed,
      "{ b }",
      "{\ }"{swap, name=s}
    ]
    & 
    \hotype{A}
    \ar[
      d,
      "{ p }"
    ]
    \ar[
      from=s,
      to=t,
      Rightarrow,
      dashed,
      "{ \eta }"
    ]
    \\
    X 
    \ar[
      r,
      "{ c }"{swap}
    ]
      & 
    \hotype{B}
  \end{tikzcd}
  }
  \right\}_{\!\!\big/\mathrm{hmtp}}
  \hspace{2pt}
  \overset{
    \scalebox{.7}{$
    \def\arraystretch{.9}
    \begin{array}{l}
      \Sigma \in \mathrm{Cof}
      \\
      p \in \mathrm{Fib}
    \end{array}
    $}
  }{
    \simeq
  }
  \hspace{6pt}
  \left\{
  \adjustbox{
    raise=3pt
  }{
  \begin{tikzcd}[column sep=large]
    \Sigma 
    \ar[
      d,
      "{ \phi }"{swap},
      "{\ }"{name=t}
    ]
    \ar[
      r,
      dashed,
      "{ b }",
      "{\ }"{swap, name=s}
    ]
    & 
    \hotype{A}
    \ar[
      d,
      "{ p }"
    ]
    \ar[
      from=s,
      to=t,
      equals
    ]
    \\
    X 
    \ar[
      r,
      "{ c }"{swap}
    ]
      & 
    \hotype{B}
  \end{tikzcd}
  }
  \right\}_{\!\!\big/\mathrm{hmtp}}
$$

\vspace{.1cm}

\smallskip

\noindent
{\bf Pointed homotopy theory} (cf. \cite[\S 3]{James84}).
To reflect the condition that {\it solitonic fields are localized} in that they {\it vanish at infinity} we

-- equip domain spaces $X$ with a {\it point at infinity}, $\infty_X \,\in \, X$,

 -- equip classifying spaces $\hotype{B}$ with a {\it point representing zero}, $0_{\hotype{B}} \in \hotype{B}$,

-- require maps $f \,:\, (X,\infty_X) \xrightarrow{\;} (\hotype{B}, 0_{\hotype{B}})$ to respect these base points

  so that 
  maps literally 
  vanish at infinity 
$$
\hspace{-14pt}
\adjustbox{raise=4pt}{
  \begin{tikzcd}[
    row sep=8pt,
    column sep=34pt
  ]
    X 
     \ar[
       rr, 
       "{c}"
      ]
      && \hotype{B}
    \\
    \{\infty_X\}
    \mathrlap{\,.}
    \ar[
      rr
    ]
    \ar[
      u,
      shorten <=-2,
      hook
    ]
    &&
    \{0_{\hotype{B}}\}
    \mathrlap{\,.}
    \ar[
      u,
      shorten <=-2,
      hook
    ]
  \end{tikzcd}
  }
$$

\vspace{.1cm}

For instance, to make fields on $\mathbb{R}^n$ vanish at infinity, we adjoin its would-be ``point at infinity'' to it (jargon: ``one-point compactification'') to obtain $\mathbb{R}^n_{\cpt} \,\simeq\, S^n$. On the other hand, if we want fields on some $X$ without a vanishing condition, we may adjoin a {\it disjoint} point-at-infinity, then pointed maps $X_{\plus} \xrightarrow{} \hotype{B}$ are ordinary $X \xrightarrow{} \hotype{B}$. For example, 
$$
  \begin{tikzcd}[
    row sep=-6pt
  ]
  \mathclap{
    \adjustbox{
      scale=.7
    }{
      \color{darkblue}
      \bf
      \begin{tabular}{c}
        based loop space
      \end{tabular}
    }
  }
  &
  \mathclap{
    \adjustbox{
      scale=.7
    }{
      \color{darkblue}
      \bf
      \begin{tabular}{c}
        free loop space
      \end{tabular}
    }
  }
  &
  \mathclap{
    \adjustbox{
      scale=.7
    }{
      \color{darkblue}
      \bf
      \begin{tabular}{c}
        maps out of contractible
      \end{tabular}
    }
  }
  \\
  \mathrm{Maps}^{\ast}\!(
    \mathbb{R}^1_{\cpt}
    ,\,
    \hotype{B}
  )
  \;=\;
  \Omega \hotype{B}
  \,,
  &
  \mathrm{Maps}^{\ast}\!(
    S^1_{\plus}
    ,\,
    X
  )
  =:
  \mathcal{L}\, \hotype{B}
  \,,
  &
  \mathrm{Maps}^{\ast}(
    \mathbb{R}^1_{\plus}
    ,\,
    \hotype{B}
  )
  \;=\;
  \hotype{B} \,.
  \end{tikzcd}
$$
Given a pair of pointed spaces $(X, \infty_X)$, $(Y, \infty_Y)$, in their product space $X \times Y$ any point should be regarded as being at infinity which is so with respect to either factor space; this yields the {\it smash product}:

\vspace{4mm} 
\noindent
$$
  X \wedge Y
  \,:=\,
  \frac{
    X \times Y
    \mathclap{\phantom{\vert_{\vert}}}
  }{
\mathclap{\phantom{\vert^{\vert}}}
    \{\infty_X\}
    \!\times\! Y
    \,\cup\,
    X \!\times\! \{\infty_Y\}
  }
$$
  to which the sub-space 
  $\mathrm{Maps}^{\ast}(-,-)$
  of
  pointed maps 
  is again adjoint:
$$
  \def\arraystretch{.9}
  \begin{array}{l}
  \big\{
  P
  \xrightarrow{\mathrm{pntd}}
  \mathrm{Maps}^{\!\ast/}(X,Y)
  \big\}
  \;\simeq\;
  \big\{
  P \wedge X
  \xrightarrow{\mathrm{pntd}}
  Y
  \big\}
  \,.
  \end{array}
$$

\vspace{.1cm}
\noindent For example,
$$
S^n \wedge S^m \,\simeq\, \mathbb{R}^n_{\cpt} \wedge \mathbb{R}^m_{\cpt} \,\simeq\, (\mathbb{R}^n \times \mathbb{R}^m)_{\cpt} \simeq S^{n+m}\,,
$$
so that, for instance:
$$
  \mathrm{Maps}^{\ast}\!\big(
    X \wedge S^1
    ,\,
    \hotype{B}
  \big)
  \,\simeq\,
  \mathrm{Maps}^{\ast}\!
  \Big(
  S^1
  ,\,
  \mathrm{Maps}^{\!\ast/}\big(
    X
    ,\,
    \hotype{B}
  \big)
  \Big)
  \;=:\;
  \Omega
  \,
  \mathrm{Maps}^{\ast}\!\big(
    X
    ,\,
    \hotype{B}
  \big)
  \,.
$$

\smallskip 
\noindent {\bf The differential character map}
$\mathbf{ch}_{\hotype{A}}$,
at the heart of flux-quantization
in the generality of flux densities with non-linear Bianchi identities:
\begin{itemize}[
 leftmargin=.5cm
]
\item[--] takes maps into a classifying space $\hotype{A}$ (classifying {\bf charges}),

\item[--] to maps into the moduli $\infty$-stack of closed $\mathfrak{l}\hotype{A}$-valued differential forms (classifying corresponding {\bf flux densities}),

\item[--] thereby allowing 
{\bf gauge potentials} to relate local flux densities to global charges.
\end{itemize}
\vspace{2mm} 
$$
  \begin{tikzcd}[
    column sep={between origins, 80pt}
  ]
    &
    \mathcolor{gray}{
      X
    }
    \ar[
      dl,
      gray,
      "{
        \scalebox{.7}{
          charges
        }
      }"{sloped}
    ]
    \ar[
      dr,
      gray,
      "{
        \scalebox{.7}{
          fluxes
        }
      }"{sloped}
    ]
    \\
    \hotype{A}
    \ar[
      dr,
      "{
        \mathbf{ch}_{\hotype{A}}
      }"{swap, sloped, pos=.4},
      "{
        \scalebox{.7}{
          \color{darkblue}
          \bf
          character
        }
      }"{sloped}
    ]
    \ar[
      rr,
      Rightarrow,
      shorten <=25pt,
      shorten >=8pt,
      darkorange,
      "{
        \scalebox{.7}{
          gauge potentials
        }
      }"{pos=.6,sloped}
    ]
    &&
    \mathcolor{gray}{
    \Omega^1_{\mathrm{dR}}\big(
      -;
      \,
      \mathfrak{l}\hotype{A}
    \big)_{\mathrm{clsd}}
    }
    \ar[
      dl,
      gray,
      "{
        \eta^{\scalebox{.6}{$\shape$}}
      }"
    ]
    \\
    &
    \shape
    \,
    \Omega^1_{\mathrm{dR}}\big(
      -;
      \,
      \mathfrak{l}\hotype{A}
    \big)_{\mathrm{clsd}}
  \end{tikzcd}
$$

\smallskip

\noindent
At a high level, this $\mathbf{ch}_{\hotype{A}}$ is readily described: It is the smooth differential-form model for the {\bf $\mathbb{R}$-rationalization} of $\hotype{A}$, followed by derived extension of scalars $\mathbb{Q} \to \mathbb{R}$
--- as indicated in the following paragraphs.

\smallskip

However, under the hood, this construction makes use of a fair bit of model category-theoretic rational-homotopy theory which we do not have space nor inclination to review here (all details in \cite{FSS23-Char}), whence the following should be ignored by readers without serious background in (rational) homotopy theory --- or else taken as motivation to learn it! (Start at \cite[\S 1]{FSS23-Char}.)
\;\;
Here is how it goes:

\vspace{.3cm}

\noindent
{\bf Fundamental theorem of homotopy theory.}
Regarding (classifying) spaces up to (weak) homotopy equivalence means equivalently to regard them as their {\it $\infty$-groupoids} 
(Kan simplicial sets)
$\mathrm{Sing}(-)$ of points, paths, 2-paths, etc., in that there is a Quillen equivalence
\cite[Ex. 1.13]{FSS23-Char}
\vspace{2mm} 
$$
\adjustbox{
  bgcolor=lightolive
}{
$
  \begin{tikzcd}
    \mathrm{TopSp}_{\mathrm{Qu}}
    \ar[
      from=rr,
      shift right=7pt,
      "{
       \phantom{A}
      }"{swap}
    ]
    \ar[
      rr,
      phantom,
      "{
        \simeq_{\mathrlap{{}_{\mathrm{Qu}}}}
      }"
    ]
    \ar[
      rr,
      shift right=7pt,
      "{
        \mathrm{Sing}
      }"{swap}
    ]
    &&
    \Delta\mathrm{Set}_{\mathrm{Qu}}
  \end{tikzcd}
$
}
$$
\vspace{.1cm}

\newpage 
\noindent {\bf Fundamental theorem of dg-algebraic rational homotopy theory.}
Sending simplicial sets to their dgc-algebras of simplex-wise $\mathcolor{purple}{\mathbb{Q}}$-polynomial differential forms (``piecewise linear'', PL)
is the left adjoint in a Quillen adjunction 
\cite[Prop. 5.5]{FSS23-Char}

$$
\adjustbox{
  bgcolor=lightolive
}{
$
  \begin{tikzcd}[column sep=60pt]
    \big(
      \mathrm{dgcAlgs}^{\geq 0}
    \big)
      ^{\mathrm{op}}
      _{\mathrm{proj}}
     \ar[
       from=rr,
       shift right=8pt,
       "{
         \Omega
           ^\bullet
           _{\mathrm{P}\mathcolor{purple}{\mathbb{Q}}\mathrm{LdR}}
       }"{swap}
     ]
     \ar[
       rr,
       phantom,
       "{
         \bot_{\mathrlap{{}_{\mathrm{Qu}}}}
       }"
     ]
     \ar[
       rr,
       shift right=8pt,
       "{
         \mathrm{Hom}\big(
           (-)
           ,\,
           \Omega_{\mathrm{P}\mathcolor{purple}{\mathbb{Q}}\mathrm{LdR}}(\Delta^\bullet)
         \big)
       }"{swap}
     ]
     &&
    \Delta\mathrm{Sets}_{\mathrm{Qu}}
  \end{tikzcd}
$
}
$$
whose derived adjunction-unit models rationalization of 
(connected, nilpotent, $\mathbb{Q}$-finite) homotopy types $\hotype{A}$ 
\cite[Prop. 5.6]{FSS23-Char}.
$$
  \begin{tikzcd}
    \hotype{A} 
    \ar[
      rr,
      "{
        \eta
          ^{\mathcolor{purple}{\mathbb{Q}}}
          _{\hotype{A}}
      }"
    ]
    &&
    L^{\mathcolor{purple}{\mathbb{Q}}}\hotype{A}
  \end{tikzcd}
$$

\noindent {\bf For $\mathbb{R}$-rational homotopy.}
The analogous Quillen adjunction with $\mathcolor{purple}{\mathbb{R}}$-polynomial forms
$$
\adjustbox{
  bgcolor=lightolive
}{
$
  \begin{tikzcd}[column sep=60pt]
    \big(
      \mathrm{dgcAlgs}^{\geq 0}
    \big)
      ^{\mathrm{op}}
      _{\mathrm{proj}}
     \ar[
       from=rr,
       shift right=8pt,
       "{
         \Omega
           ^\bullet
           _{\mathrm{P}\mathcolor{purple}{\mathbb{R}}\mathrm{L}}
       }"{swap}
     ]
     \ar[
       rr,
       phantom,
       "{
         \bot_{\mathrlap{{}_{\mathrm{Qu}}}}
       }"
     ]
     \ar[
       rr,
       shift right=8pt,
       "{
         \mathrm{Hom}\big(
           (-)
           ,\,
           \Omega_{\mathrm{P}\mathcolor{purple}{\mathbb{R}}\mathrm{LdR}}(\Delta^\bullet)
         \big)
       }"{swap}
     ]
     &&
    \Delta\mathrm{Sets}_{\mathrm{Qu}}
  \end{tikzcd}
$
}
$$
models rationalization followed by derived extension of scalars from $\mathbb{Q}$ to $\mathbb{R}$ (no longer a localization but still denoted like one)
\cite[Prop. 5.8]{FSS23-Char}.
$$
  \begin{tikzcd}
    \hotype{A}
    \ar[
      rr,
      "{
        \eta
          ^{\mathcolor{purple}{\mathbb{Q}}}
          _{\hotype{A}}
      }"
    ]
    &&
    L^{\mathcolor{purple}{\mathbb{Q}}} \hotype{A}    
    \ar[
      r
    ]
    &
    L^{\mathcolor{purple}{\mathbb{R}}} \hotype{A}
  \end{tikzcd}
$$
Now with $\mathbb{R}$-coefficients, we may equivalently use simplex-wise {\it smooth} differential forms ({\it piecewise smooth}, PS) 
$$
\adjustbox{
  bgcolor=lightolive
}{
$
  \begin{tikzcd}[column sep=60pt]
    \big(
      \mathrm{dgcAlgs}^{\geq 0}
    \big)
      ^{\mathrm{op}}
      _{\mathrm{proj}}
     \ar[
       from=rr,
       shift right=8pt,
       "{
         \Omega
           ^\bullet
           _{\mathrm{P}\mathcolor{purple}{\mathbf{S}}\mathrm{dR}}
       }"{swap}
     ]
     \ar[
       rr,
       phantom,
       "{
         \bot_{\mathrlap{{}_{\mathrm{Qu}}}}
       }"
     ]
     \ar[
       rr,
       shift right=8pt,
       "{
         \mathrm{Hom}\big(
           (-)
           ,\,
           \Omega_{\mathrm{P}\mathcolor{purple}{\mathbf{S}}\mathrm{dR}}(\Delta^\bullet)
         \big)
       }"{swap}
     ]
     &&
    \Delta\mathrm{Sets}_{\mathrm{Qu}}
  \end{tikzcd}
$
}
$$
In fact, we may equivalently use smooth differential forms on simplices times any $\mathbb{R}^n$ 
\cite[Prop. 5.10]{FSS23-Char}.
$$
\adjustbox{
  bgcolor=lightolive
}{
$
  \begin{tikzcd}[column sep=65pt]
    \big(
      \mathrm{dgcAlgs}^{\geq 0}
    \big)
      ^{\mathrm{op}}
      _{\mathrm{proj}}
     \ar[
       from=rr,
       shift right=8pt,
       "{
         \Omega
           ^\bullet
           _{\mathrm{P}\mathcolor{purple}{\mathbf{S}}\mathrm{dR}}
       }"{swap}
     ]
     \ar[
       rr,
       phantom,
       "{
         \bot_{\mathrlap{{}_{\mathrm{Qu}}}}
       }"
     ]
     \ar[
       rr,
       shift right=8pt,
       "{
         \mathrm{Hom}\big(
           (-)
           ,\,
           \Omega_{\mathrm{P}\mathcolor{purple}{\mathbf{S}}\mathrm{dR}}(
             \mathcolor{purple}{\mathbb{R}^n} 
               \times
             \Delta^\bullet
            )
         \big)
       }"{swap}
     ]
     &&
    \Delta\mathrm{Sets}_{\mathrm{Qu}}
  \end{tikzcd}
$
}
$$

\medskip

\noindent
{\bf Taking values in deformations of flux densities.}
Via the minimal Sullivan model $\mathrm{CE}(\mathfrak{l}\hotype{A})$ of $\hotype{A}$, this derived adjunction 
takes values in closed smooth $\mathfrak{l}\hotype{A}$-valued differential forms 
\cite[(9.9)]{FSS23-Char}
$$
  \Omega^1_{\mathrm{dR}}\big(
    \mathbb{R}^n \times \Delta^\bullet
    ,\,
    \mathfrak{l}\hotype{A}
  \big)_{\mathrm{clsd}}
  \;:=\;
  \mathrm{Hom}\Big(
    \mathrm{CE}(\mathfrak{l}\hotype{A})
    ,\,
    \Omega_{\mathrm{dR}}(
      \mathbb{R}^n \times \Delta^\bullet
    )
  \Big)
$$
which is the value on $\mathbb{R}^n$ of the homotopy-constant $\infty$-stack that is the {\it shape} 
$\shape(-)$ of 
the sheaf of closed forms \cite[Prop. 3.3.48]{SS21-EquBund}
\vspace{1mm} 
$$
\shape
\;
\Omega^1_{\mathrm{dR}}\big(
  -
  ;\,
  \mathfrak{l}\hotype{A}
\big)_{\mathrm{clsd}}
  \;\;
  \in
  \;\;
  \mathrm{Sh}_\infty\big(\mathrm{CartSp}\big).
$$
In total, regarding also $\hotype{A} \in \mathrm{Sh}_\infty(\ast) \xrightarrow{\mathrm{Disc}} \mathrm{Sh}_\infty(\mathrm{CartSp})$, this establishes the {\it differential character} map as promised
\cite[Def. 9.2]{FSS23-Char}
\vspace{1mm} 
$$
\adjustbox{
  margin=4pt,
  bgcolor=darkblue!10
}{
$
  \begin{tikzcd}
    \hotype{A}
    \ar[
      rr,
      "{
        \mathbf{ch}_{\hotype{A}}
      }"
    ]
    &&
    \shape\, 
    \Omega^1_{\mathrm{dR}}\big(
      -
      ;\,
      \mathfrak{l}\hotype{A}
    \big)_{\mathrm{clsd}}
  \end{tikzcd}
$
}
$$

%%%%%%%%%%%%%%%%%%%%%%%%%%%%%%%
\section{Background on TED Cohomotopy}
%%%%%%%%%%%%%%%%%%%%%%%%%%%%%%%

\vspace{-1mm} 
\noindent
{\bf Gauge potentials in twistorial Cohomotopy --- and the Green-Schwarz mechanism.}
  Consider the 
  Whitehead $L_\infty$-algebra 
  of the twistor fibration
  $
    \mathbb{C}P^3
    \xrightarrow{ t_{\mathbb{H}} }
    \mathbb{H}P^1
    \simeq
    S^4
    \,,
  $
  \vspace{1mm} 
$$
  \mathrm{CE}\big(
    \mathfrak{l}_{{}_{S^4}}
    \mathbb{C}P^3
  \big)
  =
  \mathbb{R}_{\mathrm{d}}
  \left[
  \def\arraystretch{1}
  \begin{array}{c}  
    f_2
    \\
    h_3
    \\
    g_4
    \\
    g_7
  \end{array}
  \right]
  \!\Big/\!
  \left(
  \def\arraystretch{1}
  \begin{array}{ccl}  
    \mathrm{d}\, f_2 &=& 0
    \\
    \mathrm{d}\, h_3 &=& g_4 + 
    \mathcolor{darkblue}{f_2 \, f_2}
    \\
    \mathrm{d}\, g_4 &=& 0
    \\
    \mathrm{d}\, g_7 &=& 
    \tfrac{1}{2} g_4 \, g_4
  \end{array}
  \right)
  \!,
$$

\vspace{-2mm} 
\noindent and bigons 
 parameterized
  like this:
$
  \begin{tikzcd}[
    column sep={between origins, 40pt},
    row sep=0pt
  ]
  &
  {}
  \ar[
    dd,
    -Latex,
    shorten=-4pt,
    gray,
    shift left=2pt,
    "{
      s
    }"
  ]
  \\
  \phantom{-}
  \ar[
    rr,
    -,
    bend left=45,
    shift right=7.4pt,
    "{ 0 }"{gray, description},
  ]
  \ar[
    rr,
    -,
    bend right=45,
    shift left=7.4pt,
    "{ 1 }"{
      gray, description,
      yshift=1pt
    }
  ]
  &&
  \phantom{-}
  \\
  {}
  \ar[
    rr,
    -Latex,
    gray,
    bend right=45,
    shift left=11pt,
    shorten=3pt,
    "{ t }"{swap, yshift=-2pt},
    "{ 0 }"{pos=-.02, description},
    "{ 1 }"{pos=1.02, description}
  ]
  &{}& 
  {}
  \end{tikzcd}
$

\smallskip

\noindent
{\bf Theorem} (\cite[p 23]{GSS24-SuGra}\cite[\S 4.1]{GSS24-FluxOnM5}).
Given a manifold $U_i$ (generically: a coordinate chart):

\vspace{1mm} 
\noindent
{\bf (i)}
Closed $\mathfrak{l}_{{}_{S^4}}\mathbb{C}P^3$-valued differential forms are in natural 
bijection with {\bf flux densities} of this form:
$$
  \left\{
  \adjustbox{raise=4pt}{$
  \begin{tikzcd}[
    row sep=30pt
  ]
    U_i
    \ar[
      d,
      "{
        (F_2, H_3, G_4, G_7)
      }"{description}
    ]
    \\
    \Omega^1_{\mathrm{dR}}\big(
      -;
      \mathfrak{l}_{{}_{S^4}}
      \mathbb{C}P^3
    \big)_{\mathrm{clsd}}
  \end{tikzcd}
  $}
  \right\}
  \begin{tikzcd}[column sep=70pt]
    {}
    \ar[
      r,
      ->>,
      shift left=5pt,
      "{ p_0 }"
    ]
    \ar[
      from=r,
      hook',
      shift left=5pt,
      "{ i_0 }"
    ]
    \ar[
      r,
      phantom,
      shift right=20pt,
      "{
        p_0 
          \circ 
        i_0 
          \,=\, 
        \mathrm{id}
      }"{scale=.72}
    ]
    \ar[
      r,
      phantom,
      shift left=20pt,
      "{
        i_0 
          \circ 
        p_0 
          \,=\, 
        \mathrm{id}
      }"{scale=.72}
    ]
    &
    {}
  \end{tikzcd}
  \left\{
  \def\arraystretch{1.4}
  \def\arraycolsep{2pt}
  \begin{array}{c}
    F_2
    \,\in\,
    \Omega^2_{\mathrm{dR}}(U_i)
    \\
    H_3
    \,\in\,
    \Omega^3_{\mathrm{dR}}(U_i)
    \\
    G_4
    \,\in\,
    \Omega^4_{\mathrm{dR}}(U_i)
    \\
    G_7
    \,\in\,
    \Omega^7_{\mathrm{dR}}(U_i)
  \end{array}
  \,
  \middle\vert
  \,
  \adjustbox{
    bgcolor=lightolive
  }{$
  \def\arraystretch{1.5}
  \def\arraycolsep{2pt}
  \begin{array}{ccl}
    \mathrm{d}\, F_2
    &=&
    0
    \\
    \mathrm{d}\, H_3
    &=&
    G_4 \,+\, 
    \mathcolor{darkblue}{F_2\, F_2}
    \\
    \mathrm{d}\, G_4
    &=&
    0
    \\
    \mathrm{d}\, G_7
    &=&
    \tfrac{1}{2} \, G_4\, G_4
  \end{array}  
  $}
  \right\}
$$

\vspace{.1cm}

\noindent
{\bf (ii)} Given one of these, 
its set of coboundaries (null-concordances) naturally
retracts onto the set of {\bf gauge potentials} of this form:
$$ 
  \def\arraystretch{1}
  \def\arraycolsep{4pt}
  \begin{array}{ccc}
  \left\{
  \hspace{-3pt}
  \begin{tikzcd}[
    row sep=30pt,
    column sep=13pt
  ]
    U_i
    \ar[
      d,
      "{
        (F_2, H_3, G_4, G_7)
      }"{description, name=t}
    ]
    \ar[
      r,
      "{\ }"{swap, name=s}
    ]
    \ar[
      from=s,
      to=t,
      dashed,
      end anchor={[xshift=16pt]},
      Rightarrow,
      darkorange,
      "{
        (
        \widehat F_2, 
        \widehat H_3, 
        \widehat G_4, 
        \widehat G_7)
      }"{xshift=-3pt}
    ]
    &
    \ast
    \ar[
      d,
      "{ 0 }"
    ]
    \\
    \Omega^1_{\mathrm{dR}}\big(
      -;
      \mathfrak{l}_{{}_{S^4}}
      \mathbb{C}P^3
    \big)_{\mathrm{clsd}}
    \ar[
      r,
      shorten <=-3pt,
      "{
        \eta^{\scalebox{.6}{$\shape$}}
      }"
    ]
    &
    \shape
    \,
    \Omega^1_{\mathrm{dR}}\big(
      -;
      \mathfrak{l}_{{}_{S^4}}
      \mathbb{C}P^3
    \big)_{\mathrm{clsd}}
  \end{tikzcd}
  \hspace{-3pt}
  \right\}
  &
  \begin{tikzcd}
    {}
    \ar[
      r,
      ->>,
      shift left=5pt,
      "{ p_1 }"
    ]
    \ar[
      from=r,
      hook',
      shift left=5pt,
      "{ i_1 }"
    ]
    \ar[
      r,
      phantom,
      shift right=20pt,
      "{
        p_1 
          \circ 
        i_1 
          \,=\, 
        \mathrm{id}
      }"{scale=.72}
    ]
    &
    {}
  \end{tikzcd}
  &
  \left\{
  \def\arraystretch{1.4}
  \begin{array}{l}
    A_1 \,\in\,
    \Omega^1_{\mathrm{dR}}\big(U_i\big)
    \\
    B_2 \,\in\,
    \Omega^2_{\mathrm{dR}}\big(U_i\big)
    \\
    C_3 \,\in\,
    \Omega^3_{\mathrm{dR}}\big(U_i\big)
    \\
    C_6 \,\in\,
    \Omega^6_{\mathrm{dR}}\big(U_i\big)
  \end{array}
  \middle\vert
  \,
  \adjustbox{
    bgcolor=lightolive
  }{$
  \def\arraystretch{1.4}
  \def\arraycolsep{2pt}
  \begin{array}{ccl}
    \mathrm{d}\, A_1 
      &=&
    F_2
    \\
    \mathrm{d}\, B_2 &=&
    H_3 - C_3 - 
    \mathcolor{darkblue}{A_1 \, F_2}
    \\
    \mathrm{d}\, C_3
    &=&
    G_4
    \\
    \mathrm{d}\, G_6
    &=&
    G_7 
      - 
    \tfrac{1}{2} 
    C_3\, G_4
  \end{array}
  $}
  \right\}
  \\
  \\[-5pt]
  \left(
  \def\arraystretch{1.4}
  \begin{array}{ccl}
    \widehat F_2
    &:=&
    t\, F_2 \,+\, \mathrm{d}t\, A_1
    \\
    \widehat H_3
    &:= &
    t\, H_3 \,+\, \mathrm{d}t\, B_2
    \,+\,
    \mathcolor{darkblue}{
      (t^2 - t) A_1 F_2 
    }
    \\
    \widehat{G}_4
    &:=&
    t\, G_4 \,+\, \mathrm{d}t\, C_3
    \\
    \widehat{G}_7
    &:=&
    t^2\, G_7 \,+\,
    2t\mathrm{d}t\, C_6
  \end{array}
  \right)
  &
  \begin{tikzcd}
    \ar[
      r,
      |->,
      shift left=5pt,
    ]
    \ar[
      from=r,
      |->,
      shift left=5pt,
    ]
    &
    {}
  \end{tikzcd}
  &
  \left(
  \def\arraystretch{1.5}
  \def\arraycolsep{4pt}
  \begin{array}{ccl}
    A_1 &:=&
    \int_{[0,1]} \widehat F_2
    \\
    B_2 &:=&
    \int_{[0,1]} 
    \Big(
      \widehat H_3
      -
      \mathcolor{darkblue}{
      \big(
        \int_{[0,-]}
        \widehat F_2
      \big)
      \widehat F_2
      }
    \Big)
    \\
    C_3 
      &:=&
    \int_{[0,1]} \widehat G_4
    \\
    C_6 &:=&
    \int_{[0,1]} 
    \Big(
      \widehat{G}_7
      -
      \tfrac{1}{2}\big(
        \int_{[0,-]}
        \widehat{G}_4
      \big)
      \widehat G_4
    \Big)
  \end{array}
  \right)
  \end{array}
$$

\smallskip

\noindent {\bf (iii)} Given a pair of these, the set of higher coboundaries (2nd-order concordances) between them naturally
retracts onto the set of {\bf gauge transformations} of this form:
$$
\hspace{-2cm} 
  \def\arraycolsep{2pt}
  \begin{array}{ccc}
  \left\{
  \hspace{-2pt}
  \begin{tikzcd}[
    column sep=28pt
  ]
    0
    \ar[
      rr,
      bend left=40,
      Rightarrow,
      "{\ }"{swap,name=s}
    ]
    \ar[
      rr,
      phantom,
      shift left=13,
      "{
        \scalebox{.74}{$
        (
          \widehat{F}_2,
          \widehat{H}_3,
          \widehat{G}_4,
          \widehat{G}_7
        )
        $}
      }"{pos=.8}
    ]
    \ar[
      rr,
      bend right=40,
      Rightarrow,
      "{\ }"{name=t}
    ]
    \ar[
      rr,
      phantom,
      shift right=13,
      "{
        \scalebox{.8}{$
        (
          \widehat{F}'_2,
          \widehat{H}'_3,
          \widehat{G}'_4,
          \widehat{G}'_7
        )
        $}
      }"{pos=.8}
    ]
    \ar[
      from=s,
      to=t,
      Rightarrow,
      dashed,
      color=darkorange,
      "{
        \scalebox{.8}{$
          \left(
          \def\arraystretch{.9}
          \def\arraycolsep{-1pt}
          \begin{array}{l}
            \doublehat{F}_2
            ,\, \doublehat H_3
            \\
            \doublehat G_4,\,
            \doublehat G_7
          \end{array}
          \right)
        $}
      }"{swap}
    ]
    &&
    (F_2, H_3, G_4, G_7)
  \end{tikzcd}
  \hspace{-3pt}
  \right\}
  &
    \hspace{-2cm}
  \begin{tikzcd}
    {}
    \ar[
      r,
      ->>,
      shift left=5pt,
      "{ p_2 }"
    ]
    \ar[
      from=r,
      hook',
      shift left=5pt,
      "{ i_2 }"
    ]
    \ar[
      r,
      phantom,
      shift right=20pt,
      "{
        p_2
          \circ 
        i_2 
          \,=\, 
        \mathrm{id}
      }"{scale=.72}
    ]
    &
    {}
  \end{tikzcd}
  &
%\quad 
  \left\{
  \def\arraystretch{1.4}
  \begin{array}{l}
    \alpha_0
    \,\in\,\Omega^0_{\mathrm{dR}}(U_i)
    \\
    \beta_1
    \,\in\,\Omega^1_{\mathrm{dR}}(U_i)
    \\
    \gamma_2
    \,\in\,\Omega^2_{\mathrm{dR}}(U_i)
    \\
    \gamma_5
    \,\in\,\Omega^5_{\mathrm{dR}}(U_i)
  \end{array}
  \middle\vert
  \,
  \adjustbox{
    bgcolor=lightolive
  }{$
  \def\arraystretch{1.4}
  \begin{array}{ccl}
    \mathrm{d}
    \,\alpha_0
    &=&
    A'_1 - A_1
    \\
    \mathrm{d}
    \,\beta_1
    &=&
    B'_2 - B_2 + 
    \gamma_2 + 
    \mathcolor{darkblue}{\alpha_0 \, F_2}
    \\
    \mathrm{d}
    \,\gamma_2
    &=&
    C_3' - C_3
    \\
    \mathrm{d}
    \,\gamma_5
    &=&
    C'_6 - C_6  - 
    \tfrac{1}{2}C'_3 \, C_3
  \end{array}
  $}
  \right\}
  \\
  \\[-5pt]
  \hspace{1.9cm}
  \left(
  \def\arraystretch{1.4}
  \def\arraycolsep{2pt}
  \begin{array}{ccl}
    \doublehat F_2
    &:=&
    t\, F_2
    \,+\,
    \mathrm{d}t\, A_1
    \,+\,
    s\, \mathrm{d}t
    \big(
      A'_1 
      -
      A_1
    \big)
    \,-\,
    \mathrm{d}s
    \,
    \mathrm{d}t
    \,
    \alpha_0
    \\
    \doublehat H_3
    &:=&
    t\, H_3 \,+\, \mathrm{d}t\, B_2
    \,+\,
    s\, \mathrm{d}t\, 
    (B'_2  - B_2)
    \,-\,
    \mathrm{d}s\, 
    \mathrm{d}t
    \,
    \beta_1
    \\
    &&
    \,+\,
    \mathcolor{darkblue}{
    (t^2 - t) A_1
    F_2
    +
    (t^2 - t) s(A'_1 - A_1)
    F_2}
    \\
    &&
    \;\;\;
    \mathcolor{darkblue}{
    +
    (t^2 - t)
    \mathrm{d}s\,
    \alpha_0 F_2
    }
    \\
    \doublehat G_4
    &:=&
    t\, G_4
    \,+\,
    \mathrm{d}t\, C_3
    \,+\,
    s\, \mathrm{d}t
    \big(
      C'_3 
      -
      C_3
    \big)
    \,-\,
    \mathrm{d}s
    \,
    \mathrm{d}t
    \,
    \gamma_2
    \\
    \doublehat G_7
    &:=&
    t^2\, G_7
    \,+\,
    2t\mathrm{d}t\, C_6
    \,+\,
    2 s t \mathrm{d}t
    (C'_6 - C_6)
    \\
    &&
    \,-\,
    2 \mathrm{d}s \, t\mathrm{d}t
    \big(
      \gamma_5 
        \,+\,
      \tfrac{1}{2} \gamma_2\, C_3
    \big)
  \end{array}
  \right)
  &
  \begin{tikzcd}
    \ar[
      r,
      |->,
      shift left=5pt,
    ]
    \ar[
      from=r,
      |->,
      shift left=5pt,
    ]
    &
    {}
  \end{tikzcd}
  &
  \left(
  \def\arraystretch{1.9}
  \def\arraycolsep{1pt}
  \begin{array}{ccl}
    \alpha_0
    &:=&
    \int_{s \in [0,1]}
    \int_{t \in [0,1]}
    \doublehat F_2
    \\
    \beta_1 
    &:=&
    \int_{s \in [0,1]}
    \int_{t \in [0,1]}
    \Big(
    \doublehat H_3 
    \,-\,
    \mathcolor{darkblue}{
    \big(
      \int_{t'\in [0,-]}
      \doublehat F_2
    \big)
    \doublehat F_2
    }
    \Big)
    \\
    \gamma_2 &:=&
    \int_{s \in [0,1]}
    \int_{t \in [0,1]}
    \doublehat G_4    
    \\
    \gamma_5
    &:=&
    \int_{s \in [0,1]}
    \int_{t \in [0,1]}
    \Big(
      \doublehat G_7
      -
      \tfrac{1}{2}\big(
        \int_{t'\in [0,-]}
        \doublehat G_4
      \big)
      \doublehat G_4
    \Big)
    \\
    &&
    -
    \tfrac{1}{2} \gamma_2 \, C_3
  \end{array}
  \hspace{-2pt}
  \right)
  \end{array}
$$

\medskip

\noindent
Notice the expression for
flux density subject to an (abelian) Green-Schwarz mechanism:

$$
\adjustbox{
  margin=4pt,
  bgcolor=lightolive
}{$
  H_3
  \;=\;
  \mathrm{d}\,B_2
  \,+\,
  \mathcolor{darkblue}{A_1 F_2}
  \,+\,
  C_3
$}.
$$

\medskip 
\begin{proof}
With the blue terms discarded, this is the statement of \cite[p 23]{GSS24-SuGra}\cite[\S 4.1]{GSS24-FluxOnM5}. We compile the full argument.

\smallskip

\noindent
To see that $p_1$ is well-defined:

\vspace{1mm} 
-- for $C_3, C_6$ this is \cite[(70)]{GSS24-SuGra},

-- for $A_1$ it works just as for $C_3$,

-- for $B_2$ we compute, in generalization of \cite[below (138)]{GSS24-FluxOnM5}, like this:
$$
  \def\arraystretch{2}
  \begin{array}{ccl}
    \mathrm{d}\, B_2
    &\defneq&
    \mathrm{d}
    \int_{[0,1]} 
    \Big(
      \widehat H_3
      -
      \big(
        \int_{[0,-]}
        \widehat F_2
      \big)
      \widehat F_2
    \Big)
    \\
    &=&
    \grayunderbrace{
    \iota_1^\ast
    \Big(
      \widehat H_3
      -
      \big(
        \textstyle{\int_{[0,-]}}
        \widehat F_2
      \big)
      \widehat F_2
    \Big)
    }{
      H_3 - A_1\, F_2
    }
    \,-\,
    \grayunderbrace{
    \iota_0^\ast
    \Big(
      \widehat H_3
      -
      \big(
        \textstyle{\int_{[0,-]}}
        \widehat F_2
      \big)
      \widehat F_2
    \Big)
    }{
      = 0
    }
    \,-\,
    \int_{[0,1]} 
    \grayunderbrace{
    \mathrm{d}
    \Big(
      \widehat H_3
      -
      \big(
        \textstyle{\int_{[0,-]}}
        \widehat F_2
      \big)
      \widehat F_2
    \Big)}{
      \widehat{G}_4
    }
    \\[-6pt]
    &=&
    H_3 \,-\, A_1 \, F_2 \;-\; C_3
    \mathrlap{\,.}
  \end{array}
$$

\noindent
To see that $i_1$ is well-defined:
\begin{itemize}[
  leftmargin=.5cm,
  itemsep=2pt,
  topsep=1pt
]
\item[--] for $\widehat{G}_4, \widehat{G}_7$ this is \cite[(72)]{GSS24-SuGra},

\item[--] for $\widehat{F}_2$ it works just as for $\widehat{G}_4$,

\item[--] for $\widehat{H}_3$
we compute, in generalization of \cite[further below (138)]{GSS24-FluxOnM5}, as follows:
$$
  \left.
  \def\arraystretch{1.5}
  \begin{array}{l}
    \mathrm{d}\big(
      t H_3
      + 
      \mathrm{d}t\, B_2
      +
      \mathcolor{darkblue}{
      (t^2 - t)A_1 F_2
      }
    \big)
    \\
    \;=\;
    \mathrm{d}t\, H_3
    +
    t G_4 + 
    \mathcolor{darkorange}{t F_2 F_2}
    \\
    \phantom{\;=\;}
    -
    \mathrm{d}t\,  H_3
    +
    \mathrm{d}t\, C_3
    +
    \mathcolor{darkorange}{
    \mathrm{d}t\, A_1 F_2
    }
    \\
    \phantom{\;=\;}
    +
    \mathrm{d}
    \big(
    (\mathcolor{darkgreen}{t^2}
    -
    \mathcolor{darkorange}{t})A_1 F_2
    \big)
  \end{array}
  \right\}
  \;\;
  \mbox{hence indeed:}
  \;\;
    \mathrm{d}\, \widehat H_3
    \;=\;
    \grayunderbrace{
    t \, G_4 + \mathrm{d}t\, C_3
    }{
      \widehat G_4
    }
    \,+\,
    \grayoverbrace{
    \grayunderbrace{
      (t F_2 + \mathrm{d}t \, A_1)
    }{ \widehat F_2 }
    \grayunderbrace{
    (t F_2 + \mathrm{d}t \, A_1)    
    }{\widehat F_2}
    }{
      =\,
      \mathcolor{darkgreen}{
      \mathrm{d}(
        t^2 A_1 F_2
      )
      }
    }
$$
Moreover, it is immediate from inspection that $\iota_1^\ast \widehat H_3 \,=\, H_3$ and $\iota^\ast_0 \widehat H_3 = 0$.
\end{itemize}
\smallskip

\noindent
To see that $p_1 \circ i_1 = \mathrm{id}$:

\begin{itemize}[
  leftmargin=.5cm,
  itemsep=2pt,
  topsep=1pt
]
\item[--] for $C_3, C_6$ this is \cite[below (72)]{GSS24-SuGra},

\item[--] for $A_1$ this works just as for $C_3$,

\item[--] for $B_2$ we immediately compute:
$$
  \textstyle{\int_{[0,1]}}
  \Big(
    \widehat H_3
    -
    \big(
      \textstyle{\int_{[0,-]}}
      \widehat F_2
    \big)
    F_2
  \Big)
  \;=\;
  \grayunderbrace{
    \textstyle{\int_{[0,1]}} 
     \mathrm{d}t\, B_2
  }{ B_2 }
  \,-\,
  \textstyle{\int_{[0,1]}}
  \grayunderbrace{
  t A_1 \, \mathrm{d}t \, A_1
  }{= 0}
  \;=\;
  B_2
  \mathrlap{\,.}
$$
\end{itemize}

\noindent
To see that $p_2$ is well-defined:

\begin{itemize}[
  leftmargin=.5cm,
  itemsep=2pt,
  topsep=1pt
]
\item[--] for $\doublehat G_4, \doublehat G_7$ this is \cite[(74-5)]{GSS24-SuGra},

\item[--] for $\doublehat F_2$ this works just as for $\doublehat F_2$,

\item[--] for $\doublehat H_3$ we compute, in generalization of \cite[below (140)]{GSS24-FluxOnM5},
as follows:
$$
\hspace{-3mm}
  \def\arraystretch{1.8}
  \begin{array}{ccl}
    \mathrm{d} \beta_1
    &\defneq&
    \mathrm{d}
    \int_{s \in [0,1]}
    \int_{t \in [0,1]}
    \Big(
    \doublehat H_3
    -
      \big(
        \int_{t'\in[0,-]} 
        \doublehat F_2
      \big)
      \doublehat F_2
    \Big)
    \\
    &=&
    \iota_{s=1}^\ast 
    \int_{t \in [0,1]}
    \big(
      \doublehat H_3
      -
      \cdots
    \big)
    -
    \iota_{s=0}^\ast 
    \int_{t \in [0,1]}
    \big(
      \doublehat H_3
      -
      \cdots
    \big)
    -
    \int_{s \in [0,1]}
    \mathrm{d}
    \int_{t \in [0,1]}
    \big(
      \doublehat H_3
      -
      \cdots
    \big)
    \\
    &=&
    \int_{t \in [0,1]}
    \iota_{s=1}^\ast 
    \big(
      \doublehat H_3
      -
      \cdots
    \big)
    -
    \int_{t \in [0,1]}
    \iota_{s=0}^\ast 
    \big(
      \doublehat H_3
      -
      \cdots
    \big)
    -
    \int_{s \in [0,1]}
    \iota_{t=1}^\ast
    \big(
      \doublehat H_3
      -
      \cdots
    \big)
    +
    \int_{s \in [0,1]}
    \int_{t \in [0,1]}
    \mathrm{d}
    \big(
      \doublehat H_3
      -
      \cdots
    \big)
    \\
    &=&
    \int_{t \in [0,1]}
    \big(
      \widehat H'_3
      -
      \cdots
    \big)
    -
    \int_{t \in [0,1]}
    \big(
      \widehat H_3
      -
      \cdots
    \big)
    \,+\,
    \big(
    \int_{s \in [0,1]}
    \int_{t \in [0,1]}
    \doublehat F_2
    \big)
    F_2
    +
    \int_{s \in [0,1]}
    \int_{t \in [0,1]}
      \doublehat G_4
    \\
    &=&
    B'_2
    -
    B_2
    +
    \alpha_0 \, F_2
    +
    \gamma_2 
    \mathrlap{\,.}
  \end{array}
$$
\end{itemize}

\noindent
To see that $i_2$ is well-defined:

\vspace{1mm} 
\begin{itemize}[
  leftmargin=.5cm,
  itemsep=2pt,
  topsep=1pt
]
\item[--] for $\gamma_2, \gamma_5$ this is \cite[(76)]{GSS24-SuGra},

\item[--] for $\alpha_0$ this works just as for $\gamma_2$,

\item[--] for $\beta_1$ we compute as follows:
$$
  \def\arraystretch{1.6}
  \def\arraycolsep{2pt}
  \begin{array}{rcl}
    \mathrm{d}\, 
    (
      t\, H_3
      \,+\,
      \mathrm{d}t\, B_2
      \,+\,
      s\, \mathrm{d}t\, (B'_2 - B_2)
      \,-\,
      \mathrm{d}s\, \mathrm{d}t
      \,
      \beta_1
    )
    &=&
    \grayoverbrace{
    t\, G_4 
      \,+\, 
    \mathrm{d}t\, C_3
      \,+\,
    s\, \mathrm{d}t(C'_3 - C_3)
      \,-\,
    \mathrm{d}s\, \mathrm{d}t\, 
    \gamma_2
    }{ 
      \adjustbox{
        scale=.7
      }{$
        \doublehat G_4
      $}
    }
    \\
    &&
    \,+\, t\, F_2 F_2
    \,+\, \mathrm{d}t\, A_1 \, F_2
    \,+\, s\, \mathrm{d}t(A'_1 - A_1)F_2
    \,-\, \mathrm{d}s\, \mathrm{d}t\,
    \alpha_0 \, F_2
    \\
    \\[-5pt]
    \mathrm{d}\left(
      \begin{array}{l}
      (
        \mathcolor{darkgreen}{t^2}
        -
        \mathcolor{darkorange}{t}
      )
      A_1 \, F_2
      \,+\,
      (
        \mathcolor{darkgreen}{t^2}
        -
        \mathcolor{darkorange}{t}
      )
      s(A'_1 - A_1) F_2
      \\
      \;\;\;
      \,+\,
      (
        \mathcolor{darkgreen}{t^2}
        -
        \mathcolor{darkorange}{t}
      )
      \, \mathrm{d}s\, \alpha_0\,
      F_2
      \end{array}
    \right)
    &=&
    \grayoverbrace{
    \mathcolor{darkgreen}{t^2} 
    \, F_2 \, F_2
    \,+\,
    \mathcolor{darkgreen}{
      2t\mathrm{d}t
    }\, A_1 F_2
    \,+\,
    \mathcolor{darkgreen}{
      2t\mathrm{d}t
    }\, s(A'_1 - A_1)
    \,+\,
    \mathcolor{darkgreen}{
      2t\mathrm{d}t
    }\, \mathrm{d}s\, \alpha_0\, F_2
    }{
      \scalebox{.7}{$
        \doublehat F_2
        \, 
        \doublehat F_2
      $}
    }
    \\[-10pt]
    &&
    \,-\,
    \mathcolor{darkorange}{t}
    \,
    F_2 \, F_2
    \,-\,
    \mathcolor{darkorange}{\mathrm{d}t}
    \,
    A_1\, F_2
    \,-\,
    \mathcolor{darkorange}{\mathrm{d}t}
    \,
    s(A'_1 - A_1) F_2
    \,-\,
    \mathcolor{darkorange}{\mathrm{d}t}
    \,
    \mathrm{d}s\, \alpha_0\, F_2
    \\
    \hline
    \\[-15pt]
    \mathrm{d}\, 
    \doublehat H_3
    \hspace{3cm}\,
    &=&
    \hspace{3cm}
    \doublehat G_4
    \,+\,
    \doublehat F_2
    \,
    \doublehat F_2
    \,.
  \end{array}
$$
Moreover, it is immediate from inspection that 
$\iota_{s=0}^\ast \doublehat H_3 \,=\, \widehat H_3$ , 
$\iota_{s=1}^\ast \doublehat H_3 \,=\, \widehat H_3'$
and $\iota_{t=0}^\ast \,=\, 0$,
$\iota_{t=1}^\ast \,=\, H_3$.
\end{itemize}

\medskip

\noindent
To see that $p_2 \circ i_2 \,=\, \mathrm{id}$, we directly compute,

\noindent
first
$$
  \def\arraystretch{1.8}
  \begin{array}{l}
  \textstyle{\int_{s \in [0,1]}}
  \textstyle{\int_{t \in [0,1]}}
  \doublehat G_4
  \;=\;
  \textstyle{\int_{s \in [0,1]}}
  \textstyle{\int_{t \in [0,1]}}
  (- \mathrm{d}s\, \mathrm{d}t \,\gamma_2)
  \;=\;
  \gamma_2
  \\
  \textstyle{\int_{s \in [0,1]}}
  \textstyle{\int_{t \in [0,1]}}
  \doublehat F_2
  \;=\;
  \textstyle{\int_{s \in [0,1]}}
  \textstyle{\int_{t \in [0,1]}}
  (- \mathrm{d}s\, \mathrm{d}t \,\alpha_0)
  \;=\;
  \alpha_0
  \end{array}
$$
then
$$
  \def\arraystretch{1.8}
  \begin{array}{l}
  \textstyle{\int_{s \in [0,1]}}
  \textstyle{\int_{t \in [0,1]}}
  \Big(
    \doublehat G_7
   \mathcolor{darkblue}{
    -
    \tfrac{1}{2}
  \big(
    \textstyle{\int_{t' \in [0,t]}}
    \doublehat G_4
  \big)
  \doublehat G_4
  }
  \Big)
  \,-\,
  \tfrac{1}{2}
  \,\gamma_2 \, C_3
  \\[5pt]
  \;=\;
  \textstyle{\int_{s \in [0,1]}}
  \textstyle{\int_{t \in [0,1]}}
  \doublehat G_7
  \mathcolor{darkblue}{
  \,-\,
  \tfrac{1}{2}
  \textstyle{\int_{s \in [0,1]}}
  \textstyle{\int_{t \in [0,1]}}
  \big(
  t \, C_3 + s t (C'_3 - C_2)
  +
  t \mathrm{d}s \, \gamma_2
  \big)
  \big(
    t\, G_4
    \,+\,
    \mathrm{d}t\, C_3
    \,+\,
    s\, \mathrm{d}t
    (C'_3 - C_3)
    -
    \mathrm{d}s
    \,
    \mathrm{d}t
    \,
    \gamma_2
  \big)
  }
  \\
  \;\;\;\;\;
  - \tfrac{1}{2} \, \gamma_2\, C_3
  \\
  \;=\;
  \big(
    \gamma_5
    \,+\,
    \tfrac{1}{2}
    \gamma_2 C_3
  \big)
  \grayunderbrace{
  \mathcolor{darkblue}{
  -
  \tfrac{1}{2}
  C_3\gamma_2
  -
  \tfrac{1}{4}(C'_3 - C_3) \gamma_2
  +
  \tfrac{1}{2}
  \,
  \gamma_2 \, C_3
  +
  \tfrac{1}{4}
  \,
  \gamma_2
  \,
  (C'_3 - C_3) 
  }
  }{
  0
  }
  \,-\,
  \tfrac{1}{2}\, \gamma_2\, C_3
  \\[-10pt]
  \;=\;
  \gamma_5 
  \end{array}
$$
and analogously
$$
  \def\arraystretch{1.8}
  \begin{array}{l}
    \int_{s \in [0,1]}
    \int_{t \in [0,1]}
    \big(
    \int_{t'\in [0,-]}
    \doublehat F_2
    \big)
    \doublehat F_2
    \\
    \;=\;
    \int_{s \in [0,1]}
    \int_{t \in [0,1]}
    \big(
      t A_1 + st(A'_1 - A_1) 
      \,+\,
      t\mathrm{d}s \, \alpha_0
    \big)
    \big(
      t\, F_2 
      \,+\, \mathrm{d}t\, A_1
      \,+\, s\, \mathrm{d}t(A'_1 - A_1)
      \,-\, \mathrm{d}s\, \mathrm{d}t
      \, \alpha_0
    \big)
    \\
    \;=\;
    \tfrac{1}{2}A_1 \alpha_0
    +
    \tfrac{1}{4}(A'_1-A_1)\alpha_0
    -
    \tfrac{1}{2}\alpha_0 A_1
    -
    \tfrac{1}{4}\alpha_0(A'_1-A_1)
    \\
    \;=\; 0
  \end{array}
$$
so that also
$$
    \textstyle{\int_{s \in [0,1]}}
    \textstyle{\int_{t \in [0,1]}}
    \Big(
      \doublehat H_3
      -
      \big(
       \textstyle{\int_{t' \in [0,-]}}
       \doublehat F_2
      \big)
      \doublehat F_2
    \Big)
    \;=\;
    \textstyle{\int_{s \in [0,1]}}
    \textstyle{\int_{t \in [0,1]}}
    (- \mathrm{d}s\, \mathrm{d}t\ \beta_1)
    \;=\;
    \beta_1
    \,.
$$

\vspace{-.4cm}
\end{proof}

\bigskip 
\noindent
{\bf Cocycles in differential 2-Cohomotopy and the abelian Chern-Simons invariant on the 3-Sphere.}
Notice that the Bianchi identities encoded by 2-Cohomotopy
are the characteristic property of the abelian Chern-Simons term:
$$
  \mathrm{CE}\big(
    \mathfrak{l}S^2
  \big)
  \;\simeq\;
  \mathbb{R}_{\mathrm{d}}
  \left[
  \def\arraystretch{1.1}
  \def\arraycolsep{1pt}
  \begin{array}{c}
    f_2
    \\
    h_3
  \end{array}
  \right]
  \Big/
  \left(
   \def\arraystretch{1.1}
  \def\arraycolsep{1pt}
  \begin{array}{ccl}
    \mathrm{d}\, f_2
    &=& 
    0
    \\
    \mathrm{d}\, h_3
    &=&
    f_2 f_2
  \end{array}
  \right)
  \;\;\;\;\;
  \Rightarrow
  \;\;\;\;\;
  \Omega^1_{\mathrm{dR}}\big(
    X
    ;\,
    \mathfrak{l}S^2
  \big)_{\mathrm{clsd}}
  \;\simeq\;
  \left\{
  \def\arraystretch{1.3}
  \begin{array}{ccl}
    F_2 \,\in\,
    \Omega^2_{\mathrm{dR}}(X)
    \\
    H_3 \,\in\,
    \Omega^3_{\mathrm{dR}}(X)
  \end{array}
  \middle\vert
  \;
  \adjustbox{
    bgcolor=lightgray
  }{$
  \def\arraystretch{1.3}
  \def\arraycolsep{2pt}
  \begin{array}{ccl}
    \mathrm{d}\, F_2 &=& 0
    \\
    \mathrm{d}\, H_3 &=&
    F_2\, F_2
  \end{array}
  $}
  \right\}.
$$

\medskip 
\noindent
 We may bring this out more concretely:

\medskip 
\noindent {\bf Gauge-field configurations on $\mathbb{R}^3$ flux-quantized in 2-Cohomotopy} and vanishing in a neighborhood of infinity are 
cocycles in differential 2-Cohomotopy on $\mathbb{R}^3_{\cpt}$, hence dashed homotopies as shown in the following diagram (cf. \cite[\S 3.3]{SS24-Flux}):

\newpage 
 $$
  \begin{tikzcd}[row sep=15pt, 
    column sep=70pt
  ]
    & 
    \mathbb{R}^3_{\cpt}
    \ar[
      dr,
      dashed,
      "{
        \scalebox{.7}{
          \color{darkgreen}
          \bf
          \def\arraystretch{.9}
          \begin{tabular}{c}
            charge in
            \\
            2-Cohomotopy
          \end{tabular}
        }
      }"{sloped},
      "{
        n
      }"{pos=.7, swap, name=s}
    ]
    \ar[
      dl,
      dashed,
      "{
        \scalebox{.7}{
          \color{darkgreen}
          \bf
          \def\arraystretch{.9}
          \begin{tabular}{c}
            flux density
            \\
            valued in \scalebox{1.3}{$\mathfrak{l}S^2$}
          \end{tabular}
        }
      }"{sloped},
      "{
        (F_2, H_3)
      }"{pos=.7,name=t}
    ]
    \ar[
      from=s,
      to=t,
      Rightarrow,
      shorten=4pt,
      dashed,
      "{ (\mathbf{A}_1, \mathbf{B}_2) }"{swap},
      "{
        \scalebox{.7}{
          \color{darkorange}
          \bf
          \def\arraystretch{.9}
          \begin{tabular}{c}
            gauge potentials
            \\
            in diff. 2-Cohomotopy
          \end{tabular}
        }
      }"
    ]
    \\
    \Omega^1_{\mathrm{dR}}\big(
      -;\,
      \mathfrak{l}S^2
    \big)_{\mathrlap{\mathrm{clsd}}}
    \;\;\;\;
    \ar[
      dr,
      "{
        \eta^{\scalebox{.6}{$\shape$}}
      }"{swap}
    ]
    &&
\;\;    \shape\, S^2
    \ar[
      dl,
      "{
        \mathbf{ch}
      }"
    ]
    \\[-6pt]
    &
    \shape
    \,
    \Omega^1_{\mathrm{dR}}\big(
      -;\,
      \mathfrak{l}S^2
    \big)_{\mathrlap{\mathrm{clsd}}}
  \end{tikzcd}
$$

\noindent
\colorbox{lightolive}{{\bf Theorem.} For each $[n] \in \pi^2\big(\mathbb{R}^3_{\cpt}\big) \,\simeq\, \mathbb{Z}$ this exists with $H_3 = 0$ and $[n] = \int_{\mathbb{R}^3} A_1\, F_2$ the Chern-Simons invariant.}

\medskip

To see this, first consider:

\noindent {\bf Lemma.} 
{\it On a smooth manifold $X$, 
every cocycle $\alpha$ in \emph{rational} 3-Cohomotopy is represented by a globally defined differential form $H_3$:}
$$
  \adjustbox{raise=3pt}{
  \begin{tikzcd}[
    row sep=2pt
  ]
    X
    \ar[
      rr,
      "{\ }"{swap, name=s}
    ]
    \ar[
      dr,
      dashed,
      "{
        H_3
      }"{swap},
      "{\ }"{name=t}
    ]
    \ar[
      from=s,
      to=t,
      Rightarrow,
      dashed, color=darkorange
    ]
    &&
    \shape\,
    \Omega^1_{\mathrm{dR}}\big(
      -;\,
      \mathfrak{l}S^3
    \big)_{\mathrlap{\mathrm{clsd}}}
    \\
    &
    \Omega^1_{\mathrm{dR}}\big(
      -;\,
      \mathfrak{l}S^3
    \big)_{\mathrlap{\mathrm{clsd}}}
    \ar[
      ur,
      shorten=-2pt,
      "{ 
        \eta^{\scalebox{.6}{$\shape$}}
      }"{swap}
    ]
  \end{tikzcd}
  }
$$

\vspace{-.3cm}
\begin{proof}[Proof of the Lemma]
 Since $\mathfrak{l}S^3 \,\simeq\, \mathfrak{l}B^3 \mathbb{Q}$ this is 
 just the degree=3 case of the statement that cocycles in de Rham hyper-cohomology have global representatives on smooth manifolds (using partitions of unity).
\end{proof}

\begin{proof}[Proof of the Theorem]
Stereographic projection provides a  homeomorphism $\mathbb{R}^3_{\cpt} \xrightarrow{\sim} S^3$ which is smooth away from the point at infinity. We may slightly deform this to a smooth degree=1 map that is constant on a neighborhood of infinity. Since $\pi^2(S^3) \simeq \pi_3(S^2) \simeq \mathbb{Z}$ we may find a smooth map $n : S^3 \xrightarrow{} S^2$, with compact support away from the base point, so that $\mathbb{R}^3_{\cpt} \to S^3 \xrightarrow{n} S^2$ represents the charge $[n]$.

Now the 2-cohomotopical character map for charges on $S^3$,
shown in black,
factors 
{\color{darkblue}as shown in blue} (by naturality of rationalization),
which furthermore factors 
{\color{darkorange}as shown in orange} (by the above Lemma):
\begin{equation}
  \label{2CohomotopyCharacterOn3Sphere}
  \begin{tikzcd}[
    row sep=20pt,
    column sep=23pt
  ]
    \mathbb{R}^3_{\cpt}
    \ar[r]
    &[-40pt]
    S^3
    \ar[
      r,
      "{
        \eta^{
          \scalebox{.7}{$
            \shape
          $}
        }
      }",
      "{\ }"{swap, pos=.3, name=s1}
    ]
    \ar[
      d,
      darkorange,
      "{
        n \cdot \mathrm{dvol}_{S^3}
      }"{swap, name=t1},
    ]
    \ar[
      from=s1,
      to=t1,
      end anchor={[xshift=6pt]},
      Rightarrow,
      darkorange
    ]
    &
    \shape \, S^3 
    \ar[
      r,
      "{ n }",
      "{\ }"{swap, pos=.3, name=s2}
    ]
    \ar[
      d,
      darkblue,
      "{
        \mathbf{ch}_{S^3}
      }"{swap, name=t2}
    ]
    \ar[
      from=s2,
      to=t2,
      Rightarrow,
      end anchor={[xshift=6pt]},
      darkblue
    ]
    &
    \shape \, S^2
    \ar[
      d,
      "{
        \mathbf{ch}_{S^2}
      }"
    ]
    \\
    &
    \mathcolor{darkorange}{
    \Omega^1_{\mathrm{dR}}
      (
        -; \mathfrak{l}S^3
      )_{\mathrm{clsd}}
    }
    \ar[
      r,
      darkorange,
      "{
        \eta^{
          \scalebox{.7}{$\shape$}
        }
      }"
    ]
    &
    \mathcolor{darkblue}{
    \shape\,
    \Omega^1_{\mathrm{dR}}(-;\mathfrak{l}S^3)_{\mathrm{clsd}}
    }
    \ar[
      r,
      darkblue,
      "{
        (\mathfrak{l}n)_\ast
      }"
    ]
    &
    \shape\, 
    \Omega^1_{\mathrm{dR}}(-;\mathfrak{l}S^2)_{\mathrm{clsd}}
    \mathrlap{\,.}
  \end{tikzcd}
\end{equation}
Hence, to get a differential cocycle as desired, it is sufficient to exhibit gauge potentials $(A_1, B_2)$ encoding a {\color{olive}concordance}   filling the following diagram:
\hspace{-4mm}
\begin{equation}
  \label{AConcordance}
  \begin{tikzcd}[column sep=large]
    \mathbb{R}^3_{\cpt}
    \ar[r]
    &[-40pt]
    S^3
    \ar[
      r,
      "{
        \eta^{\scalebox{.6}{$\shape$}}
      }"
    ]
    \ar[
      dr,
      "{
        n\cdot 
        \mathrm{dvol}_{S^3}
      }"{description, name=s1}
    ]
    \ar[
      dd,
      "{
        (F_2, H_3 = 0)
      }"{description, name=t1}
    ]
    \ar[
      from=s1,
      to=t1,
      Rightarrow,
      olive,
      "{
        (A_1, B_2)
      }"
    ]
    &
    \shape S^3
    \ar[
      r,
      "{
        n
      }"
    ]
    &
    \shape\,
    S^2
    \ar[
      dd,
      "{
        \mathbf{ch}_{S^2}
      }"
    ]
    \ar[
      dl,
      Rightarrow,
      shorten=7pt,
      start anchor={[xshift=-10pt]},
      end anchor={[xshift=+10pt, yshift=-4pt]},
      "{
        \scalebox{.7}{
          \eqref{2CohomotopyCharacterOn3Sphere}
        }
      }"
    ]
    \\
    &
    &
    \Omega^1_{\mathrm{dR}}(
      -;\,
      \mathfrak{l}S^3
    )_{\mathrm{clsd}}    
    \ar[
      dr,
      "{
        (\mathfrak{l}n)_\ast
        \circ
        \,
        \eta^{\scalebox{.7}{$\shape$}}
      }"{description}
    ]
    \\
    &
    \Omega^1_{\mathrm{dR}}(
      -;\,
      \mathfrak{l}S^2
    )_{\mathrm{clsd}}    
    \ar[
      rr,
      "{
        \eta^{\scalebox{.6}{$\shape$}}
      }"{swap}
    ]
    &&
    \shape
    \,
    \Omega^1_{\mathrm{dR}}(
      -;\,
      \mathfrak{l}S^2
    )_{\mathrm{clsd}}
    \mathrlap{\,.}
  \end{tikzcd}
\end{equation}
\vspace{-1mm} 

\noindent
However, since $H^2_{\mathrm{dR}}(S^3) = 0$
and by  the {\it Whitehead integral formula}
(cf. \cite[p 134]{GriffithsMorgan81}\cite[p 228]{BottTu82}\cite[p 19]{FSS19-Hopf})
there exists:
\begin{equation}
    \label{WhiteheadIntegralFormula}
\left\{\!\!
\def\arraystretch{1.4}
\begin{array}{ccll}
A_1 
&\in&
\Omega^1_{\mathrm{dR}}(S^3)
\\
  B_2 
  &\in&
  \Omega^2_{\mathrm{dR}}(S^3)
\end{array}
\;\;\mbox{s.t.}\;\;
\adjustbox{
  bgcolor=lightgray
}{$
\def\arraystretch{1.4}
\begin{array}{l}
  \mathrm{d}\, A_1
  \;=\;
  F_2
  \;:=\;
  (-n)^\ast \mathrm{dvol}_{S^2}
  \\
    \mathrm{d}\, B_2
    \;=\;
    n \cdot \mathrm{dvol}_{S^3}
    +
    A_1 F_2
\end{array}
$}
  \right.
\end{equation}

\noindent
From this we get the desired {\color{olive}concordance} \eqref{AConcordance} \footnote{
  To note that $F_2 F_2 = 0$ over $S^3$, by degree reasons, and that this holds also after pullback to the concordance domain $[0,1] \times S^3$, while $\widehat{F}_2 \widehat{F}_2 = 2 t\mathrm{d}t \, A_1 F_2$ is generally non-vanishing there.
}:
\vspace{-.2cm}
$$
  (0, n \cdot \mathrm{dvol}_{S^3})
  \Rightarrow
  (F_2,0)
  \;\;
  :
  \;
  \left\{\!\!
  \def\arraystretch{1.5}
  \begin{array}{l}
    \widehat F_2
    :=
    t\, F_2 \,+\,
    \mathrm{d}t\, A_1\,,
    \\
    \widehat H_3
    :=
    (1-t) \,
    n \cdot \mathrm{dvol}_{S^3}
    \,-\,
    \mathrm{d}t
    \,
    B_2
    \,+\,
    (t^2 - t)
    A_1 F_2
    \,,
  \end{array}
  \right.
  \;\; 
  \def\arraystretch{1.4}
  \begin{array}{l}
    \big(
      \widehat{F}_2, \widehat{H}_3
    \big)\vert_{t=0}  
      \,=\, 
    (0,n \cdot \mathrm{dvol}_{S^3})
    \\
    \big(
      \widehat{F}_2, \widehat{H}_3
    \big)\vert_{t=1} 
      \,=\, 
    (F_2, 0)
    \\
    \mathrm{d}\, \widehat{F}_2 \,=\, 0
    \,,\;
    \mathrm{d}\, \widehat{H}_3 \,=\, \widehat{F}_2\, \widehat{F}_2
    \,.
  \end{array}
$$
\vspace{-.5cm}

\end{proof}

%\newpage 
\noindent
{\bf Cartesian M5-Probes charged in Cohomotopy.} The equations of motion for a(n orbifolded) cartesian M5-probe demand that the flux $H_3 = \mathrm{const}$ \cite[Ex. 3.14]{GSS24-FluxOnM5},
and thus its solitonic vanishing-at-infinity implies $H_3 = 0$. The above theorem says that such solutions still support non-vanishing cohomotopical charge, in fact that the vanishing of $H_3$ forces the charge to be carried by the Chern-Simons invariant of the auxiliary gauge field $\mathbf{A}_1$ that is brought in by the cohomotopical flux quantization.


\bigskip 
\begin{thebibliography}{100}

\bibitem{AGMOOY00}
Aharony, O., Gubser, S., Maldacena, J., Ooguri, H., Oz, Y., 
{\it\color{darkblue}Large $N$ Field Theories, String Theory and Gravity}, Phys. Rept. {\bf 323}  (2000), 
183-386,
[\href{https://doi.org/10.1016/S0370-1573(99)00083-6}{\tt doi:10.1016/S0370-1573(99)00083-6}],  \newline 
[\href{https://arxiv.org/abs/hep-th/9905111}{\tt arXiv:hep-th/9905111}].



\bibitem{AnLiYang22}
An, Y.-S., Li L., Yang, F.-G., Yang, R.-Q., {\it\color{darkblue}Interior Structure and Complexity Growth Rate of Holographic Superconductor from M-Theory}, 
J. High Energ. Phys. {\bf 2022} 133 (2022), 
[\href{https://doi.org/10.1007/JHEP08(2022)133}{\tt doi:10.1007/JHEP08(2022)133}],
[\href{https://arxiv.org/abs/2205.02442}{\tt arXiv:2205.02442}].

\bibitem{ASW84}
Arovas, D. P., Schrieffer, R., Wilczek, F., 
{\it\color{darkblue}Fractional Statistics and the Quantum Hall Effect}, 
Phys. Rev. Lett. {\bf 53} (1984) 722, 
[\href{https://doi.org/10.1103/PhysRevLett.53.722}{\tt doi:10.1103/PhysRevLett.53.722}].

\bibitem{ASWZ85}
Arovas, D. P., Schrieffer, R., Wilczek, F.,  Zee, A., 
{\it\color{darkblue}Statistical mechanics of anyons}, 
Nucl. Phys. B {\bf 251} (1985), 117-126,
[\href{https://doi.org/10.1016/0550-3213(85)90252-4}{\tt doi:10.1016/0550-3213(85)90252-4}].

\bibitem{BaeLee21}
Bae, J.-B., Lee, S.,  
{\it\color{darkblue}Emergent Supersymmetry on the Edges}, SciPost Phys. {\bf 11} 091 (2021), 
[\href{https://arxiv.org/abs/2105.02148}{\tt arXiv:2105.02148}],
\newline  
[\href{https://doi.org/10.21468/SciPostPhys.11.5.091}{\tt doi:10.21468/SciPostPhys.11.5.091}].



\bibitem{BakulevShirkov10}
Bakulev, A. P., Shirkov, D., 
{\it\color{darkblue}Inevitability and Importance of Non-Perturbative Elements in Quantum Field Theory}, Proceedings of the 6th Mathematical Physics Meeting, Belgrade (2010), 27–54, [\href{https://arxiv.org/abs/1102.2380}{\tt arXiv:1102.2380}],
[{\tt ISBN:978-86-82441-30-4}].

\bibitem{Balachandran91}
Balachandran, A. P., Srivastava, A. M., {\it\color{darkblue}Chern-Simons Dynamics and the Quantum Hall Effect},
\newline
[\href{https://arxiv.org/abs/hep-th/9111006}{\tt arXiv:hep-th/9111006}].


\bibitem{BarkeshliJianQi13}
Barkeshli, M., Jian, C.-M., Qi, X>-L., 
{\it\color{darkblue}Twist defects and projective non-Abelian braiding statistics}, Phys. Rev. B {\bf 87} (2013) 045130, 
[\href{https://doi.org/10.1103/PhysRevB.87.045130}{\tt doi:10.1103/PhysRevB.87.045130}],
[\href{https://arxiv.org/abs/1208.4834}{\tt arXiv:1208.4834}].



\bibitem{BeardleyNakamura24}
Beardsley, J., Nakamura, S., {\it\color{darkblue}Projective Geometries and Simple Pointed Matroids as $\mathbb{F}_1$-modules},
\newline
[\href{https://arxiv.org/abs/2404.04730}{\tt arXiv:2404.04730}].


\bibitem{Beterov24}
Beterov, I. I., 
{\it\color{darkblue}Progress and Prospects in the Field of Quantum Computing}, Optoelectron. Instrument. Proc. {\bf 60} (2024),
74–83,  
[\href{https://doi.org/10.3103/S8756699024700043}{\tt doi:10.3103/S8756699024700043}].

\bibitem{BhattacharjeeMacphersonMollerNeumann06}
Bhattacharjee, M., Macpherson, D., Möller, R. G., Neumann, P. M., 
{\it\color{darkblue}Notes on Infinite Permutation Groups}, Lecture Notes in Mathematics {\bf 1698}, Springer (2006), 67-76,
[\href{https://doi.org/10.1007/BFb0092558}{\tt doi:10.1007/BFb0092558}].



\bibitem{BottTu82}
Bott, R., Tu, L., 
{\it\color{darkblue}Differential Forms in Algebraic Topology}, Graduate Texts in Mathematics {\bf 82}, Springer (1982),
[\href{https://link.springer.com/book/10.1007/978-1-4757-3951-0}{\tt doi:10.1007/978-1-4757-3951-0}].


\bibitem{BourjailyEspahbodi08}
Bourjaily, J. L., Espahbodi S.,  
{\it\color{darkblue}Geometrically Engineerable Chiral Matter in M-Theory},
[\href{https://arxiv.org/abs/0804.1132}{\tt arXiv:0804.1132}].

\bibitem{SS21-FracD}
Burton, S., Sati, H., Schreiber, U., {\it\color{darkblue}Lift of fractional D-brane charge to equivariant Cohomotopy theory}, J. Geometry and Physics {\bf 161} (2021) 104034 
[\href{https://doi.org/10.1016/j.geomphys.2020.104034}{\tt doi:10.1016/j.geomphys.2020.104034}],
[\href{https://arxiv.org/abs/1909.12277}{\tt arXiv:1909.12277}].



\bibitem{ChakrabortyPietilainen95}
Chakraborty, T., Pietiläinen, P.,
{\it\color{darkblue}The Quantum Hall Effects -- Integral and Fractional}, 
Springer Series in Solid State Sciences (1995),
[\href{https://doi.org/10.1007/978-3-642-79319-6}{\tt doi:10.1007/978-3-642-79319-6}].



\bibitem{CTY16}
Chan, P. O., Teo, J. C. Y., Ryu, S.,
{\it\color{darkblue}Topological Phases on Non-orientable Surfaces: Twisting by Parity Symmetry},
New J. Phys. {\bf 18}  (2016) 035005,
[\href{https://arxiv.org/abs/1509.03920}{\tt arXiv:1509.03920}],
[\href{https://doi.org/10.1088/1367-2630/18/3/035005}{\tt doi:10.1088/1367-2630/18/3/035005}].

\bibitem{ChenEtAl07}
Chen, G. et al., 
{\it\color{darkblue}Quantum Computing Devices -- Principles, Designs, and Analysis}, Routledge (2007), \newline 
[\href{https://www.routledge.com/Quantum-Computing-Devices-Principles-Designs-and-Analysis/Chen-Church-Englert-Henkel-Rohwedder-Scully-Zubairy/p/book/9780367390372?srsltid=AfmBOooGiSmyK4mzxjmR1LBJz4zU9cgCpp_z4cPiCzEhM6SJo4b2GxRp}{\tt ISBN:9780367390372}]


\bibitem{ChoGangKim20}
Cho, G. Y., Gang, D., Kim, H.-C., 
{\it\color{darkblue}M-theoretic Genesis of Topological Phases}, J. High Energ. Phys. {\bf 2020} 115 (2020),
[\href{https://doi.org/10.1007/JHEP11(2020)115}{\tt doi:10.1007/JHEP11(2020)115}],
[\href{https://arxiv.org/abs/2007.01532}{arXiv:2007.01532}].




\bibitem{ChuLorscheidSanthanam12}
Chu, C., Lorscheid, O., Santhanam, R., {\it\color{darkblue}Sheaves and K-theory for $\mathbb{F}_1$-schemes}, Adv. Math.
{\bf 229} 4 (2012), 2239-2286, 
[\href{https://arxiv.org/abs/1010.2896}{\tt arXiv:1010.2896}],
[\href{https://doi.org/10.1016/j.aim.2011.12.023}{\tt doi:10.1016/j.aim.2011.12.023}].


\bibitem{CMI}
Clay Math Institute, 
{\it\color{darkblue}The Millennium Prize Problems},
[\href{https://www.claymath.org/millennium-problems}{\tt www.claymath.org/millennium-problems}].


\bibitem{Cohen09}
Cohen, F. R.,
{\it\color{darkblue}Introduction to configuration spaces and their applications}, in: {\it Braids}, Lecture Notes Series, Institute for Mathematical Sciences, National University of Singapore {\bf 19} (2009), 183-261, \newline  [\href{https://doi.org/10.1142/9789814291415_0003}{\tt doi:10.1142/9789814291415\_0003}].


\bibitem{QBI}
DARPA, 
{\it\color{darkblue}Quantum Benchmarjing Initiative}
(2024),
\newline
[\href{https://www.darpa.mil/work-with-us/quantum-benchmarking-initiative}{\tt www.darpa.mil/work-with-us/quantum-benchmarking-initiative}].

\bibitem{DasSarma22}
Das Sarma, S., {\it\color{darkblue}Quantum computing has a hype problem}, MIT Tech Review (March 2022),
\newline
\scalebox{.9}{
[\href{https://www.technologyreview.com/2022/03/28/1048355/quantum-computing-has-a-hype-problem}{\tt www.technologyreview.com/2022/03/28/1048355/quantum-computing-has-a-hype-problem/}]}


\bibitem{DasSarma23}
Das Sarma, S., 
{\it\color{darkblue}In search of Majorana}, Nature Physics {\bf 19} (2023), 165-170, 
[\href{https://arxiv.org/abs/2210.17365}{\tt arXiv:2210.17365}], \newline 
[\href{https://doi.org/10.1038/s41567-022-01900-9}{\tt doi:10.1038/s41567-022-01900-9}].


\bibitem{DF82}
D'Auria, R., Fr\'e, P., 
{\it\color{darkblue} Geometric Supergravity in $D=11$ and its hidden supergroup}, 
Nucl. Phys. B {\bf 201} (1982), 101-140, 
[\href{https://doi.org/10.1016/0550-3213(82)90376-5}{\tt doi:10.1016/0550-3213(82)90376-5}].



\bibitem{DonosGauntlett13a}
Donos, A., Gauntlett, J. P., Sonner, J., Withers, B., 
{\it\color{darkblue}Competing orders in M-theory: superfluids, stripes and metamagnetism}, 
J. High Energ. Phys. {\bf 2013} 108 (2013), 
[\href{https://doi.org/10.1007/JHEP03(2013)108}{\tt doi:10.1007/JHEP03(2013)108}], \newline 
[\href{https://arxiv.org/abs/1212.0871}{\tt arXiv:1212.0871}].



\bibitem{DonosGauntlett13b}
Donos, A., Gauntlett, J. P., Pantelidou, C., {\it\color{darkblue}Semi-local quantum criticality in string/M-theory}, 
J. High Energ. Phys. {\bf 2013} 103 (2013),  
[\href{https://doi.org/10.1007/JHEP03(2013)103}{\tt doi:10.1007/JHEP03(2013)103}],  
[\href{https://arxiv.org/abs/1212.1462}{\tt arXiv:1212.1462}].


\bibitem{Duff99}
 Duff, M.,  
{\it\color{darkblue}The World in Eleven Dimensions: Supergravity, Supermembranes and M-theory}, 
IoP (1999), \newline 
[\href{https://www.crcpress.com/The-World-in-Eleven-Dimensions-Supergravity-supermembranes-and-M-theory/Duff/9780750306720}{\tt ISBN:9780750306720}].


\bibitem{Dul23}
Dul, F., 
{\it\color{darkblue}General Covariance from the Viewpoint of Stacks}, Lett Math Phys {\bf 113} (2023) 30,
\newline
[\href{https://arxiv.org/abs/2112.15473}{\tt arXiv:2112.15473}],
[\href{https://doi.org/10.1007/s11005-023-01653-3}{\tt doi:10.1007/s11005-023-01653-3}].

\bibitem{Dyakonov14}
Dyakonov, M. I.,
{\it\color{darkblue}Prospects for quantum computing: extremely doubtful}, Int. J. Mod. Phys.: Conf. Ser. {\bf 33} (2014) 1460357, 
[\href{https://doi.org/10.1142/S2010194514603573}{\tt doi:10.1142/S2010194514603573}],
[\href{https://arxiv.org/abs/1401.3629}{\tt arXiv:1401.3629}].



\bibitem{Ezratty23a}
Ezratty O., 
{\it\color{darkblue}Where are we heading with NISQ?}, 
[\href{https://arxiv.org/abs/2305.09518}{\tt arXiv:2305.09518}].


\bibitem{Ezratty23b}
Ezratty O., 
{\it\color{darkblue}Where are we heading with NISQ?}, blog post (2023), 
\newline
[\href{https://www.oezratty.net/wordpress/2023/where-are-we-heading-with-nisq}{\tt www.oezratty.net/wordpress/2023/where-are-we-heading-with-nisq}].

\bibitem{FarbMargalit12}
Farb, B., Margalit, D.,
{\it\color{darkblue}A primer on mapping class groups}, Princeton University Press (2012), \newline 
[\href{https://www.jstor.org/stable/j.ctt7rkjw}{\tt doi:j.ctt7rkjw}],
[\href{https://press.princeton.edu/books/hardcover/9780691147949/a-primer-on-mapping-class-groups}{\tt ISBN:9780691147949}].


\bibitem{FerrazEtAl20}
Ferraz, A., Gupta, K. S., Semenoff, G. W., Sodano, P.  (eds), 
{\it\color{darkblue}Strongly Coupled Field Theories for Condensed Matter and Quantum Information Theory}, Springer Proceedings in Physics {\bf 239}, Springer (2020), \newline  
[\href{https://doi.org/10.1007/978-3-030-35473-2}{\tt doi:10.1007/978-3-030-35473-2}].


\bibitem{FSS14-Stacky}
Fiorenza, D., Sati, H. ,  Schreiber, U.,
{\it\color{darkblue}A higher stacky perspective on Chern-Simons theory},
in
{\it Mathematical Aspects of Quantum Field Theories}
Springer (2014), 153-211,
[\href{https://doi.org/10.1007/978-3-319-09949-1}{\tt doi:10.1007/978-3-319-09949-1}].

\bibitem{FSS15-WZWTerm}
Fiorenza, D., Sati, H., Schreiber, U.,
{\it\color{darkblue} The WZW term of the M5-brane and differential cohomotopy},
J. Math. Phys. {\bf 56} (2015) 102301, 
[\href{https://doi.org/10.1063/1.4932618}{\tt doi:10.1063/1.4932618}],
[\href{https://arxiv.org/abs/1506.07557}{\tt arXiv:1506.07557}].


\bibitem{FSS18-TDual}
Fiorenza, D., Sati, H., Schreiber, U., 
{\it\color{darkblue}T-Duality from super Lie $n$-algebra cocycles for super $p$-branes}, Adv. Theor. Math. Phys {\bf 22} 5 (2018), 
[\href{https://arxiv.org/abs/1611.06536}{\tt arXiv:1611.06536}],
[\href{https://dx.doi.org/10.4310/ATMP.2018.v22.n5.a3}{\tt doi:10.4310/ATMP.2018.v22.n5.a3}].


\bibitem{FSS19-Hopf}
Fiorenza, D., Sati, H., Schreiber, U.,
{\it\color{darkblue} Twisted Cohomotopy implies M5 WZ term level quantization},
Commun. Math. Phys. {\bf 384} (2021), 403-432,
[\href{https://doi.org/10.1007/s00220-021-03951-0}{\tt doi:10.1007/s00220-021-03951-0}],
[\href{https://arxiv.org/abs/1906.07417}{\tt arXiv:1906.07417}].


\bibitem{FSS20-H}
Fiorenza, D., Sati, H., Schreiber, U., {\it\color{darkblue} Twisted Cohomotopy implies M-theory anomaly cancellation on 8-manifolds}, 
Commun. Math. Phys. {\bf 377} (2020), 1961-2025, 
[\href{https://doi.org/10.1007/s00220-020-03707-2}{\tt doi:10.1007/s00220-020-03707-2}], \newline 
[\href{https://arxiv.org/abs/1904.10207}{\tt arXiv:1904.10207}].

\bibitem{FSS21-TwistedString}
Fiorenza, D., Sati, H., Schreiber, U.
{\it\color{darkblue}Twisted cohomotopy implies twisted String structure on M5-branes}
J. Math. Phys. 
{\bf 62} (2021) 042301,
[\href{https://arxiv.org/abs/2002.11093}{\tt arXiv:2002.11093}],
[\href{https://doi.org/10.1063/5.0037786}{\tt doi:10.1063/5.0037786}].


\bibitem{FSS22-Twistorial}
Fiorenza, D., Sati, H. ,  Schreiber, U.,
{\it\color{darkblue} Twistorial Cohomotopy Implies Green-Schwarz anomaly cancellation}, Rev. Math. Phys.
{\bf 34} 05 (2022) 2250013,
[\href{https://doi.org/10.1142/S0129055X22500131}{\tt doi:10.1142/S0129055X22500131}],
[\href{https://arxiv.org/abs/2008.08544}{\tt arXiv:2008.08544}].

\bibitem{FSS23-Char}
Fiorenza, D., Sati, H., Schreiber, U.,
{\it\color{darkblue} The Character map in Nonabelian Cohomology --- Twisted, Differential and Generalized},
World Scientific, Singapore (2023),
[\href{https://doi.org/10.1142/13422}{\tt doi:10.1142/13422}],
[\href{https://arxiv.org/abs/2009.11909}{\tt arXiv:2009.11909}].



\bibitem{FF24}
Foss-Feig, M., Pagano, G., Potter, A. C., Yao, N. Y., 
{\it\color{darkblue}Progress in Trapped-Ion Quantum Simulation}, Ann. Rev. Condensed Matter Phys. (2024),
[\href{https://doi.org/10.1146/annurev-conmatphys-032822-045619}{\tt doi:10.1146/annurev-conmatphys-032822-045619}], \newline 
[\href{https://arxiv.org/abs/2409.02990}{\tt arXiv:2409.02990}].


\bibitem{FowlerHollenberg07}
Fowler, A. G., Hollenberg, L. C. L., 
{\it\color{darkblue}Scalability of Shor’s algorithm with a limited set of rotation gates}, Phys. Rev. A {\bf 70} (2007) 032329, 
[\href{https://doi.org/10.1103/PhysRevA.103.032417}{\tt doi:10.1103/PhysRevA.103.032417}].




\bibitem{FreedmanHastingsNayakQiWalkerWang11}
Freedman, M., Hastings, M. B.,   Nayak, C., Qi, X.-L., Walker, K.,  Wang, Z., 
{\it\color{darkblue}Projective Ribbon Permutation Statistics: a Remnant of non-Abelian Braiding in Higher Dimensions}, Phys. Rev. B {\bf 83} 115132 (2011), 
[\href{https://doi.org/10.1103/PhysRevB.83.115132}{\tt arXiv:10.1103/PhysRevB.83.115132}],
[\href{https://arxiv.org/abs/1005.0583}{\tt arXiv:1005.0583}].


\bibitem{FKLW03}
Freedman, M., Kitaev, A., Larsen, M.,  Wang, Z., 
{\it\color{darkblue}Topological quantum computation}, Bull. Amer. Math. Soc. {\bf 40} (2003), 31-38, 
[\href{https://doi.org/10.1090/S0273-0979-02-00964-3}{\tt doi:10.1090/S0273-0979-02-00964-3}],
[\href{https://arxiv.org/abs/quant-ph/0101025}{\tt arXiv:quant-ph/0101025}].


\bibitem{GallierXu13}
Gallier, J., Xu, D., 
{\it\color{darkblue}A Guide to the Classification Theorem for Compact Surfaces}, Springer (2013), \newline  
[\href{https://doi.org/10.1007/978-3-642-34364-3}{\tt doi:10.1007/978-3-642-34364-3}].


\bibitem{GanorMotl98}
Ganor, O., Motl, L., 
{\it\color{darkblue}Equations of the $(2,0)$ Theory and Knitted Fivebranes}, 
J. High Energy Phys. {\bf 9805} (1998) 009,
[\href{https://doi.org/10.1016/S0550-3213(00)00148-6}{\tt doi:10.1016/S0550-3213(00)00148-6}],
[\href{https://arxiv.org/abs/hep-th/9803108}{\tt arXiv:hep-th/9803108}].




\bibitem{GauntlettSonnerWiseman10a}
Gauntlett, J. P., Sonner, J., Wiseman, T., 
{\it\color{darkblue}Holographic superconductivity in M-Theory}, 
Phys. Rev. Lett. {\bf 103} (2009) 151601, 
[\href{https://doi.org/10.1103/PhysRevLett.103.151601}{\tt doi:10.1103/PhysRevLett.103.151601}],
[\href{https://arxiv.org/abs/0907.3796}{\tt arXiv:0907.3796}].


\bibitem{GauntlettSonnerWiseman10b}
Gauntlett, J., Sonner, J., Wiseman, T., 
{\it\color{darkblue}Quantum Criticality and Holographic Superconductors in M-theory}, 
J. High Energ. Phys. {\bf 2010} 60 (2010),
[\href{https://doi.org/10.1016/j.physletb.2009.12.017}{\tt doi:10.1016/j.physletb.2009.12.017}],
[\href{https://arxiv.org/abs/0912.0512}{\tt arXiv:0912.0512}].


\bibitem{Gent23}
Gent, E, {\it\color{darkblue}Quantum Computing’s Hard, Cold Reality Check}, IEEE Spectrum (Dec. 2023),
\newline
[\href{https://spectrum.ieee.org/quantum-computing-skeptics}{\tt spectrum.ieee.org/quantum-computing-skeptics}]

\bibitem{Gill24}
Gill, S. G.,  et al.,
{\it\color{darkblue}Quantum Computing: Vision and Challenges} ,
[\href{https://arxiv.org/abs/2403.02240}{\tt arXiv:2403.02240}].

\bibitem{GSS24-SuGra}
Giotopoulos G., Sati, H.,  Schreiber, U,
{\it\color{darkblue} Flux Quantization on 11d Superspace},
J. High Energy Phys. {\bf 2024}  (2024) 82,
[\href{https://doi.org/10.1007/JHEP07(2024)082}{\tt doi:10.1007/JHEP07(2024)082}],
[\href{https://arxiv.org/abs/2403.16456}{\tt arXiv:2403.16456}].


\bibitem{GSS24-FluxOnM5}
Giotopoulos, G., Sati, H., Schreiber, U., 
{\it\color{darkblue}Flux Quantization on M5-Branes}
J. High Energy Phys. {\bf 2024} 140 (2024),
[\href{https://doi.org/10.1007/JHEP10(2024)140}{\tt doi:10.1007/JHEP10(2024)140}],
[\href{https://arxiv.org/abs/2406.11304}{\tt arXiv:2406.11304}].


\bibitem{GSS24-TDual}
Giotopoulos, G., Sati, H.,  Schreiber, U.,
{\it\color{darkblue}Super $L_\infty$ T-Duality and M-theory},
[\href{https://arxiv.org/abs/2411.10260}{\tt arXiv:2411.10260}].


\bibitem{GS21}
Grady, D., Sati, H.,
{\it\color{darkblue}Differential cohomotopy versus differential cohomology for M-theory and differential lifts of Postnikov towers}, J. Geom. Phys. {\bf 165} (2021) 104203, 
[\href{https://doi.org/10.1016/j.geomphys.2021.104203}{\tt doi:10.1016/j.geomphys.2021.104203}], \newline 
[\href{https://arxiv.org/abs/2001.07640}{\tt arXiv:2001.07640}].


\bibitem{GriffithsMorgan81}
Griffiths, P., Morgan, J., 
{\it\color{darkblue}Rational Homotopy Theory and Differential Forms}, Progress in Mathematics {\bf 16}, Birkh{\"a}user (1981, 2013), 
[\href{https://doi.org/10.1007/978-1-4614-8468-4}{\tt doi:10.1007/978-1-4614-8468-4}].


\bibitem{GromovMartinecRyu20}
Gromov, A., Martinec, E. J., Ryu, S., 
{\it\color{darkblue}Collective excitations at filling factor $5/2$: The view from superspace}, 
Phys. Rev. Lett. {\bf 125} (2020) 077601, 
[\href{https://doi.org/10.1103/PhysRevLett.125.077601}{\tt doi:10.1103/PhysRevLett.125.077601}],
[\href{https://arxiv.org/abs/1909.06384}{\tt arXiv:1909.06384}].


\bibitem{GrumblinHorowitz19}
Grumblin, E., Horowitz, M. (eds.), 
{\it\color{darkblue}Quantum Computing: Progress and Prospects}, The National Academies Press (2019),
[\href{https://doi.org/10.17226/25196}{\tt doi:10.17226/25196}],
[\href{https://nap.nationalacademies.org/catalog/25196/quantum-computing-progress-and-prospects}{\tt ISBN:9780309479691}].


\bibitem{Gubser10}
Gubser, S. S., Pufu, S. S., Rocha, F. D., 
{\it\color{darkblue}Quantum critical superconductors in string theory and M-theory}, 
Phys. Lett. B {\bf 683} (2010), 201-204, 
[\href{https://doi.org/10.1016/j.physletb.2009.12.017}{\tt doi:10.1016/j.physletb.2009.12.017}],
[\href{https://arxiv.org/abs/0908.0011}{\tt arXiv:0908.0011}]. 


\bibitem{Hannabuss18}
Hannabuss, K. C., 
{\it\color{darkblue}T-duality and the bulk-boundary correspondence}, 
J. Geom. Phys.
{\bf 124} (2018), 421-435, 
[\href{https://doi.org/10.1016/j.geomphys.2017.11.016}{\tt doi:10.1016/j.geomphys.2017.11.016}],
[\href{https://arxiv.org/abs/1704.00278}{\tt arXiv:1704.00278}].

\bibitem{Hansen74}
Hansen, V. L., 
{\it\color{darkblue}On the Space of Maps of a Closed Surface into the 2-Sphere}, Math. Scand. {\bf 35} (1974), 149-158,
[\href{https://doi.org/10.7146/math.scand.a-11542}{\tt doi:10.7146/math.scand.a-11542}],
[\href{https://www.jstor.org/stable/24490694}{\tt jstor:24490694}].

\bibitem{HartleTaylor69}
Hartle, J. B., Taylor, J. R., 
{\it\color{darkblue}Quantum Mechanics of Paraparticles}, 
Phys. Rev. {\bf 178} (1969) 2043, \newline 
[\href{https://doi.org/10.1103/PhysRev.178.2043}{\tt doi:10.1103/PhysRev.178.2043}].


\bibitem{HartnollLucasSachdev18}
Hartnoll, S., Lucas, A., Sachdev, S., 
{\it\color{darkblue}Holographic quantum matter}, 
MIT Press (2018), 
[\href{https://arxiv.org/abs/1612.07324}{\tt arXiv:1612.07324}],
[\href{https://mitpress.ublish.com/book/holographic-quantum-matter}{\tt ISBN:9780262348010}].


\bibitem{Hasebe08}
Hasebe, K., 
{\it\color{darkblue}Unification of Laughlin and Moore–Read states in SUSY quantum Hall effect}, 
Phys. Lett. A {\bf 372} 9 (2008), 1516-1520,
[\href{https://doi.org/10.1016/j.physleta.2007.09.071}{\tt doi:10.1016/j.physleta.2007.09.071}].



\bibitem{Hatcher02}
Hatcher, A.,
{\it\color{darkblue}Algebraic Topology}, Cambridge University Press (2002),
[\href{https://www.cambridge.org/gb/academic/subjects/mathematics/geometry-and-topology/algebraic-topology-1?format=PB&isbn=9780521795401}{\tt ISBN:9780521795401}].


\bibitem{HerzogKovtun07}
Herzog, C. P., Kovtun, P., Sachdev, S., Thanh, D. S., 
{\it\color{darkblue}Quantum critical transport, duality, and M-theory}, Phys. Rev. D {\bf 75} (2007) 085020, 
[\href{https://doi.org/10.1103/PhysRevD.75.085020}{\tt doi:10.1103/PhysRevD.75.085020}],
[\href{https://arxiv.org/abs/hep-th/0701036}{\tt arXiv:hep-th/0701036}]. 



\bibitem{HoefletHaenerTroyer23}
Hoefler, T., Haener, T., Troyer, M., 
{\it\color{darkblue}Disentangling Hype from Practicality: On Realistically Achieving Quantum Advantage}, 
Commun. ACM {\bf 66} 5 (2023), 82-87,
[\href{https://doi.org/10.1145/3571725}{\tt doi:10.1145/3571725}],
[\href{https://arxiv.org/abs/2307.00523}{\tt arXiv:2307.00523}].


\bibitem{HSS19}
Huerta, J., Sati, H., Schreiber, U.,
{\it\color{darkblue}Real ADE-equivariant (co)homotopy and Super M-branes}, 
Commun. Math. Phys. {\bf 371} (2019) 425,
[\href{https://doi.org/10.1007/s00220-019-03442-3}{\tt doi:10.1007/s00220-019-03442-3}],
[\href{https://arxiv.org/abs/1805.05987}{\tt arXiv:1805.05987}].

\bibitem{IqbalEtAl24}
Iqbal, M.,  Tantivasadakarn, N., Verresen, R. et al., 
{\it\color{darkblue}Non-Abelian topological order and anyons on a trapped-ion processor}, Nature {\bf 626} (2024),
505–511, 
[\href{https://doi.org/10.1038/s41586-023-06934-4}{\tt doi:10.1038/s41586-023-06934-4}].



\bibitem{Intriligator00}
Intriligator, K., 
{\it\color{darkblue}Anomaly Matching and a Hopf-Wess-Zumino Term in 6d, $\mathcal{N} = (2,0)$ Field Theories}, 
Nucl. Phys. B {\bf 581} (2000) 257-273 
[\href{https://arxiv.org/abs/hep-th/0001205}{\tt arXiv:hep-th/0001205}],
[\href{https://doi.org/10.1016/S0550-3213(00)00148-6}{\tt doi:10.1016/S0550-3213(00)00148-6}].


\bibitem{Jacek21}
Jacak, J. E.,
{\it\color{darkblue}Topological approach to electron correlations at fractional quantum Hall effect}, Ann. Phys. 
{\bf 430} (2021) 168493, 
[\href{https://doi.org/10.1016/j.aop.2021.168493}{\tt doi:10.1016/j.aop.2021.168493}].

\bibitem{James84}
James, I. M.,  
{\it\color{darkblue}General Topology and Homotopy Theory}, Springer (1984), 
[\href{https://doi.org/10.1007/978-1-4613-8283-6}{\tt doi:10.1007/978-1-4613-8283-6}].

\bibitem{JiangEtAl21}
Jiang, B., Bouhon, A., Lin, Z.-K.,  Zhou, X., Hou, B., Li, F.,  Slager, R.-J., Jiang, J.-H., 
{\it\color{darkblue}Experimental observation of non-Abelian topological acoustic semimetals and their phase transitions}, Nature Physics {\bf 17} (2021), 1239-1246, 
[\href{https://doi.org/10.1038/s41567-021-01340-x}{\tt doi;10.1038/s41567-021-01340-x}],
[\href{https://arxiv.org/abs/2104.13397}{\tt arXiv:2104.13397}].


\bibitem{Jordan10a}
Jordan, S. P., 
{\it\color{darkblue}Quantum Computation Beyond the Circuit Model}, PhD thesis, MIT (2010), 
[\href{https://arxiv.org/abs/0809.2307}{\tt arXiv:0809.2307}].

\bibitem{Jordan10b}
Jordan, S. P.,
{\it \color{darkblue} Permutational Quantum Computing}, Quantum Information and Computation {\bf 10} (2010) 470,
[\href{https://doi.org/10.26421/QIC10.5-6-7}{\tt doi:10.26421/QIC10.5-6-7}],
[\href{https://arxiv.org/abs/0906.2508}{\tt arXiv:0906.2508}].



\bibitem{Kak08}
Kak, S., 
{\it\color{darkblue} Prospects for Quantum Computing}, talk at CIFAR Nanotechnology program meeting, Halifax (November 2008), 
[\href{https://arxiv.org/abs/0902.4884}{\tt arXiv:0902.4884}].


\bibitem{Kallel01}
Kallel, S., 
{\it\color{darkblue}Configuration Spaces and the Topology of Curves in Projective Space}, 
in: {\it Topology, Geometry, and Algebra: Interactions and new directions}, 
Contemporary Mathematics {\bf 279}, AMS (2001), 151–175, 
\newline 
[\href{https://doi.org/10.1090/conm/279}{\tt doi:10.1090/conm/279}].


\bibitem{Kallel25}
Kallel, S., 
{\it\color{darkblue}Configuration spaces of points: A user’s guide}, Encyclopedia of Mathematical Physics 2nd ed. {\bf 4} (2025), 
[\href{https://doi.org/10.1016/B978-0-323-95703-8.00211-1}{\tt doi:10.1016/B978-0-323-95703-8.00211-1}],
[\href{https://arxiv.org/abs/2407.11092}{\tt arXiv:2407.11092}].

\bibitem{KatzKlemmVafa97}
Katz, S., Klemm, A., Vafa, C., 
{\it\color{darkblue}Geometric Engineering of Quantum Field Theories}, Nucl. Phys. B {\bf 497} (1997), 173-195, 
[\href{https://doi.org/10.1016/S0550-3213(97)00282-4}{\tt doi:10.1016/S0550-3213(97)00282-4}],
[\href{https://arxiv.org/abs/hep-th/9609239}{\tt arXiv:hep-th/9609239}].


\bibitem{Kitaev01}
Kitaev, A., 
{\it\color{darkblue}Unpaired Majorana fermions in quantum wires}, Phys. Uspekhi {\bf 44} 10S (2001), 131-136, 
\newline
[\href{https://iopscience.iop.org/article/10.1070/1063-7869/44/10S/S29}{\tt doi:10.1070/1063-7869/44/10S/S29}],
[\href{https://arxiv.org/abs/cond-mat/0010440}{\tt arXiv:cond-mat/0010440}].


\bibitem{Kitaev03}
Kitaev, A., 
{\it\color{darkblue}Fault-tolerant quantum computation by anyons}, 
Ann. Phys. {\bf 303} (2003), 2-30, 
\newline
[\href{https://doi.org/10.1016/S0003-4916(02)00018-0}{\tt doi:10.1016/S0003-4916(02)00018-0}],
[\href{https://arxiv.org/abs/quant-ph/9707021}{\tt arXiv:quant-ph/9707021}].


\bibitem{Kitaev06}
Kitaev, A., 
{\it\color{darkblue}Anyons in an exactly solved model and beyond}, Ann. Phys. {\bf 321} 1 (2006), 2-111, 
\newline
[\href{https://doi.org/10.1016/j.aop.2005.10.005}{\tt doi:10.1016/j.aop.2005.10.005}].


\bibitem{KoSmolinsky92}
Ko, K. H., Smolinsky, L., 
{\it\color{darkblue}The framed braid group and 3-manifolds}, Proc. Amer. Math. Soc. {\bf 115} (1992) 541-551 [\href{https://doi.org/10.1090/S0002-9939-1992-1126197-1}{\tt doi:10.1090/S0002-9939-1992-1126197-1}].




\bibitem{Kosinski93}
Kosinski, A., 
{\it\color{darkblue}Differential manifolds}, Academic Press (1993), [\href{https://www.sciencedirect.com/bookseries/pure-and-applied-mathematics/vol/138/suppl/C}{\tt ISBN:978-0-12-421850-5}].



\bibitem{LarmoreThomas80}
Larmore, L. L., Thomas, E., 
{\it\color{darkblue}On the Fundamental Group of a Space of Sections}, Math. Scand. {\bf 47} 2 (1980), 232-246,
[\href{https://www.jstor.org/stable/24491393}{\tt jstor:24491393}].


\bibitem{LauLimEtAl22}
Lau J. W. Z., Lim, K. H., Shrotriya H., Kwek, L. C., {\it\color{darkblue}NISQ computing: where are we and where do we go?}, AAPPS Bull. {\bf 32} 27 (2022), [\href{https://doi.org/10.1007/s43673-022-00058-z}{\tt doi:10.1007/s43673-022-00058-z}].

\bibitem{LeePacker96}
Lee, S. T., Packer, J. A., 
{\it\color{darkblue}The Cohomology of the Integer Heisenberg Groups}, J. Algebra {\bf 184} 1 (1996), 230-250, [\href{https://doi.org/10.1006/jabr.1996.0258}{\tt doi:10.1006/jabr.1996.0258}].


\bibitem{LidarBrun13}
Lidar, D. A., Brun, T. A. (eds.), {\it\color{darkblue}Quantum Error Correction}, Cambridge University Press (2013), 
\newline
[\href{https://www.cambridge.org/de/universitypress/subjects/physics/quantum-physics-quantum-information-and-quantum-computation/quantum-error-correction?format=HB=9780521897877}{\tt ISBN:9780521897877}], 
[\href{https://doi.org/10.1017/CBO9781139034807}{\tt doi:10.1017/CBO9781139034807}].


\bibitem{Liu24}
Liu, S.,
{\it\color{darkblue}Anyon quantum dimensions from an arbitrary ground state wave function}, Nature Commun. {\bf 15} (2024) 5134, 
[\href{https://doi.org/10.1038/s41467-024-47856-7}{\tt doi:10.1038/s41467-024-47856-7}], [\href{https://arxiv.org/abs/2304.13235}{\tt arXiv:2304.13235}].



\bibitem{LZX24}
Liu, Y., Zhao, T., Xiang, T., 
{\it\color{darkblue}Resolving Geometric Excitations of Fractional Quantum Hall States}, Phys. Rev. B {\bf 110} 195137 (2024), 
[\href{https://doi.org/10.1103/PhysRevB.110.195137}{\tt doi:10.1103/PhysRevB.110.195137}],
[\href{https://arxiv.org/abs/2406.11195}{\tt arXiv:2406.11195}].





\bibitem{Lloyd02}
Lloyd, S., 
{\it\color{darkblue} Quantum computation with abelian anyons}, Quantum Information Processing {\bf 1} 1/2 (2002), 13-18, 
[\href{https://doi.org/10.1023/A:1019649101654}{\tt doi:10.1023/A:1019649101654}],
[\href{https://arxiv.org/abs/quant-ph/0004010}{\tt arXiv:quant-ph/0004010}].

\bibitem{Lobb24}
Lobb, A.,
{\it\color{darkblue}A feeling for Khovanov homology}, Notices AMS {\bf 71} 5 (2024), 
[\href{https://doi.org/10.1090/noti2928}{\tt doi:10.1090/noti2928}].



\bibitem{LutchyEtAl10}
Lutchyn, R. M. Sau, J. D., Das Sarma, S., 
{\it\color{darkblue}Majorana Fermions and a Topological Phase Transition in Semiconductor-Superconductor Heterostructures}, Phys. Rev. Lett. {\bf 105} (2010) 077001, 
\newline
[\href{https://doi.org/10.1103/PhysRevLett.105.077001}{\tt doi:10.1103/PhysRevLett.105.077001}].

\bibitem{Manoliu98}
Manoliu, M.,
{\it\color{darkblue}Abelian Chern-Simons theory}, 
J. Math. Phys. {\bf 39} (1998), 170-206, 
[\href{https://arxiv.org/abs/dg-ga/9610001}{\tt arXiv:dg-ga/9610001}], \newline 
[\href{https://doi.org/10.1063/1.532333}{\tt doi:10.1063/1.532333}].


\bibitem{Massuyeau21}
Massuyeau, G.,
{\it\color{darkblue}Lectures on Mapping Class Groups, Braid Groups and Formality} (2021),
\newline
[\href{https://massuyea.perso.math.cnrs.fr/notes/formality.pdf}{\tt massuyea.perso.math.cnrs.fr/notes/formality.pdf}].



\bibitem{MathaiThiang15}
Mathai, V., Thiang, G. C., 
{\it\color{darkblue}T-Duality of Topological Insulators}, 
J. Phys. A: Math. Theor. {\bf 48} (2015) 42FT02, 
[\href{https://doi.org/10.1088/1751-8113/48/42/42FT02}{\tt doi:10.1088/1751-8113/48/42/42FT02}],
[\href{https://arxiv.org/abs/1503.01206}{\tt arXiv:1503.01206}].

\bibitem{MathaiThiang16}
Mathai, V., Thiang, G. C., 
{\it\color{darkblue}T-Duality Simplifies Bulk-Boundary Correspondence}, 
Commun. Math. Phys. {\bf 345} (2016), 675–701, 
[\href{https://doi.org/10.1007/s00220-016-2619-6}{\tt doi:10.1007/s00220-016-2619-6}],
[\href{https://arxiv.org/abs/1505.05250}{\tt arXiv:1505.05250}].

\bibitem{Morava23}
Morava, J., 
{\it\color{darkblue}A homotopy-theoretic context for CKM/Birkhoff renormalization},
[\href{https://arxiv.org/abs/2307.10148}{\tt arXiv:2307.10148}].

\bibitem{Morava24}
Morava, J., 
{\it\color{darkblue}Some very low-dimensional algebraic topology},
[\href{https://arxiv.org/abs/2411.15885}{\tt arXiv:2411.15885}].



\bibitem{MyersSatiSchreiber23-TQG}
Myers, D. J., Sati, H., Schreiber, U.,
{\it\color{darkblue}Topological Quantum Gates in Homotopy Type Theory}, Commun. Math. Phys. {\bf 405} 172 (2024), 
[\href{https://doi.org/10.1007/s00220-024-05020-8}{\tt 10.1007/s00220-024-05020-8}],
[\href{https://arxiv.org/abs/2303.02382}{\tt arXiv:2303.02382}].




\bibitem{NOF86}
Naito, S., Osada, K., Fukui, T., {\it\color{darkblue} Fierz Identities and Invariance of Eleven-dimensional Supergravity Action},
Phys. Rev. D {\bf 34} (1986), 536-552, 
[\href{https://doi.org/10.1103/PhysRevD.34.536}{\tt doi:10.1103/PhysRevD.34.536}];
Erratum Ibid. {\bf 35} (1986), 1536-1536, 
[\href{https://doi.org/10.1103/PhysRevD.35.1536}{\tt 
doi:10.1103/PhysRevD.35.1536}]. 

\bibitem{Nakamura20}
Nakamura, J., et al., 
{\it\color{darkblue}Direct observation of anyonic braiding statistics}, Nature Phys. {\bf 16} (2020), 931–936, \newline 
[\href{https://doi.org/10.1038/s41567-020-1019-1}{\tt doi:10.1038/s41567-020-1019-1}].



\bibitem{NielsenChuang00}
Nielsen, M. A., Chuang, I. L., 
{\it\color{darkblue}Quantum computation and quantum information}, Cambridge University Press (2000), 
[\href{https://doi.org/10.1017/CBO9780511976667}{\tt doi:10.1017/CBO9780511976667}].

\bibitem{nLabYMMassGap}
nLab, {\it\color{darkblue}Yang-Mills mass gap},
[\href{https://ncatlab.org/nlab/show/Yang-Mills+mass+gap}{\tt ncatlab.org/nlab/show/Yang-Mills+mass+gap}]

\bibitem{nLabCellStructureOnProjectiveSpaces}
nLab,
{\it\color{darkblue} Cell structure of projective spaces},
[\href{https://ncatlab.org/nlab/show/cell+structure+of+projective+spaces}{\tt ncatlab.org/nlab/show/cell+structure+of+projective+spaces}].

\bibitem{Okuyama05}
Okuyama, S., 
{\it\color{darkblue}The space of intervals in a Euclidean space}, Algebr. Geom. Topol. {\bf 5} (2005), 1555-1572, 
[\href{https://doi.org/10.2140/agt.2005.5.1555}{\tt doi:10.2140/agt.2005.5.1555}],
[\href{https://arxiv.org/abs/math/0511645}{\tt arXiv:math/0511645}].

\bibitem{Ohtsuki01}
Ohtsuki, T., 
{\it\color{darkblue}Quantum Invariants -- A Study of Knots, 3-Manifolds, and Their Sets}, World Scientific (2001), 
[\href{https://doi.org/10.2140/agt.2005.5.1555}{\tt doi:10.2140/agt.2005.5.1555}].

\bibitem{Pachos06}
Pachos, J. K., 
{\it\color{darkblue} Quantum computation with abelian anyons on the honeycomb lattice}, Int. J. Quantum Information {\bf 4} 6 (2006), 947-954,
[\href{https://doi.org/10.1142/S0219749906002328}{\tt doi:10.1142/S0219749906002328}],
[\href{https://arxiv.org/abs/quant-ph/0511273}{\tt arXiv:quant-ph/0511273}].


\bibitem{Pachos12}
Pachos, J. K., 
{\it\color{darkblue}Introduction to Topological Quantum Computation}, 
Cambridge University Press (2012), \newline 
[\href{https://doi.org/10.1017/CBO9780511792908}{\tt doi:10.1017/CBO9780511792908}].


\bibitem{PapicBalram24}
Papi{\'c}, Z., Balram, A. C., {\it\color{darkblue}Fractional quantum Hall effect in semiconductor systems}, Encyclopedia of Condensed Matter Physics 2nd ed. {\bf 1} (2024), 285-307, [\href{https://doi.org/10.1016/B978-0-323-90800-9.00007-X}{\tt doi:10.1016/B978-0-323-90800-9.00007-X}], \newline 
[\href{https://arxiv.org/abs/2205.03421}{\tt arXiv:2205.03421}].


\bibitem{Polchinski17}
Polchinski, J., 
{\it\color{darkblue}Dualities of Fields and Strings}, 
Studies in History and Philosophy of Modern Physics {\bf 59} (C) (2017), 6-20, 
[\href{https://arxiv.org/abs/1412.5704}{\tt arXiv:1412.5704}].


\bibitem{Polychronakos96}
Polychronakos, A. P., 
{\it\color{darkblue}Path Integrals and Parastatistics}, 
Nucl. Phys. B {\bf 474} (1996) 529-539
\newline
[\href{https://arxiv.org/abs/hep-th/9603179}{\tt arXiv:hep-th/9603179}],
[\href{https://doi.org/10.1016/0550-3213(96)00277-5}{\tt doi:10.1016/0550-3213(96)00277-5}].



\bibitem{Pontrjagin38}
Pontrjagin, L., 
{\it\color{darkblue}Classification of continuous maps of a complex into a sphere, 
Communication I}, Doklady Akademii Nauk SSSR {\bf 19} 3  (1938), 147-149.

\bibitem{PrangeGirvin86}
Prange, R. E., Girvin, S. M. (eds.),
{\it\color{darkblue}The Quantum Hall Effect}, Graduate Texts in Contemporary Physics, Springer (1986, 1990), 
[\href{https://doi.org/10.1007/978-1-4612-3350-3}{\tt doi:10.1007/978-1-4612-3350-3}].



\bibitem{Preskill18}
Preskill, J.,  
{\it\color{darkblue}Quantum Computing in the NISQ era and beyond}, Quantum {\bf 2} 79 (2018),
[\href{https://arxiv.org/abs/1801.00862}{\tt arXiv:1801.00862}],
[\href{https://doi.org/10.22331/q-2018-08-06-79}{\tt doi:10.22331/q-2018-08-06-79}].

\bibitem{Preskill23}
Preskill, J., 
{\it\color{darkblue}Crossing the Quantum Chasm: From NISQ to Fault Tolerance}, talk at Q2B 2023, Silicon Valley (2023),
[\href{https://ncatlab.org/nlab/files/Preskill-Crossing.pdf}{\tt ncatlab.org/nlab/files/Preskill-Crossing.pdf}].

\bibitem{Preskill24}
Preskill, J.,
{\it\color{darkblue}Beyond NISQ: The Megaquop Machine}, talk at {\it Q2B 2024 Silicon Valley} (Dec. 2024),
\newline 
[\href{https://www.preskill.caltech.edu/talks/Preskill-Q2B-2024.pdf}{\tt www.preskill.caltech.edu/talks/Preskill-Q2B-2024.pdf}].


\bibitem{PBFGP23}
Pu, S., Balram, A. C., Fremling, M.,  Gromov, A., Papi{\'c}, Z., 
{\it\color{darkblue}Signatures of Supersymmetry in the $\nu = 5/2$ Fractional Quantum Hall Effect}, 
Phys. Rev. Lett. {\bf 130} (2023) 176501,
[\href{https://doi.org/10.1103/PhysRevLett.130.176501}{\tt doi:10.1103/PhysRevLett.130.176501}],
[\href{https://arxiv.org/abs/2301.04169}{\tt arXiv:2301.04169}].




\bibitem{RigolinOrtiz12}
Rigolin, G., Ortiz, G., 
{\it\color{darkblue}The Adiabatic Theorem for Quantum Systems with Spectral Degeneracy}, Phys. Rev. A {\bf 85} 062111 (2012), 
[\href{https://doi.org/10.1103/PhysRevA.85.062111}{\tt doi:10.1103/PhysRevA.85.062111}],
[\href{https://arxiv.org/abs/1111.5333}{\tt arXiv:1111.5333}].

\bibitem{Rowell16}
Rowell, E. C., 
{\it\color{darkblue}An Invitation to the Mathematics of Topological Quantum Computation}, 
J. Phys.: Conf. Ser. {\bf 698} (2016) 012012, 
[\href{https://iopscience.iop.org/article/10.1088/1742-6596/698/1/012012}{\tt doi:10.1088/1742-6596/698/1/012012}], [\href{https://arxiv.org/abs/1601.05288}{\tt arXiv:1601.05288}].


\bibitem{Rowell22}
Rowell, E. C., 
{\it\color{darkblue}Braids, Motions and Topological Quantum Computing},
[\href{https://arxiv.org/abs/2208.11762}{\tt arXiv:2208.11762}].



\bibitem{SS20-Tad}
Sati, S., Schreiber, U.,
{\it\color{darkblue}Equivariant Cohomotopy implies orientifold tadpole cancellation},
J. Geom. Phys. {\bf 156} (2020), 103775,
[\href{https://doi.org/10.1016/j.geomphys.2020.103775}{\tt doi:10.1016/j.geomphys.2020.103775}],
[\href{https://arxiv.org/abs/1909.12277}{\tt arXiv:1909.12277}].


\bibitem{SS20-Orb}
Sati, H., Schreiber, U., 
{\it\color{darkblue}Geometric Orbifold Cohomology},
CRC Press (2026)
[{\tt ISBN:9781041147510}]
\newline
[\href{https://ncatlab.org/schreiber/show/Geometric+Orbifold+Cohomology}{\tt ncatlab.org/schreiber/show/Geometric+Orbifold+Cohomology}].


\bibitem{SS21-M5Anomaly}
Sati, H., Schreiber, U.
{\it\color{darkblue}Twisted Cohomotopy implies M5-brane anomaly cancellation}, 
Lett. Math. Phys. {\bf 111} (2021) 120,
[\href{https://doi.org/10.1007/s11005-021-01452-8}{\tt doi;10.1007/s11005-021-01452-8}],
[\href{https://arxiv.org/abs/2002.07737}{\tt 
arXiv:2002.07737}]. 

\bibitem{SS23-Mf}
Sati, H., Schreiber, U.,
{\it\color{darkblue}M/F-Theory as $M\!f$-Theory},
Rev. Math. Phys. {\bf 35} 10 (2023),
[\href{https://arxiv.org/abs/2103.01877}{\tt arXiv:2103.01877}],
[\href{https://doi.org/10.1142/S0129055X23500289}{\tt doi:10.1142/S0129055X23500289}].


\bibitem{SS21-EquBund}
Sati, H., Schreiber, U.,
{\it\color{darkblue}Equivariant Principal $\infty$-Bundles},
Cambridge University Press (2026) 
\newline
[{\tt ISBN:9781009698559}],
[\href{https://ncatlab.org/schreiber/show/Equivariant+Principal+infinity-Bundles}{\tt ncatlab.org/schreiber/show/Equivariant+Principal+infinity-Bundles}].


\bibitem{SS23-DefectBranes}
Sati, H., Schreiber, U.,
{\it\color{darkblue} Anyonic Defect Branes and Conformal Blocks in Twisted Equivariant Differential K-Theory}, 
Rev. Math. Phys. {\bf 35} 06 (2023) 2350009,
[\href{https://doi.org/10.1142/S0129055X23500095}{\tt doi:10.1142/S0129055X23500095}],
[\href{https://arxiv.org/abs/2203.11838}{\tt arXiv:2203.11838}].

\bibitem{SS23-Monadology}
Sati, H., Schreiber, U.,
{\it\color{darkblue}The Quantum Monadology},
Quantum Studies: Mathematics and Foundations {\bf 12} 25 (2025)
[\href{https://arxiv.org/abs/2310.15735}{\tt arXiv:2310.15735}],
[\href{https://doi.org/10.1007/s40509-025-00368-5}{\tt doi:10.1007/s40509-025-00368-5}].


\bibitem{SS23-Entanglement}
Sati, H., Schreiber, U.,
{\it\color{darkblue}Entanglement of Sections},
Quantum Studies: Mathematics and Foundations {\bf 13} 23 (2026)
[\href{https://arxiv.org/abs/2309.07245}{\tt arXiv:2309.07245}],
[\href{https://doi.org/10.1007/s40509-026-00397-8}{\tt doi:10.1007/s40509-026-00397-8}].


\bibitem{SS24-TopOrd}
Sati, H., Schreiber, U.,
{\it\color{darkblue}Anyonic topological order in TED K-theory},
Rev. Math. Phys. (2023) {\bf 35} 03 (2023) 2350001,
[\href{https://doi.org/10.1142/S0129055X23500010}{\tt doi:10.1142/S0129055X23500010}],
[\href{https://arxiv.org/abs/2206.13563}{\tt arXiv:2206.13563}].

\bibitem{SS24-Flux}
Sati, H., Schreiber, U.,
{\it\color{darkblue}Flux quantization},
Encyclopedia of Mathematical Physics 2nd ed. 
{\bf 4} (2025) 281-324
[\href{doi:10.1016/B978-0-323-95703-8.00078-1}{\tt doi:10.1016/B978-0-323-95703-8.00078-1}],
[\href{https://arxiv.org/abs/2402.18473}{\tt arXiv:2402.18473}]. 

\bibitem{SS24-QObs}
Sati, H., Schreiber, U.,
{\it\color{darkblue}Quantum Observables of Quantized Fluxes},
Ann. Henri Poincar{\'e}
{\bf 26} (2025) 4241–4269 (2024),
[\href{https://doi.org/10.1007/s00023-024-01517-z}{\tt doi:10.1007/s00023-024-01517-z}],
[\href{https://arxiv.org/abs/2312.13037}{\tt arXiv:2312.13037}].


\bibitem{SS24-AbAnyons}
Sati, H., Schreiber, U.,
{\it\color{darkblue}Cohomotopy, Framed Links, and Abelian Anyons},
in Proceedings of {\it Focus Program on Algebraic Topology  in memory of Fred Cohen}, Fields Institute Communications (2026, in press)
[\href{https://arxiv.org/abs/2408.11896}{\tt arXiv:2408.11896}].

\bibitem{SS25-TQBits}
Sati, H., Schreiber, U.,
{\it\color{darkblue}Topological QBits in Flux-Quantized Supergravity}, 
in: {\it Quantum Gravity and Computation},
Routledge (2026),
[\href{https://arxiv.org/abs/2411.00628}{\tt arXiv:2411.00628}],
[\href{https://www.routledge.com/Quantum-Gravity-and-Computation-Information-Pregeometry-and-Digital-Physics/Rickles-Arsiwalla-Elshatlawy/p/book/9781032900940}{\tt ISBN:9781032900940}].

\bibitem{SS25-EquTwistorial} 
Sati, H., Schreiber, U., 
{\it\color{darkblue}The Character Map in Twisted Equivariant Nonabelian Cohomology},
in: {\it Applied Algebraic Topology},
special issue of: 
Beijing J. Pure Appl. Math. (2025),
[\href{https://arxiv.org/abs/2011.06533}{\tt arXiv:2011.06533}].

\bibitem{SS25-Seifert}
Sati, H., Schreiber, U.,
{\it\color{darkblue}Anyons on M5-Probes of Seifert 3-Orbifolds via Flux-Quantization},
Lett.  Math. Phys.
{\bf 115} 36 (2025),
[\href{https://doi.org/10.1007/s11005-025-01918-z}{\tt doi:10.1007/s11005-025-01918-z}],
[\href{https://arxiv.org/abs/2411.16852}{\tt arXiv:2411.16852}].


\bibitem{SS25-ViaAlgTop}
Sati, H., Schreiber, U.,
{\it\color{darkblue}Fractional Quantum Hall Anyons via the Algebraic Topology of exotic Flux Quanta},
[\href{https://arxiv.org/abs/2505.22144}{\tt arXiv:2505.22144}].

\bibitem{SS25-FQAH}
Sati, H., Schreiber, U.,
{\it\color{darkblue}Identifying Anyonic Topological Order in Fractional Quantum Anomalous Hall Systems},
Applied Physics Letters
{\bf 128} (2026) 023101
[\href{https://arxiv.org/abs/2507.00138}{\tt arXiv:2507.00138}],
[\href{https://doi.org/10.1063/5.0305441}{\tt doi:10.1063/5.0305441}].


\bibitem{SatiValera25}
Sati, H., Valera, S., 
{\it\color{darkblue}Topological Quantum Computing}, Encyclopedia of Mathematical Physics 2nd ed {\bf 4}, Elsevier (2025) 325-345, 
[\href{https://doi.org/10.1016/B978-0-323-95703-8.00262-7}{\tt doi:10.1016/B978-0-323-95703-8.00262-7}].


\bibitem{Sau17}
Sau S., 
{\it\color{darkblue}A Roadmap for a Scalable Topological Quantum Computer}, Physics {\bf 10} 68 (2017),
\newline
[\href{https://physics.aps.org/articles/v10/68}{\tt physics.aps.org/articles/v10/68}].

\bibitem{Schreiber25-ICMAT}
Schreiber, U., 
{\it\color{darkblue}Quantum Language via Linear Homotopy Types},
ICMAT lecture notes (2025),
\newline
[\href{https://ncatlab.org/schreiber/show/Quantum+Language+via+Linear+Homotopy+Types}{\tt ncatlab.org/schreiber/show/Quantum+Language+via+Linear+Homotopy+Types}].


\bibitem{Segal73}
Segal, G., 
{\it\color{darkblue}Configuration-spaces and iterated loop-spaces}, Invent. Math. {\bf 21} (1973), 213-221, 
\newline
[\href{https://doi.org/10.1007/BF01390197}{\tt doi:10.1007/BF01390197}].



\bibitem{Stormer99}
St{\" o}rmer, H. L.,
{\it\color{darkblue}Nobel Lecture: The fractional quantum Hall effect}, Rev. Mod. Phys. {\bf 71} (1999) 875, \newline  
[\href{https://doi.org/10.1103/RevModPhys.71.875}{\tt doi:10.1103/RevModPhys.71.875}].


\bibitem{Strom11}
Strom, J., 
{\it\color{darkblue}Modern classical homotopy theory}, Graduate Studies in Mathematics {\bf 127}, 
American Mathematical Society (2011), 
[\href{http://www.ams.org/books/gsm/127}{\tt ams:gsm/127}].

\bibitem{TeoKane10}
Teo, J. C. Y., Kane, C. L.
{\it\color{darkblue}Majorana Fermions and Non-Abelian Statistics in Three Dimensions}, Phys. Rev. Lett. {\bf 104}
(2010) 046401,
[\href{https://doi.org/10.1103/PhysRevLett.104.046401}{\tt doi:10.1103/PhysRevLett.104.046401}],
[\href{https://arxiv.org/abs/0909.4741}{\tt arXiv:0909.4741}].



\bibitem{TiwariBzdusek20}
Tiwari, A.,  Bzdu{\v s}ek, T., 
{\it\color{darkblue}Non-Abelian topology of nodal-line rings in PT-symmetric systems}, Phys. Rev. B {\bf 101} (2020) 195130,
[\href{https://doi.org/10.1103/PhysRevB.101.195130}{\tt doi:10.1103/PhysRevB.101.195130}].

\bibitem{Tong16}
Tong, D., 
{\it\color{darkblue}The Quantum Hall Effect}, lecture notes (2016), 
[\href{https://arxiv.org/abs/1606.06687}{\tt arXiv:1606.06687}],
\newline
[\href{http://www.damtp.cam.ac.uk/user/tong/qhe/qhe.pdf}{\tt www.damtp.cam.ac.uk/user/tong/qhe/qhe.pdf}].


\bibitem{vonKlitzing86}
von Klitzing, K., 
{\it\color{darkblue}The quantized Hall effect}, 
Rev. Mod. Phys. {\bf 58} 519 (1986),
\newline
[\href{https://doi.org/10.1103/RevModPhys.58.519}{\tt doi:10.1103/RevModPhys.58.519}].


\bibitem{Waintal24}
Waintal, X., 
{\it\color{darkblue}The Quantum House Of Cards}, PNAS {\bf 121} 1 (2024) e2313269120,
[\href{https://arxiv.org/abs/2312.17570}{\tt arXiv:2312.17570}],
\newline
[\href{https://doi.org/10.1073/pnas.2313269120}{\tt doi:10.1073/pnas.2313269120}].


\bibitem{Waldorf24}
Waldorf, K., 
{\it\color{darkblue}Geometric T-duality: Buscher rules in general topology}, 
Ann. Henri Poincar{\'e} {\bf 25} (2024), 1285–1358,  
[\href{https://doi.org/10.1007/s00023-023-01295-0}{\tt doi:10.1007/s00023-023-01295-0}],
[\href{https://arxiv.org/abs/2207.11799}{\tt arXiv:2207.11799}].


\bibitem{WangHuEtAl20}
Wang, Y.,  Hu, Z., Sanders, B. C., Sabre, K.,
{\it\color{darkblue}Qudits and high-dimensional quantum computing}, Front. Phys. {\bf 8} 479 (2020),
[\href{https://doi.org/10.3389/fphy.2020.589504}{\tt doi:10.3389/fphy.2020.589504}],
[\href{https://arxiv.org/abs/2008.00959}{\tt arXiv:2008.00959}].


\bibitem{WangHazzard25}
Wang, Z., Hazzard, K. R. A., 
{\it\color{darkblue}Particle exchange statistics beyond fermions and bosons}, Nature {\bf 637} (2025), 314-318,
[\href{https://doi.org/10.1038/s41586-024-08262-7}{\tt doi:10.1038/s41586-024-08262-7}],
[\href{https://arxiv.org/abs/2308.05203}{\tt arXiv:2308.05203}]. 



\bibitem{Wen89}
Wen, X.-G., 
{\it\color{darkblue}Vacuum Degeneracy of Chiral Spin State in Compactified Spaces}, Phys. Rev. B {\bf 40} 7387 (1989), 
[\href{https://doi.org/10.1103/PhysRevB.40.7387}{\tt doi:10.1103/PhysRevB.40.7387}].

\bibitem{Wen95}
Wen, X.-G., 
{\it\color{darkblue}Topological orders and Edge excitations in FQH states}, 
Adv. Phys.
{\bf 44} 5 (1995) 405,  \newline 
[\href{https://doi.org/10.1080/00018739500101566}{\tt doi:10.1080/00018739500101566}],
[\href{https://arxiv.org/abs/cond-mat/9506066}{\tt arXiv:cond-mat/9506066}].


\bibitem{WenDagottoFradkin90}
Wen, X.-G., Dagotto, E., Fradkin, E.,
{\it\color{darkblue}Anyons on a torus}, Phys. Rev. B {\bf 42} (1990) 6110, 
\newline
[\href{https://doi.org/10.1103/PhysRevB.42.6110}{\tt doi:10.1103/PhysRevB.42.6110}].


\bibitem{Williams20}
Williams, L., 
{\it\color{darkblue}Configuration Spaces for the Working Undergraduate}, Rose-Hulman Undergrad. Math. J.
{\bf 21} 1 (2020) 8,
[\href{https://scholar.rose-hulman.edu/rhumj/vol21/iss1/8}{\tt rhumj:vol21/iss1/8}],
[\href{https://arxiv.org/abs/1911.11186}{\tt arXiv:1911.11186}].

\bibitem{Witten16}
Witten, E., 
{\it\color{darkblue}Three Lectures On Topological Phases Of Matter}, 
La Rivista del Nuovo Cimento {\bf 39} (2016), 313-370, 
[\href{https://doi.org/10.1393/ncr/i2016-10125-3}{\tt doi:10.1393/ncr/i2016-10125-3}],
[\href{https://arxiv.org/abs/1510.07698}{\tt arXiv:1510.07698}].



\bibitem{Wootton10}
Wootton, J. R., 
{\it\color{darkblue} Dissecting Topological Quantum Computation}, PhD thesis, Leeds (2010), 
[\href{https://etheses.whiterose.ac.uk/1163}{\tt etheses:1163}].

\bibitem{WoottonPachos11}
Wootton, J. R., Pachos, J. K., 
{\it\color{darkblue} Universal Quantum Computation with Abelian Anyon Models}, Electron. Notes Theor. Comput. Sci. {\bf 270} 2 (2011), 209-218, 
[\href{https://doi.org/10.1016/j.entcs.2011.01.032}{\tt doi:10.1016/j.entcs.2011.01.032}],
[\href{https://arxiv.org/abs/0904.4373}{\tt arXiv:0904.4373}].


\bibitem{WuSoluyanovBzdusek19}
Wu, Q., Soluyanov, A. A., Bzdu{\v s}ek, T., 
{\it\color{darkblue}Non-Abelian band topology in noninteracting metals}, Science {\bf 365} (2019), 1273-1277,
[\href{https://doi.org/10.1126/science.aau8740}{\tt doi:10.1126/science.aau8740}],
[\href{https://arxiv.org/abs/1808.07469}{\tt arXiv:1808.07469}].

\bibitem{YeEtAl11}
Ye, C., Peng, S.-G., Zheng ,C.,  Long G.-L., 
{\it\color{darkblue}Quantum Fourier Transform and Phase Estimation in Qudit System}, Commun. Theor. Phys. {\bf 55} 790 (2011), 
[\href{https://iopscience.iop.org/article/10.1088/0253-6102/55/5/11}{\tt doi:10.1088/0253-6102/55/5/11}].

\bibitem{Zaanen15}
Zaanen, J., Liu, Y., Sun, Y.-W., Schalm, K., {\it\color{darkblue}Holographic Duality in Condensed Matter Physics}, 
Cambridge University Press (2015), 
[\href{https://doi.org/10.1017/CBO9781139942492}{\tt doi:10.1017/CBO9781139942492}].


\bibitem{WenEtAl19}
Zeng, B., Chen, X., Zhou, D.-L., Wen, X.-G.,
{\it\color{darkblue}Quantum Information Meets Quantum Matter -- From Quantum Entanglement to Topological Phases of Many-Body Systems}, Quantum Science and Technology (QST), Springer (2019), 
[\href{https://arxiv.org/abs/1508.02595}{\tt arXiv:1508.02595}],
[\href{https://doi.org/10.1007/978-1-4939-9084-9}{\tt doi:10.1007/978-1-4939-9084-9}].

\bibitem{Zhang92}
Zhang, S. C. 
{\it\color{darkblue}The Chern-Simons-Landau-Ginzburg theory of the fractional quantum Hall effect}, 
Int. J. Mod. Phys. B {\bf 06} 01 (1992), 25-58, 
[\href{https://doi.org/10.1142/S0217979292000037}{\tt doi:10.1142/S0217979292000037}].


\end{thebibliography}
\end{document}